\documentclass[aps,rmp,amsmath,amssymb,longbibliography]{revtex4-1}

\usepackage{bm}

\usepackage{graphicx}
\usepackage{epsfig}
\usepackage{amsthm}
\usepackage{longtable}
\usepackage{booktabs}
\usepackage{epstopdf}
\usepackage[mathlines]{lineno}

\usepackage{grffile}
\usepackage{float}

\usepackage[usenames]{color}
\usepackage{colortbl}
\newcommand\T{\rule{0pt}{2.2ex}}
\newcommand\B{\rule[-1.6ex]{0pt}{0pt}}

\begin{document}

\title{Neutrinoless Double-Electron Capture}

\author{K.~Blaum}
\author{S.~Eliseev}
\affiliation{Max-Planck-Institut f\"ur Kernphysik, Saupfercheckweg 1 69117 Heidelberg, Germany}
\author{F.~A.~Danevich}
\author{V.~I.~Tretyak}
\affiliation{Institute for Nuclear Research of NASU, Prospekt Nauky 47 03028  
Kyiv, Ukraine}
\author{Sergey~Kovalenko}
%\affiliation{Universidad T\'{e}cnica Federico Santa Mar\'{\i}a and Centro Cient\'{\i}fico Tecnol\'{o}gico de Valpara\'{\i}so, Casilla 110-V, Valpara\'{\i}so, Chile}
\affiliation{Departamento de Ciencias F\'{\i}sicas, Universidad Andres Bello, Sazi\'{e} 2212, Santiago, Chile}

\author{M.~I.~Krivoruchenko}
\affiliation{Institute for Theoretical and Experimental Physics, NRC "Kurchatov Institute", B. Cheremushkinskaya 25 117218 Moscow, Russia}
\affiliation{National Research Centre "Kurchatov Institute",
pl. Akademika Kurchatova 1 123182 Moscow, Russia}
\author{Yu.~N.~Novikov}
\affiliation{Petersburg Nuclear Physics Institute, NRC "Kurchatov Institute", Gatchina, 188300 St. Petersburg, Russia}
\affiliation{Department of Physics, St. Petersburg State University, 199034 St. Petersburg, Russia}
\author{J.~Suhonen}
\affiliation{Department of Physics, University of Jyvaskyla, P.O. Box 35, Jyvaskyla, FI-40014, Finland}

%\date{\today{}}

\begin{abstract}
Double-beta processes play a key role in the exploration of
neutrino and weak interaction properties, and in the searches for
effects beyond the Standard Model. During the last half century
many attempts were undertaken to search for double-beta decay with
emission of two electrons, especially for its neutrinoless mode
($0\nu2\beta^-$), the latter being still not observed.
Double-electron capture (2EC) was not in focus so far because of
its in general lower transition probability. However, the rate of
neutrinoless double-electron capture ($0\nu2$EC) can experience a
resonance enhancement by many orders of magnitude in case the
initial and final states are energetically degenerate. In the
resonant case, the sensitivity of the $0\nu2$EC process
can approach the sensitivity of the $0\nu2\beta^-$ decay
in the search for the Majorana mass of neutrinos, right-handed currents,
and other new physics. We present an overview of the main experimental and
theoretical results obtained during the last decade in this field.
The experimental part outlines search results of 2EC processes and
measurements of the decay energies for possible resonant $0\nu$2EC
transitions. An unprecedented precision in the determination of
decay energies with Penning traps has allowed one to refine the
values of the degeneracy parameter for all previously known
near-resonant decays and has reduced the rather large
uncertainties in the estimate of the $0\nu2$EC half-lives. The
theoretical part contains an updated analysis of the electron
shell effects and an overview of the nuclear structure models, in
which the nuclear matrix elements of the $0\nu2$EC decays are
calculated. One can conclude that the decay probability of
$0\nu$2EC can experience a significant enhancement in several
nuclides.
\end{abstract}

%\pacs{81.05.Uw,68.37.-d,73.20-r}

\maketitle

\renewcommand{\baselinestretch}{1.0}
\renewcommand{\theequation}{\thesection.\arabic{equation}}

\tableofcontents{}

%\newpage

\section{Introduction}
\setcounter{equation}{0}
\label{intro}

In 1937, Ettore Majorana found an evolution equation for a truly neutral spin-1/2 fermion \cite{Majorana1937}.
His work was motivated by the experimental observation of a free neutron by James Chadwick in 1932.
Majorana also conjectured that his equation applies to a hypothetical neutrino introduced by Wofgang Pauli
to explain the continuous spectrum of electrons in the $\beta $ decay of nuclei.
In the mid-1950s, \textcite{Reines:1956,Reines:1959} discovered a particle with neutrino properties.
At the time of the establishment of the Standard Model (SM) of electroweak interactions,
three families of neutrinos were known. While the neutron is a composite fermion consisting of quarks,
neutrinos acquired the status of elementary particles with zero masses in the SM framework.
The discovery of neutrino oscillations by the Super-Kamiokande collaboration %(Fukuda \textit{et al.},~1998) %
\cite{SK:1998}
showed that neutrinos are mixed and massive.
A comprehensive description of the current state of neutrino physics can be found
in the monograph of \textcite{Bilenky:2019}.

Whether neutrinos are truly neutral fermions is one of the fundamental questions in modern particle physics, astrophysics and cosmology. The fermions described by the Dirac equation can also be electrically neutral. However, even in this case, there is a conserved current in Dirac�s theory, which ensures a constant number of particles (minus antiparticles). Majorana fermions, like photons, do not have such a conserved current. In Dirac's theory, particles and antiparticles are independent, while Majorana fermions are their own antiparticles and, via CPT, it follows that they must have zero charge. Truly neutral spin-1/2 fermions are referred to as Majorana fermions and those with a conserved current are referred to as Dirac fermions. Majorana fermions in bispinor basis are vectors of the real vector space $\mathbb{R}^4$. They can also be described by two-component Weyl spinors in the complex space $\mathbb{C}^2$. In both representations, the superposition principle for Majorana fermions holds over the field of real numbers. Majorana fermions belong to the fundamental real representation of the Poincar\'e group. A Dirac fermion of mass $m$ can be represented as a superposition of two Majorana fermions of masses $m$ and $-m$, respectively.

The neutron has a non-vanishing magnetic moment, so it cannot be a pure Majorana particle.
On the other hand, there is no fundamental reason to claim that it is a pure Dirac particle.
In theories with nonconservation of the baryon number, the mass eigenstates include a mixture of baryons and antibaryons.
At the phenomenological level, the effect is modeled by adding a Majoranian mass term to the effective Lagrangian.
As a result, the neutron experiences oscillations $n \leftrightarrow \bar{n}$, whereby the nuclei decay with nonconservation of the baryon number
\cite{CBDover1983,CBDover1983b,Krivoruchenko:1995mn,Krivoruchenko1996,Gal2000,VKopeliovich2010,DGPhillips2016}.
%(Dover~\textit{et al.},~1983,~1985; Gal,~2000; Kopeliovich and Potashnikova,~2010; Krivoruchenko, 1996a,b; and Phillips II~\textit{et al.},~2016).
Experimental limits for the period of the $n \leftrightarrow \bar{n}$ oscillations in the vacuum,
$\tau_{\mathrm{vac}} > 2.7\times10^8$ s %(Abe~\textit{et al.}, 2015),
\cite{Abe2015},
constrain the neutron Majorana mass to $\Delta m \sim 1/\tau_{\mathrm{vac}} < 0.8\times 10^{-33} m$,
where $m = 939.57$ MeV/c$^2$ is the neutron Dirac mass. Thus, under the condition of nonconservation of the baryon number,
Majorana's idea on the existence of truly neutral fermions can be partially realized in relation to the neutron.
In contrast, neutrinos can be pure Majorana fermions or pure Dirac fermions, or a mixture of these two extreme cases.
It is noteworthy that none of the variants of neutrino masses are possible within the SM.
The neutrino mass problem leads us to physics beyond the SM.
\textcolor{black}{
Alternative examples of Majorana particles include
weakly interacting dark matter candidates
and Majorana zero modes in solid state systems \cite{Elliott:2015}.}

Searches for neutrinoless double-beta ($0\nu 2 \beta^-$) decay,  neutrinoless double-electron capture ($ 0\nu$2EC) by nuclei and other lepton number violating (LNV) processes provide the possibility to shed light on the question of the nature of neutrinos: whether they are Majorana or Dirac particles.
By virtue of the black-box theorem \cite{Hirsch2006,Schechter1982}, %(Hirsch~\textit{et al.}, 2006; Schechter and Valle, 1982),
the observation of the $0\nu 2\beta$ decay  would prove that neutrinos
have a finite Majorana mass.
\textcolor{black}{
The massive Majorana neutrinos lead to a violation of the conservation of the total lepton number $L$. In the quark sector of SM, the baryon charge, $B$, is a similar quantum number. Vector currents of $B$ and $L$ are classically conserved. Left handed fermions are associated with
the $SU(2)$ electroweak gauge fields $W^{\pm}$ and $Z^0$, so that vector currents of $B$ and $L$, as 't Hooft
first pointed out \cite{tHooft:1976}, are sensitive to the axial anomaly.
Through electroweak instantons, this leads to nonconservation of $B$ and $L$, while the difference $(B - L)$ is conserved. The violating amplitude is exponentially suppressed. Another example is given by sphaleron solutions of classical field equations
of SM, that preserve $(B+L)$ but violate $B$ and $L$ individually, which can be relevant for cosmological implications \cite{White:2016}.
The conservation of $(B-L)$ within SM is not supported by any fundamental principle analogous to local gauge symmetry,
so $B$ and $L$ can be broken beyond SM explicitly.
}
Experimental observation of the proton decay and/or $n \leftrightarrow \bar{n}$ oscillations
could prove nonconservation of $B$,
while observation of the $0\nu 2 \beta^-$ decay and/or the $ 0\nu$2EC process could prove nonconservation of $L$ with constant $B$.
Moreover, these processes are of interest for determining
the absolute neutrino mass scale, the type of neutrino mass hierarchy,
and the character of CP violation in lepton sector.
Due to the exceptional value of LNV processes, vast literature is devoted to
physics of $0\nu 2 \beta^-$ decay and the underlined nuclear structure models
(for reviews see
%Avignone III~\textit{et al.} (2008); Bilenky and Petcov (1987); Ejiri~\textit{et al.} (2019);
%Engel and Men\'endez (2017); \textcite{Raduta2015}; \textcite{Suhonen2007}; Vergados~\textit{et al.} (2012); \textcite{Zdesenko:2002}).
\textcite{Vergados:2012xy,Raduta2015,Engel:2017,Suhonen2019,Suhonen2007,Bilenky1987,Zdesenko:2002,Avignone:2008}).

The $0\nu 2 \beta^-$ decay was first discussed by \textcite{Furry39}. The process is shown in Fig. 1.
A nucleus with the mass number $A$ and charge $Z$ experiences $0\nu 2\beta^-$ decay accompanied by
the exchange of a Majorana neutrino between the nucleons:
\begin{equation} \label{eq:NLDBD-1}
(A,Z) \to (A,Z+2)^{++} + e^{-} + e^{-} ,
\end{equation}
where $(A,Z+2)^{++}$ is the doubly ionized atom in the final state.
There are many models beyond SM that provide alternative mechanisms of the $0\nu 2 \beta^-$-decay,
some of which are discussed in Sec.~II.

In 1955, the related $0\nu2$EC process
\begin{equation} \label{eq:0Nu2EC-1}
e^{-}_b + e^{-}_b + (A,Z) \to (A,Z-2)^{**}
\end{equation}
was discussed by Winter \cite{WINTER55}. Here $e^{ - } _b$ are bound electrons.
The nucleus and the electron shell of the neutral atom $ (A,Z-2)^{**}$ are in excited states.
An example of the mechanism related to the Majorana neutrino exchange is shown in Fig.~2.
Subsequent de-excitation of the nucleus occurs via gamma-ray radiation or $\beta$ decays.
De-excitation of the electron shell is associated with the emission of Auger electrons or gamma rays in
a cascade formed by filling of electron vacancies. In the absence of special selection rules,
dipole radiation dominates in X-rays.
Since the dipole moment of electrons is much higher than that of nucleons in the nucleus,
the de-excitation of the electron shell goes faster.
For atoms with a low value of $Z$, the Auger electron emission is more likely.
With increase in the atomic number, the radiation of X-ray photons becomes dominant.
The de-excitation of high electron orbits is due to Auger-electron emission for all $Z$.

Estimates show that the sensitivity of the $0\nu2$EC process to the Majorana
neutrino mass is many orders of magnitude lower than that of the $0\nu2\beta^-$ decay.
Winter pointed out that degeneracy of the energies of the parent atom $(A,Z)$
and the daughter atom $(A,Z-2)^{**}$ gives rise to resonant enhancement of the decay.
In the early 80s, a number of other authors also remarked on
the possible resonant enhancement of the $0\nu2$EC process \cite{Georgi1981,VOLO82}. %(Georgi~\textit{et al.}, 1981; Voloshin~\textit{et al.}, 1982).
The resonances in 2EC were considered, however, as an unlikely coincidence.

%%%%%%%%%%%%%%%%%%%%%%%%%%%%%%%%%%%%%%%%%%%%%%%%%%%%%%%%%%%%%%%%%%%%%%%%%%%%
%%%%%%%%%%%%%%%%%%%%%%%%%%%%%%%%%%%%%%%%%%%%%%%%%%%%%%%%%%%%%%%%%%%%%%%%%%%%
%%%%%%%%%%%%%%%%%%%%%%%%%%%%%%%%%%%%%%%%%%%%%%%%%%%%%%%%%%%%%%%%%%%%%%%%%%%%
\begin{figure}[t]
\includegraphics[angle = 270,width=0.146\textwidth]{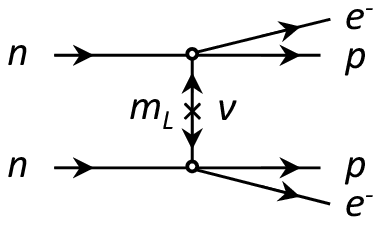}
\caption[fig]{\label{fig2:1} Schematic representation of neutrinoless double-beta decay: Two neutrons in the nucleus experience $\beta$ decay
accompanied by the exchange of a Majorana neutrino.
Neutrons, protons, electrons and neutrinos are represented by solid lines.
Arrows show the flow of baryon charge of protons and neutrons and the flow of lepton charge of electrons and neutrinos.
The cross denotes the Majorana neutrino mass term, $m_L$, that causes the helicity-flip of the intermediate neutrino and violates lepton number by two units.}
\end{figure}
%%%%%%%%%%%%%%%%%%%%%%%%%%%%%%%%%%%%%%%%%%%%%%%%%%%%%%%%%%%%%%%%%%%%%%%%%%%%
%%%%%%%%%%%%%%%%%%%%%%%%%%%%%%%%%%%%%%%%%%%%%%%%%%%%%%%%%%%%%%%%%%%%%%%%%%%%
%%%%%%%%%%%%%%%%%%%%%%%%%%%%%%%%%%%%%%%%%%%%%%%%%%%%%%%%%%%%%%%%%%%%%%%%%%%%by increasing the heart rate

To compensate for the low probability of the $0\nu$2EC process by a resonance effect, it is necessary to determine
the energy difference of atoms with high accuracy. The decay probability is proportional to the Breit-Wigner factor
$\Gamma_f / (\Delta^2 + \Gamma_f^2/4)$,
where $\Gamma_f$ is the electromagnetic decay width of the daughter atom and $ \Delta = M_{A,Z} - M^{**}_{A,Z-2}$ is the degeneracy parameter
equal to the mass difference of the parent and the daughter atoms.
The maximum increase in probability is achieved for $\Delta = 0$ when the decay amplitude approaches the unitary limit.
Taking $\Delta \sim 10$ keV for the typical splitting of the masses of the %initial and final
atoms and $\Gamma_f \sim 10$ eV for the typical decay width of the excited electron shell of the daughter atom, one finds the maximum enhancement of $\sim 10^6$.
The degeneracy parameter $\Delta \lesssim \Gamma_f$ gives the half-life of a nuclide with respect to $0\nu$2EC
comparable to the half-life of nuclides with respect to $0\nu2\beta^-$ decay.

The near-resonant $0\nu2$EC process was analyzed in detail by \textcite{BERN83}.
The authors developed a non-relativistic formalism of the resonant $0\nu2$EC in atoms and
specified a dozen of nuclide pairs for which degeneracy is not excluded.
The $0\nu2$EC process became the subject of a detailed theoretical study by \textcite{Sujkowski:2004}.
A list of the near-resonant $0\nu2$EC nuclide pairs is also provided by \textcite{Karpeshin2008}.
The problem acquired an experimental character:
the difference between masses of the parent and daughter atoms, i.e. $Q$-values, known
to an accuracy of about 10 keV, which is too far from the accuracy required to identify the unitary limit.
The determination of the degeneracy parameter has acquired fundamental importance.

In the 1980s, there was no well-developed technique to measure the masses of nuclides with relative uncertainty of about $10^{-9}$
sufficient to find resonantly enhanced $0\nu2$EC processes.
The presently state-of-the-art technique high-precision Penning-trap mass spectrometry was still in its \textcolor{black}{infancy}. Its triumphal advance in the field of high-precision mass measurements on radioactive nuclides began with the installation of the ISOLTRAP facility at CERN in the late 1980s
\cite{Bollen1987,Mukherjee2008,Kluge2013}. The last decades was marked by a rising variety of high-precision Penning-trap facilities in Europe, the USA and Canada \cite{Blaum2006, Blaum13}.  This lead to a tremendous development of Penning-trap mass-measurement techniques \cite{Kretzschmar2007,George2007,Eliseev2007,Kretzschmar2013,Blaum13,Eliseev2013,Eliseev2014} and finally made it possible to routinely carry out mass measurements on a broad variety of nuclides with a relative uncertainty of about 10$^{-9}$.
%Nowadays, Penning-trap mass spectrometry is the method of choice, if the mass of a particular nuclide is to be determined with a very high precision. This technique is superior in achievable sensitivity, accuracy and resolving power to all other mass-measurement methods due to the very idea which forms the basis of a Penning trap: one confines a single particle to a minute volume by the superposition of an extremely stable strong static homogeneous magnetic field and a weak static electric quadrupole potential.
The mass of the ion is determined via the measurement of its free cyclotron frequency in a pure magnetic field, the most precisely measurable quantity in physics.

These factors motivated a new study of the near-resonant $0\nu2$EC process.
A relativistic formalism for calculating electron shell effects was developed and an updated realistic list of nuclide pairs for which the measurement of $ Q_{\mathrm {2EC}} $ values ​
​has high priority was compiled \cite{KRIV11}.
Also a significantly refreshed database of the nuclides and their excited states is now available, thirty years since the previous publication by \textcite{BERN83}.
An overview of the investigation of the resonant $0\nu$2EC is given by \textcite{Eliseev2012}
including the persistent experimental attempts to search for appropriate candidates for this extraordinary phenomenon.

%%%%%%%%%%%%%%%%%%%%%%%%%%%%%%%%%%%%%%%%%%%%%%%%%%%%%%%%%%%%%%%%%%%%%%%%%%%%
%%%%%%%%%%%%%%%%%%%%%%%%%%%%%%%%%%%%%%%%%%%%%%%%%%%%%%%%%%%%%%%%%%%%%%%%%%%%
%%%%%%%%%%%%%%%%%%%%%%%%%%%%%%%%%%%%%%%%%%%%%%%%%%%%%%%%%%%%%%%%%%%%%%%%%%%%
\begin{figure}[t]
\includegraphics[angle = 270,width=0.146\textwidth]{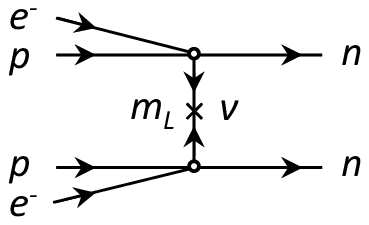}
\caption[fig]{\label{fig2:2} Schematic representation of neutrinoless double-electron capture:
Two protons in the nucleus each capture a bound electron from the electron shell and
turn into two neutrons by the exchange of a Majorana neutrino.
Notations are the same as in Fig.~1.}
\end{figure}
%%%%%%%%%%%%%%%%%%%%%%%%%%%%%%%%%%%%%%%%%%%%%%%%%%%%%%%%%%%%%%%%%%%%%%%%%%%%
%%%%%%%%%%%%%%%%%%%%%%%%%%%%%%%%%%%%%%%%%%%%%%%%%%%%%%%%%%%%%%%%%%%%%%%%%%%%
%%%%%%%%%%%%%%%%%%%%%%%%%%%%%%%%%%%%%%%%%%%%%%%%%%%%%%%%%%%%%%%%%%%%%%%%%%%%

The advancements of the experiments in search of the $0\nu$2EC
process are lower than of those searching for $0\nu
2\beta^-$ decay. While the sensitivity of the $0\nu 2\beta^-$
experiments approaches half-life limits
$T_{1/2}\sim10^{24}-10^{26}$ yr, which constrains the effective
Majorana neutrino mass of electron neutrino to $| m_{\beta \beta} | \lesssim 0.1-0.7$ eV, the results of the
best $0\nu$2EC experiments are yet on the level of
$T_{1/2}\sim10^{19}-10^{22}$ yr. The reasons for this difference
are rather obvious: there is usually a much lower relative abundance of the
isotopes of interest (typically lower than 1\%), and additionally
a more complicated effect signature due to the emission of a
gamma-quanta cascade (instead of a clear $0\nu2\beta^-$ peak at
the decay energy). The second circumstance results in a lower
detection efficiency for the most energetic peak in a $0\nu$2EC
energy spectrum. Furthermore, the energy of the most energetic
$0\nu$2EC peak is generally lower than in the  $0\nu 2\beta^-$
processes, yet the higher the energy of a certain process the
better the suppression of the radioactive background. As a result
the scale of the 2EC experiments is
substantially smaller than that of the $0\nu 2\beta^-$ ones. At
the same time, there is a motivation to search for the
neutrinoless EC$\beta^+$ and $2\beta^+$ decays owing to the
potential to clarify the possible contribution of the right-handed
currents to the $0\nu 2\beta^{-}$ decay rate \cite{Hirsch:1994},
and the appealing possibility of the resonant $0\nu$2EC processes.
The complicated effect signature expected in resonant $0\nu$2EC
transitions becomes an advantage: the detection of several gamma
quanta with well-known energies could be a strong proof of the
pursued effect.

The above mentioned aspects of the phase space, degeneracy, abundance factors, etc. play an important role
in determining the half-lives of the $0\nu$EC$\beta^+ $ and $0\nu $2EC processes.
\textcolor{black}{A further ingredient affecting the decay half-lives are the involved
nuclear matrix elements (NMEs), see the reviews \cite{Maalampi:2013,Suhonen2012a,Suhonen2019}. These
NMEs have been calculated in various nuclear-theory frameworks for a
number of nuclei.}
In this review we use these NMEs, as well as NMEs which have been calculated just for this review, to estimate the
half-lives of those $0\nu$2EC transitions which are of interest due to their (possibly) favorable resonance conditions.

\section{Double-electron capture and physics beyond the Standard Model}
%%%%%%%%%%%%%%%%%%%%%%%%%%%%%%%%%%%%%%%%%%%%%%%%%%%%%%%%%%%%%%%%%%%%%%%%%%%%
%%%%%%%%%%%%%%%%%%%%%%%%%%%%%%%%%%%%%%%%%%%%%%%%%%%%%%%%%%%%%%%%%%%%%%%%%%%%
\setcounter{equation}{0}
\label{sec:Double-electron capture and physics beyond the Standard Model}

The underlying quark-level physics behind the $0\nu$2EC process (see Eq.~(\ref{eq:0Nu2EC-1}))
is basically the same as for the $0\nu2\beta^-$, $0\nu 2\beta^+$ and $0\nu$EC$\beta^+$ decays.
In Figs.~\ref{fig2:1} and \ref{fig2:2}, we show the mechanism of exchange of light or heavy Majorana neutrinos
that arises beyond the Standard Model within the Weinberg effective Lagrangian approach \cite{Weinberg:1979}.
In the Weinberg scenario, the Majorana mass occurs from  LNV operator of dimension 5,
which provides conditions for the existence of the processes $0\nu$2EC and $0\nu2\beta^-$.
A violation of the lepton number can also occur from quark-lepton effective Lagrangians of higher dimensions,
corresponding to other possible mechanisms of the $0\nu 2$EC process.
The neutrinoless 2EC can be accompanied by the emission of one or more very light particles,
other than neutrinos in $2\nu$2EC. A well-known example is the Majoron, $J$,
as the Goldstone boson of a spontaneously broken $U(1)_{L}$-symmetry of lepton number.
Passing to the hadronic level one meets two possibilities of hadronization of the quark-level underlying process, known from $0\nu2\beta^-$ decay:
direct nucleon and pionic mechanisms.
Below we consider the above-mentioned aspects of  the $0\nu$2EC process in more detail.

First of all, the underlying quark level mechanisms of the neutrinoless 2EC can be classified
according to the possible exotic final states:
%\begin{itemize}

$\bullet$ No exotic particles in the final state: the reaction $0\nu$2EC is shown in Eq.~(\ref{eq:0Nu2EC-1}). %[3mm]

$\bullet$ The reaction $0\nu $2EC$nJ$:
%\begin{eqnarray}
%\label{eq:0Nu2ECnJ-1}
$
e^{-}_b + e^{-}_b + (A,Z) \to (A,Z-2)^{**}+ n J,
$
%\end{eqnarray}
with $n$ being the number of Majorons or Majoron-like exotic particles in the final state.

Both kinds of reaction can be further classified by the typical distance between particles
involved in the underlying quark-lepton process, depending on the masses of  the intermediate particles
\cite{Pas:1999fc,Pas:2000vn,Prezeau:2003xn,Cirigliano:2017,Cirigliano:2018}, as illustrated in Fig.~\ref{fig:mechanisms-1}.

$\bullet$  Long-range mechanisms with the Weinberg $d = 5$ operator of Fig.~\ref{fig:mechanisms-1}~(a)
and an effective $d=6$ operator in the upper vertex of Fig.~\ref{fig:mechanisms-1}~(b).
%\cite{Pas:1999fc}.

$\bullet$  Short-range mechanisms with a dimension 9 effective operator in the vertex of Fig.~\ref{fig:mechanisms-1}~(c).
%\cite{Pas:2000vn}.

The effective operators in the low-energy limit originate from diagrams with heavy exotic particles in the internal lines.

The diagrams of the $0\nu$2EC$nJ$ decays are derived from those in Fig.~\ref{fig:mechanisms-1}
by inserting one or more scalar Majoron lines either into the blobs of effective operators or into
the central neutrino line of Fig.~\ref{fig:mechanisms-1}~(a).

\subsection{Quark-level mechanisms of $0\nu$2EC}
\label{sec:Quark-Level mechanisms of $0Nu2EC}

\begin{figure}[t]
\includegraphics[angle = 270,width=0.62\textwidth]{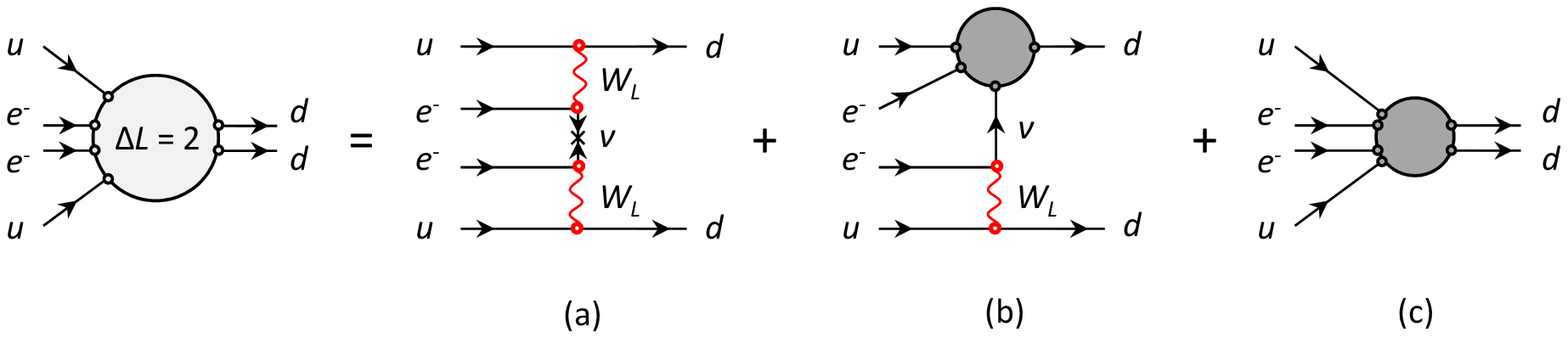}
\caption[fig]{\label{fig:mechanisms-1} A decomposition of the generic $\Delta L= 2$ vertex into the long-range  (a), (b) and
the short-range (c) quark-level contributions to $0\nu2$EC.
The diagram (a) is the conventional Majorana neutrino mass mechanism. The blobs in (b) and (c) denote the effective $\Delta L= 2$ vertices.}
\end{figure}

Let us consider in more detail the short- and long-ranged mechanisms of $0\nu 2$EC.
The corresponding diagrams for the $0\nu 2$EC process are shown in  Fig.~\ref{fig:mechanisms-1}.
The blobs in Figs.~\ref{fig:mechanisms-1}~(b,c) represent the $\Delta L = 2$ effective vertices beyond the SM.
At the low-energy scales $\mu \sim 100$ MeV, typical for $0\nu 2$EC process, the blobs are essentially point-like,
being generated by the exchange of a heavy particle with the characteristic masses $M_{H}$ much larger than the
$0\nu 2$EC-scale, i.e. $M_{H} \gg \mu $.
Integrating them out one finds the effective Lagrangian terms describing the vertices at the scale  $\mu \ll \Lambda \sim M_{H}$
for any kind of underlying high-scale physics beyond SM.
These vertices can be written in the following form \cite{Pas:1999fc,Pas:2000vn}
\begin{eqnarray}
\label{eq:Dim-6}
\mathcal{L}_{ql}^{(6)}&=&\frac{G_F}{\sqrt{2}}
\left( - j_{CC}^{\mu}J_{CC\, \mu}+
\sum_{i}C^{X}_{i}(\mu)\mathcal{O}^{(6)X}_{i}(\mu)\right),\\[3mm]
\label{eq:Dim-9}
{\cal L}^{(9)}_{ql} &=& \frac{G_F^2}{2 m_p} \,
              \sum_{i, XY} C_{i}^{XY}(\mu)\cdot \mathcal{O}^{(9)XY}_{i}(\mu).
\end{eqnarray}
%:{SK_16.03.2019-6} END
The first and second lines correspond to Figs.~\ref{fig:mechanisms-1}~(b) and (c), respectively.
The proton mass $m_{p}$ is introduced to match the conventional notations.
%The characteristic scale of $0\nu2$EC amounts $\mu\sim$100 MeV.
The complete set of the $\Delta L=2$ operators for $d = 6$ and $9$ is as follows \cite{Gonzalez:2015ady,Arbelaez:2016zlt,Arbelaez:2016uto}:
%and dimension-6 operators
\begin{eqnarray}
\label{eq:LROperBasis-1}
\mathcal{O}_{1}^{(6) X} &=& 4 (\bar{d} P_{X} u)  \left(\overline{\nu^{c}} P_{L} e\right), \\
%\left(\bar{e} P_{R} \nu^{C}\right), \\
%
\label{eq:LROperBasis-2}
\mathcal{O}_{2}^{(6) X} &=& 4  (\bar{d} \sigma^{\mu\nu}P_{X} u) \left(\overline{\nu^{c}} \sigma^{\mu\nu}P_{L} e\right)\\
%\left(\bar{e}\sigma^{\mu\nu}  P_{R} \nu^{C}\right),\\
%
\label{eq:LROperBasis-3}
\mathcal{O}_{3}^{(6) X} &=& 4  (\bar{d} \gamma_{\mu}P_{X} u) \left(\overline{\nu^{c}}\gamma^{\mu} P_{R} e\right).
%\left(\bar{e} \gamma^{\mu} P_{R} \nu^{C}\right).
%
\end{eqnarray}
\begin{eqnarray}
\label{eq:SROperBasis-1}
\mathcal{O}^{(9)XY}_{1}&=& 4 ({\bar d}P_{X}u) ({\bar d}P_{Y}u) \ j,\\
\label{eq:SROperBasis-2}
\mathcal{O}^{(9)XY}_{2}&=& 4 ({\bar d}\sigma^{\mu\nu}P_{X}u)
                         ({\bar d}\sigma_{\mu\nu}P_{X}u) \ j,\\
\label{eq:SROperBasis-3}
\mathcal{O}^{(9)XY}_{3}&=& 4 ({\bar d}\gamma^{\mu}P_{X}u)
                        ({\bar d}\gamma_{\mu}P_{Y}u) \  j,\\
\label{eq:SROperBasis-4}
\mathcal{O}^{(9)XY}_{4}&=& 4 ({\bar d}\gamma^{\mu}P_{X}u)
                         ({\bar u}\sigma_{\mu\nu}P_{Y}d) \ j^{\nu},\\
\label{eq:SROperBasis-5}
\mathcal{O}^{(9)XY}_{5}&=& 4 ({\bar d}\gamma^{\mu}P_{X}u) ({\bar d}P_{Y}u) \ j_{\mu},
\end{eqnarray}
where $X,Y = L,R$ and the leptonic currents are
%\begin{eqnarray}\label{eq:Curr}
%
$
j = \overline{e^{c}}(1\pm \gamma_{5})e$, $j_{\mu} = \overline{e^{c}}\gamma_{\mu}\gamma_{5} e .
$
%\end{eqnarray}
%:{SK_16.03.2019-5} I supplied the currents with the subscript CC
The first term in Eq.~(\ref{eq:Dim-6}) describes the SM low-energy  4-fermion effective
interaction of the Charged Current (CC):
\begin{eqnarray}\label{eq:V-A}
&&j_{CC}^{\mu} = \bar{\nu}\gamma^{\mu}(1 - \gamma_{5}) e, \ \ \
J_{CC\, \mu} = \bar{d}\gamma_{\mu}(1 - \gamma_{5}) u.
\end{eqnarray}
The $SU(3)_{c}\times U(1)_{em}$-symmetric operators in
Eqs.~(\ref{eq:LROperBasis-1}) - (\ref{eq:SROperBasis-5})
are written in the mass-eigenstate basis. They originate from the
$SU(3)_{c}\times SU(2)_{W}\times U(1)_{Y}$ gauge invariant operators
after the electroweak symmetry breaking
(see, e.g., the papers of \textcite{Bonnet:2012kh,Lehman:2014jma,Graesser:2016bpz}).

The diagrams in Figs.~\ref{fig:mechanisms-1}~(a,b) are of second-order in the Lagrangian (\ref{eq:Dim-6}).
The effect of $\Delta L = 2$ is introduced in the diagrams (a) and (b) by the Majorana neutrino mass term
and by the $d = 6$ effective operators  (\ref{eq:LROperBasis-1}) - (\ref{eq:LROperBasis-3}), respectively.
The diagram in Fig.~\ref{fig:mechanisms-1}~(a) is the conventional Majorana neutrino mass mechanism
with the contribution to the $0\nu2$EC amplitude
%(cf. (\ref{LNVP2}))
\begin{eqnarray}\label{eq.MNM-1}
V_{\alpha\beta} & \sim& m_{\beta\beta} \equiv \sum_{i} U^{2}_{ei}m_{\nu_{i}},
\end{eqnarray}
where $U$ is the Pontecorvo - Maki - Nakagawa - Sakata (PMNS) mixing matrix and
$m_{\beta\beta}$
{\color{black}is the effective electron neutrino Majorana mass parameter
%,the very same parameter, which is
well-known from the analysis of  $0\nu2\beta^-$ decay.}
% entering the $0\nu2\beta^-$ decay amplitude.
The contribution of the diagram of Fig.~\ref{fig:mechanisms-1}~(b) is independent of the neutrino mass,
but proportional to the momentum $q$ flowing in the neutrino propagator. This is the so-called $\hat{q}$-type contribution.
%:>>>>>

{\textcolor{black} The following comment is in order.
The diagrams in Fig.~\ref{fig:mechanisms-1}~(a,b) show two possible mechanisms of both $0\nu 2$EC and $0\nu2\beta^-$  (with the inverted final to initial states) processes. The first mechanism Fig.~\ref{fig:mechanisms-1}~(a) contributes to the amplitude of these processes with terms proportional to the effective Majorana mass parameter $m_{\beta\beta}$ defined in (\ref{eq.MNM-1}). The contribution of the second mechanism Fig.~\ref{fig:mechanisms-1}~(b) has no explicit dependence on $m_{\beta\beta}$. This is because the upper vertex in Fig.~\ref{fig:mechanisms-1}~(b) breaks lepton number in two units as necessary for this process to proceed without the need of  the $\Delta L=2$ Majorana neutrino mass insertion into the neutrino line.
Note that $m_{\beta\beta}=0$ is compatible with the neutrino oscillation data in the case of Normal neutrino mass Ordering. This result shows that the mechanism Fig.~\ref{fig:mechanisms-1}~(a), proportional to $m_{\beta\beta}$, can be negligible in comparison with the mechanism in Fig.~\ref{fig:mechanisms-1}~(b).  Therefore, even if $m_{\beta\beta}$ turns out to be very small, both $0\nu2\beta^-$ decay  and $0\nu 2$EC process can be observable due to the latter mechanism. This possibility has been studied in the literature for $0\nu2\beta^-$ decay (see, e.g., the papers of \textcite{Pas:1999fc,Pas:2000vn,Arbelaez:2016zlt,Arbelaez:2016uto,Gonzalez:2015ady}).  }

%%%%%%%%%%%%%%%%%%%%%%%%%%%%%%%%%%%%%%%%%%%%%%%%%%%%%%%%%%%%%%%%%%%%%%%%%%%%
%%%%%%%%%%%%%%%%%%%%%%%%%%%%%%%%%%%%%%%%%%%%%%%%%%%%%%%%%%%%%%%%%%%%%%%%%%%%
%%%%%%%%%%%%%%%%%%%%%%%%%%%%%%%%%%%%%%%%%%%%%%%%%%%%%%%%%%%%%%%%%%%%%%%%%%%%
%%%%%%%%%%%%%%%%%%%%%%%%%%%%%%%%%%%%%%%%%%%%%%%%%%%%%%%%%%%%%%%%%%%%%%%%%%%%
%%%%%%%%%%%%%%%%%%%%%%%%%%%%%%%%%%%%%%%%%%%%%%%%%%%%%%%%%%%%%%%%%%%%%%%%%%%%
%%%%%%%%%%%%%%%%%%%%%%%%%%%%%%%%%%%%%%%%%%%%%%%%%%%%%%%%%%%%%%%%%%%%%%%%%%%%

\begin{figure}[t]
\begin{center}
\includegraphics[angle = 270,width=.15\textwidth]{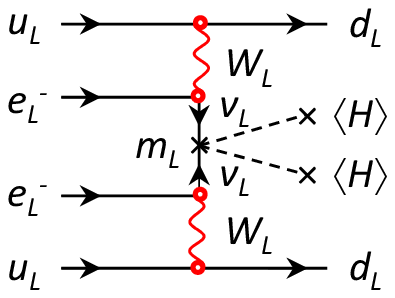}
\caption{
The contribution of the Weinberg operator to the $0\nu2$EC process in the flavor basis of the neutrino states.
}
\label{fig:Weinberg-1}
\end{center}
\end{figure}

The SM gauge invariant Weinberg dimension-5 effective operator generating the neutrino mass mechanism is given by \cite{Weinberg:1979}
%
%{\textcolor{black}
\begin{eqnarray}
\mathcal{L}^{(5)}_{W}&=& \kappa\frac{\left(\overline{L^{c}} \cdot H \right) \left(L \cdot H\right)}{\Lambda}=
\frac{\kappa}{\Lambda} \overline{\nu^{c}} \nu H^{0} H^{0} +... \nonumber \\
&\xrightarrow[]{SSB}& \kappa\frac{\langle H^{0}\rangle^{2}}{\Lambda} \overline{\nu^{c}} \nu  +... \label{eq:Weinberg-1}\;.
\end{eqnarray}
In the above equation we mean $L\cdot H = L_{i}H_{j}\epsilon_{ij}$ the singlet combination of two $SU(2)_{W}$ doublets $L(\overline{L^{c}})$ and $H$.%}
After the electroweak spontaneous symmetry breaking (SSB) with the Higgs vacuum expectation value $\langle H\rangle \neq 0$
neutrinos acquire a Majorana mass $m_{\nu} = -2\kappa \langle H\rangle^{2}/\Lambda$, with $\kappa$ being a dimensionless parameter.
In the flavor basis of neutrino states the contribution of the Weinberg operator to an LNV process
such as $0\nu2$EC is displayed in Fig.~\ref{fig:Weinberg-1}. The summation of multiple insertions of the Weinberg operator
into the bare neutrino propagator entails the renormalized neutrino propagator with Majorana mass $m_{\nu}$.
The operator (\ref{eq:Weinberg-1}) is unique. Other operators of the effective Lagrangian are suppressed by higher powers of the unification scale $\Lambda$.
The study of the neutrinoless 2EC process and $2\beta^-$ decays
could be the most direct way of testing physics beyond the Standard Model.
In terms of the naive dimensional counting one can expect the dominance of the Weinberg operator,
with the dimension 5, over the operators of dimensions 6 and 9 in Eqs.~(\ref{eq:LROperBasis-3}) - (\ref{eq:SROperBasis-1}). However,
in order to set $m_{\nu}$ at eV-scale one should provide
{\textcolor{black}a very small coupling $\kappa \sim 10^{-11}$ for the phenomenologically interesting case of $\Lambda \sim O$(1 TeV).
However, the smallness  of any dimensionless coupling requires explanation. Typically in this case one expects the presence of some underlying physics, for example, symmetry.}
%Numerically this gives suppression of the order
%$\sim (\langle H\rangle/\Lambda)^{-10}$. %equivalent to a huge effective dimensionality of the Weinberg operator.
The situation changes with the increase of the LNV scale up to $\Lambda \sim 10^{13 - 14}$, with $\kappa \sim 1$,
where the contribution of the Weinberg operator to $0\nu2$EC dominates. %over the operators (\ref{eq:SROperBasis-1}) - (\ref{eq:LROperBasis-3}).
The final count depends on the concrete high-scale underlying LNV model:
not all operators appear in the low-energy limit and $\kappa$ is a small suppression factor allowing TeV-scale $\Lambda$.
The latter can stem from loops or the ratio of the SSB scales in multi-scale models (for a recent analysis see, e.g., the paper of \textcite{Helo:2016vsi}).

In this review the mechanism of the neutrino Majorana masses is discussed in detail, for which numerical evaluation of the neutrinoless 2EC half-lives of near-resonant nuclides with the known NMEs will be given.
In the case of high-dimensional operators, as well as for the d = 5 mechanism with the unknown NMEs,
normalized estimates will be given, which take into account the factorization of nuclear effects in the $0\nu$2EC amplitude.
Keeping the above comments in mind, we also discuss mechanisms based on the operators of Eqs.~(\ref{eq:LROperBasis-3}) - (\ref{eq:SROperBasis-1}),
leading to the contributions shown in Figs.~\ref{fig:mechanisms-1}~(b,c).

The blobs in the diagrams of Figs.~\ref{fig:mechanisms-1}~(b,c) can be opened up (ultraviolet completed) in terms of all possible types of renormalizable interactions
consistent with the SM gauge invariance. These are the high-scale models, which lead to the $0\nu2$EC process.
A list of all the possible ultraviolet completions for $0\nu2\beta^-$ decay is given by
\textcite{Bonnet:2012kh}. %The same (?) models contribute to $0\nu2$EC.

The Wilson coefficients $C_{i}$ in Eqs.~(\ref{eq:Dim-6}), (\ref{eq:Dim-9}) are calculable in terms of the
parameters (couplings and masses) of a particular underlying model at the scale $\Lambda \sim M_{H}$, called
``matching scale''.  Note that some of $C_{i}(\Lambda)$ may vanish.
In order to make contact with $0\nu2$EC one needs to estimate
$C_{i}$ at a scale $\mu_{0}$ close to the typical $0\nu2$EC-energy scale.
%The QCD corrections, such as shown in Fig.~\ref{fig:qcdc}, lead
%to running of the coefficients between the matching $\Lambda$ and
%$\mu_{0}$ scales.
The coefficients $C_{i}$ run from the scale $\Lambda$ down to $\mu_{0}$ due to the QCD corrections.
Also the $d=9$ operators undergo the RGE-mixing with each other leading to the mixing of the corresponding Wilson coefficients.
%:SK
%Because of vast numerical difference of the NME of some operators, their mixing may have a crucial impact on the relation between
%the process rate and the parameters of underlying high-scale models.

%%%%%%%%%%%%%%%%%%%%%%%%%%%%%%%%%%%%%%%%%

The general parameterization of the $0\nu2$EC amplitude derived from the diagrams in Fig.~\ref{fig:mechanisms-1},
taking into account the leading order QCD-running \cite{Gonzalez:2015ady,Cirigliano:2018,Liao:2019gex,Ayala:2020gtv}, reads:
\begin{equation}
V_{\alpha \beta } = G_{F}^{2}\cos ^{2}\theta _{C}K_Z \left( \sum_{i=1}^{3}\beta_{i}^{X}(\mu _{0},\Lambda )C_{i}^{X}(\Lambda ) \right.
+ \left. \sum_{i=1}^{5}\beta
_{i}^{XY}(\mu _{0},\Lambda )C_{i}^{XY}(\Lambda )\right) \mathcal{A}_{\alpha \beta }. \label{eq:ECECRate-1}
\end{equation}
The parameters $\beta_{i}^{X}$ and $\beta_{i}^{XY}$ incorporate the QCD-running of the Wilson coefficients
and the matrix elements of
the operators in Eqs.(\ref{eq:LROperBasis-1}) - (\ref{eq:LROperBasis-3}) combined with $j_{CC}^{\mu}J_{CC\,\mu}$ and
the operators in Eqs.~(\ref{eq:SROperBasis-1}) - (\ref{eq:SROperBasis-5}).
The wave functions of the captured electrons with quantum numbers $\alpha$ and $\beta$
enter the coefficients $\mathcal{A}_{\alpha \beta }$ defined by Eqs.~(IV.20) - (IV.23).
In Eq.~(\ref{eq:ECECRate-1}) the summation over the different chiralities $X,Y =
L,R$ is implied.  It is important to note that the Wilson coefficients
$C_{i}(\Lambda)$ entering Eqs.~(\ref{eq:Dim-6}) and (\ref{eq:Dim-9}) are linked to the
matching scale $\Lambda$, where they are calculable in terms of the Lagrangian parameters of a particular high-scale underlying model.
\textcolor{black}{
The decay amplitude (\ref{eq:ECECRate-1}) is supplemented by the overlap amplitude $K_Z$
of the electron shells of the initial and final atoms.
In this review, we discuss mainly the light Majorana neutrino exchange mechanism of Fig.~\ref{fig:mechanisms-1}~(a).
}

\textcolor{black}{The $0\nu2$EC NMEs are currently known only for the Majorana neutrino exchange mechanisms coupled to left- and right-handed currents.
Calculations of the NMEs corresponding to the other long- and short-range mechanisms
of Figs.~\ref{fig:mechanisms-1}~(b) and (c), respectively, for all operators
(\ref{eq:LROperBasis-3}) - (\ref{eq:SROperBasis-1}) are still in progress.}

%%%%%%%%%%%
%%%%%%%%%%%%

\subsection{Examples of underlying high-scale models}
\label{sec:UV}

We give three examples of popular high-scale models that can underlie the $0\nu2$EC process.
In the low-energy limit their contribution is described by the effective Lagrangians (\ref{eq:Dim-6}) and/or (\ref{eq:Dim-9}). %\\[3mm]

$\bullet$ {\it Left-Right symmetric models:}
A well-known example of a high-scale model leading to $\Delta L =2 $ processes, such as $0\nu2\beta$-decay and
$0\nu2$EC as well as generating Majorana mass for neutrinos, is the Left-Right symmetric extension of the SM. The Left-Right Symmetric Model (LRSM) is based on the gauge group $\mathcal{G}$ spontaneously broken via the chain
\begin{eqnarray}\label{eq:LRSM-G}
&\mathcal{G} = SU(3)_{C}\times SU(2)_{L} \times SU(2)_{R}\times U(1)_{B-L}&\\
\nonumber
%\xRightarrow[v_{R}]{}
&\Downarrow\ \ v_{R}&\\
\nonumber
&SU(3)_{C}\times SU(2)_{L}\times U(1)_{Y}
%\xRightarrow[v_{SM}]{}
&\\
\nonumber
&\ \ \Downarrow\ \ v_{SM} &\\
\nonumber
&SU(3)_{C}\times U(1)_{em}.&
\end{eqnarray}
%
%:{SK_16.03.2019-4} I modified significantly this fragment until the label %:{SK_16.03.2019-4} END
%
where $v_{R}\equiv \langle \Delta_{R}\rangle \gg \langle \Phi \rangle \equiv v_{SM}$ are the vacuum expectation values (VEV) of a Higgs $SU(2)_{R}$-triplet, $\Delta_{R}$ and a Higgs  bi-doublet, $\Phi$, respectively. The bi-doublet belongs to  the doublet representation of both $SU(2)_{L}$ and $SU(2)_{R}$.
There is also a Higgs $SU(2)_{L}$-triplet,
$\Delta_{L}$, with the VEV $v_{L} \equiv \langle \Delta_{L}\rangle$. Left- and right-handed leptons and quarks
\textcolor{black}{belong to the doublet representations} of the $SU(2)_{L}$ and $SU(2)_{R}$ gauge groups, respectively.
The $SU(3)_{C}\times SU(2)_{L} \times SU(2)_{R}\times U(1)_{B-L}$ assignments of the LRSM fields are
\begin{eqnarray}\label{eq:LRSM-assign-1}
L_{{L(R)}}&=&\left(\begin{array}{c}
\nu_{i}\\
\ell^{-}_{i} %
\end{array}\right)_{L(R)}\sim \left[1, 2(1),1(2); -1\right], \\
\nonumber
Q_{L(R)}&=&\left(\begin{array}{c}
u_{i}\\
d_{i} %
\end{array}\right)_{L(R)}\sim \left[3,2(1), 2(1); -1/3\right],\\
\nonumber
\Delta_{L(R)}&=&
\left(\begin{array}{cc}
\frac{\Delta^{+}}{\sqrt{2}}& \Delta^{++}\\
\Delta^{0}&  \frac{-\Delta^{+}}{\sqrt{2}}
\end{array}\right)_{L(R)}\sim \left[1,3(1),1(3); 2\right],\\
\nonumber
\Phi &=&
\left(\begin{array}{cc}
\Phi_{1}^{0}& \Phi_{1}^{+}\\
\Phi_{2}^{-}&  \Phi_{2}^{0}
\end{array}\right)\sim \left[1,2,2; 0\right],
\end{eqnarray}
where $i=1,2,3$ is the generation index. Previously introduced VEVs are related with the VEVs of the electrically neutral components $\langle \Delta_{L,R}\rangle\equiv \langle \Delta_{L,R}^{0}\rangle$,
$\langle \Phi\rangle^{2}= \langle \Phi^{0}_{1}\rangle^{2}+ \langle \Phi^{0}_{2}\rangle^{2}\equiv v_{SM}$. There are two charged gauge bosons $W^{\pm}_{L,R}$ and two neutral gauge bosons $Z_{L,R}$
%corresponding to $SU(2)_{L,R}$ groups
with masses of the order
$M_{W_{R}}, M_{Z_{R}} \sim g_{R} v_{R}$, $M_{W_{L}},M_{Z_{L}} \sim g_{L} v_{SM}$.
Note that in the scenario with the manifest Left-Right symmetry the $SU(2)_{L,R}$-gauge couplings obey $g_{L}=g_{R}$.
Since the bosons $W_{R}, Z_{R}$ have not been experimentally observed, the scale of the left-right symmetry breaking
$v_{R}$
must be sufficiently large, above few TeV.  On the other hand the VEV of the ``left'' triplet $v_{L}$ must be small, since it affects the SM relation $\rho=1$, which is in good agreement with the experimental measurements setting an upper limit
$v_{L}\lesssim 8$~GeV.  From the scalar potential of the LRSM follows $v_{L}\sim v_{SM}^{2}/v_{R}$, which satisfies the above upper limit for $v_{R}\gtrsim10$~TeV.
%, typically less than
%The angle of  $W_{L}-W_{R}$-mixing $\zeta$ is given by $\tan\zeta = M_{W_{L}}/M_{W_{R}}$.
%
%:{SK_16.03.2019-4}

\begin{figure}
\includegraphics[angle = 270,width=0.5\textwidth]{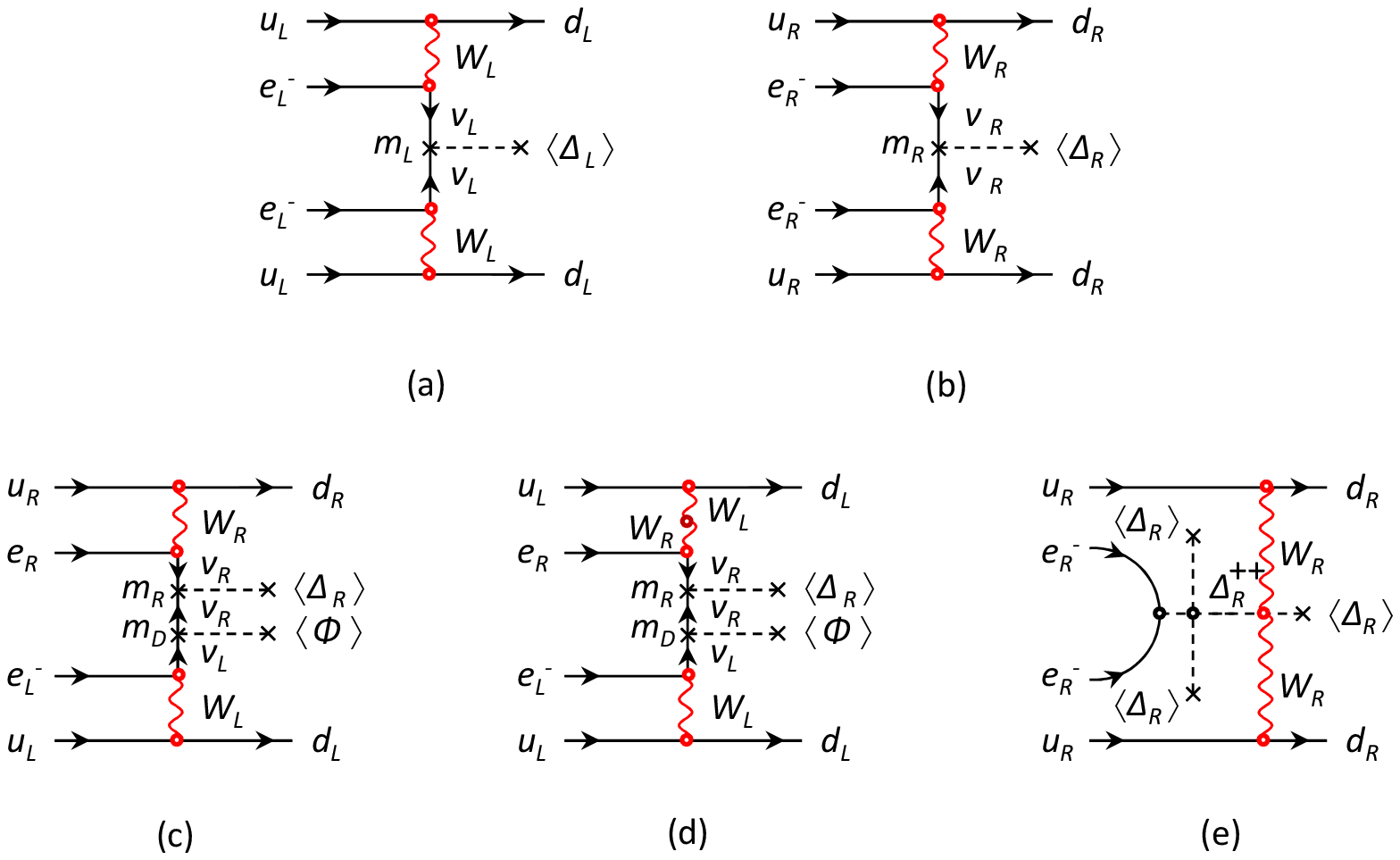}
\caption[fig]{\label{fig:Fig-LRSM-1} Possible flavor-basis contributions to $0\nu2$EC within LRSM.}
\end{figure}

The spontaneous gauge symmetry breaking (\ref{eq:LRSM-G}) generates a $6\times 6$ neutrino seesaw-I mass matrix
given in the basis $(\nu_{L}, \nu_{R}^{\ C})^{T}$ by
\begin{eqnarray}\label{eq:NuMass-1}
M^{\nu} &=& \left(\begin{array}{cc}
m_{L}&m_{D}\\
m^{T}_{D}& m_{R}%
\end{array}\right)
\end{eqnarray}
with $m_{L,R} \sim y_{L,R} v_{L,R}$ and $m_{D} \sim y_{\Phi} v_{SM}$ being $3\times 3$ block matrices  in the generation space.
The matrix (\ref{eq:NuMass-1}) is diagonalized to
%
%\begin{eqnarray}\label{eq:diag-1}
%
$
U^{T} M^{\nu} U = \mbox{Diag}(m_{\nu_{i}}; m_{N_{i}})
$
%
%\end{eqnarray}
by an orthogonal mixing matrix
\begin{eqnarray}\label{eq:NuMassMix-1}
U &=& \left(\begin{array}{cc}
U_{L}&U_{D}\\
U^{T}_{D}& U_{R}%
\end{array}\right),
\end{eqnarray}
with $U_{L,R,D}$ being $3\times 3$ block matrices  in the generation space.  The neutrino mass spectrum consists of three light $\nu_{1,2,3}$ and three heavy $N_{1,2,3}$ Majorana neutrino states with the masses  $m_{\nu_{1,2,3,}}$ and $m_{N_{1,2,3}}$, respectively.\\
The possible contributions to $0\nu2$EC within LRSM are shown in Fig.~\ref{fig:Fig-LRSM-1}.\\
The diagram of Fig.~\ref{fig:Fig-LRSM-1}~(a) shows the conventional long-range light Majorana neutrino exchange mechanisms with the contribution shown in Eq.~(\ref{eq.MNM-1}).
The diagrams in Figs.~\ref{fig:Fig-LRSM-1}~(b,e) are short-range mechanisms with two heavy right-handed bosons $W_{R}$ and heavy neutrino $\nu_{R}$ or doubly-charged Higgs $\Delta^{++}$ exchange. In the low-energy limit they are reducing to the effective operators  $\mathcal{O}^{(9)RR}_{3}$ in Eq.~(\ref{eq:SROperBasis-3}) depicted in Fig.~3~(c).
The diagrams in  Figs.~\ref{fig:Fig-LRSM-1}~(c) and (d), containing light virtual neutrinos, represent the long-range mechanism of $0\nu2$EC.\\
In the low-energy limit the upper parts of Figs.~\ref{fig:Fig-LRSM-1}~(c) and (d) with heavy particles $W_{R}$ and $ \nu_R$
reduce to the $d=6$ effective operators
$\mathcal{O}_{3}^{(6) R}$ and $\mathcal{O}_{3}^{(6) L}$, respectively.  Note that these contributions to the $0\nu2$EC amplitude do not depend on the light neutrino mass $m_{\nu}$, but on its momentum $\mathbf{q}$ flowing in the neutrino propagator.
Technically this happens because different chiralities of the lepton vertices project the $\hat{q}$-term out of the neutrino propagator:
$P_{L}(\hat{q} + m_{\nu})P_{R}=P_{L}\hat{q}$. On the contrary, the diagram
Fig.~\ref{fig:Fig-LRSM-1}~(a), with the same chiralities in both vertices, is proportional to $m_{\nu}$ due to $P_{L}(\hat{q} + m_{\nu})P_{L}=P_{L}m_{\nu}$.
This is consistent with the fact that in the latter case the source of LNV is the Majorana neutrino mass $m_{\nu}$ and in the limit $m_{\nu}\rightarrow 0$ the corresponding contribution must vanish. On the other hand in the former case,
Figs.~\ref{fig:Fig-LRSM-1}~(c), (d),
 the LVN source is  the operator in the upper vertex and $m_{\nu}$ is not needed to allow for the $\Delta L = 2$ process to proceed.
These are the so called $\hat{q}$-type contributions.

%%%%%%%%%%%%%%%%%%%%%%%%%%%%%%%%%%%%%%%%%%%%%%%%%%%%%%%%%%%%%%%%%%%%%%%%%%%%
%%%%%%%%%%%%%%%%%%%%%%%%%%%%%%%%%%%%%%%%%%%%%%%%%%%%%%%%%%%%%%%%%%%%%%%%%%%%
%%%%%%%%%%%%%%%%%%%%%%%%%%%%%%%%%%%%%%%%%%%%%%%%%%%%%%%%%%%%%%%%%%%%%%%%%%%%
%%%%%%%%%%%%%%%%%%%%%%%%%%%%%%%%%%%%%%%%%%%%%%%%%%%%%%%%%%%%%%%%%%%%%%%%%%%%
%%%%%%%%%%%%%%%%%%%%%%%%%%%%%%%%%%%%%%%%%%%%%%%%%%%%%%%%%%%%%%%%%%%%%%%%%%%%
%%%%%%%%%%%%%%%%%%%%%%%%%%%%%%%%%%%%%%%%%%%%%%%%%%%%%%%%%%%%%%%%%%%%%%%%%%%%

The Wilson coefficients $C^{K}_{n}$ in Eqs.~(\ref{eq:Dim-9}) and (\ref{eq:ECECRate-1}) at the matching high-energy scale $\Lambda \sim M_{R}$ corresponding to the diagrams in Figs.~\ref{fig:Fig-LRSM-1}(b)-(e) are given by
\begin{widetext}
\begin{eqnarray}\label{eq:WC-WR-1}
\mbox{Fig. \ref{fig:Fig-LRSM-1}~(b)} \sim y_{R} \langle \Delta_{R}\rangle
&\rightarrow& C^{RR, N_{R}}_{3} \ = \sum_{i=1}^{3} U_{L\, ei}^{2} \frac{m_{p}}{m_{N_{i}}}
\left(\frac{M_{W_{L}}}{M_{W_{R}} }\right)^{4}
%\sim  y_{R} \langle \Delta_{R}\rangle,
\\
\mbox{Fig. \ref{fig:Fig-LRSM-1}~(c)} \sim  y_{\Phi} y_{R}\langle\Phi\rangle \langle \Delta_{R}\rangle &\rightarrow&
C^{R, \hat{q}}_{3}=
\sum_{i=1}^{3} U_{L\, ei} U_{D, ei}
\left(\frac{M_{W_{L}}}{M_{W_{R}} }\right)^{2},\\
%\sim  y_{R} \langle \Delta_{R}\rangle
%\frac{m_{N_{i}}m_{p}}{m^{2}_{\Delta_{R}}}}
%\left(\frac{M_{W_{L}}}{M_{W_{R}} }\right)^{4}
%\sim  y_{R} \langle \Delta_{R}\rangle,
\mbox{Fig. \ref{fig:Fig-LRSM-1}~(d)} \sim  \zeta_{W} y_{\Phi} y_{R}\langle\Phi\rangle \langle \Delta_{R}\rangle &\rightarrow&
C^{L, \hat{q}}_{3} =
\sum_{i=1}^{3} U_{L\, ei} U_{D, ei}
\tan\zeta_{W}, \\
\mbox{Fig. \ref{fig:Fig-LRSM-1}~(e)} \sim  \lambda_{\Delta_{R}} g_{R}^{2} \langle \Delta_{R}\rangle &\rightarrow&
C^{RR, \Delta^{++}_{R}}_{3} =
\sum_{i=1}^{3} U_{R\, ei}^{2}
\frac{m_{N_{i}}m_{p}}{m^{2}_{\Delta_{R}}}
\left(\frac{M_{W_{L}}}{M_{W_{R}} }\right)^{4}.
\end{eqnarray}
\end{widetext}
Here, $\zeta_{W}$ is the angle of  $W_{L}-W_{R}$-mixing.
%:The angle of  $W_{L}-W_{R}$-mixing $\zeta$ is given by $\tan\zeta = M_{W_{L}}/M_{W_{R}}$.
For convenience we showed the correspondence of the flavor-basis diagrams in Fig.~\ref{fig:Fig-LRSM-1} to the particular combinations of  the parameters of  the  LRSM Lagrangian -- quartic $\Delta_{R}$ coupling $\lambda_{\Delta_{R}}$, gauge coupling $g_{R}$ and VEV $\langle \Delta_{R}\rangle $ -- and, then, give the corresponding Wilson coefficients.
%
%A representative case is the so called $\lambda$-contribution within the Left-Right symmetric models. The latter class of models also generate the short-range contributions represented by the diagram in
%Fig.~\ref{fig:mechanisms-1}(d) via the heavy sterile neutrino $N$ exchange, like in the diagram Fig.~\ref{fig:mechanisms-1}(b) but with $\nu$ and $W_{L}$ replaced by $N$ and $W_{R}$, respectively.
%\\[3mm]
%

\begin{figure}
\includegraphics[angle = 270,width=0.50\textwidth]{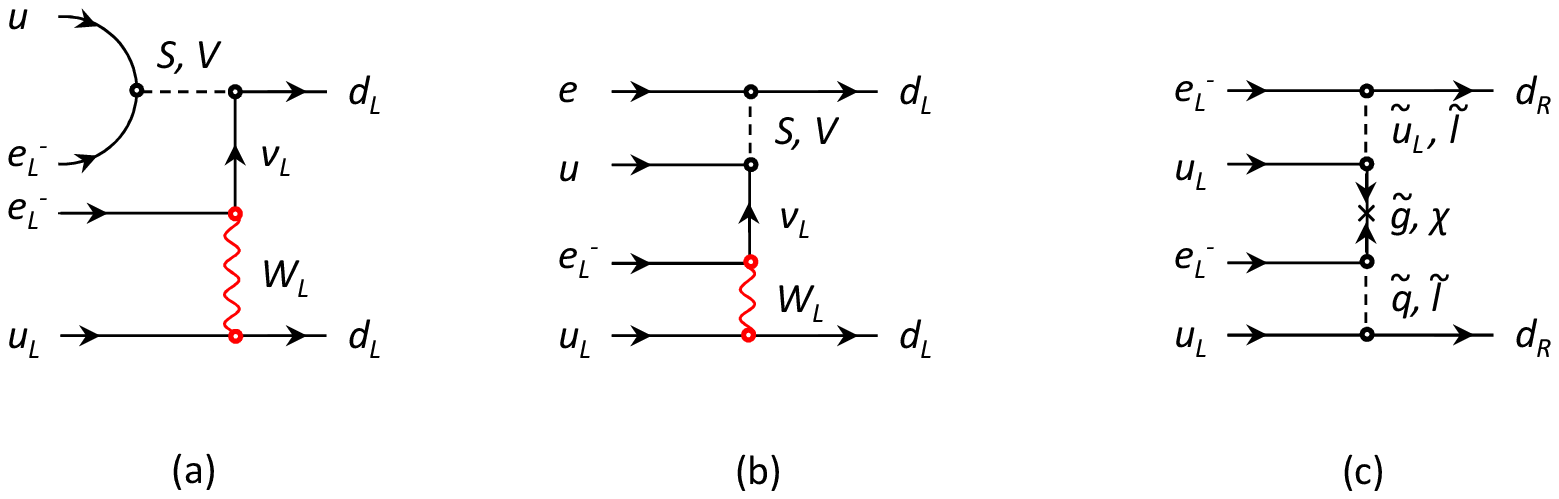}
\caption[fig]{\label{fig:Fig-LQ-1} Possible flavor-basis contributions to $0\nu2$EC within Leptoquark
(a,b) and RPV MSSM (a,c) models. The cross $\times$ in diagram (c) denotes the Majorana mass of a gluino $\tilde{g}$ or neutralino $\chi$.}
\end{figure}

$\bullet$ {\it Leptoquark models:}
Leptoquarks (LQ) are exotic scalar or vector particles coupled to lepton-quark pairs in such a way $\bar{L}\cdot LQ \cdot Q$.
They appear in various high-scale contexts,
for example, in Grand Unification, extended Technicolor, Compositeness etc.
For a generic LQ theory all the renormalizable interactions were specified by \textcite{BRW:1987}.
Current experimental limits \cite{PDG}
allow them to be relatively light at the TeV scale. The SM gauge symmetry allows LQ to mix with the SM Higgs.
This mixing generates $\Delta L=2$ interactions with the chiral structure leading to the long-range $\hat{q}$-type contribution not
suppressed by the smallness of the Majorana mass $m_{\nu}$ of the light virtual neutrino displayed in the diagram of Fig.~\ref{fig:Fig-LQ-1}(a) with $S$ or $V$ being scalar or vector LQ. In the low-energy limit, the upper part of this diagram with heavy LQ
reduces to the point-like vertex described by the operator $\mathcal{O}_{1}^{(6) X}$ in (\ref{eq:LROperBasis-1}).
The chirality structure of this vertex combined with the SM vertex in the bottom part  render $\hat{q}$-type contribution to $0\nu2$EC.

$\bullet$ {\it R-parity violating  supersymmetric models:}
The TeV-scale supersymmetric (SUSY) models offer a natural explanation of the GUT-SM scale hierarchy, introducing superpartners to each SM particle, so that they form supermultiplets (chiral superfields):
$(q, \tilde{q})$, $(l, \tilde{l})$,
$(g,\tilde{g})$  etc. Here $\tilde{q}$ and $\tilde{l}$ are scalar squarks and sleptons while $\tilde{g}$ is a spin-1/2 gluino.  The SUSY framework requires at least two electroweak Higgs doublets $H_{U}$ and $H_{D}$.
A class of SUSY models, the so called $R$-parity violating (RPV) SUSY models, allow for LNV interactions described by a superpotential
\begin{eqnarray}\label{eq:Superpotential-1}
W_{RPV} &=& \lambda'_{ijk} L_{i}Q_{j} \bar{D}_{k} + \epsilon_{i} L_{i} H_{U},
\end{eqnarray}
where $Q$,  $D$ and $H_{U}$ conventionally denote here
the chiral superfields of the left-chiral electroweak doublet quarks, the right-chiral electroweak singlet down quark and the up-type electroweak  Higgs doublet, respectively.

The RPV SUSY models with the interactions  (\ref{eq:Superpotential-1}) contribute to the long- and short range mechanisms of $0\nu2$EC process.
%, analogously to $0\nu2\beta$ decay.
The corresponding diagrams are shown in Figs.~\ref{fig:Fig-LQ-1}.
{The first two diagrams generate the long-range $\hat{q}$-type contribution, while the last one with the gluino $\tilde{g}$ or
neutralino $\chi$ exchange entail the short-range contribution. In  the diagrams of Fig.~\ref{fig:Fig-LQ-1}(a,b) the source of LNV is located
in the vertices of the upper part, while in Fig.~\ref{fig:Fig-LQ-1}(c) it is given by the Majorana mass of the neutralino $m_{\chi}$ and/or
gluino $m_{\tilde{g}}$.
In the low-energy limit the upper part of the diagrams in Fig.~\ref{fig:Fig-LQ-1}(a,b) lead} to the operator
$\mathcal{O}_{1}^{(6) L}$ while in this limit the  diagram in Fig.~\ref{fig:Fig-LQ-1}(c),
where all internal particles are heavy,
collapses to a point-like short-range contribution given by a linear combination of d = 9
operators $\mathcal{O}_{1}^{(9) LL}$ and $\mathcal{O}_{2}^{(9) LL}$.

\subsection{Hadronization of quark-level interactions}
\label{sec:Hadronization}
\begin{figure}
\includegraphics[angle = 270,width=0.50\textwidth]{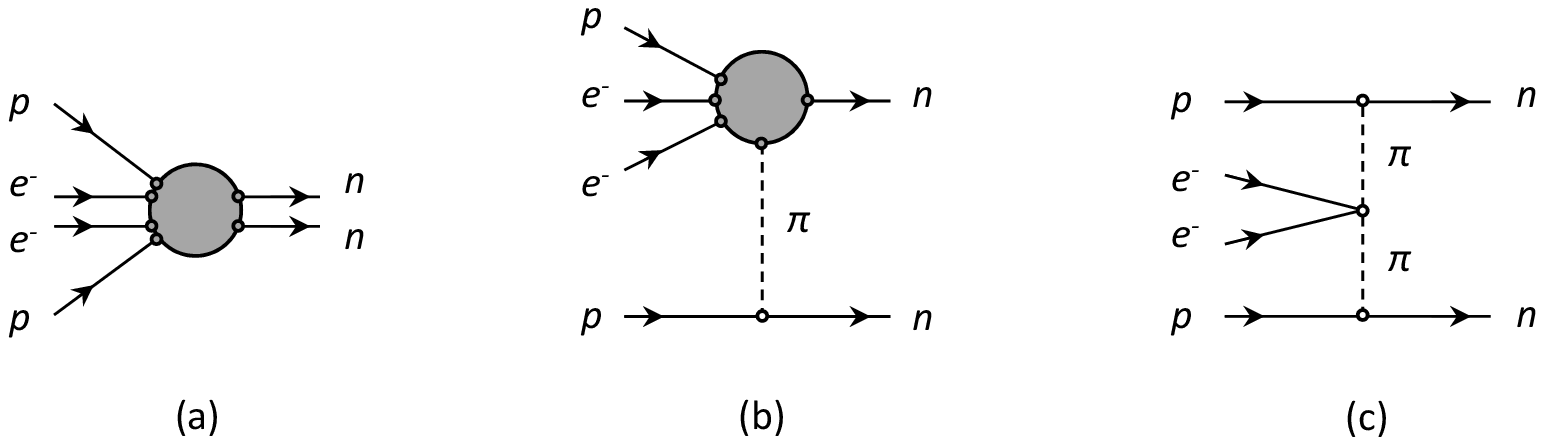}
\caption[fig]{\label{fig:Hadronization} Hadron-level diagrams for $0\nu2$EC: (a) the conventional two-nucleon mechanism, (b) the one-pion exchange and   (c) the two-pion exchange mechanisms.}
\end{figure}

Let us comment on the calculation of the structure coefficients $\beta^{X}_{i}, \beta^{XY}_{i}$ in the amplitude (\ref{eq:ECECRate-1}) depending on the NMEs and nucleon structure.
%%%%%%%%%%%%%%%%%%%%%%%%%%%%%%%%%%%%%%%%%%%%%%%%%%%%%%%%%%%%%%%%%%%%%%%%%%%%
%%%%%%%%%%%%%%%%%%%%%%%%%%%%%%%%%%%%%%%%%%%%%%%%%%%%%%%%%%%%%%%%%%%%%%%%%%%%
%%%%%%%%%%%%%%%%%%%%%%%%%%%%%%%%%%%%%%%%%%%%%%%%%%%%%%%%%%%%%%%%%%%%%%%%%%%%
%%%%%%%%%%%%%%%%%%%%%%%%%%%%%%%%%%%%%%%%%%%%%%%%%%%%%%%%%%%%%%%%%%%%%%%%%%%%
%%%%%%%%%%%%%%%%%%%%%%%%%%%%%%%%%%%%%%%%%%%%%%%%%%%%%%%%%%%%%%%%%%%%%%%%%%%%
%%%%%%%%%%%%%%%%%%%%%%%%%%%%%%%%%%%%%%%%%%%%%%%%%%%%%%%%%%%%%%%%%%%%%%%%%%%%
The Lagrangians (\ref{eq:Dim-6}) and (\ref{eq:Dim-9}) can, in principle,  be applied to any LNV processes with whatever hadronic states: quarks, mesons, nucleons, other baryons,
as well as nuclei. The corresponding amplitude, such as in Eq.~(\ref{eq:ECECRate-1}),  involves
the hadronic matrix elements of the operators (\ref{eq:SROperBasis-1}) and (\ref{eq:LROperBasis-3}). The Wilson coefficients $C^{XY}_{i}$ are calculated in terms of the parameters of the high-scale model and are independent of the low-energy scale non-perturbative hadronic dynamics. This is the celebrated property of the operator product expansion, expressing interactions of some high-scale
renormalizable model in the form of Eqs.~(\ref{eq:Dim-6}) and (\ref{eq:Dim-9}) below a certain scale $\mu$.
In the case of  $0\nu2\beta^-$ decay, $0\nu$2EC and other similar nuclear processes, the corresponding NMEs of the operators (\ref{eq:SROperBasis-1}) and (\ref{eq:LROperBasis-3}) are %used for calculations
calculated in the framework of
%using
the approach based on non-relativistic impulse approximation
(for a detailed description see, e.g., the paper of \textcite{Takasugi1985}).  This implies, as the first step, reformulating the quark-level theory
%, initially formulated  with the quark states,
in terms of the nucleon degrees of freedom, which the existing nuclear structure models operates with.  This is the so-called hadronization procedure.
%
%As we already mentioned the underlying quark-level effective interactions (\ref{eq:Dim-6}) and (\ref{eq:Dim-9}) must be properly hadronized, to wit,
%the quark fields are replaced with the corresponding interpolating fields of nucleons or pions.
In the absence of  firm theory of hadronization one is left to resort upon general principles and particular models. Imbedding two initial(final) quarks into two different protons (neutrons) is conceptually a more simple option illustrated in Fig.~\ref{fig:Hadronization}(a). This is the conventional two-nucleon mechanism relying on the nucleon form factors as a phenomenological representation of  the nucleon structure. On the other hand, putting one initial and one final quark into a charged pion while the other initial quark is put into a proton and one final quark into a neutron, as in Fig.~\ref{fig:Hadronization}(b),  we deal with 1-pion mechanism. The 2-pion mechanism, displayed in Fig.~\ref{fig:Hadronization}~(c), treats all the quarks to be incorporated in two charged pions. In both cases the pions are virtual and interact with nucleons via the ordinary pseudoscalar pion-nucleon coupling
$\bar{N}i\gamma_{5} \mbox{\boldmath$\tau$} \mbox{\boldmath$\pi$}  N$. One may expect a priori dominance of the pion mechanism for the reason that it extends the region of the nucleon-nucleon interaction due to the smallness of the pion mass leading to a long-range potential. As a result, the suppression caused by the short-range nuclear correlation can be significantly alleviated in comparison to the conventional two-nucleon mechanism.
Nevertheless, one should consider all these mechanisms to contribute to the process in question with corresponding relative amplitudes. The latter is  as yet unknown. In principle, it can be evaluated in particular hadronic models. These kind of studies are still missing in literature and consensus on the dominance of one of these two mechanisms is pending.
It is a common trend to posit the analysis on one of these two hadronization scenarios. Note that, for the long-range  contributions described by the effective Lagrangian (\ref{eq:Dim-6}) the above-mentioned advantage of the pion mechanism is absent, and one can, in a sense, safely resort to the conventional two nucleon mechanism.
The light Majorana exchange contribution to $0\nu2$EC, on which we focus in the rest of this review,  is of this kind.  {\textcolor{black}This limitation is explained by the fact that for the moment there are no yet Nuclear Matrix Elements calculated in the literature for other mechanisms different from this one. }

The procedure of hadronization
%based on Eqs.~(\ref{eq:2Nucl-matching-1})-(\ref{eq:2pion-matching-1})
is essentially the same as for $0\nu\beta\beta$-decay and described in the literature. For more details on this approach to hadronization we refer readers to the original papers of \textcite{Takasugi1985,Faessler:1998qv,Faessler:2007nz} and the recent review by \textcite{Graf:2018ozy}.

{\textcolor{black} Recently there has been developed another approach, which resorts to matching the high-scale quark-level theory to Chiral Perturbation Theory. The latter is believed to provide a low-energy description of QCD in terms of nucleon and pion degrees of freedom. It is expected that the parameters of the low-energy effective theory can be determined from experimental measurements or from the lattice QCD.  This approach leads to quite different picture of hadronization and numerical results in comparison with the conventional approach sketched above.
Surprisingly,  contrary to the conventional approach  short-range nucleon-nucleon interactions
%of the type (\ref{eq:2N-Lagrangian-1})
should be introduced for theoretical self-consistency even in the case of the long-range light neutrino exchange mechanism in Fig.~\ref{fig:mechanisms-1}(a).
For the detailed description of this approach we refer the reader to the original papers of
\textcite{Prezeau:2003xn,Graesser:2016bpz,Cirigliano:2019vdj,Cirigliano:2017ymo,Cirigliano:2018hja,Cirigliano:2017,Cirigliano:2018} and the recent review of \textcite{Cirigliano:2020yhp}.
}

%%%%%%%%%%%%%%%%%%%%%%%%%%%%%%%%%%%%%%%%%%%%%%%%%%%%%%%%%%%%%%%%%%%%%%%%%%%%
%%%%%%%%%%%%%%%%%%%%%%%%%%%%%%%%%%%%%%%%%%%%%%%%%%%%%%%%%%%%%%%%%%%%%%%%%%%%
%%%%%%%%%%%%%%%%%%%%%%%%%%%%%%%%%%%%%%%%%%%%%%%%%%%%%%%%%%%%%%%%%%%%%%%%%%%%
%%%%%%%%%%%%%%%%%%%%%%%%%%%%%%%%%%%%%%%%%%%%%%%%%%%%%%%%%%%%%%%%%%%%%%%%%%%%
%%%%%%%%%%%%%%%%%%%%%%%%%%%%%%%%%%%%%%%%%%%%%%%%%%%%%%%%%%%%%%%%%%%%%%%%%%%%

%
%\subsection{Concluding remarks}
%
To conclude, neutrinoless double-electron capture $0\nu$2EC, the same as $0\nu\beta\beta$-decay, is a $\Delta L =2$ lepton number violating process. Moreover, at the level of nucleon sub-process it is virtually equal to $0\nu\beta\beta$-decay.  Consequently, the underlying $\Delta L =2$ physics driving both these processes is the same. Obviously there are many formal differences in the form of the effective operators representing this physics at low energy sub-GeV scales.  We specified a complete basis of the  $0\nu$2EC effective operators in Eqs.~(\ref{eq:Dim-6})-(\ref{eq:SROperBasis-5}) and exemplified high-energy scale models presently popular in the literature, which can be reduced to these operators in the low-energy limit. Akin to  $0\nu\beta\beta$-decay there are basically three types of mechanisms of  $0\nu$2EC shown in Fig.~\ref{fig:mechanisms-1}: (a) the conventional neutrino exchange mechanism with the amplitude proportional to the effective Majorana neutrino  mass $m_{\beta\beta}$ defined in (\ref{eq.MNM-1}); (b) neutrino exchange mechanism  independent of Majorana neutrino mass, when lepton number violation necessary for
$0\nu$2EC to proceed is gained from a
$\Delta L =2$ vertex; Both (a) and (b) are long-range mechanics induced by the exchange of a very light particle, a neutrino. On the other hand the diagram (d) represents a short-range mechanism induced by the exchange of heavy particles with masses much larger than the typical scale ($\sim$ few MeV) of  $0\nu$2EC.
Despite the underlying physics of both  $0\nu$2EC and $0\nu\beta\beta$-decay is the same, their nuclear matrix elements (NME) are very different. We will discuss the nuclear structure aspects and atomic physics involved  in the calculations of the $0\nu$2EC NMEs in the subsequent sections.
Here it is worth noting that so far  only the NMEs for the Majorana neutrino exchange mechanism Fig.~\ref{fig:mechanisms-1}(a) have been calculated in the literature. The similar calculations for NMEs of other mechanisms Fig.~\ref{fig:mechanisms-1}(b,c)  are still pending.

%%%%%%%%%%%%%%%%%%%%%%%%%%%%%%%%%%%%%%%%%%%%%%%%%%%%%%%%%%%
%%%%%%%%%%%%%%%%%%%%%%%%%%%%%%%%%%%%%%%%%%%%%%%%%%%%%%%%%%%
%%%%%%%%%%%%%%%%%%%%%%%%%%%%%%%%%%%%%%%%%%%%%%%%%%%%%%%%%%%

%%%%%%%%%%%%%%%%%%%%%%%%%%%%%%%%%%%%%%%%%%%%%%%%%%%%%%%%%%%%%%%%%%%%%%%%%%%%
\section{Phenomenology of neutrinoless 2EC}% half-lives}
%%%%%%%%%%%%%%%%%%%%%%%%%%%%%%%%%%%%%%%%%%%%%%%%%%%%%%%%%%%%%%%%%%%%%%%%%%%%
\setcounter{equation}{0}

\begin{figure}[t]
\begin{center}
\includegraphics[angle = 270,width=0.209\textwidth]{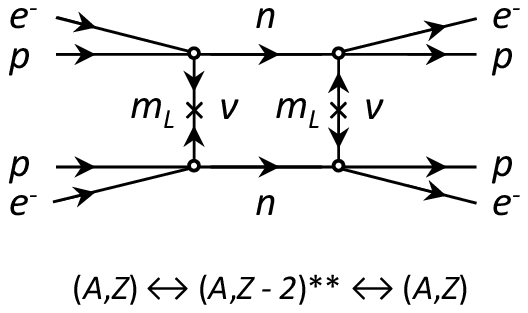}
\caption{
\textcolor{black}{Oscillations of atoms induced by $0\nu2$EC and $0\nu2\beta$ transitions.
The notations are the same as in Figs.~\ref{fig2:1} and \ref{fig2:2}.
The intermediate atom $(A,Z-2)^{**}$ is coupled to the continuum
through the emission of a photon and/or Auger electrons. These channels generate
a finite width $\Gamma_f$ in Eq.~(\ref{HAMI}),
they are also responsible for the non-Hermitian character of $H_{\mathrm{eff}}$.}
}
\label{fig4:12}
\end{center}
\end{figure}

The diagrams in Figs.~\ref{fig2:1} and \ref{fig2:2} can be combined as shown in Fig. \ref{fig4:12}.
In the initial state, there is an atom $(A,Z)$. The electron lines belong to the electron shells,
and the proton and neutron lines belong to the initial and intermediate nuclei, respectively. As a result of neutrinoless double-electron capture,
an atom $(A, Z-2)^{**}$ is formed, generally in an excited state.
\textcolor{black}{
In what follows, $(A, Z)^*$ denotes an atom with the excited electron shell,
and $(A, Z)^{**}$ means that the nucleus is also excited.
The intermediate atom can decay by emitting a photon or Auger electrons,
but it can also experience $0\nu2\beta^-$ transition and
evolve back to the initial state. As a result, LNV oscillations $(A,Z) \leftrightarrow (A, Z-2)^{**}$ occur in the two-level system.
These oscillations are affected by the coupling of the $(A, Z-2)^{**}$ atom to the continuum,
which eventually leads to the decay of $(A, Z)$.
The Hamiltonian of the system is not Hermitian because $(A, Z-2)^{**}$ has a finite width. }

The LNV oscillations of atoms are discussed by \textcite{KRIV11,Simkovic:2009zz,Bernabeu:2017ape}.
The formalism of LNV oscillations allows to find a relationship between
the half-life $T_{1/2}$ of the initial atom $(A,Z)$,
the amplitude of neutrinoless double-electron capture $V_{\alpha\beta}$,
and the decay width of the intermediate atom $(A,Z-2)^{**}$, which has an electromagnetic origin.

%%%%%%%%%%%%%%%%%%%%%%%%%%%%%%%%%%%%%%%%%%%%%%%%%%%%%%%%%%%%%%%%%%%%%%%%%%%%
\subsection{Underlying formalism}
%%%%%%%%%%%%%%%%%%%%%%%%%%%%%%%%%%%%%%%%%%%%%%%%%%%%%%%%%%%%%%%%%%%%%%%%%%%%

\textcolor{black}{
The evolution of a system of mixed states, each of which may be unstable due to the coupling with the continuum,
can described by an effective non-Hermitian Hamiltonian \cite{Weisskopf:1930}. In the case under consideration,
the Hamiltonian takes the form}
\begin{eqnarray}
H_{\mathrm{eff}} &=& \left(
\begin{array}{cc}
M_i & V_{\alpha\beta} \\
V^*_{\alpha\beta} & M_f - \frac{i}{2}\Gamma_f
\end{array}
\right),  \label{HAMI}
\end{eqnarray}
where $M_i$ and $M_f$ are the masses of the initial and final atoms.
The width $\Gamma_f $ of the final excited atom with two vacancies $\alpha$ and $\beta$
is of electromagnetic origin.
The off-diagonal matrix elements are due to a violation of lepton number conservation.
They can be chosen real by changing the phase of one of the states;
thus, we set $ V_{\alpha\beta} = V^*_{\alpha\beta} $.
\textcolor{black}{The real and imaginary parts of the Hamiltonian do not commute.}

Let us find the evolution operator
\begin{equation}
U(t) = \exp(-i H_{\mathrm{eff}} t).
\end{equation}
According to Sylvester's theorem, the function of a finite-dimensional $ n \times n $ matrix $ A $
is expressed in terms of the eigenvalues $ \lambda_k $ of the matrix $ A $, which are solutions of the characteristic equation
$\det(A - \lambda) = 0$, and a polynomial of $ A $:
\begin{equation}
f(-iAt) = \sum_{k} f(-i\lambda_k t) \prod_{l\neq k}\frac{\lambda_l - A}{\lambda_l - \lambda_k},
\end{equation}
where the sum runs over $1 \leq k \leq n$, the product runs over $1 \leq l \leq n$, $l \neq k$,
and the eigenvalues are assumed to be pairwise distinct.
The matrix function $f(-iAt)$ evolves with the time $t$ like the superposition of $n$ terms $f(-i\lambda_k t) $
with the matrix coefficients which are projection operators onto the $k$-th eigenstates of $A$.

The eigenvalues of the Hamiltonian (\ref{HAMI}) are equal to
$
\lambda_{\pm} = M_{+} \pm \Omega,
$
where $M_{\pm} = (M_i \pm M_f)/2 \mp i\Gamma_f/4$, and $\Omega = \sqrt{V^2_{\alpha\beta} + M_-^2}$.
The values of $\lambda_{\pm}$ are complex,
so the norm of the states is not preserved in time. A series expansion around $V=0$ yields
\begin{eqnarray}
\lambda_{+} &\approx& M_i + \Delta M - \frac{i}{2}     \Gamma_i ,  \label{LP} \\
\lambda_{-} &\approx& M_f - \Delta M - \frac{i}{2} (\Gamma_f - \Gamma_i), \label{LM}
\end{eqnarray}
with $\Delta M = (M_i - M_f)\Gamma_i/{\Gamma_f}$, $\Gamma_i = V^2_{\alpha\beta} R_f$, and
\begin{eqnarray}
R_f = \frac{\Gamma_f }{(M_i - M_f)^2 + \frac{1}{4} \Gamma_f^2}. \label{DF}
\label{G1}
\end{eqnarray}
The initial state decays at the rate
$\Gamma_{i} \ll \Gamma_f$. The width $\Gamma_i$ is maximal for complete degeneracy of the atomic masses:
\begin{equation}
\Gamma_i^{\max} = \frac{4V^2_{\alpha\beta}}{\Gamma_f}.
\end{equation}

A simple calculation gives
\begin{eqnarray}
U(t) = \exp(-iM_{+}t) \left( \cos(\Omega t) - i\frac{H_{\mathrm{eff}} - M_{+}}{\Omega}\sin(\Omega t) \right).
\label{EVOLMATR}
\end{eqnarray}

The decay widths of single-hole excitations of atoms are known experimentally
and tabulated for $10 \leq Z \leq 92$ and principal quantum numbers $1\leq n \leq 4$ by \textcite{CAMP01}.
The width of a two-hole state $\alpha\beta$ is represented by the sum of the widths of the one-hole states
$\Gamma_f =  \Gamma_{\alpha}+ \Gamma_{\beta}$.
The de-excitation width of the daughter nucleus is much smaller and can be neglected.
The values $\Gamma_f$ are used in estimating the decay rates $\Gamma_i$.

The transition amplitude from the initial to the final state
for small time $t$, according to Eq.~(\ref{EVOLMATR}), is equal to
\begin{equation} \label{transition}
\langle f|U(t)|i\rangle = -iV_{\alpha\beta}t +...
\end{equation}
This equation is valid for $t \lesssim 1/|M_{+}|$ and also over a wider range $t \lesssim 1/|\Omega|$,
given that the real part of the phase can be made to vanish via redefinition of the Hamiltonian $H_{\mathrm{eff}} \to H_{\mathrm{eff}} - \Re(M_{+})$.
The value of $V_{\alpha\beta}$ can be evaluated by means of field-theoretical methods
that allow one to find the amplitude (\ref{transition}) from first principles.
\textcolor{black}{Formalism described in this subsection reproduces
results of \textcite{BERN83} with respect to $0\nu$2EC decay rates.}

%%%%%%%%%%%%%%%%%%%%%%%%%%%%%%%%%%%%%%%%%%%%%%%%%%%%%%%%%%%%%%%%%%%%%%%%%%%%
%%%%%%%%%%%%%%%%%%%%%%%%%%%%%%%%%%%%%%%%%%%%%%%%%%%%%%%%%%%%%%%%%%%%%%%%%%%%
%%%%%%%%%%%%%%%%%%%%%%%%%%%%%%%%%%%%%%%%%%%%%%%%%%%%%%%%%%%%%%%%%%%%%%%%%%%%
\vspace{-3mm}
%%%%%%%%%%%%%%%%%%%%%%%%%%%%%%%%%%%%%%%%%%%%%%%%%%%%%%%%%%%%%%%%%%%%%%%%%%%%
%%%%%%%%%%%%%%%%%%%%%%%%%%%%%%%%%%%%%%%%%%%%%%%%%%%%%%%%%%%%%%%%%%%%%%%%%%%%
%%%%%%%%%%%%%%%%%%%%%%%%%%%%%%%%%%%%%%%%%%%%%%%%%%%%%%%%%%%%%%%%%%%%%%%%%%%%

%%%%%%%%%%%%%%%%%%%%%%%%%%%%%%%%%%%%%%%%%%%%%%%%%%%%%%%%%%%%%%%%%%%%%%%%%%%
%\subsection{Decay amplitude in Majorana neutrino exchange mechanism}
\subsection{Decay amplitude of the light Majorana neutrino exchange mechanism}
%%%%%%%%%%%%%%%%%%%%%%%%%%%%%%%%%%%%%%%%%%%%%%%%%%%%%%%%%%%%%%%%%%%%%%%%%%%%

The total lepton number violation is due to the Majorana masses of the neutrinos. It is
assumed that the left electron neutrino is a superposition of three left
Majorana neutrinos:
\begin{equation}
\nu_{e L} = \sum^3_{i=1} U_{e k} \chi_{k L},
\end{equation}
where $U$ is the PMNS mixing matrix. In the Majorana bispinor representation,
$\chi_{k L} = \frac{1}{2}(1 - \gamma_5)\chi_{k}$ and $\chi_{k}^{*} =\chi_{k}.$
The vertex describing the creation and annihilation of a neutrino has the standard form
\begin{equation}
{\cal H}(x)=\frac{G_F \cos{\theta_C}}{\sqrt{2}} j_{\mu}(x) J^{\mu}_h(x) + \mathrm{h. c.},
\end{equation}
where ${\theta_C}$ is the Cabibbo angle. The lepton and quark charged currents
are defined by Eq.~(\ref{eq:V-A}). In terms of the composite fields,
the hadron charged current is given by
\begin{equation}
J^{\mu}_h (x)=\bar{n}(x) \gamma^{\mu}  (g_{V} - g_{A} \gamma_{5}) p(x),
\end{equation}
where $n(x)$ and $p(x)$ are the neutron and the proton field operators and
$g_{{V}}=1$ and $g_{{A}}=1.27$ are the vector and axial-vector coupling constants, respectively.
An effective theory could also include $\Delta$-isobars, meson fields and their vertices
for decaying into lepton pairs and interacting with nucleons and each other.

%%%%%%%%%%%%%%%%%%%%%% wave function of electrons

As a result of the capture of electrons, the nucleus $(A,Z)$ undergoes a $0^{+} \to J^{\pi}$ transition.
Conservation of total angular momentum requires that the captured electron pair be in the state $J$.
In weak interactions, parity is not conserved; thus, it is not required that the parity of the electron pair
be correlated with the parity of the daughter nucleus.

The wave function of a relativistic electron in a central potential has the form
\begin{equation}
\Psi _{\alpha m_\alpha}(\mathbf{r})= \left(
\begin{array}{l}
 f_{\alpha }(r) \Omega_{\alpha m_{\alpha}}  (\mathbf{n})\\
i g_{\alpha}(r)  \Omega_{\alpha^{\prime} m_{\alpha}}  (\mathbf{n})
\end{array}
\right),  \label{DIWF}
\end{equation}
where $\alpha =(njl)$, $\alpha ^{\prime }=(njl^{\prime }),$ $l^{\prime}=2j-l$.
The radial wave functions are defined in agreement with \textcite{BERE};
$\Omega_{\alpha m_{\alpha}}  (\mathbf{n}) \equiv
 \Omega_{j m_{\alpha}}^{l}(\mathbf{n})$,
$\Omega_{\alpha^{\prime} m_{\alpha}}  (\mathbf{n})=
 \Omega_{j m_{\alpha}}^{l^{\prime}}(\mathbf{n})$
are spherical spinors in the notations of \textcite{Varshalovich1988}.
The normalization condition for $\Psi _{\alpha m_\alpha}(\mathbf{r})$ is given by
\begin{equation} \label{normWF}
\int d\mathbf{r}\Psi _{\alpha m_{\alpha}}^{\dagger}(\mathbf{r})\Psi _{\beta m_{\beta}}(\mathbf{r}) = \delta_{\alpha\beta}\delta_{m_{\alpha}m_{\beta}}.
\end{equation}
If the captured electrons occupy the states $\alpha \equiv (n 2j l)_1$ and $\beta \equiv (n 2j l)_2$,
we must take the superposition of products of their wave functions:
\begin{equation} \label{WF2}
\psi _{\alpha \beta }^{JM}(\mathbf{r}_{1},\mathbf{r}_{2})= \sum_{m_{\alpha}
m_{\beta }}C_{j_{\alpha }m_{\alpha }j_{\beta }m_{\beta }}^{JM} \Psi _{\alpha
m_{\alpha }}(\mathbf{r}_{1})\Psi _{\beta m_{\beta }}(\mathbf{r}_{2}),
\end{equation}
where $j_{\alpha }$ and $j_{\beta}$ are the total angular momenta,
$m_\alpha$ and $m_\beta$ are their projections on the direction of the $z$ axis,
and $\Psi _{\alpha m_{\alpha}}(\mathbf{r})$ and $\Psi _{\beta m_{\beta }}(\mathbf{r})$
are the relativistic wave functions of the bound electrons
in an electrostatic mean field of the nucleus and the surrounding electrons.
The identity of the fermions implies that the wave function of two fermions is antisymmetric;
thus, the final expression for the wave function takes the form
\begin{equation}
\Psi _{\alpha \beta }^{JM}(\mathbf{r}_{1},\mathbf{r}_{2})= \mathcal{N}_{\alpha \beta }
(\psi _{\alpha \beta }^{JM}(\mathbf{r}_{1}, \mathbf{r}_{2})
-(-)^{j_{\alpha}+j_{\beta }-J}\psi _{\beta \alpha }^{JM}(\mathbf{r}_{1},\mathbf{r}_{2})), \label{TWOE}
\end{equation}
where $\mathcal{N}_{\alpha \beta }$ equals $ {1}/{\sqrt{2}}$ for $\alpha \neq \beta$ and ${1}/2$ for $\alpha = \beta$.

As a consequence of the identity $C_{jm_{1}jm_{2}}^{JM}=(-1)^{J-2j}C_{jm_{2}jm_{1}}^{JM}$,
the wave function of two electrons with equal quantum numbers $ \alpha = \beta $ is symmetric under the permutation
$ m_{\alpha} \leftrightarrow m_{\beta} $ provided their angular momenta are combined to the total angular momentum $J = 2j$~mod(2).
In such a case, the antisymmetrization (\ref{TWOE}) yields zero, which means that the states $J = 2j$~mod(2) are nonexistent.
The antisymmetrization (\ref{TWOE}) of the states $J = 2j + 1$~mod(2) leads to a doubling of the initial wave function.
To keep the norm, the additional factor $1/\sqrt{2}$ is thus required for $\alpha = \beta$.

%%%%%%%%%%%%%%%%%%%%%% V from QFT

The derivation of the equation for $V_{\alpha\beta}$ is analogous to the corresponding derivation of the $0\nu2\beta$ decay amplitude, as described in the review of \cite{Bilenky1987}. The specificity is that a transition from a discrete level
to a quasi-discrete level is considered. Accordingly, the delta function expressing the energy conservation
is replaced by a time interval that can be identified with the parameter $t$ in Eq.~(\ref{transition}). We thus write
\begin{equation}
\langle f|\mathcal{U}(t)|i\rangle = -iV_{\alpha\beta}t+...,
\end{equation}
where $\mathcal{U}(t)$ is Dyson's $U$-matrix. The amplitude takes the form
\begin{eqnarray}
V_{\alpha\beta} &=&  i K_Z m_{\beta\beta} \sqrt{2} \mathcal{N}_{\alpha \beta }
                               \left(\frac{G_F \cos \theta_C}{\sqrt{2}}\right)^2
                               \int \frac{d \mathbf{q}}{(2\pi)^3} d{\mathbf{r}}_1 d{\mathbf{r}}_2
                               \frac{1}{\sqrt{2J + 1}}\sum_M \frac{e^{-i \mathbf{q}({\mathbf{r}}_1-{\mathbf{r}}_2)}}{2 q_0} \label{LNVP1}\\
&\times&
\left[
T^{JM}_{\mu\nu \alpha\beta}({\mathbf{r}}_1,{\mathbf{r}}_2)N^{\mu\nu}_{JM \alpha\beta}({\mathbf{r}}_1,{\mathbf{r}}_2) -
(-1)^{j_{\alpha} + j_{\beta} - J} T^{JM}_{\mu\nu \beta\alpha}({\mathbf{r}}_1,{\mathbf{r}}_2)N^{\mu\nu}_{JM \beta\alpha}({\mathbf{r}}_1,{\mathbf{r}}_2)
\right],
\nonumber
\end{eqnarray}
where
\begin{eqnarray*}
T^{JM}_{\mu\nu \alpha\beta}({\mathbf{r}}_1,{\mathbf{r}}_2) &=& \sum_{m_\alpha m_\beta} C^{JM}_{j_\alpha m_\alpha j_\beta m_\beta}
\left[ {\bar{\Psi}_{\alpha m_\alpha}^c} ({\mathbf{r}}_1) \gamma_\mu (1 + \gamma_5 ) \gamma_\nu {\Psi_{\beta  m_\beta }}   ({\mathbf{r}}_2) \right], \\
N^{\mu\nu}_{JM \alpha\beta}({\mathbf{r}}_1,{\mathbf{r}}_2) &=& \sum_n \left[
                       \frac{ \langle JM|J^\mu_h({\mathbf{r}}_1)|n\rangle
                              \langle n|J^\nu_h({\mathbf{r}}_2)|00 \rangle }
                              {q_0 + E_n - M_i - \varepsilon_\beta}
+                       \frac{\langle JM|J^\nu_h({\mathbf{r}}_2)|n\rangle
                              \langle n|J^\mu_h({\mathbf{r}}_1)|00 \rangle }
                              {q_0+ E_n - M_i - \varepsilon_\alpha} \right].
\end{eqnarray*}
Here, $q_0 \approx |\mathbf{q}|$, $\langle JM| = \langle f|$ and $ |00\rangle = |i \rangle$ are the states of the final and initial nuclei, respectively;
$\varepsilon_\gamma = m - \varepsilon_\gamma^*$, $\varepsilon_\gamma^*$ is the one-hole excitation energy of the initial atom.
The sum is taken over all excitations of the intermediate atom $(A,Z - 1)$.
In the Majorana bispinor representation, $\Psi_{\alpha m_\alpha}^c = \Psi_{\alpha m_\alpha}^{*}$.
The amplitude $V_{\alpha\beta}$ is a scalar under rotation. By virtue of identities
\begin{eqnarray*}
T^{JM}_{\mu\nu \alpha\beta}({\mathbf{r}}_1,{\mathbf{r}}_2) &=& - (-1)^{j_{\alpha} + j_{\beta} - J} T^{JM}_{\nu\mu \beta\alpha}({\mathbf{r}}_2,{\mathbf{r}}_1), \\
N^{\mu\nu}_{JM \alpha\beta}({\mathbf{r}}_1,{\mathbf{r}}_2) &=& N^{\nu\mu}_{JM  \beta\alpha}({\mathbf{r}}_2,{\mathbf{r}}_1),
\end{eqnarray*}
$V_{\alpha\beta}$ is also invariant under permutations of $\alpha$ and $\beta$.
For $\alpha = \beta$ and $J = 2j_{\alpha} + 1$ mod(2), the second term in the square brackets of Eq.~(\ref{LNVP1}) doubles the result,
whereas for $\alpha = \beta$ and $J = 2j_{\alpha}$ mod(2), $V_{\alpha\beta} = 0$.
The factor $\mathcal{N}_{\alpha \beta }$ provides the correct normalization.

\textcolor{black}{
In the processes associated with the electron capture, shell electrons of the parent atom appear
in a superposition of the stationary states of the daughter atom. The overlap amplitude of two atoms with atomic numbers $Z$ and $Z + \Delta Z$
can be evaluated for $\Delta Z \ll Z$ in a simple non-relativistic shell model to give \cite{Krivoruchenko:2019b}
\begin{eqnarray}
K_Z \approx  \exp\left(-{\frac {3^{5/3} 2^{1/3}}{80}} \frac {\Delta Z^2}{Z^{1/3}} \right).
\label{OTA}
\end{eqnarray}
The overlap factors for $^{96}$Ru, $^{152}$Gd and $^{190}$Pt atoms, e.g., equal
$K_Z = 0.895$, $0.906$, and $0.912$, respectively. The result is not very sensitive to the charge.
Valence-shell electrons are involved in the formation of chemical bonds and give an important contribution to $K_Z $.
We limit ourselves to estimating the core-shell electrons contribution which weakly depends on the environment.
}

The weak charged current of a nucleus for a low-energy transfer can be written in the form
\begin{equation}\label{NRIA}
J^{\mu} (0,\mathbf{r}) = \sum_{a} \tau^-_a [g_V g^{\mu 0} + g_A
(\sigma_k)_a g^{\mu k} ] \delta(\mathbf{r}-{\mathbf{r}}_a).
\end{equation}
This approximation neglects the contribution of the exchange currents.
The short-term contribution of some higher-dimensional operators is dominated by the pion exchange mechanism
(see, e.g., the paper of \textcite{Faessler:2007nz}).

The neutrino momentum enters the energy denominators of Eq. (\ref{LNVP1}). The typical value of $q_0$ is of the order of the Fermi momentum, $p_F = 270$ MeV. The remaining quantities in the energy denominators are of the order of the nucleon binding energy in the nucleus $\sim 8$ MeV, i.e., substantially lower. The energy denominators can therefore be taken out from the square bracket such that
the sum over the excited states can be performed using the completeness condition $\sum_n | n \rangle \langle n |$ = 1.
This approximation is called the closure approximation.
The integral over $\mathbf{q}$ entering Eq.~(\ref{LNVP1}) with good accuracy is inversely proportional to the distance between two nucleons.
The decay amplitude can finally be written in the form (cf. \cite{KRIV11})
\begin{eqnarray}
 V_{\alpha \beta} \approx G_F^2 \cos ^2 \theta_C K_Z
m_{\beta\beta} \frac{g^2_A}{4 \pi R} \sqrt{2 J_f + 1} M^{\mathrm{2EC}} {\cal A}_{\alpha\beta} .
\label{LNVP2}
\end{eqnarray}
Here, the electron and nuclear parts of the amplitude are assumed to factorize.
Such an approximation is well justified in the case of $K$ capture given the approximate constancy of the electron wave functions inside the nucleus.
The still-probable capture of an electron from the $p_{1/2}$ state is determined by the lower dominant component of the electron wave function inside the nucleus, which is also approximately constant. The factorization is also supported by the fact of localization of nucleons involved in the decay near the nuclear surface.

The decay amplitude due to the operators of higher dimension of Fig.~3~(b,c) has the form of Eq.~(\ref{LNVP2}) with the replacement
\begin{equation} \label{T12dgt5}
m_{\beta \beta }\frac{g_{A}^{2}}{4\pi R}\sqrt{2J_f +1}M^{\mathrm{2EC}%
}\rightarrow \sum_{i=1}^{3}\beta _{i}^{X}(\mu _{0},\Lambda
)C_{i}^{X}(\Lambda )+\sum_{i=1}^{5}\beta _{i}^{XY}(\mu _{0},\Lambda
)C_{i}^{XY}(\Lambda ).
\end{equation}
The value of $ \mathcal{A}_ {\alpha \beta}$ entering Eq.~(\ref{LNVP2}) is the product of electron wave functions, whose bispinor indices are contracted in a way depending on the type of nuclear transition and the type of operator responsible for the decay.

For neutrino exchange mechanism,
the explicit expressions for ${\cal A}_{\alpha\beta}$ of low-$J$ nuclear transitions $0^+ \to 0^{\pm},1^{\pm}$
in terms of the upper and lower radial components of the electron wave functions are given by \textcite{KRIV11}.
For $j_{\alpha} = j_{\beta} = 1/2$ and arbitrary $n_{\alpha}$ and $n_{\beta}$, one gets
\begin{eqnarray}
{\mathcal A}_{\alpha\beta}(0^+ \to 0^+)    &=&  \langle F^{(+) }_{\alpha \beta}(r_a,r_b) \rangle, \label{Aab1}\\
{\mathcal A}_{\alpha\beta}(0^+ \to 0^-)    &=&  \langle H^{(+) }_{\alpha \beta}(r_a,r_b) \rangle, \label{Aab2}\\
{\mathcal A}_{\alpha\beta}(0^+ \to 1^+)    &\approx&  \langle F^{(-)2}_{\alpha \beta}(r_a,r_b) \rangle ^{1/2}, \label{Aab3}\\
{\mathcal A}_{\alpha\beta}(0^+ \to 1^-)    &\approx&  \langle (H^{(-)}_{\alpha \beta}(r_a,r_b) - H^{(-)}_{\alpha \beta}(r_b,r_a))^2/4\rangle ^{1/2}. \label{Aab4}
\end{eqnarray}
The functions $F^{(\pm)}$ and $H^{(\pm)}$ depend on the radial variables $r_a$ and $r_b$
and quantum numbers $\alpha$ and $\beta$ of the captured electrons. For $l_{\alpha} = l_{\beta} = 0$, one finds
$4\pi F_{\alpha \beta }^{(\pm)}(r_{a},r_{b}) = \mathcal{N}_{\alpha \beta}(f_{\alpha }(r_{a})f_{\beta }(r_{b}) \pm f_{\beta }(r_{a}) f_{\alpha }(r_{b}))$
and $4\pi H^{(\pm)}_{\alpha\beta}(r_{a},r_{b}) = \mathcal{N}_{\alpha \beta}(g_{\alpha}(r_{a}) f_{\beta}(r_{b}) \pm g_{\beta}(r_{a}) f_{\alpha}(r_{b}))$.
Computation of electron radial wave functions  $f_{\alpha}(r)$ and $g_{\alpha}(r)$ is discussed in Sec.~IV.
Nuclear structure models for matrix elements $M^{\mathrm{2EC}}$ entering Eq.~(\ref{LNVP2}) are discussed in Sec.~V.

%%%%%%%%%%%%%%%%%%%%%%%%%%%%%%%%%%%%%%%%%%%%%%%%%%%%%%%%%%%%%%%%%%%%%%%%%%%%%%%%%%%%%%%%%%
\subsection{Comparison of $0\nu$2EC and $0\nu 2\beta^- $ decay half-lives}
%%%%%%%%%%%%%%%%%%%%%%%%%%%%%%%%%%%%%%%%%%%%%%%%%%%%%%%%%%%%%%%%%%%%%%%%%%%%%%%%%%%%%%%%%%

Here, we obtain estimates for half-lives of the $0\nu$2EC
and $0\nu 2\beta ^{-}$ decay, starting from the expressions of the paper of
\textcite{Suhonen:1998ck}. The inverse $0\nu 2\beta^{-}$ half-life can be written in the form
\begin{equation}
\left( T_{1/2}^{0\nu 2\beta }\right) ^{-1}=\left( \frac{|m_{\beta \beta }|}{m%
}\right) ^{2}|{M}_{2\beta }|^{2}G_{2\beta },
\end{equation}%
where ${M}_{2\beta }$ is the nuclear matrix element of the $0\nu 2\beta ^{-}$
decay, $m$ is the electron rest mass,
$G_{2\beta }=g_{\mathrm{A}}^{4} K_Z^2 g^{(0\nu )}r_{A}^{-2}\mathcal{I}\ $
is the phase-space factor, with $r_{A}=mR$, $R=1.2A^{1/3}\,\text{fm}$
being the nuclear radius.
$K_Z$ describes the overlap of the electron shells of the parent and daughter atoms including the possible ionization of the latter.
In what follows, we neglect the electron shell effects and set $K_Z = 1$.
The factor $g^{(0\nu)}=2.80\times 10^{-22}\,$y$^{-1}$
includes all the fundamental constants and other numerical coefficients entering the
half-life. The phase-space integral reads
\begin{equation}
\mathcal{I}=\int_{1}^{\tilde{Q}+1}F_{0}(Z_{f},\varepsilon
_{1})F_{0}(Z_{f},\varepsilon _{2})p_{1}p_{2}\varepsilon _{1}\varepsilon
_{2}d\varepsilon _{1},  \label{eq:phspint}
\end{equation}%
where $\varepsilon_{1,2} $ are the total energies and $p_{1,2}$ the momenta of the
emitted electrons, scaled by the electron mass. Here $\tilde{Q}=Q/m$ is the
normalized $Q$ value of the decay. The quantities $F_{0}(Z_{f},\varepsilon )$
are the Fermi functions taking into account the Coulomb interaction between
the emitted electrons and the final nucleus with charge number $Z_{f}$. The
integral $\mathcal{I}$ can be integrated analytically by noticing that $%
\varepsilon _{2}=\tilde{Q}+2-\varepsilon _{1}$ and using the Primakoff-Rosen
approximation $F_{0}(Z_{f},\varepsilon )=(\varepsilon /p)F_{0}^{\mathrm{%
(PR)}}(Z_{f}).$ This leads in a good approximation to $\mathcal{I}%
\approx 10\pi ^{2}\alpha ^{2}Z_{f}^{2}(\tilde{Q}+1)^{5}/3$ (cf.
\cite{Suhonen:1998ck}) and to the corresponding phase-space factor $G_{2\beta }=g_{%
\mathrm{A}}^{4}Z_{f}^{2}A^{-2/3}(\tilde{Q}+1)^{5}~5\times 10^{-20}\,\mathrm{y%
}^{-1}$. Combining with the rest of the $0\nu 2\beta $ observables, the
inverse half-life can be written as
\begin{equation}
\left( T_{1/2}^{0\nu 2\beta }\right) ^{-1}=g_{\mathrm{A}}^{4}\left( \frac{%
|m_{\beta \beta }|}{m}\right) ^{2}|{M}_{2\beta }|^{2}Z_{f}^{2}A^{-2/3}(%
\tilde{Q}+1)^{5}\,5.0\times 10^{-20}\text{y}^{-1}.  \label{eq:T0vbb}
\end{equation}

The inverse $0\nu $2EC half-life is given by
\begin{equation}
\left( T_{1/2}^{0\nu 2\text{EC}}\right) ^{-1}=\Gamma _{i}/\ln 2\equiv G_{%
\mathrm{2EC}}R_f,  \label{eq:T0v2EC}
\end{equation}%
where $G_{\mathrm{2EC}}=V_{\alpha \beta }^{2}/\ln 2,$ and $R_f$ is
defined by Eq. (\ref{DF}). For $J_{f}^{\pi}=0^+$ and $K_Z = 1$,
we find
in the non-relativistic approximation for two-electron capture from the lowest $K$ shell
\begin{equation}
\left( T_{1/2}^{0\nu 2\text{EC}}\right) ^{-1}=g_{\mathrm{A}}^{4}\left( \frac{%
|m_{\beta \beta }|}{m}\right) ^{2}|M_{\mathrm{2EC}}|^{2}Z_{i}^{6}A^{-2/3}%
\alpha ^{2}m R_f\,5.1\times 10^{-25}\text{y}^{-1},
\label{eq:T0v2EC2}
\end{equation}%
where $M_{\mathrm{2EC}}$ is the $0\nu $2EC nuclear matrix element.

We can now find the ratio of the two processes. Adopting the simplification $%
Z_{f}\approx Z_{i}\equiv Z$ and assuming $\mathrm{M}_{2\beta }\approx M_{\mathrm{2EC}}$,
one finds for the half-life ratio
\begin{equation}
\frac{T_{1/2}^{0\nu 2\text{EC}}}{T_{1/2}^{0\nu 2\beta }}\approx \left( \frac{%
20}{Z}\right) ^{4}\frac{(\tilde{Q}+1)^{5}}{\alpha ^{2}m R_f}.
\label{eq:ratio}
\end{equation}%
Given that $\Gamma _{f}\sim \alpha ^{2}m=27.2$ eV, one immediately derives
that the two processes have comparable half-lives for $\alpha ^{2}m R_f \sim 1$
which is the case for $|M_{i}-M_{f}| \lesssim \Gamma _{f}$.

%%%%%%%%%%%%%%%%%%%%%%%%%%%%%%%%%%%%%%%%%%%%%%%%%%%%%%
%%%%%%%%%%%%%%%%%%%%%%%%%%%%%%%%%%%%%%%%%%%%%%%%%%%%%%

%%%%%%%%%%%%%%%%%%%%%%%%%%%%%%%%%%%%%%%%%%%%%%%%%%%%%%%%%%%%%%%%%%%%%%%%%%%%
%%%%%%%%%%%%%%%%%%%%%%%%%%%%%%%%%%%%%%%%%%%%%%%%%%%%%%%%%%%%%%%%%%%%%%%%%%%%
\section{Electron shell effects}
%%%%%%%%%%%%%%%%%%%%%%%%%%%%%%%%%%%%%%%%%%%%%%%%%%%%%%%%%%%%%%%%%%%%%%%%%%%%
%%%%%%%%%%%%%%%%%%%%%%%%%%%%%%%%%%%%%%%%%%%%%%%%%%%%%%%%%%%%%%%%%%%%%%%%%%%%
\setcounter{equation}{0}

The selection of atoms with near-resonant $0\nu2$EC transitions
requires an accurate value of the double-electron ionization potentials of the atoms
and the electron wave functions in the nuclei.
The electron shell models are based on the Hartree-Fock and post-Hartree-Fock methods,
density functional theory, and semiempirical methods of quantum chemistry.
Analytical parametrizations of the non-relativistic wave functions of electrons in neutral atoms,
obtained with the use of the Roothaan-Hartree-Fock method and covering almost the entire periodic table,
are provided by \textcite{Clementi1974,McLean1981,Snijders1981,Bunge1993}.
The various feasible $0\nu2$EC decays are expected to occur in medium-heavy and heavy atoms,
for which relativistic effects are important.
With the advent of personal computers,
physicists acquired the opportunity to use advanced software packages,
such as G{\lowercase{\scshape RASP}}2K \cite{GRASP1989,Grant2007}, DIRAC \footnote{See T Saue, L Visscher, H J Aa Jensen and R Bast,
with contributions from V Bakken, K G Dyall, S Dubillard et al, (2018)
\textit{DIRAC, a relativistic ab initio electronic structure program, Release DIRAC18 (2018)}
(available at $\mathrm{https://doi.org/10.5281/zenodo.2253986}$, see also $\mathrm{http://www.diracprogram.org}$).}, %\cite{Dirac2008},
CI-MBPT \cite{Kozlov2015} and others,
for applications of relativistic computational methods in modeling complex atomic systems.

Quantum electrodynamics (QED) of electrons and photons is known to be a self-consistent theory within infinite renormalizations.
One could expect the existence, at least, of a similarly formally consistent theory of electrons, photons and nuclei,
regarded as elementary particles, which would be a satisfactory idealization for most practical purposes.

In quantum field theory, the relativistic bound states of two particles are described by the Bethe-Salpeter equation \cite {BS1951,HM1952}.
In the nonrelativistic limit, this equation leads to the Schr\"odinger wave equation, but it also includes
additional anomalous solutions that do not have a clear physical interpretation:
First, there are bound states corresponding to excitations of the time-like component of the relative four momenta of the particles.
Such states have no analogs in the nonrelativistic potential scattering theory. None of the particles observed experimentally have been identified with the anomalous solutions so far.
Second, some solutions appear with a negative norm.
Third, the Bethe-Salpeter kernel, evaluated at any finite order of perturbation theory, breaks crossing symmetry and gauge invariance of
QED \cite{Nakanishi1969,Itzykson1980}. The anomalous solutions do not arise when retardation effects are neglected.

Applications of the Bethe-Salpeter equation to
 hydrogen atom \cite{Salpeter1952} and positronium \cite{Itzykson1980}
appeared to be successful because of the non-relativistic character of the bound-state problems and the possibility to account for the retardation effects with the help of perturbation theory in terms of the small parameter $1/c$.

%%%%%%%%%%%%%%%%%%%%%%%%%%%%%%%%%%%%%%%%%%%%%%%%%%%%%%%%%%%%%%%%%%%%%%%%%%%%
%%%%%%%%%%%%%%%%%%%%%%%%%%%%%%%%%%%%%%%%%%%%%%%%%%%%%%%%%%%%%%%%%%%%%%%%%%%%
%%%%%%%%%%%%%%%%%%%%%%%%%%%%%%%%%%%%%%%%%%%%%%%%%%%%%%%%%%%%%%%%%%%%%%%%%%%%
%\begin{figure} [H] %
%\begin{center}
%\includegraphics[angle = -90,width=0.136\textwidth]{figSdiagram.eps}
%\vspace{-10cm}
%\caption{
%An instantaneous photon exchange diagram of the non-covariant perturbation theory.
%A photon is represented by a wavy line and an electron is represented by a solid line.
%}
%\label{figSdiagram}
%\end{center}
%\end{figure}
%\vspace{-5cm}
%%%%%%%%%%%%%%%%%%%%%%%%%%%%%%%%%%%%%%%%%%%%%%%%%%%%%%%%%%%%%%%%%%%%%%%%%%%%
%%%%%%%%%%%%%%%%%%%%%%%%%%%%%%%%%%%%%%%%%%%%%%%%%%%%%%%%%%%%%%%%%%%%%%%%%%%%
%%%%%%%%%%%%%%%%%%%%%%%%%%%%%%%%%%%%%%%%%%%%%%%%%%%%%%%%%%%%%%%%%%%%%%%%%%%%

A successful attempt at generalization of the series expansion
around the instantaneous approximation to multielectron atoms is presented in the papers of \textcite{Sucher1980,Broyles1988},
where the progress was achieved by using non-covariant perturbation theory in %terms of
the coupling constant, $\alpha$, of electrons with the transverse part of the electromagnetic vector potential
and
the magnitude of $Z$ diagrams describing creation and annihilation of electron-positron pairs.
Such a perturbation theory appears well-founded
because the transverse components of the electromagnetic vector potential interact
with spatial components of the electromagnetic current, which contain the small term $1/c$,
whereas $Z$ diagrams contribute to observables in higher orders of $1/c$.
Compared with lowest-order Coulomb photon exchange diagrams, $Z$ diagrams are
suppressed by the factors $ {u}^{\dagger}v^{\prime} \sim 1/c$ or ${v}^{\prime \dagger}u \sim 1/c$ in each of the photon vertices
due to overlapping of the small and large bispinor components and the factor $ 1/c^2 $ originating from the propagator of the positron.
As a result, %the diagram shown in Fig.~\ref{figZdiagram} is
$Z$ diagrams are of the order $ O (\alpha^2/c^4) $.
The second-order correction to the energy due to the Fermi-Breit potential is of the same magnitude.
The exchange of transverse photons leading to the Fermi-Breit potential
contributes to the interaction potential of order $ \alpha /c^2 $,
such that the magnitude of the $Z$ diagrams %shown in Fig.~\ref{figZdiagram}
is suppressed.
The scheme adopted  by \textcite{Sucher1980,Broyles1988} is equivalent to neglecting the Dirac sea at the zeroth-order approximation,
except for the two-body problem, in which the bound-state energy equation is the same as in the paper of \textcite{Salpeter1952}.
The non-sea approximation is widely used to study nuclear matter in the Dirac-Brueckner-Hartree-Fock method \cite{DBHF1983,DBHF1984,DBHF1987}.
Let us remark that the negative-energy fermion states are required to ensure causality and
guarantee Lorentz invariance of the $T$-product and hence of the $S$-matrix.

%%%%%%%%%%%%%%%%%%%%%%%%%%%%%%%%%%%%%%%%%%%%%%%%%%%%%%%%%%%%%%%%%%%%%%%%%%%%
%%%%%%%%%%%%%%%%%%%%%%%%%%%%%%%%%%%%%%%%%%%%%%%%%%%%%%%%%%%%%%%%%%%%%%%%%%%%
%%%%%%%%%%%%%%%%%%%%%%%%%%%%%%%%%%%%%%%%%%%%%%%%%%%%%%%%%%%%%%%%%%%%%%%%%%%%
%\begin{figure} [!t] %
%\begin{center}
%\includegraphics[angle = -90,width=0.39\textwidth]{figZdiagram.eps}
%\vspace{-10cm}
%\caption{
%$Z$ diagram of the non-covariant perturbation theory, which generates a three-body potential $\sim 1/c^4$.
%}
%\label{figZdiagram}
%\end{center}
%\end{figure}
%\vspace{-5cm}
%%%%%%%%%%%%%%%%%%%%%%%%%%%%%%%%%%%%%%%%%%%%%%%%%%%%%%%%%%%%%%%%%%%%%%%%%%%%
%%%%%%%%%%%%%%%%%%%%%%%%%%%%%%%%%%%%%%%%%%%%%%%%%%%%%%%%%%%%%%%%%%%%%%%%%%%%
%%%%%%%%%%%%%%%%%%%%%%%%%%%%%%%%%%%%%%%%%%%%%%%%%%%%%%%%%%%%%%%%%%%%%%%%%%%%

The self-consistent non-relativistic expansion becomes possible because the corrections related to the finite speed of light are small.
The neglect of retardation allows for the formulation of an equation for the Bethe-Salpeter wave function integrated over the relative energy of the particles \cite{Broyles1988}.
The non-covariant wave function obtained in this manner yields the wave function of the equivalent many-body non-covariant Schr\"odinger equation.
%In the approach of Refs.~\cite{Sucher1980,Broyles1988},
%the instantaneous part of the photon propagator provides a zeroth-order approximation for interactions,
%whereas retardation effects are estimated using perturbation theory in terms of the small parameter $1/c$.
Gauge invariance of QED ensures Lorentz invariance of the theory, %as it will be discussed in Sec. 7.2,
but in the intermediate stages of the computation, it is necessary to work with Lorentz-noncovariant and gauge-dependent expressions.

In the Feynman gauge, the product of two photon vertices and the photon propagator
\begin{equation}
D_{\mu \nu}(k) = \frac{- g_{\mu \nu}}{k^2}
\end{equation}
is represented as follows:
\begin{equation} \label{FG}
e\gamma_1^{\mu} D_{\mu \nu}(k) e\gamma_2^{\nu} = - \gamma_1^{0} \gamma_2^{0}\frac{e^2}{k^2}
\left(1 - \mbox{\boldmath$\alpha$}_1 \cdot \mbox{\boldmath$\alpha$}_2\right),
\end{equation}
where $k=(\omega,\mathbf{k})$ is the photon momentum,
$\mbox{\boldmath$\alpha$}_i = \gamma_i^{0}\mbox{\boldmath$\gamma$}_i$
are velocity operators for the particles $i=1,2$, and $\gamma^{\mu}$ are the Dirac $\gamma$-matrices.
The corresponding interaction potential obtained in the static limit,
\begin{eqnarray}
V_{CG}(\mathbf{r}) &\equiv& \int \frac{d\mathbf{k}}{(2\pi)^3} \exp(i\mathbf{k}\mathbf{r})
e\gamma_1^{0}\gamma_1^{\mu} D_{\mu \nu}(\omega = 0,\mathbf{k}) e\gamma_2^{0}\gamma_2^{\nu} \nonumber \\
&=&  \frac{e^2}{4\pi r}\left (1 - \mbox{\boldmath$\alpha$}_1 \cdot \mbox{\boldmath$\alpha$}_2\right),
\label{CGpot}
\end{eqnarray}
acquires familiar form of electrostatic interaction energy of charges plus  magnetostatic interaction energy of  electric currents. The correction to the Coulomb potential entering Eq.~(\ref{CGpot}) was first derived quantum-mechanically by \textcite{Gaunt1929}. The expansion of $D_{\mu \nu}(k)$ in higher powers of $\omega$ describes retardation effects, which are expected to be most pronounced for inner orbits.
A typical splitting of the energy levels  is $\omega \sim \alpha^2 Z^2 m$, and
a typical momentum of electrons is $|\mathbf{k}| \sim \alpha Z m$, such that
for light and medium-heavy atoms, the expansion parameter
$\omega^2/\mathbf{k}^2 \sim (\alpha Z)^2$ is still sufficiently small.
The potential of Eq.~(\ref{CGpot}) is known as the Coulomb-Gaunt potential.
Such a potential can be used to approximate the lowest-order interaction of electrons,
although the magnetostatic energy $\sim 1/c^2$ is of the same order as the retardation corrections to the Coulomb potential.

In the Coulomb gauge, the photon propagator $D_{\mu \nu}(k) $ takes the form
\begin{eqnarray}
D_{0 0}(k) &=& \frac{1}{\mathbf{k}^2},                        \nonumber \\
D_{i j}(k) &=& \frac{\delta_{ij} - k_i k_j/\mathbf{k}^2}{k^2},  \label{CG} \\
D_{i 0}(k) &=& D_{0 j}(k) = 0, \nonumber
\end{eqnarray}
where $i,j=1,2,3$. The Coulomb gauge breaks Lorentz covariance but appears natural in the problem of quantization of the electromagnetic field,  since it allows for explicitly solving the constraint equations.
The photon propagator appears split in two pieces,
the first of which corresponds to the instantaneous interaction;
the second describes the static terms $\sim 1/c^2$ plus retardation effects $\sim 1/c^4$.
The potential of a zero-order approximation contains contributions from the time-like components
and the space-like component of the propagator in the limit of $\omega = 0$.
The product of two photon vertices and the propagator (\ref{CG}) is represented by
\begin{equation}
e\gamma_1^{\mu} D_{\mu \nu}(k) e\gamma_2^{\nu} = \gamma_1^{0} \gamma_2^{0}\left(
  \frac{e^2}{\mathbf{k}^2}
+ \frac{e^2}{k^2} \left(
\mbox{\boldmath$\alpha$}_1 \cdot \mbox{\boldmath$\alpha$}_2 -
\frac{ \mbox{\boldmath$\alpha$}_1 \cdot \mathbf{k} \mbox{\boldmath$\alpha$}_2 \cdot \mathbf{k}}{\mathbf{k}^2}
\right)
\right).                             \label{expCG}
\end{equation}

%Suppose $\Psi_{a}$ and $\Psi_{b}$ are wave functions of electrons with energies $\epsilon_a^*$ and $\epsilon_b^*$, respectively, satisfying the Dirac equation in a self-consistent mean field of an atom, $\omega = \epsilon_b^* - \epsilon_a^*$ and $\mathbf{k} = \mathbf{k}_b - \mathbf{k}_a$.
%The current conservation $\Psi^{\dagger}_{b} \omega \Psi_{a} = \Psi^{\dagger}_{b} \mbox{\boldmath$\alpha$} \cdot \mathbf{k} \Psi_{a}$ then makes Eqs.~(\ref{FG}) and (\ref{expCG}) %after substituting $\mbox{\boldmath$\alpha$}_{1,2} \cdot \mathbf{k} \to \omega$
%equivalent to order $\sim 1/c^2$, which is a manifestation of gauge invariance.

The interaction potential corresponding to the static limit of Eq.~(\ref{expCG}) becomes
\begin{equation}
V_{CB}(\mathbf{r}) = \frac{e^2}{4\pi r}\left(1 - \frac{\mbox{\boldmath$\alpha$}_1 \cdot \mbox{\boldmath$\alpha$}_2 + \mbox{\boldmath$\alpha$}_1 \cdot \mathbf{n} \mbox{\boldmath$\alpha$}_2 \cdot \mathbf{n}}{2} \right),
\label{CBpot}
\end{equation}
where $\mathbf{n} = \mathbf{r}/|\mathbf{r}|$.
Equation (\ref{CBpot}) can be recognized as the sum of the classical Coulomb and Darwin potentials \cite{Darwin1920},
which demonstrates the essentially classical origin of $V_{CB}$.
The no-sea approximation is thus sufficient to ensure the correct expression for $V_{CB}$.
The potential $V_{CB}$ given by Eq.~(\ref{CBpot}) was derived first quantum-mechanically by \textcite{Breit1929};
it is known as the Coulomb-Breit potential.
Starting with $V_{CB}$ is a natural choice,
since the retardation effects in this case are of the order of $1/c^4$.

The techniques of \textcite{Sucher1980,Broyles1988} can be considered as the starting point for developing
a systematic $1/c$ expansion around the instantaneous approximation in the bound-state problem for
light and medium-heavy atoms in analogy with positronium and the hydrogen atom.
Heavy atoms for which the expansion parameter $1/c$ is not small are somewhat beyond the scope of
perturbational treatment.
Since theoretical estimates of accuracy are difficult, it is required to compare model predictions with empirical data wherever possible.
%In the nonrelativistic theory, the Thomas-Fermi model is known to be exact in the limit of $Z \to \infty$ \cite{Thirring1980}.
%In the $0\nu2$EC decays, medium-heavy and heavy atoms are of interest.
%One can hope that in the relativistic case, a self-consistent description of the mean field produced by the nucleus and the shell electrons allows for one to sum up the leading series in $\alpha Z$.
%Since theoretical estimates of accuracy are difficult, we base our conclusions on a comparison of the model predictions with empirical data.
%In what follows, we employ models that offer practical insights into problems of atomic physics.

The relativistic approach is based on the Dirac-Coulomb Hamiltonian,
\begin{equation}
H=\sum_{i}\left( \mbox{\boldmath$\alpha$} \mathbf{p}_i + \beta m - \frac{\alpha Z}{r_i} \right) + \sum_{i<j} V(\mathbf{r}_i - \mathbf{r}_j),
\end{equation}
where the sum runs over electrons, $r_i=|\mathbf{r}_i|$.
The potential $V(\mathbf{r})$ is, as a rule, taken to be the Coulomb-Breit potential,
which already accounts for relativistic effects $\sim 1/c^2$ at the lowest order of the perturbation expansion.
Neglecting the Dirac sea could require the projection of the potential onto positive energy states.
The electron wave function is constructed as a Slater determinant of one-electron orbitals.
Solutions to the eigenvalue problem are sought using the Dirac-Hartree-Fock approximation.

In the non-relativistic Coulomb problem, physical quantities are determined by a one-dimensional parameter,
which is the Bohr radius, $a_{B} = 1/(\alpha m) = 5.2 \times 10^4$ fm.
In the relativistic problem, the Bohr radius acts as a scale, which determines the normalization of electron wave functions and
the integral characteristics, such as the interaction energies of holes in the electron shell.
In addition to the Bohr radius, there are other scales of the relativistic problem.
The electron Compton wavelength  $\lambda_e =  1/m $ determines the distance from the nucleus,
at which the electron should be considered in a relativistic manner.
On a scale smaller than $1/m$, the non-relativistic wave function is markedly different from the upper component of the Dirac wave function;
thus, the effects associated with the finite size of the nucleus must be calculated on the basis of the relativistic Dirac equation.
The third scale in the hierarchy is the distance $\alpha Z/m = 2.8 Z$ fm, at which the Coulomb potential becomes greater than the electron mass.
The smallest (fourth) scale is the nuclear radius $R=1.2 A^{1/3}$ fm. %The hierarchy is shown schematically in Fig.~\ref{scales}; it is valid for light and heavy atoms.
The size of the $^{238}$U nucleus, e.g., is approximately 35 times smaller than $\alpha Z/m = 260$ fm,
approximately 50 times smaller than the Compton wavelength $ 1/m = 390$ fm,
and approximately 75 times smaller than $1/(\alpha m Z) = 580$ fm.
The depth of the potential extending from $0$ to $\alpha Z/m$ is too small to produce bound states of electrons in the negative continuum.

%%%%%%%%%%%%%%%%%%%%%%%%%%%%%%%%%%%%%%%%%%%%%%%%%%%%%%%%%%%%%%%%%%%%%%%%%%%%
%\begin{figure} [htb] %
%\begin{center}
%\includegraphics[angle = -90,width=0.786\textwidth]{scales.bis.eps}
%\caption{
%The hierarchy of scales typical for a multielectron atom.
%Shown in ascending order are the following:
%(i) the radius $R$ of a uranium nucleus;
%(ii) the distance $r_1 = \alpha Z/m$ at which the energy of the Coulomb interaction of the electron with the nucleus
%becomes comparable to the electron mass;
%(iii) the Compton wavelength of an electron $\lambda_e = 1/m$;
%(iv) the radius $a_K = a_B/Z = 1/(\alpha Z m)$ of the $K$ orbit;
%(v) the Bohr radius, which coincides with the radius of the outermost orbit of the electron shell.
%The scales are given for $Z=92$.
%}
%\label{scales}
%\end{center}
%\end{figure}
%\vspace{-5cm}
%%%%%%%%%%%%%%%%%%%%%%%%%%%%%%%%%%%%%%%%%%%%%%%%%%%%%%%%%%%%%%%%%%%%%%%%%%%%

According to the Thomas-Fermi model, the majority of the shell electrons are at a distance $1/(\alpha m Z^{1/3})$
from the nucleus, and the total binding energy of the electrons scales as 20.8$Z^{7/3}$ eV.
The potential energy of the interaction of electrons with each other is 1/7 of the energy of the interaction of electrons with the nucleus.
The numerical smallness of the electron-electron interaction shows that the Coulomb wave functions of electrons
can be used as a first approximation to the self-consistent mean-field solutions.

The single-electron ionization potentials (SEIP) of innermost orbits,
which are of specific interest to the $0\nu2$EC problem,
increase quadratically with $Z$ from 13.6 eV in the hydrogen atom up to 115.6 keV in the uranium atom.
The radii of the outer orbits and hence the size of atoms do not depend on $Z$.
The SEIPs of outermost orbits are of the order of a few eV for all $Z$.
The greatest overlap with the nucleus is achieved for electrons of the innermost orbits.
In the $(A, Z) \to (A, Z-2)$ transitions associated with the $0\nu2$EC decays, we are interested in the electron wave functions inside the parent nucleus $(A, Z)$, whereas the energy balance is provided by the energy of the excited electron shells of the daughter nucleus $(A, Z-2)$.
The SEIPs of all orbitals across the entire periodic table are given by \textcite{LARK},
where
experimental data on the binding energy of electron subshells and
data obtained from Hartree-Fock atomic calculations are combined within a general semi-empirical method.

The double-electron ionization potentials (DEIPs) are additive to first approximation.
A more accurate estimate of DEIPs takes account of interaction energy of electron holes,
relaxation energy and other specific effects.
In the innermost orbitals, the Coulomb interaction energy of two holes is of the order $ \alpha Z/ a_B \sim \alpha^2 Z m$.
This energy grows linearly with $Z$ and reaches a value of $\sim 1$ keV in heavy atoms.
Relaxation energy for a medium-heavy atom of $^{101}$Ru reaches a value of 400 eV \cite{Niskanen2011}.
The two-hole excitation energy of the daughter atoms differs from the corresponding DEIP
by the sum of the energies of two outermost occupied orbits, approximately 10 eV.

The required accuracy of two-hole excitation energies is dictated
by the typical width of vacancies of electron shells, which is approximately 10 eV.
This accuracy is required to specify the $0\nu2$EC transitions in the unitary limit.
The best achieved accuracy in the $Q$-value measurements with Penning traps is on the order of 10 eV for heavy systems, and furthermore,
the DEIP calculations for heavy atoms are successful to within several tens of eV.
To realistically calculate the excitation energies and the short-distance components of electron wave functions,
we use the G{\lowercase{\scshape RASP}}2K software package \cite{GRASP1989,Grant2007}, which is well-founded theoretically
and successful in the description of a wide range of atomic physics data.

%%%%%%%%%%%%%%%%%%%%%%%%%%%%%%%%%%%%%%%%%%%%%%%%%%%%%%%%%%%%%%%%%%%%%%%%%%%%
\subsection{Interaction energy of electron holes}
%%%%%%%%%%%%%%%%%%%%%%%%%%%%%%%%%%%%%%%%%%%%%%%%%%%%%%%%%%%%%%%%%%%%%%%%%%%%

The average electron velocity $ v/c \sim \alpha Z / n $ increases with the nuclear charge and
becomes large in heavy atoms. \textcolor{black}{
In a uranium atom, an electron on the $K$ shell, localized at an average distance $a_B/Z\sim 600$ fm from the nucleus, moves at a speed of $v \sim 0.7\,c$.}
A fully relativistic description is thus required to construct
accurate electron wave functions inside the nucleus.

The wave function of a relativistic electron in a central potential is defined in Eq.~(\ref{DIWF}).
We consider transitions between nuclei with good quantum numbers.
In what follows, $J_i$ and $J_f$ are the total angular momenta of the parent and daughter nuclei and
$j_i$ and $j_f$ are the total angular momenta of electron shells in the initial and final states, respectively.
The daughter nucleus $(A,Z-2)$ inherits the electron shell of the parent nucleus $(A,Z)$
with two electron holes formed by the electron capture
and possible excitations of spectator electrons into vacant orbits.
The total angular momentum of captured electrons, $J$, is in the interval
$\max(|J_f - J_i|,|j_f - j_i|) \leq J \leq \min(J_f + J_i,j_f + j_i)$.

Let $\mathbf{J}_{i}^{\textrm{tot}} = \mathbf{J}_i + \mathbf{j}_i$ and $\mathbf{J}_f^{\textrm{tot}} = \mathbf{J}_f + \mathbf{j}_f$
be the total angular momenta of atoms in the initial and final states, respectively.
The reaction involves the nucleus and two electrons,
whereas $Z - 2$ electrons are spectators that can be excited due to the nuclear recoil and/or
the non-orthogonality of the initial- and final-state electron wave functions.
Total angular momentum conservation implies
$\mathbf{J}_f^{\textrm{tot}} = \mathbf{J}_i^{\textrm{tot}}$ as well as $\mathbf{J}_f = \mathbf{J}_i + \mathbf{J}$ and $\mathbf{j}_f + \mathbf{J} = \mathbf{j}_i$.

The atomic-state wave function with definite $j_i$ is a superposition of configuration states,
which are anti-symmetric products of the one-electron orbitals (\ref{DIWF}).
For $J_i^{\textrm{tot}} \neq 0$, the atomic-state wave function is further superimposed with the wave function of the nucleus.
The atomic states are split into levels with typical energy separation of fractions of electronvolts.
Transitions between these levels produce radiation in the short- and mid-wavelength infrared range.
Such effects lie beyond the energy scale in which we are interested, 10 eV.
Since, at room temperature, atoms are in their ground states, each time, we select the lowest-energy eigenstate.
In the $0\nu2$EC decays, the spin of the initial nucleus is zero,
in which case the configuration space reduces, and the calculations simplify.

The capture from the $s_{1/2}$ and $p_{1/2}$ orbits occurs with the dominant probability,
which restricts the admissible values of $J^{\pi}$ to $0^{\pm},\;1^{\pm}$.
The higher orbits, relevant to the daughter nuclei with $J_f \geq 2$, thus may be disregarded.
The spin $J$ is the suitable quantum number for the classification of transitions.

After capturing the pair, the atomic-state wave function is still a superposition of configuration states,
onto which the states with various $j_f$ are further superimposed.
The typical level splitting is a fraction of an electronvolt, whereas the radiation width of the excited electron shell is about 10 eV.
This is the case for overlapping resonances. %, which have their own specifics.
The influence of the coherent overlap on the $0\nu2$EC decay has not been discussed thus far.
We sum all the contributions decoherently.

The two-electron wave function has the total angular momentum $J$, projection $M$, and a definite parity.
This can be arranged by weighting the product of wave functions of one-electron orbitals (\ref{DIWF})
with the Clebsch-Gordan coefficients, as done in Eq.~(\ref{WF2}).
\textcolor{black}{The Pauli principle} says that the wave function must be antisymmetric under exchange of two electrons.
The normalized antisymmetric two-electron wave function takes the form shown in Eq.~(\ref{TWOE}).
The interaction energy of electron holes in the static approximation can be found from
\begin{equation}
\epsilon = \int d\mathbf{r}_{1} d\mathbf{r}_{2}
\Psi _{\beta \delta }^{JM \dagger}(\mathbf{r}_{1},\mathbf{r}_{2})
V(\mathbf{r}_{1} - \mathbf{r}_{2})
\Psi_{\beta \delta }^{JM}(\mathbf{r}_{1},\mathbf{r}_{2}), \label{EINT}
\end{equation}
where $V(\mathbf{r})$ is the Coulomb-Gaunt potential (\ref{CGpot}) or the Coulomb-Breit potential (\ref{CBpot}).

The interaction energy (\ref{EINT}) is given by the matrix element of the two-particle operator.
In such cases, the angular variables are explicitly integrated out and the problem is reduced to the calculation
of a two-dimensional integral in the radial variables (see, e.g., \cite{Grant2007}).
We present results of this reduction needed to demonstrate the independence of the interaction energy from the gauge.

%%%%%%%%%%%%%%%%%%%%%%%%%%%%%%%%%%%%%%%%%%%%%%%%%%%%%%%%%%%%%%%%%%%%%%%%%%%%
\subsubsection{Electrostatic interaction}
%%%%%%%%%%%%%%%%%%%%%%%%%%%%%%%%%%%%%%%%%%%%%%%%%%%%%%%%%%%%%%%%%%%%%%%%%%%%

The interaction energy in the static approximation splits into the sum of the electrostatic and magnetostatic energies:
$\epsilon = \epsilon_E + \epsilon_M$.
The Coulomb part, $\epsilon_E$, does not depend on the gauge condition. Equation~(\ref{EINT}) can be written in the form
\begin{eqnarray}
\epsilon_{E} &=& \int d\mathbf{r}_{1} d\mathbf{r}_{2}
\Psi _{\beta \delta }^{JM \dagger}(\mathbf{r}_{1},\mathbf{r}_{2})
\frac{\alpha}{r}
\Psi_{\beta \delta }^{JM}(\mathbf{r}_{1},\mathbf{r}_{2}) \nonumber \\
&=& 2\mathcal{N}^2_{\beta \delta}
\left(
\mathcal{K}_{\beta \delta \beta \delta }^{JM}-(-)^{j_{\beta
}+j_{\delta}-J} \mathcal{K}_{\beta \delta \delta \beta }^{JM}
\right),
\label{estat}
\end{eqnarray}
where $4\pi \alpha = e^2$, $r = |\mathbf{r}_{1} - \mathbf{r}_{2}|$ and
\begin{equation}
\mathcal{K}_{\alpha \gamma \beta \delta }^{JM}=\sum_{m_{\alpha }m_{\gamma
}m_{\beta}m_{\delta }} C_{j_{\alpha }m_{\alpha }j_{\gamma }m_{\gamma}}^{JM}
C_{j_{\beta }m_{\beta }j_{\delta }m_{\delta }}^{JM}
\mathcal{K}^{\alpha m_{\alpha }\gamma m_{\gamma } }_{\beta m_{\beta } \delta m_{\delta }}
\label{estatbis1}
\end{equation}
with
\begin{eqnarray*}
\mathcal{K}^{\alpha m_{\alpha }\gamma m_{\gamma } }_{\beta m_{\beta } \delta m_{\delta }}
&=&\int d\mathbf{r}_{1}d\mathbf{r}_{2}\left. \Psi _{\alpha m_{\alpha
}}^{\dagger}(\mathbf{r}_{1})\Psi _{\beta m_{\beta }}(\mathbf{r}_{1})\right.
\frac{\alpha}{r}\left. \Psi _{\gamma m_{\gamma
}}^{\dagger} (\mathbf{r}_{2})\Psi _{\delta m_{\delta }}(\mathbf{r}_{2})\right.
\label{EC1}
\end{eqnarray*}

The Hermitian product of spherical spinors weighted with a spherical harmonic and integrated over angles can be represented as
\begin{equation}
\int d\Omega _{\mathbf{n}}
\Omega _{\alpha m_{\alpha}}^{\dagger}(\mathbf{n})
\Omega_{\beta m_{\beta}} (\mathbf{n})Y_{lm}(\mathbf{n})= C_{j_{\beta }m_{\beta
}lm}^{j_{\alpha}m_{\alpha }}\mathcal{A}_{\alpha \beta }^{l},
\label{spinors}
\end{equation}
where $|j_{\alpha} - j_{\beta}| \leq l \leq j_{\alpha} + j_{\beta}$,
\begin{equation}
\mathcal{A}_{\alpha \beta }^{l}=(-)^{1/2+j_{\beta }+l_{\alpha }+l}
\sqrt{\frac{[l][l_{\beta }][j_{\beta }]}{4\pi }} C_{l_{\beta
}0l0}^{l_{\alpha}0}\left\{
\begin{array}{cll}
1/2 & l_{\beta } & j_{\beta } \\
l & j_{\alpha } & l_{\alpha }
\end{array}
\right\},
\end{equation}
and $[x]=2x+1$ . By introducing the unit matrix $( \mbox{\boldmath$\sigma$} \mathbf{n} )^2=1$
between the spherical spinors of Eq. (\ref{spinors})
and taking into account the identity $ \mbox{\boldmath$\sigma$} \mathbf{n}
\Omega _{\alpha m_{\alpha}}(\mathbf{n}) = - \Omega _{\alpha^{\prime} m_{\alpha}}(\mathbf{n})$, one obtains
$\mathcal{A}_{\alpha \beta }^{l} = \mathcal{A}_{\alpha^{\prime} \beta^{\prime} }^{l}$.
The angular integral of the Hermitian product of the electron wave functions and a spherical harmonic leads to the expression
\begin{equation}
\int d\Omega _{\mathbf{n}}
\Psi _{\alpha m_{\alpha}}^{\dagger}(\mathbf{n})
\Psi_{\beta m_{\beta}} (\mathbf{n})Y_{lm}(\mathbf{n})=
C_{j_{\beta }m_{\beta}lm}^{j_{\alpha}m_{\alpha }}
\mathcal{F}_{\alpha \beta }^{l}(r),
\label{bispinors}
\end{equation}
where $\mathcal{F}_{\alpha \beta }^{l}(r)$ is defined by
\begin{equation}
\mathcal{F}_{\alpha \beta }^{l}(r)=(f_{\alpha }(r)f_{\beta }(r)
+g_{\alpha }(r) g_{\beta}(r))\mathcal{A}_{\alpha \beta }^{l}.
\end{equation}

The electrostatic interaction integral takes the form \cite{KRIV11}
\begin{equation}
\mathcal{K}_{\alpha \gamma \beta \delta }^{JM} = \alpha \sum_{l} \frac{4\pi }{2l+1}
\mathcal{C}_{\alpha \gamma \beta \delta }^{Jl}\int
r_{1}^{2}dr_{1}r_{2}^{2}dr_{2}\frac{r_{<}^{l}}{r_{>}^{l+1}} \mathcal{F}_{\alpha
\beta }^{l}(r_{1})\mathcal{F}_{\gamma \delta }^{l}(r_{2}),
\label{elstenergy}
\end{equation}
where $r_i = |\mathbf{r}_i|$, $r_{<}$ is the lesser and $r_{>}$ the greater of $r_1$ and $r_2$, and
\begin{equation}
\mathcal{C}_{\alpha \gamma \beta \delta }^{Jl}=(-1)^{j_{\gamma }+j_{\beta }+J}
\sqrt{[j_{\alpha }][j_{\gamma }]}\left\{
\begin{array}{lll}
l & j_{\delta } & j_{\gamma }  \\
J & j_{\alpha } & j_{\beta }
\end{array}
\right\}.
\end{equation}
The interaction energy %(\ref{estat})
is invariant under rotations; thus, $\mathcal{K}_{\alpha \gamma \beta \delta }^{JM}$ does not depend on the spin projection $M$.

%%%%%%%%%%%%%%%%%%%%%%%%%%%%%%%%%%%%%%%%%%%%%%%%%%%%%%%%%%%%%%%%%%%%%%%%%%%%
\subsubsection{Retardation correction in the Feynman gauge}
%%%%%%%%%%%%%%%%%%%%%%%%%%%%%%%%%%%%%%%%%%%%%%%%%%%%%%%%%%%%%%%%%%%%%%%%%%%%

The time-like component of the free photon propagator in the Feynman gauge,
expanded in powers of the small parameter ${\omega^2}/{\mathbf{k}^2} \sim (\alpha Z)^2 \sim 1/c^2$, takes the form
\begin{equation}
D_{0 0}(\omega,\mathbf{k}) = \frac{1}{\mathbf{k}^2}+\frac{\omega^2}{\mathbf{k}^4} + \ldots .
\end{equation}
The second term provides the lowest-order retardation correction to the Coulomb interaction energy of electrons:
\begin{equation}
\Delta \epsilon_{E} = - \alpha \int d\mathbf{r}_{1} d\mathbf{r}_{2}
\Psi _{\beta \delta }^{JM \dagger}(\mathbf{r}_{1},\mathbf{r}_{2})
\frac{\omega^2 r}{2}
\Psi_{\beta \delta }^{JM}(\mathbf{r}_{1},\mathbf{r}_{2}),
\label{ereta}
\end{equation}
\textcolor{black}{
where $\omega$ is the energy of virtual photon, and $-r/2$ is the Fourier transform of ${4\pi}/{\mathbf{k}^4}$,
obtained using the analytical continuation in $z$ of the expression
\[
\int \frac{d\mathbf{k}}{(2\pi )^{3}}\frac{4\pi }{|\mathbf{k}|^{z}}e^{i%
\mathbf{kr}}=\frac{r^{z -3}\Gamma ((3 - z)/2)}{2^{z - 2}\sqrt{\pi }\Gamma (z/2)}.
\]
$\Delta \epsilon_{E}$} is of the same order as the magnetostatic interaction energy.

The angular variables can be integrated out, similar to the case of the instantaneous Coulomb interaction.
We write Eq.~(\ref{ereta}) in the form
\begin{equation}
\Delta \epsilon_{E} = 2\mathcal{N}^2_{\beta \delta}
\left(
\Delta \mathcal{K}_{\beta \delta \beta \delta }^{JM}-(-)^{j_{\beta
}+j_{\delta}-J} \Delta  \mathcal{K}_{\beta \delta \delta \beta }^{JM}
\right),
\label{eretstat}
\end{equation}
where
\begin{eqnarray}
\Delta \mathcal{K}_{\alpha \gamma \beta \delta }^{JM} &=& \frac{\alpha}{2}
\sum_{l} \frac{4\pi }{2l+1}
\mathcal{C}_{\alpha \gamma \beta \delta }^{Jl} (\epsilon^{*}_{\alpha} - \epsilon^{*}_{\beta})(\epsilon^{*}_{\gamma} - \epsilon^{*}_{\delta}) \nonumber \\
&\times& \int r_{1}^{2}dr_{1}r_{2}^{2}dr_{2} \frac{r_{<}^{l}}{r_{>}^{l+1}}
\left(
\frac{r_{<}^2}{2l + 3} - \frac{r_{>}^2}{2l - 1}
\right)
\mathcal{F}_{\alpha\beta }^{l}(r_{1})
\mathcal{F}_{\gamma \delta }^{l}(r_{2}).
\label{retard}
\end{eqnarray}
One can observe that
only the exchange interaction $\alpha = \delta \neq \gamma = \beta$ contributes to $\Delta \mathcal{K}_{\alpha \gamma \beta \delta }^{JM} \neq 0$, whereas
$\Delta \mathcal{K}_{\alpha \gamma \beta \delta }^{JM} = 0$ for the direct interaction $\alpha = \beta \neq \gamma = \delta$ and
$\Delta \mathcal{K}_{\alpha \gamma \beta \delta }^{JM} = 0$ if the electrons occupy the same shell $\alpha = \beta = \gamma = \delta$.
The retardation corrections of higher orders can be calculated in a similar manner.
In the Coulomb gauge, the retardation corrections to the electrostatic interaction energy vanish.

%%%%%%%%%%%%%%%%%%%%%%%%%%%%%%%%%%%%%%%%%%%%%%%%%%%%%%%%%%%%%%%%%%%%%%%%%%%%
\subsubsection{Magnetostatic  interaction in the Feynman gauge}
%%%%%%%%%%%%%%%%%%%%%%%%%%%%%%%%%%%%%%%%%%%%%%%%%%%%%%%%%%%%%%%%%%%%%%%%%%%%

The magnetostatic part of the interaction energy (\ref{CGpot}) can be represented in a form similar to Eq.~(\ref{estat}):
\begin{eqnarray}
\epsilon_{M} &=& - \alpha \int d\mathbf{r}_{1} d\mathbf{r}_{2}
\Psi _{\beta \delta }^{JM \dagger}(\mathbf{r}_{1},\mathbf{r}_{2})
\frac{ \mbox{\boldmath$\alpha$}_1 \mbox{\boldmath$\alpha$}_2 }{r}
\Psi_{\beta \delta }^{JM}(\mathbf{r}_{1},\mathbf{r}_{2}) \nonumber \\
&=& 2\mathcal{N}^2_{\beta \delta}
\left(
\mathcal{M}_{\beta \delta \beta \delta }^{JM}-(-)^{j_{\beta
}+j_{\delta}-J} \mathcal{M}_{\beta \delta \delta \beta }^{JM}
\right),
\label{mstat}
\end{eqnarray}
where
\begin{equation}
\mathcal{M}_{\alpha \gamma \beta \delta }^{JM}=\sum_{m_{\alpha }m_{\gamma
}m_{\beta}m_{\delta }} C_{j_{\alpha }m_{\alpha }j_{\gamma }m_{\gamma}}^{JM}
C_{j_{\beta }m_{\beta }j_{\delta }m_{\delta }}^{JM}
\mathcal{M}^{\alpha m_{\alpha }\gamma m_{\gamma } }_{\beta m_{\beta } \delta m_{\delta }},
\label{estatbis2}
\end{equation}
with
\begin{eqnarray*}
\mathcal{M}^{\alpha m_{\alpha }\gamma m_{\gamma } }_{\beta m_{\beta } \delta m_{\delta }}
=-\int d\mathbf{r}_{1}d\mathbf{r}_{2}
\left. \Psi _{\alpha m_{\alpha}}^{\dagger}(\mathbf{r}_{1})  \mbox{\boldmath$\alpha$} \Psi _{\beta  m_{\beta  }}(\mathbf{r}_{1})\right. \frac{\alpha}{r}
\left. \Psi _{\gamma m_{\gamma}}^{\dagger}(\mathbf{r}_{2})  \mbox{\boldmath$\alpha$} \Psi _{\delta m_{\delta }}(\mathbf{r}_{2})\right..
\label{EC2}
\end{eqnarray*}

The angular integrals are calculated with the use of equation
\begin{equation}
\int d\Omega _{\mathbf{n}}\Omega _{\alpha m_{\alpha}}^{\dagger}(\mathbf{n})
\mbox{\boldmath$\sigma$}
\Omega_{\beta m_{\beta}} (\mathbf{n})Y_{lm}(\mathbf{n})=
\mathbf{e}^{\mu} \sum_{j}
C_{j_{\beta }m_{\beta}j\kappa}^{j_{\alpha}m_{\alpha }}
C_{1\mu lm}^{j\kappa}
\mathcal{B}_{\alpha \beta }^{jl},
\end{equation}
where $\mathbf{e}^{\mu}$ are basis vectors of the cyclic coordinate system \cite{Varshalovich1988},
the sum runs over $j=l,l\pm 1$ for $\kappa= m_{\alpha } - m_{\beta}$ and $\mu =  m_{\alpha } - m_{\beta} - m$, and
\begin{equation}
\mathcal{B}_{\alpha \beta }^{jl}=(-)^{ j_{\alpha} - j_{\beta} + l_{\alpha }+l_{\beta }}
\sqrt{\frac{ 3 [j] [j_{\beta }] [l ] [l_{\beta }]}{2\pi }}
C_{l_{\beta}0l0}^{l_{\alpha}0}
\left\{
\begin{array}{ccc}
l_{\alpha } &  j_{\alpha } & 1/2\\
l_{\beta  } &  j_{\beta  } & 1/2\\
l                 &  j                 & 1
\end{array}
\right\}.
\end{equation}

The transition current projection on a spherical harmonic can be found to be
\begin{equation}
\int d\Omega _{\mathbf{n}} \Psi_{\alpha m_{\alpha}}^{\dagger}(\mathbf{r})
\mbox{\boldmath$\alpha$}
\Psi_{\beta m_{\beta}}(\mathbf{r}) Y_{lm}(\mathbf{n}) = -i\mathbf{e}^{\mu}
\sum_{j}
C_{j_{\beta }m_{\beta}j\kappa}^{j_{\alpha}m_{\alpha }}
C_{1\mu lm}^{j\kappa}
\mathcal{G}_{\alpha \beta }^{jl}(r),
\end{equation}
where
\begin{equation}
\mathcal{G}_{\alpha \beta }^{jl}(r) =
g_{\alpha}(r) f_{\beta} (r)\mathcal{B}_{\alpha^{\prime} \beta }^{jl} -
f_{\alpha}(r) g_{\beta}(r) \mathcal{B}_{\alpha \beta^{\prime}}^{jl}.
\end{equation}
The sum over $j$ runs within the limits $|j_{\alpha} - j_{\beta}| \leq j \leq j_{\alpha} + j_{\beta}$,
whereas $l$ are constrained by $|j-1| \leq l \leq j+1$.

The interaction integral over the radial variables takes the form
\begin{equation}
\mathcal{M}_{\alpha \gamma \beta \delta }^{JM}=- e^{2}\sum_{jl} \frac{4\pi }{2l+1}
(-1)^{j-l}\mathcal{C}_{\alpha \gamma \beta \delta }^{Jj}
%\mathcal{D}_{\alpha \gamma \beta \delta }^{Jjl}
\int r_{1}^{2}dr_{1}r_{2}^{2}dr_{2}\frac{r_{<}^{l}}{r_{>}^{l+1}} \mathcal{G}_{\alpha
\beta }^{jl}(r_{1})\mathcal{G}_{\gamma \delta }^{jl}(r_{2}).
\end{equation}
%where $r_{<}=\min (r_{1},r_{2})$, $r_{>}=\max (r_{1},r_{2})$.

%%%%%%%%%%%%%%%%%%%%%%%%%%%%%%%%%%%%%%%%%%%%%%%%%%%%%%%%%%%%%%%%%%%%%%%%%%%%
\subsubsection{Magnetostatic  interaction in the Coulomb gauge}
%%%%%%%%%%%%%%%%%%%%%%%%%%%%%%%%%%%%%%%%%%%%%%%%%%%%%%%%%%%%%%%%%%%%%%%%%%%%

In the Coulomb-Breit potential, the magnetostatic interaction energy is given by the expression
\begin{eqnarray}
\epsilon_{M}^{\prime} &=& - \alpha \int d\mathbf{r}_{1} d\mathbf{r}_{2}
\Psi _{\beta \delta }^{JM \dagger}(\mathbf{r}_{1},\mathbf{r}_{2})
\frac{  \mbox{\boldmath$\alpha$}_1  \mbox{\boldmath$\alpha$}_2
+  (\mbox{\boldmath$\alpha$}_1  \mathbf{n})( \mbox{\boldmath$\alpha$}_2 \mathbf{n})}{2r}
\Psi_{\beta \delta }^{JM}(\mathbf{r}_{1},\mathbf{r}_{2}).
\label{mstatbreit}
\end{eqnarray}
Using identity
$
{{n}^i {n}^j}/{r} =
  {\delta^{ij}}/{r}
- \nabla^{i}  \nabla^{j} r,
$
we integrate the derivative term by parts. The result can be written in the form
$
\epsilon_{M}^{\prime} = \epsilon_{M} + \Delta  \epsilon_{M},
$
with $\epsilon_{M}$ given by Eq.~(\ref{mstat}) and
\begin{equation}
\Delta \epsilon_{M} = 2\mathcal{N}^2_{\beta \delta} \left(
                                 \Delta \mathcal{M}_{\beta \delta \beta \delta }^{JM}
 - (-1)^{j_{\beta}+j_{\delta}-J} \Delta \mathcal{M}_{\beta \delta \delta \beta }^{JM} \right).
\end{equation}
The variance of the magnetostatic interaction energy is determined by
\begin{equation}
\Delta \mathcal{M}_{\alpha \gamma \beta \delta }^{JM}=\sum_{m_{\alpha }m_{\gamma
}m_{\beta}m_{\delta }} C_{j_{\alpha }m_{\alpha }j_{\gamma }m_{\gamma}}^{JM}
C_{j_{\beta }m_{\beta }j_{\delta }m_{\delta }}^{JM}
\Delta \mathcal{M}^{\alpha m_{\alpha }\gamma m_{\gamma } }_{\beta m_{\beta } \delta m_{\delta }},
\end{equation}
with
\begin{eqnarray*}
\Delta \mathcal{M}^{\alpha m_{\alpha }\gamma m_{\gamma } }_{\beta m_{\beta } \delta m_{\delta }}
=- \frac{\alpha}{2} \int d\mathbf{r}_{1}d\mathbf{r}_{2}
\left( \mbox{\boldmath$\nabla$} \, \Psi _{\alpha m_{\alpha}}^{\dagger}(\mathbf{r}_{1})  \mbox{\boldmath$\alpha$} \Psi _{\beta m_{\beta }}(\mathbf{r}_{1})\right)
r
\left(  \mbox{\boldmath$\nabla$} \, \Psi _{\gamma m_{\gamma}}^{\dagger} (\mathbf{r}_{2}) \mbox{\boldmath$\alpha$}  \Psi _{\delta m_{\delta }}(\mathbf{r}_{2})\right).
\label{DMC}
\end{eqnarray*}

The divergence of the transition current projected onto a spherical harmonic can be found to be
\begin{equation}
\int d\Omega_{\mathbf{n}}
\left(  \mbox{\boldmath$\nabla$} \, \Psi_{\alpha m_{\alpha}}^{\dagger}(\mathbf{r})  \mbox{\boldmath$\alpha$} \Psi _{\beta m_{\beta }}(\mathbf{r})\right) Y_{lm}(\mathbf{n}) =-i
 C_{j_{\beta }m_{\beta}lm}^{j_{\alpha}m_{\alpha }}
\mathcal{H}_{\alpha \beta }^{l}(r),
\label{forH}
\end{equation}
where
\begin{equation*}
\mathcal{H}_{\alpha \beta }^{l}(r)=
\left(
- \left( \frac{d}{dr}  + \frac{2 - \kappa_{\alpha} + \kappa_{\beta}}{r}\right) g_{\alpha }(r) f_{\beta }(r)
+ \left( \frac{d}{dr}  + \frac{2 + \kappa_{\alpha} - \kappa_{\beta}}{r}\right) f_{\alpha }(r) g_{\beta}(r) \right)
\mathcal{A}_{\alpha \beta }^{l}.
\end{equation*}

The variance of the magnetostatic interaction energy becomes
\begin{eqnarray}
\Delta \mathcal{M}_{\alpha \gamma \beta \delta }^{JM} &=& \frac{\alpha}{2}
\sum_{l} \frac{4\pi }{2l+1}
\mathcal{C}_{\alpha \gamma \beta \delta }^{Jl} \nonumber \\
&\times& \int r_{1}^{2}dr_{1}r_{2}^{2}dr_{2} \frac{r_{<}^{l}}{r_{>}^{l+1}}
\left(
\frac{r_{<}^2}{2l + 3} - \frac{r_{>}^2}{2l - 1}
\right)
\mathcal{H}_{\alpha\beta }^{l}(r_{1})
\mathcal{H}_{\gamma \delta }^{l}(r_{2}).
\end{eqnarray}

%%%%%%%%%%%%%%%%%%%%%%%%%%%%%%%%%%%%%%%%%%%%%%%%%%%%%%%%%%%%%%%%%%%%%%%%%%%%
\subsubsection{Gauge invariance of the interaction energy of electron holes}
%%%%%%%%%%%%%%%%%%%%%%%%%%%%%%%%%%%%%%%%%%%%%%%%%%%%%%%%%%%%%%%%%%%%%%%%%%%%

The wave function $\Psi_{\alpha m_{\alpha}}(\mathbf{r}) $ is assumed to satisfy
the Dirac equation in a mean-field potential $U(\mathbf{r})$
created by the nucleus and surrounding electrons.
The divergence of the transition current between states with energies $\epsilon_{\alpha}^{*}$ and $\epsilon_{\beta}^{*}$ equals
\begin{equation}
 \mbox{\boldmath$\nabla$} \, \Psi _{\alpha m_{\alpha}}^{\dagger} \mbox{\boldmath$\alpha$} \Psi_{\beta m_{\beta}} =
-i(\epsilon_{\alpha}^{*} - \epsilon_{\beta}^{*}) \Psi _{\alpha m_{\alpha}}^{\dagger}  \Psi_{\beta m_{\beta}}.
\end{equation}
We substitute this expression into Eq.~(\ref{forH}). A comparison with Eq.~(\ref{bispinors}) gives
\begin{eqnarray}
\mathcal{H}_{\alpha\beta }^{l}(r) =
 (\epsilon_{\alpha}^{*} - \epsilon_{\beta}^{*})
\mathcal{F}_{\alpha \beta }^{l}(r),
\label{symmetricf}
\end{eqnarray}
and similarly for  $\mathcal{H}_{\gamma \delta }^{l}(r)$. The contribution of the direct interaction to $\Delta \epsilon_M$ vanishes,
such that $\Delta \epsilon_M = 0$ for electrons of the same shell $\alpha = \beta = \gamma = \delta$,
whereas the exchange interaction for $\alpha = \delta \neq \gamma = \beta$ contributes to $\Delta \epsilon_M \neq 0$.

The function $\mathcal{F}_{\alpha \beta }^{l}(r)$ entering Eq.~(\ref{symmetricf}) appeared earlier
in the interaction energy integrals (\ref{elstenergy}) and (\ref{retard}).
As a consistency check, we observe that $\Delta \epsilon_M$
is equal to the lowest-order retardation correction $\Delta \epsilon_E$ to the Coulomb potential in the Feynman gauge.
We thus conclude that the two-electron interaction energy to the order $1/c^2$ does not depend on the gauge condition.
The gauge independence of the interaction energy of two electrons is thus demonstrated
without assuming a specific type of mean-field potential $U(\mathbf{r})$.
%The above examination refines the heuristic argument after Eq.~(\ref{expCG}).
In positronium, the calculation of bound-state energies is performed to the order $O(\alpha^3)$
using an $O(\alpha^2)$ approximation for the Bethe-Salpeter kernel, which is sufficient for gauge invariance to the order of $O(\alpha^3)$.

For noble gas atoms Ne, Ar, Kr, Xe, and Rn, the difference between magnetostatic interaction energies in the Feynman and Coulomb gauges equals:
0.01 eV, 0.10 eV, 1.39 eV,  5.76 eV, and 26.96 eV for SEIP and
0.02 eV, 0.25 eV, 3.04 eV, 12.16 eV, and 55.72 eV for DEIP, respectively \cite{Niskanen2011}.
The variance does not exceed 60 eV.
The origin of this variance can be attributed to the retardation part of the Coulomb interaction energy
in the Feynman gauge. One can expect that the atomic structure models consistently determine
the energy conditions for the $0\nu2$EC decays with an accuracy of several tens of eV or better.

The proof of gauge invariance of QED of electrons and photons is based on the Ward-Green-Fradkin-Takahashi (WGFT) identity
\cite{Ward1950,Green1953,Fradkin1955,Takahashi1957}.
The diagrams without self-energy insertions into electron lines
are known to be gauge invariant on shell (see, e.g., \cite{BjorkenDrell,BogoliubovShirkov1980}).
The electron self-energy part, or the mass operator $\Sigma$, depends on the gauge, which can be demonstrated explicitly
in a one-loop calculation \cite{Itzykson1980} and to all orders of perturbation theory \cite{JohnsonZumino1959}.
The off-shell Green's functions depend on the gauge.
A complete proof of the gauge invariance of the physical cross-sections in QED of electrons and photons
is given first by \textcite{Bialynicki1970}.
The equivalence of covariant Lorentz gauge and non-covariant Coulomb gauge
also implies the invariance of QED with respect to Lorentz transformations.
There is currently no proof of the gauge invariance of QED of multielectron atoms in higher orders of the $1/c^2$ expansion.
The difficulties are caused by the existence of bound states as asymptotic states of the theory.
Proofs of \textcite{BjorkenDrell,BogoliubovShirkov1980,Bialynicki1970} do not apply to diagrams whose fermion
lines belong to wave functions of bound states.
We believe that the uncertainties inherent in the exitation energies of multielectron atoms
are entirely related to complexity in modeling the atomic systems.
In the shell model discussed above,
the Feynman and Coulomb gauges provide identical results up to the order of $\alpha^{2}\mathrm{Ry}$.

%%%%%%%%%%%%%%%%%%%%%%%%%%%%%%%%%%%%%%%%%%%%%%%%%%%%%%%%%%%%%%%%%%%%%%%%%%%%
%%%%%%%%%%%%%%%%%%%%%%%%%%%%%%%%%%%%%%%%%%%%%%%%%%%%%%%%%%%%%%%%%%%%%%%%%%%%
\subsection{Double-electron
ionization potentials in Auger spectroscopy}
\begin{figure} [!t] %
\begin{center}
\includegraphics[angle = -90,width=0.382\textwidth]{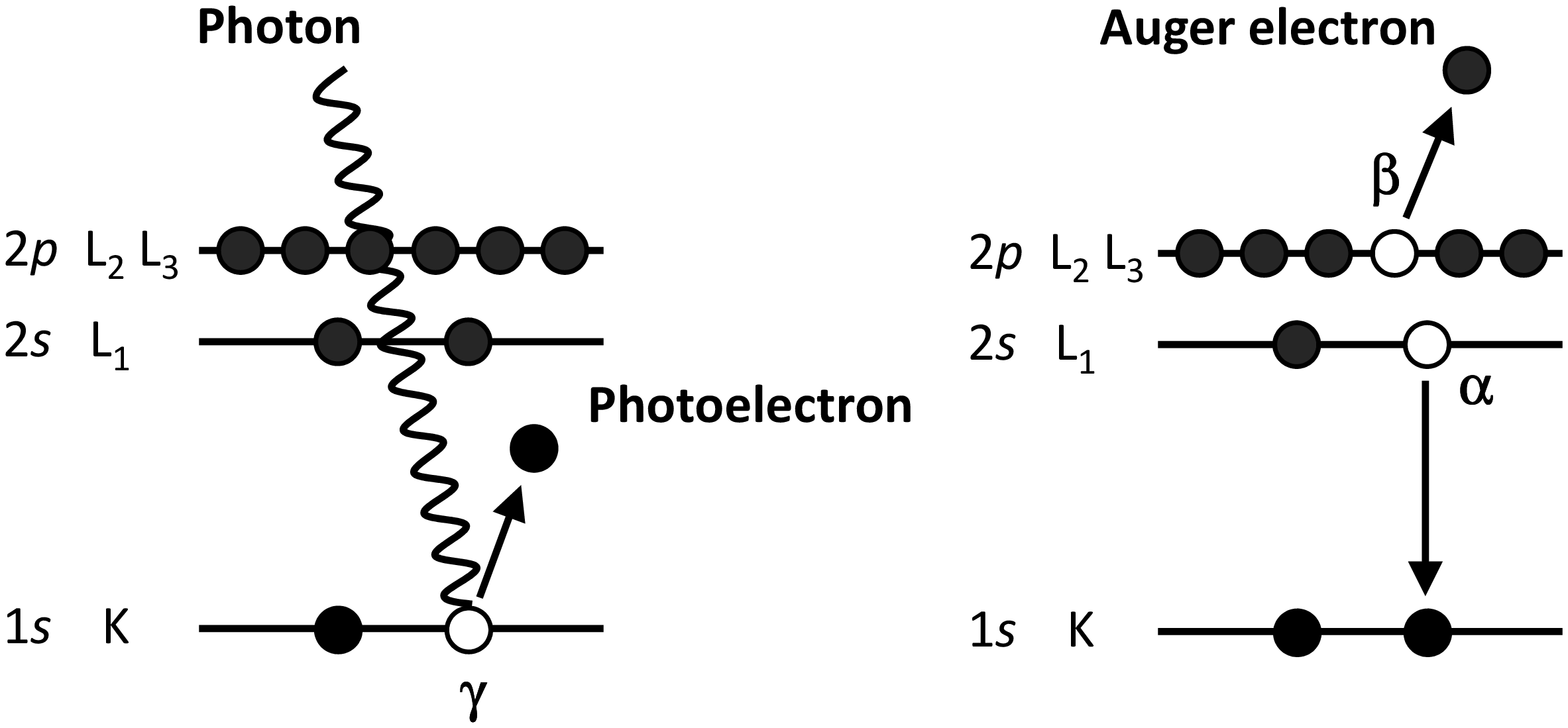}
\vspace{-2mm}
\caption{
Schematic representation of Auger electron knockout by X-ray photons ($q^2=0$) or by virtual photons ($q^2\ne 0$) emitted by electrons irradiating the surface of a sample. In the first stage, the $K$ shell electron is excited to a continuum state (left panel). The resulting vacancy, $\gamma$, is filled with an electron from the $L_1$ shell ($\alpha$), which is accompanied by the radiation of a virtual photon to knock out an electron, $\beta$, from the $L_2$  or $L_3$ shell (right panel). The excitation energy $\varepsilon^*_{\gamma}$ goes to the formation of the vacancies $\alpha$ and $\beta$ and the kinetic energy
$\varepsilon_{\mathrm{kin}}^{\mathrm{A}}$ of the Auger electron. By measuring the energy of the knocked-out electrons, it is possible to determine the peaks at energy values of the corresponding transitions.}
\label{auger}
\end{center}
\end{figure}

%%%%%%%%%%%%%%%%%%%%%%%%%%%%%%%%%%%%%%%%%%%%%%%%%%%%%%%%%%%%%%%%%%%%%%%%%%%%
%%%%%%%%%%%%%%%%%%%%%%%%%%%%%%%%%%%%%%%%%%%%%%%%%%%%%%%%%%%%%%%%%%%%%%%%%%%%
To determine the energy released in the $0\nu2$EC process, it is necessary to know
the energy of the excited electron shell of the neutral daughter atom with two
core-level vacancies and two extra valence electrons inherited from the electron shell of the parent atom.
The binding energy of valence electrons usually does not exceed several eVs,
which is lower than the required accuracy of 10 eV; thus, these two electrons are of no interest.
To estimate the excitation energy, if this simplifies the task, they can be removed from the shell.
The resulting atoms with a charge of +2 can be created in the laboratory
by irradiating the substance composed of these atoms by electrons or X-rays.
Among the knocked-out electrons, one can observe electrons that arise from the so-called Auger process,
schematically shown in Fig.~\ref{auger}.
The narrow structures in the energy distribution of the knocked-out electrons correspond to transitions between
the atomic levels. The study of these structures allows for the estimation of the excitation energy of the electron
shell with two vacancies relevant for the $0\nu2$EC decays.

When the surface of the substance is bombarded with photons or electrons with
energy sufficient for ionization of one of the inner shells of the atom,
a primary vacancy occurs ($\gamma$), as shown in the left panel of Fig.~\ref{auger}.
This vacancy is filled in a short time by an electron from a higher orbit, e.g.,
$L_{1}$, as illustrated in the right panel in Fig.~\ref{auger}.
During the transition to a lower orbit, the electron interacts with the neighboring electrons
via the Coulomb force and transmits to one of them energy sufficient for its knocking out to the continuum state.
The resulting atom has two secondary vacancies, $\alpha $ and $\beta $, plus one ejected Auger electron.
Let $\epsilon _{\gamma }^{\ast }$ be the binding energy of the first knocked-out electron (photoelectron).
The energy of the shell with one vacancy equals $\epsilon _{\gamma }^{\ast }$.
If $\epsilon _{\alpha}^{\ast }$ and $\epsilon _{\beta }^{\ast }$ are the energies of single vacancies,
the energy of the shell with two vacancies is the sum of single excitation energies, $\epsilon _{\alpha }^{\ast}+\epsilon _{\beta }^{\ast }$,
plus the Coulomb interaction of holes, relativistic and relaxation effects,
which we denote by $ \Delta \epsilon_ {\alpha \beta} ^ {\ast} $. The kinetic energy of the photoelectron equals
\begin{equation}
\epsilon _{\mathrm{kin}}^{\mathrm{}}=\omega -\epsilon _{\gamma }^{\ast
}-\phi ,
\end{equation}
where $\omega $ is the photon energy and $\phi $ is the work function. In solid-phase systems $\phi$ is equal to a few eVs
and in vapor-phase systems $\phi = 0$.
The energy of the Auger electron is also determined by conservation of energy:
\begin{equation}
\epsilon _{\mathrm{kin}}^{\mathrm{A}}=\epsilon _{\gamma }^{\ast }-\epsilon
_{\alpha \beta }^{\ast }-\phi ,
\end{equation}
where
\begin{equation} \label{east}
\epsilon _{\alpha \beta }^{\ast }=\epsilon _{\alpha }^{\ast }+\epsilon
_{\beta }^{\ast }+\Delta \epsilon _{\alpha \beta }^{\ast }.
\end{equation}

These equations show that the photoelectron energy spectrum contains information about the single-hole excitation energies,
whereas the Auger electron energy spectrum allows for the measurement of the two-hole excitation energies of electron shells.
These energies occur in the energy balance of the $0\nu $2EC transitions.
We consider the experimental values of excitation energies $\epsilon _{\alpha \beta }^{\ast }$
to estimate the probability of $0\nu2$EC decays as the preferable choice.
As can be seen also from Fig.~\ref{auger}, not all combinations of vacancies are available for the measurement.
Auger electrons are associated with vacancies $n_{\alpha }\geqslant 2$ and $n_{\beta }\geqslant 2$.
When experimental data are not available, we perform calculations using the G{\lowercase{\scshape RASP}}2K package.

%%%%%%%%%%%%%%%%%%%%%%%%%%%%%%%%%%%%%%%%%%%%%%%%%%%%%%%%%%%%%%%%%%%%%%%%%%%%
%%%%%%%%%%%%%%%%%%%%%%%%%%%%%%%%%%%%%%%%%%%%%%%%%%%%%%%%%%%%%%%%%%%%%%%%%%%%
%%%%%%%%%%%%%%%%%%%%%%%%%%%%%%%%%%%%%%%%%%%%%%%%%%%%%%%%%%%%%%%%%%%%%%%%%%%%
%%%%%%%%%%%%%%%%%%%%%%%%%%%%%%%%%%%%%%%%%%%%%%%%%%%%%%%%%%%%%%%%%%%%%%%%%%%%
%%%%%%%%%%%%%%%%%%%%%%%%%%%%%%%%%%%%%%%%%%%%%%%%%%%%%%%%%%%%%%%%%%%%%%%%%%%%
%%%%%%%%%%%%%%%%%%%%%%%%%%%%%%%%%%%%%%%%%%%%%%%%%%%%%%%%%%%%%%%%%%%%%%%%%%%%

\begin{table}[!ht]
\caption{
Total binding energies of neutral noble gas atoms in eVs.
Column 2 shows our calculations using the G{\lowercase{\scshape RASP}}2K package.
The last three columns show the results of \textcite{CCLu1971,DESCLAUX1973,Huang1976}.
}
\label{tab:table24}
\centering
\scriptsize

\begin{tabular}{crrrr}
\hline
\hline
El  &   G{\lowercase{\scshape RASP}}2K &    \textcite{CCLu1971}  &  \textcite{DESCLAUX1973}  &   \textcite{Huang1976}   \\
\hline
   Ne  &      3501.1  &     3472.0 &    3501.4  &    3500.8 \\
   Ar  &     14380.1  &  14072.6   & 14382.3    &   14382.1 \\
   Kr  &     75823.8  &  75739.0   &  75845.8   &   75851.7 \\
   Xe  &    202379.5  & 202402.7   & 202465.3   &  202498.4 \\
   Rn  &    640906.9  & 641899.1   & 641348.1   &  641591.6 \\
\hline
\hline
\end{tabular}
\end{table}
The main question of interest is whether it is possible to calculate the two-hole excitation energy of atoms
with an accuracy of 10 eV,
which is attainable experimentally in Penning-trap mass spectrometry and
is typical of the natural widths of two-hole excitations.

To estimate the magnitude of uncertainty, we first consider the total binding energy of inert gas atoms.
Table~\ref{tab:table24} summarizes the results of our calculations performed using the G{\lowercase{\scshape RASP}}2K software package.
These results are compared to \textcite{CCLu1971}, \textcite{DESCLAUX1973}, and \textcite{Huang1976}.
For light atoms, the variance is negligible.
For the medium-heavy nucleus Xe, the variance does not exceed 120 eV.
For Rn, the variance does not exceed 1 keV; the case of heavy atoms should be treated with caution.

%%%%%%%%%%%%%%%%%%%%%%%%%%%%%%%%%%%%%%%%%%%%%%%%%%%%%%%%%%%%%%%%%%%%%
\begin{table}[h]
\caption{
Single-electron ionization potentials for the noble gas series from Ne to Rn in eVs.
The second and sixth columns list the hole quantum numbers:
$n$ is the principal quantum number, $j$ is the total angular momentum, and $l$ is the orbital momentum.
Columns 3 and 7 present results of our calculations using the G{\lowercase{\scshape RASP}}2K package.
Columns 4 and 8 list the results of \textcite{LARK}.
}
\label{tab:table25}
\centering
\scriptsize

\begin{tabular}{|c|crr|c|crr|}
\hline
\hline
El  &   $n2jl$  & G{\lowercase{\scshape RASP}}2K &  \textcite{LARK} & El  &   $n2jl$  & G{\lowercase{\scshape RASP}}2K &  \textcite{LARK} \\
\hline
 Ne     &$   1    1    0  $&      869.3 &    870.1&        &$   3    1    0  $&     1151.4 &   1148.7 \\
 Ar     &$   1    1    0  $&     3205.8 &   3206.0&        &$   3    1    1  $&     1005.5 &   1002.1 \\
        &$   2    1    0  $&      327.0 &    326.3&        &$   3    3    1  $&      943.3 &    940.6 \\
        &$   2    1    1  $&      250.3 &    250.7&        &$   4    1    0  $&      222.6 &    213.3 \\
        &$   2    3    1  $&      248.2 &    248.6&        &$   4    1    1  $&      169.2 &    145.5 \\
 Kr     &$   1    1    0  $&    14325.9 &  14325.6&        &$   4    3    1  $&      156.8 &    145.5 \\
        &$   2    1    0  $&     1930.7 &   1921.0&Rn      &$   1    1    0  $&    98390.5 &   98397.0\\
        &$   2    1    1  $&     1732.3 &   1727.2&        &$   2    1    0  $&    18061.8 &   18048.0\\
        &$   2    3    1  $&     1679.3 &   1674.9&        &$   2    1    1  $&    17335.6 &   17328.0\\
        &$   3    1    0  $&      295.9 &    292.1&        &$   2    3    1  $&    14614.2 &   14610.0\\
        &$   3    1    1  $&      225.1 &    221.8&        &$   3    1    0  $&     4489.7 &    4473.0\\
        &$   3    3    1  $&      217.0 &    214.5&        &$   3    1    1  $&     4164.4 &    4150.0\\
 Xe     &$   1    1    0  $&    34562.4 &  34564.4&        &$   3    3    1  $&     3542.8 &    3529.0\\
        &$   2    1    0  $&     5458.8 &   5452.8&        &$   4    1    0  $&     1104.9 &    1090.0\\
        &$   2    1    1  $&     5107.5 &   5103.7&        &$   4    1    1  $&      961.3 &     944.0\\
        &$   2    3    1  $&     4786.9 &   4782.2&        &$   4    3    1  $&      803.9 &     790.0\\
\hline
\hline
\end{tabular}
\end{table}
%%%%%%%%%%%%%%%%%%%%%%%%%%%%%%%%%%%%%%%%%%%%%%%%%%%%%%%%%%%%%%%%%%%%%

An estimate of the uncertainties in the energy of single-hole excitations can be obtained from Table~\ref{tab:table25},
where the results of the G{\lowercase{\scshape RASP}}2K package are compared with the data of \textcite{LARK},
\textcolor{black}{
obtained within the framework of a general semi-empirical method that takes into account experimental data on the binding energy of electron subshells
and results of Hartree-Fock atomic calculations.}
For heavy atoms, the mismatch is basically on the order of 10 eV or less, and it is always less than 20 eV.
The claimed accuracy of Larkins' data is a few eVs for light atoms and 10 eV for heavy atoms.

The energies of two-hole excitations of the inert gas atoms of Ne and Ar are collected in Table \ref{tab:table14};
these data can further be supplemented with the results of calculating the energy of two-hole excitations of Kr, Xe and Rn.
The results for $2 \leq n_{\alpha} \leq 3$ and $2 \leq n_{\beta} \leq 3$ are compared with the semi-empirical values
extracted from the energy spectrum of Auger electrons \cite{LARK}.
The second column of these tables lists the total angular momentum $J$ of the two holes;
then, the sums of the energies of the single-particle excitations and the energy of two-hole excitation, determined on the basis of Auger spectroscopy data,
are presented. The next column, 5, reports two-hole excitation energy according to our calculations using the G{\lowercase{\scshape RASP}}2K package.
The last column contains quantum numbers of the pair $(n, 2j, l)$.
\textcolor{black}{We remark that  $n$ and $l$ are integers, $j$ are half-integers, and $2j$ are odd.}
Mixing occurs if some states of the pair arise in two or more combinations.
The mixing matrix is presented in column 6. For mixed states, column 5 lists the energy eigenvalues.
\textcolor{black}{
For example, for the Ar atom, the vacancies
$| (210) (211)\rangle \equiv | 2s_{1/2} 2p_{1/2} \rangle $ and
$| (210) (231)\rangle \equiv | 2s_{1/2} 2p_{3/2} \rangle $ with $J=1$ are mixed, whereby the eigenstates have the form
\begin{eqnarray*}
| J = 1;~631.2 ~\mathrm{eV}\rangle &=& 0.612 | (210) (211)\rangle + 0.791 |(210) (231)\rangle \\
| J = 1;~606.8 ~\mathrm{eV}\rangle &=& 0.791 | (210) (211)\rangle - 0.612 |(210) (231)\rangle .
\end{eqnarray*}}
The energy of these states are 631.2 eV and 606.8 eV, respectively. \textcite{LARK} does not take into account such mixing.
In the above example, the difference between the energies of the holes with and without the mixing is small.
In the case of Xe,
the deviation from the semi-empirical values of \textcite{LARK} does not exceed 10 eV, whereas for Rn, the deviation does not exceed 40 eV.
The deviation is negligible for light atoms.

%%%%%%%%%%%%%%%%%%%%%%%%%%%%%%%%%%%%%%%%%%%%%%%%%%%%%%%%%%%%%%%%%%%%%
\begin{table}[]
\caption{
Double-electron ionization potentials for the noble gas atoms Ne and Ar in eVs. Column 2 lists the total angular momentum of the pair.
Column 3 presents the sum of excitation energies of the single-hole states. The values of $\epsilon_{\alpha}^*$ and $\epsilon_{\beta}^*$ are from \textcite{LARK}.
Column 4 reports the double-electron ionization potentials extracted from the Auger electrons spectroscopy data from \textcite{LARK}. The Auger transitions
allow for the determination of the excitation energies of \textcolor{black}{two-hole states} with $n_{\alpha},n_{\beta} \ge 2$. Column 5 presents the results of our calculations using the G{\lowercase{\scshape RASP}}2K package.
The principal quantum number $n$,  the total angular momentum $j$ and the orbital momentum $l$ of electron holes, $\alpha$ and $\beta$,
are reported in the last columns. Column 6 presents the mixing matrix of \textcolor{black}{two-hole states}, which gives the energy eigenstates with definite $J$.
The energy levels in  columns 3 and 4 neglect mixing and are ordered in the coincidence with column 5, i.e., for a unit-mixing matrix.
}
\label{tab:table14}
\centering
\scriptsize

\begin{tabular}{rcrrrrc}
\hline
\hline
El  &   $J$ &    $\epsilon_{\alpha}^{*}+\epsilon_{\beta}^{*}$  &  $\epsilon_{\alpha\beta}^{\mathrm{A}}$ ~  &   $\epsilon_{\alpha\beta}^{*}$ ~    &            ~~~ $U(\cdot|\alpha\beta)$ ~~                                         &$    (n2jl)_{\alpha}   \;    (n2jl)_{\beta}  $\\
\hline
Ne  &    0 &         1740.2  &               &            1862.1    &             1.000 ~~~                                                         &$    1    1    0   \;    1    1    0  $\\
Ar  &    0 &         6412.0  &               &            6653.5    &             1.000 ~~~                                                         &$    1    1    0   \;    1    1    0  $\\
    &    0 &         3532.3  &               &            3588.7    &             1.000 ~~~                                                         &$    1    1    0   \;    2    1    0  $\\
    &    1 &         3532.3  &               &            3579.8    &             1.000 ~~~                                                         &$    1    1    0   \;    2    1    0  $\\
    &    0 &         3456.7  &               &            3513.3    &             1.000 ~~~                                                         &$    1    1    0   \;    2    1    1  $\\
    &    1 &         3456.7  &               &            3522.0    &             0.669            ~0.743                                       &$    1    1    0   \;    2    1    1  $\\
    &      &         3454.6  &               &            3511.9    &             0.743            -0.669                                       &$    1    1    0   \;    2    3    1  $\\
    &    2 &         3454.6  &               &            3510.1    &             1.000 ~~~                                                         &$    1    1    0   \;    2    3    1  $\\
    &    0 &          652.6  &          695.4&             695.0    &             1.000 ~~~                                                         &$    2    1    0   \;    2    1    0  $\\
    &    0 &          577.0  &          606.7&             607.6    &             1.000 ~~~                                                         &$    2    1    0   \;    2    1    1  $\\
    &    1 &          577.0  &          630.0&             631.2    &             0.612            ~0.791                                       &$    2    1    0   \;    2    1    1  $\\
    &      &          574.9  &          605.9&             606.8    &             0.791            -0.612                                       &$    2    1    0   \;    2    3    1  $\\
    &    2 &          574.9  &          604.6&             605.3    &             1.000 ~~~                                                         &$    2    1    0   \;    2    3    1  $\\
    &    0 &          501.4  &          553.8&             557.0    &             0.669            ~0.743                                       &$    2    1    1   \;    2    1    1  $\\
    &      &          497.2  &          539.3&             538.5    &             0.743            -0.669                                       &$    2    3    1   \;    2    3    1  $\\
    &    1 &          499.3  &          538.9&             538.0    &             1.000 ~~~                                                         &$    2    1    1   \;    2    3    1  $\\
    &    2 &          499.3  &          545.8&             545.1    &             0.881            -0.474                                       &$    2    1    1   \;    2    3    1  $\\
    &      &          497.2  &          537.4&             536.3    &             0.474            ~0.881                                       &$    2    3    1   \;    2    3    1  $\\
\hline
\hline
\end{tabular}
\end{table}
%%%%%%%%%%%%%%%%%%%%%%%%%%%%%%%%%%%%%%%%%%%%%%%%%%%%%%%%%%%%%%%%%%%%%

The energies of the electron shells of noble gas atoms with double-$K$ holes are calculated by \textcite{Niskanen2011}.
We obtain good agreement for Ne, Ar, Kr, and Xe, but
the results are noticeably different for Rn. \textcite{Niskanen2011} give $\epsilon_{{KK}}^*$ = 198912.6 eV,
whereas our calculations yield the value of 198568.3 eV, which is 344 eV less.
In the paper of \textcite{Niskanen2011}, QED corrections due to the electron self-energy and the vacuum polarization
are not taken into account. If these effects are neglected in our calculations, then a value of 198977.8 eV is obtained;
this value differs by only 65 eV from the result of \textcite{Niskanen2011}.
The uncertainty in two-hole excitation energies of heavy atoms is thus found to be about 60 eV,
which is higher than the $0\nu2$EC target of 10 eV.

The $0\nu2$EC half-lives are estimated in Sec.~V and Sec.~VIII by neglecting the mixing. The two-hole excitation energies of atoms with arbitrary $Z$
are determined using the data for Ne, Ar (Table \ref{tab:table14}), and also Kr, Xe and Rn
by means of interpolation $\epsilon _{\alpha \beta }^{\ast } = aZ^b$ between the neighboring noble gas atoms
with the same vacancies.

\subsection{Section summary}

The probability of capture of orbital electrons by the nucleus depends on the value of the electron wave functions inside the nucleus.
The values of upper and lower components of the Dirac electron wave functions of neutral atoms inside the nuclei are tabulated by \textcite{Band1986}, where the electron screening is accounted for using the relativistic Dirac-Fock-Slater and Dirac-Fock potentials. The $ns_{1/2}$ waves are dominant. The $np_{1/2}$ waves can be found to be enhanced compared to the $np_{3/2}$ waves. The nonrelativistic solutions of the $np_{3/2}$ waves are close to the relativistic ones (see also \cite{Grant2007}, Fig. 1.2). The electron capture from the $np_{3/2}$ orbits is therefore suppressed. The results of \textcite{Band1986} are in good agreement with results provided by the G{\lowercase{\scshape RASP}}2K software package.

The resonant enhancement of the $0\nu$2EC decays occurs when the excitation energies
of the parent and intermediate daughter atoms are degenerate with an accuracy of about 10 eV.
This scale characterizes the typical excitation width of the atomic shells.
Accuracy of about 10 eV is achievable on Penning traps when measuring mass difference of ionized atoms.
To identify the resonant $0\nu$2EC with the same high accuracy, information about double-hole excitations
of electron shells is required.

The two-hole interactions provide a dominant contribution to the energy of excited electron
shells but not the only contribution. Open vacancies in the occupation numbers affect the energy of all atomic levels.
The G{\lowercase{\scshape RASP}}2K package calculates the structure of electron shells
based on the Green's function method of QED, which offers a simple and clear description of various approximations.

Accuracy of up to 10 eV is readily achievable when determining theoretically single-hole excitation energies.
We demonstrated independence on the gauge of single-hole excitation energies in the order $\alpha^2$Ry.
One of the challenges of the atomic theory is the proof of gauge invariance of QED of multielectron atoms in all orders of perturbation theory.
Double-hole excitation energies can be determined theoretically with an accuracy of 60 eV or better.
The upper bound of the possible error is still higher than that required to identify the resonant $0\nu$2EC unambiguously. In calculations
with atomic shell structure, the uncertainties are associated with complexity of the bound-state problem for multielectron atoms. In many cases,
the resonant parameter of $0\nu$2EC can be extracted from the experimental data on Auger spectroscopy.
When normalization to the experimental values is not possible, quantum chemistry codes such as G{\lowercase{\scshape RASP}}2K can be used
to get the missing information.

%%%%%%%%%%%%%%%%%%%%%%%%%%%%%%%%%%%%%%%%%%%%%%%%%%%%%%%%%%%%%%%%%%%%%%%%%%%%
\section{Nuclear matrix elements}
%%%%%%%%%%%%%%%%%%%%%%%%%%%%%%%%%%%%%%%%%%%%%%%%%%%%%%%
\label{sec:NME}
%%%%%%%%%%%%%%%%%%%%%
\setcounter{equation}{0}

\textcolor{black}{In this section we describe how nuclear structure affects the half-life of the $0\nu$2EC process and review the available
calculations of the related NMEs. We also add new NME calculations in order to complement the list of the evaluated $0\nu$2EC cases.}

%\subsection{Survey of the calculated near-resonant neutrinoless 2EC half-lives \label{subsec:survey}}
\subsection{Overview of the calculated nuclear matrix elements in 0$\nu$2EC \label{subsec:survey}}

Nuclear structure is heavily involved in the decay amplitude (\ref{LNVP2})
through the appropriate nuclear matrix elements (NMEs) \cite{Suhonen2012a}. These
NMEs have been computed in various theory frameworks as described below.
A representative list of the calculations is displayed in Table~\ref{tab:sNMEs}.
In the table the available estimated lowest and highest limits for the
near-resonant $0\nu2$EC half-lives (last two columns) are listed. In the evaluation of the
half-lives Eq.~(\ref{T12abs}) was adopted and the NMEs $M^{0\nu}$ from column 4 of Table~\ref{tab:NME} were used, in
addition to the degeneracy parameters taken from Tables~\ref{tab:ground_states} and
\ref{tab:excited_states}. Also the theory frameworks used to derive these NMEs are
given (column 5), along with the corresponding reference (column 6).
The $Q_{\mathrm{2EC}}$-value measurements for the evaluation of the degeneracy parameters
have been performed by the use of modern Penning-trap techniques (see Sec.~VI).

In practically all the listed cases the decay rates are suppressed by the rather sizable
magnitude of the ratio $(\Delta/\Gamma_{\alpha \beta})^2$,
where $\Gamma_{\alpha \beta} \sim 10$ eV is the typical de-excitation width of the excited electron shells
with the electron vacancies $\alpha$ and $\beta$.
Decays to $0^+$ states are favored over the decays to $2^+$ or $1^-,2^-,3^-$ etc. states due to
the involved nuclear wave functions and/or higher-order transitions.
\textcolor{black}{A further suppression stems from nuclear deformation. This
suppression is typically a few tens of percent (Ejiri, 2019) but can be
even stronger, factors of 2-3, for large deformations \cite{Delion:2017}.}
The radial wave functions of electrons in low-lying atomic states on the surface of medium-heavy nuclei
are from \textcite{Band1986} .
In relativistic theory, the electron capture from the $ns_{1/2}$ states is dominant,
the amplitude of electron capture from the $np_{1/2}$ states is suppressed by about an order of magnitude,
while the amplitude of electron capture from the $j \geq 3/2$ state appears to be suppressed by several orders of magnitude
\cite{Kolhinen2010,KRIV11}.

There are some favorable values of degeneracy parameters listed
in Table~\ref{tab:excited_states}, like for the transitions
$^{106}\textrm{Cd}\to\,^{106}\textrm{Pd}((2,3)^-)$ and
$^{156}\textrm{Dy}\to\,^{156}\textrm{Gd}(1^-,2^+)$ but the
associated nuclear matrix elements are not yet evaluated.
At the moment the most favorable case with a half-life estimate of
$\gtrsim 5 \times 10^{28}$ years is the case
$^{152}\textrm{Gd}\to\,^{152}\textrm{Sm}(0^+_{\rm gs})$
which corresponds to a decay transition to the ground state.

\textcolor{black}{All the presently identified favorable $0\nu$2EC cases are in the regions of relatively strong
nuclear deformation so that a proper handling of this degree of freedom
poses a challenge to the nuclear-theory frameworks.}

%\subsection{Outline of the calculation frameworks \label{subsec:frameworks}}
%\subsection{Overview of the Basic calculation schemes \label{subsec:frameworks}}
\subsection{Overview of the calculation frameworks \label{subsec:frameworks}}

In the analyses of the near-resonant $0\nu2$EC decay transitions the adopted many-body
frameworks include the quasiparticle random-phase approximation (QRPA) in its
higher-QRPA versions: multiple-commutator model (MCM), the deformed QRPA,
the microscopic interacting boson model (IBM-2) and the energy-density functional
(EDF) method. The MCM and deformed QRPA frameworks compute the $0\nu2$EC-decay NME
explicitly, including the contributions by the virtual states of the intermediate
nucleus. The other two models, IBM-2 and EDF, resort to the closure approximation
where the sum over the intermediate states, with the appropriate energy denominator,
has been suppressed by assuming an average excitation energy in the denominator and
then using the closure over the complete set of intermediate virtual states.
All these models are briefly described below.

\begin{table}[H]
\vspace{2mm} \caption{The ``sQRPA'' and ``dQRPA'' of column 5
denote the spherical pnQRPA and deformed QRPA outlined in the
beginning of Sec.~V.B.1 and in Sec.~V.B.2 , respectively.
Furthermore, the symbol ``IBM-2'' denotes the microscopic
interacting boson model of Sec.~V.B.3 and the symbol ``EDF''
denotes the energy-density functional method of Sec.~V.B.4. In the
case of the multiple-commutator model (MCM of Sec.~V.B.1) we have
chosen to quote the results obtained by the use of the UCOM
short-range correlations (UCOM s.r.c.) which are the very
realistic ones \cite{Kortelainen2007}. The UCOM s.r.c. have also
been used in the sQRPA, dQRPA and EDF calculations. The ``qp
estimate'' in the fifth column denotes the procedure outlined in
Sec.~V.C. The last two columns give the minimum and maximum
half-lives (in years) calculated using Eq.~(\ref{T12abs}) with \textcolor{black}{
$K_Z$ of Eq.~(\ref{OTA}),} $m_{\beta\beta}=100$~meV and $g_A=1.27$. The excitation energies
are given in keV.} \label{tab:sNMEs}
\label{tab:NME}
\centering
\renewcommand{\arraystretch}{1.2}
%\addtolength{\tabcolsep}{-2.2pt}
\scriptsize
\resizebox{0.8541\textwidth}{!}
{\begin{tabular}{|c|c|c|c|c|c|c|c|c|c|}
\hline \hline
Transition                               & $J^{\pi}_f$ & $M^*_{A,Z-2} - M_{A,Z - 2}$ & $M^{0\nu}$ & Model     & Ref.
& $T_{1/2}^{\min}$ & $T_{1/2}^{\max}$\\
\hline
\hline
$^{\;\,74}_{\;\,34}$Se$\to ^{\;\,74}_{\;\,32}$Ge$^{**}$&  2$^{+}_2$    & $1204.205 \pm 0.007$          & $3.22 \times 10^{-4}$& MCM  & \textcite{Kolhinen2010}  & $1.2\times10^{46}$ & $1.2\times10^{46}$ \\
$^{\;\,96}_{\;\,44}$Ru$ \to ^{\;\,96}_{\;\,42}$Mo$^{**}$& [0$^+$]        & $2712.68 \pm 0.10$          & 5.57        & MCM  & \textcite{Suhonen2012}  &
$2.2\times10^{32}$ & $6.8\times10^{32}$ \\
$^{106}_{\;\,48}$Cd$ \to ^{106}_{\;\,44}$Pd$^{**}$& [0$^+$]              & $2717.59 \pm 0.21$          & 3.38 - 3.48 & MCM  & \textcite{Suhonen2011}  &
$5.3\times10^{31}$ & $7.0\times10^{31}$ \\
$^{112}_{~50}$Sn$ \to ^{112}_{~48}$Cd$^{**}$  &   (0$^{+ }$)         & $2988  \pm   8    $             & 4.76         &  estimate   & \textcite{Rahaman2009} & $3.5\times10^{35}$ & $3.8\times10^{35}$ \\
$^{124}_{\;\,54}$Xe$ \to ^{124}_{\;\,52}$Te$^{**}$ & 0$^+_4$            & $2153.29 \pm 0.03  $          & 1.11 - 1.30  & MCM  & \textcite{Suhonen2013}  & $7.3\times10^{35}$ & $7.3\times10^{35}$ \\
                                               &                    &                               & 0.297        & IBM-2 & \textcite{Kotila2014}  &
$1.2\times10^{37}$ & $1.2\times10^{37}$ \\
$^{136}_{\;\,58}$Ce$ \to ^{136}_{\;\,56}$Ba$^{**}$& 0$^+_4$            & $2315.32 \pm0.07   $          & 0.68         & MCM   & \textcite{136Ce}  &
$3.3\times10^{32}$ & $4.3\times10^{32}$ \\
$^{152}_{\;\,64}$Gd$ \to ^{152}_{\;\,62}$Sm$^{*}$  &   0$^{+}_1$          & 0                             & 7.21 - 7.59         & sQRPA & \textcite{SKF11}  & $6.8\times10^{27}$ & $6.8\times10^{28}$  \\
                                             &                      &                               & 2.67 - 3.23         & dQRPA & \textcite{Fang2012}  &
$4.3\times10^{28}$ & $4.3\times10^{29}$ \\
                                             &                      &                               & 2.445               & IBM-2 & \textcite{Kotila2014} & $6.0\times10^{28}$ & $6.0\times10^{29}$  \\
                                             &                      &                               & 0.89 - 1.07           & EDF   & \textcite{Rodriguez2010} & $3.8\times10^{29}$ & $3.8\times10^{30}$   \\
$^{156}_{\;\,66}$Dy$ \to ^{156}_{\;\,64}$Gd$^{**}$ & 0$^+_1$            & 0                             & 3.175        & IBM-2 & \textcite{Kotila2014}  & $7.3\times10^{34}$ & $7.3\times10^{34}$ \\
                                             & 0$^+_2$              & $1049.487 \pm 0.002$          & 1.749        & IBM-2 & \textcite{Kotila2014}  &
$4.8\times10^{34}$ & $4.8\times10^{34}$ \\
                                             & 0$^+_3$              & $1168.186 \pm 0.007$          & 0.466        & IBM-2 & \textcite{Kotila2014}  &
$5.0\times10^{35}$ & $5.0\times10^{35}$ \\
                                             & 0$^+_4$              & $1715.192 \pm 0.005$          & 0.311        & IBM-2 & \textcite{Kotila2014}  &
$7.5\times10^{34}$ & $7.5\times10^{34}$ \\
                                             & 0$^+_5$              & $1851.239 \pm 0.007$          & 0.346        & IBM-2 & \textcite{Kotila2014}  &
$4.8\times10^{33}$ & $4.8\times10^{33}$ \\
                                             & 0$^+_6$              & $1988.5\pm   0.2$          & 0.3             & estimate & This work  &
$1.1\times10^{28}$ & $9.5\times10^{32}$ \\
$^{164}_{\;\,68}$Er$ \to ^{164}_{\;\,66}$Dy$^{*}$  &   0$^{+}_1$          & 0                             & 5.94 - 6.12  & sQRPA & \textcite{SKF11}  &
$3.3\times10^{31}$ & $3.5\times10^{31}$ \\
                                             &                      &                               & 2.27 - 2.64  & dQRPA & \textcite{Fang2012}  &
$2.0\times10^{32}$ & $2.1\times10^{32}$ \\
                                             &                      &                               & 3.952        & IBM-2 & \textcite{Kotila2014}  &
$7.8\times10^{31}$ & $8.0\times10^{31}$ \\
                                             &                      &                               & 0.50 - 0.64  & EDF   & \textcite{Rodriguez2010}  &
$3.8\times10^{33}$ & $4.0\times10^{33}$ \\
$^{180}_{\;\,74}$W$ \to ^{180}_{\;\,72}$Hf$^{*}$   &   0$^{+}_1$          & 0
& 5.56 - 5.79  & sQRPA & \textcite{SKF11}  &
$1.4\times10^{29}$ & $1.8\times10^{29}$ \\
                                             &                      &
& 1.79 - 2.05  & dQRPA & \textcite{Fang2012}  &
$1.2\times10^{30}$ & $1.6\times10^{30}$ \\
                                             &                      &
& 4.672        & IBM-2 & \textcite{Kotila2014}  &
$2.1\times10^{29}$ & $2.8\times10^{29}$ \\
                                             &                      &                               & 0.38 - 0.58    & EDF   & \textcite{Rodriguez2010}  &
$2.0\times10^{31}$ & $2.5\times10^{31}$ \\
$^{184}_{\;\,76}$Os$\to ^{184}_{\;\,74}$W$^{**}$&   0$^{+}_2$          & $1002.48  \pm 0.04 $          & 0.631        & EDF   & \textcite{184Os}  &
$6.8\times10^{33}$ & $6.8\times10^{33}$ \\
                                             &   (0)$^{+}_3$          & $1322.152 \pm 0.022$          & 0.504        & EDF   & \textcite{184Os}  &
$6.0\times10^{30}$ & $1.2\times10^{31}$ \\
                                             &   2$^{+}_5$          & $1431.02 \pm 0.05 $           & 0.14        & MCM  & This work  &
$4.5\times10^{32}$ & $4.5\times10^{37}$ \\
                                             &   (0)$^{+}_5$          & $1713.47  \pm 0.10 $          & 0.163        & EDF   & \textcite{184Os}  &
$1.7\times10^{35}$ & $1.7\times10^{35}$ \\
%\hline
$^{190}_{\;\,76}$Pt$\to ^{190}_{\;\,74}$Os$^{**}$&   1$^{+}$                & $1326.9 \pm 0.5 $        & 1.1         & MCM  & This work  &
$3.3\times10^{26}$ & $1.6\times10^{30}$ \\
                                                 &   $(0_3)^{+}$            & $1382.4 \pm 0.2 $        & 4.7         & MCM  & This work  &
$1.0\times10^{30}$ & $6.5\times10^{30}$ \\
\hline
\hline
$^{148}_{\;\,64}$Gd$ \to ^{148}_{\;\,62}$Sm$^{**}$&    [0$^+$]           & $3004 \pm 3$                  & 0.071         & qp estimate   & This work  & $5.3\times10^{28}$ & $1.5\times10^{35}$ \\
                                               &    [1$^+$]           &                               & 0.031        & qp estimate  & This work  &
$8.3\times10^{28}$ & $2.5\times10^{35}$ \\
%                                               &    2$^+$           &                               & 0.002        & qp estimate  & This work  & $1.7\times10^{29}$ & $1.7\times10^{30}$ \\
$^{150}_{\;\,64}$Gd$ \to ^{150}_{\;\,62}$Sm$^{**}$&    0$^+$           & $1255.400 \pm 0.022$          & 0.066         & qp estimate  & This work  &
$4.0\times10^{29}$ & $1.6\times10^{36}$ \\
%                                               &    1$^+$           &                               & 0.029        & qp estimate  & This work  & $1.7\times10^{29}$ & $1.7\times10^{30}$ \\
%                                               &    2$^+$           &                               & 0.002        & qp estimate  & This work  & $1.7\times10^{29}$ & $1.7\times10^{30}$ \\
$^{154}_{\;\,66}$Dy$ \to ^{154}_{\;\,64}$Gd$^{**}$&    [0$^+$]           &  $3153.1 $               & 0.068        & qp estimate  & This work  &
$6.8\times10^{34}$ & $2.8\times10^{35}$ \\
                                                  &    [0$^+$]           &  $3154.8 \pm 0.4$        & 0.068        & qp estimate  & This work  &
$6.3\times10^{34}$ & $2.8\times10^{35}$ \\
                                               &    (1$^+$)           &  $3264.42 \pm 0.21$           & 0.030        & qp estimate  & This work  &
$7.5\times10^{28}$ & $3.8\times10^{35}$ \\
%                                               &    2$^+$           &                               & 0.002        & qp estimate  & This work  & $1.7\times10^{29}$ & $1.7\times10^{30}$ \\
$^{194}_{\;\,80}$Hg$ \to ^{194}_{\;\,78}$Pt$^{**}$&    (0$^+$)           & $2450 \pm 5$                  & 0.017        & qp estimate  & This work  & $2.5\times10^{29}$ & $1.2\times10^{36}$ \\
                                                  &    (1$^+$)           &                               & 0.004        & qp estimate  & This work  & $1.4\times10^{30}$ & $6.3\times10^{36}$ \\
                                                  &    (0$^+$)           & $2472 \pm 5$                  & 0.017        & qp estimate  & This work  & $2.5\times10^{29}$ & $1.2\times10^{36}$ \\
                                                  &    (1$^+$)           &                               & 0.004        & qp estimate  & This work  & $1.4\times10^{30}$ & $2.8\times10^{36}$ \\
                                                  &    [1$^+$]           &  $2500 \pm 10$                & 0.004        & qp estimate  & This work  & $1.4\times10^{30}$ & $6.5\times10^{36}$ \\
$^{202}_{\;\,82}$Pb$ \to ^{202}_{\;\,78}$Hg$^{**}$&    0$^+_2$         &  $1411.37 \pm 0.12$           & 0.011        & qp estimate  & This work  &
$1.7\times10^{30}$ & $1.6\times10^{37}$ \\
                                                  &    (1$^+)$           &  $1347.92 \pm 0.07$           & 0.003        & qp estimate  & This work  & $4.0\times10^{33}$ & $4.8\times10^{36}$ \\
%                                               &    2$^+$           &                               & - 0.002        & qp estimate  & This work  & $1.7\times10^{29}$ & $1.7\times10^{30}$ \\
\hline \hline
 \end{tabular}}
 \centering
 \end{table}

The NME for the near-resonant $0\nu2$EC decay to the $0^+$ final states
is written as a linear combination of the
Gamow-Teller (GT), Fermi (F) and tensor (T) NMEs as
\begin{equation} \label{eq:0nume}
M^{\rm 2EC} = M_{\rm GT}^{\rm 2EC} - \left( \frac{g_{\rm V}}{g_{\rm A}}
\right)^{2} M_{\rm F}^{\rm 2EC}+ M_{\rm T}^{\rm 2EC} \, .
\end{equation}
In the MCM and deformed QRPA frameworks the transitions through the virtual states
of the intermediate nucleus are treated explicitly. Then
the double Fermi, Gamow--Teller, and tensor nuclear matrix
elements can be written as
\begin{align} \label{eq:bbF}
M_{\rm F}^{\rm 2EC} & =  \sum_{k} (J^{+}_{f} || \sum_{mn}h_{\rm F}(r_{mn},E_{k}) || 0^{+}_{i}) \,,  \\
\label{eq:bbGT}
M_{\rm GT}^{\rm 2EC} & =  \sum_{k} (J^{+}_{f} || \sum_{mn}h_{\rm GT}(r_{mn},E_{k})
( \mbox{\boldmath{$\sigma$}}_{m}\cdot\mbox{\boldmath{$\sigma$}}_{n} ) || 0^{+}_{i}) \,, \\
\label{eq:bbT}
M_{\rm T}^{\rm 2EC} & =  \sum_{k} (J^{+}_{f} || \sum_{mn}h_{\rm T}(r_{mn},E_{k})
S^{\rm T}_{mn} || 0^{+}_{i}) \,,
\end{align}
where the tensor operator reads
\begin{equation} \label{eq:S-T}
S^{\rm T}_{mn} = 3[(\mbox{\boldmath{$\sigma$}}_{m}
\cdot\hat{\mathbf{r}}_{mn})(\mbox{\boldmath{$\sigma$}}_{n}
\cdot\hat{\mathbf{r}}_{mn})] -  \mbox{\boldmath{$\sigma$}}_{m}
\cdot\mbox{\boldmath{$\sigma$}}_{n} \,.
\end{equation}
The summations over $k$ in Eqs. (\ref{eq:bbF}), (\ref{eq:bbGT}) and
(\ref{eq:bbT}) run over all the states of the intermediate odd-odd nucleus,
$r_{mn}=\vert\mathbf{r}_m-\mathbf{r}_n\vert$ is the relative
distance between the two decaying protons, labeled $m$ and $n$, and
$\hat{\mathbf{r}}_{mn}=(\mathbf{r}_m-\mathbf{r}_n)/r_{mn}$. The neutrino potentials
$h_{\rm K}(r_{mn},E_{k})$, $K=\textrm{F,~GT,~T}$, are given by \textcite{Suhonen2012a}.
The ground state of the initial even-even nucleus is denoted by $0^{+}_{i}$ and the
positive-parity final state in the daughter even-even nucleus is denoted by $J^{+}_{f}$.
In the closure approximation the intermediate energies $E_k$ in the above equations
are replaced by one single energy $E$ and the summation over $k$ is replaced by
a unit operator.

In general, for the near-resonant $0\nu2$EC decay to the final
$J^{+}_{f}=0^+,1^+,2^+$ states the NMEs
can be written in the QRPA framework in the form
\begin{align} \label{eq:NMEs}
M_{K}^{\rm 2EC}(0^+_i\to J_f^+) & = (-1)^{J_f}
\sum_{J^{\pi},k_{1},k_{2}}\sum_{J_1,J',J''} \sum_{pp'nn'}
\sqrt{\frac{[J'][J''][J_1]}{[J_f]}} \nonumber \\
&  \times \left\{ \begin{array}{ccc} j_{n} & j_{p} & J \\
j_{n'} & j_{p'} & J_1 \\ J'' & J' & J_f \end{array} \right\}
( nn':J'' \vert\vert  {\mathcal O}_K^{(J_f)} \vert\vert pp':J' )
\\ &  \times
( J^{+}_{f} \vert \vert  \big[ c^{\dag}_{n'}
\tilde{c}_{p'}\big]_{J_1} \vert \vert J^{\pi}_{k_{1}} )
\langle J^{\pi}_{k_{1}} \vert J^{\pi}_{k_{2}} \rangle
( J^{\pi}_{k_{2}} \vert \vert  \big[ c^{\dag}_{n}
\tilde{c}_{p}\big]_J \vert \vert 0^{+}_{i}) \, , \nonumber
\end{align}
where $[x] = 2x+1$ and
$k_1$ and $k_2$ label the different QRPA solutions for a given multipole
$J^{\pi}$, stemming from the parent and daughter nuclei of the near-resonant $0\nu2$EC decay.
The operators ${\mathcal O}_K^{(0_f)}$ for the $0^+$ final states
in the reduced two-particle matrix element
denote the Fermi (K=F), Gamow-Teller (K=GT) and
tensor (K=T) parts of the double-beta operator, given in
Eqs.~(\ref{eq:bbF})--(\ref{eq:bbT}). In all the discussed theory frameworks
the two-particle matrix element contains also the appropriate short-range correlations,
higher-order nucleonic weak currents and nucleon form factors, as
given by \textcite{Simkovic:1999re}. The last line of Eq.~(\ref{eq:NMEs}) contains
the one-body transition densities between
the initial/final ground state ($0^+_i$/$0^+_f$) and the intermediate states
$J^{\pi}_{k}$, and they can be obtained in the QRPA framework as discussed below.
The term between the one-body transition densities is the overlap between the two
sets of intermediate states emerging from the two QRPA calculations based on the
parent and daughter even-even ground states and its expression for the spherical
nuclei has been given by \textcite{Suhonen2012a} and for deformed nuclei by
\textcite{Simkovic2004}.

Here it should be noted that typically only the $J_f^{+}=0^+$ final states have
been considered in the near-resonant $0\nu2$EC-decay calculations, as given by
Eqs.~(\ref{eq:bbF}) - (\ref{eq:bbT}) and the $J_f=0$ special case of Eq.~(\ref{eq:NMEs}).
The simplest procedure to reach the positive-parity $J_f^{+}=1^+,2^+$ states is to
use a generalized GT-type of operator:
\begin{equation}
M_{\rm GT}^{\rm 2EC}(J_f^+=1^+,2^+) =  \sum_{k} (J_{f}^+ || \sum_{mn}h_{\rm GT}(r_{mn},E_{k})
\big\lbrack\mbox{\boldmath{$\sigma$}}_{m}\mbox{\boldmath{$\sigma$}}_{n}
\big\rbrack_{J_f} || 0^{+}_{i}) \,,
\label{eq:ECEC-T}
\end{equation}
together with the expression (\ref{eq:NMEs}).
In Sec.~\ref{subsec:simple} we compute this NME for several cases of interest
in an \textcolor{black}{approximate} way avoiding
the vast complications involved in the use of detailed nuclear wave functions for
high-excited states in heavy daughter nuclei of the near-resonant $0\nu2$EC processes.
Furthermore, to reach the negative-parity states $J_f^{-}=0^-,1^-,2^-$ one
would need more complex nuclear transition operators and these are not thoroughly
examined yet \cite{Vergados2011}. In this work we then skip the estimation of
the order of magnitude of the related NMEs.

\textcolor{black}{
Here it has to be remarked that in the very recent studies \cite{Cirigliano:2018hja} and \cite{Cirigliano:2019vdj} it was found that in addition to the long-range NME \ref{eq:0nume} there is a notable contribution from a short-range operator affecting the Fermi part of the NME. According to preliminary studies in these works for very light nuclei the value of the NME of the neutrinoless double beta decay could change considerably by the inclusion of the new short-range term. It remains to be seen how strong is the effect for the medium-heavy and heavy nuclei which actually double beta decay.
}

\subsubsection{Multiple-commutator model \label{subsubsec:MCM}}

The nuclear states of odd-odd nuclei can be described within the
spherical proton-neutron quasiparticle random-phase approximation (pnQRPA)
framework. The solution of the pnQRPA
equations can be written as (see, e.g. \cite{Suhonen2007})
\begin{align} \label{eq:pnQRPA-st}
\vert \omega  \,M\rangle &= q^{\dagger}(\omega,M)\vert\textrm{QRPA} \rangle \nonumber \\
&= \sum_{pn} \left(X_{pn}^{\omega}
\left[a_p^{\dagger}a_{n}^{\dagger}\right]_{JM}-Y_{pn}^{\omega}
\left[a_p^{\dagger}a_n^{\dagger}\right]_{JM}^{\dagger}\right)\vert\textrm{QRPA} \rangle \,,
\end{align}
where the shorthand $\omega = J^{\pi}_k$ for the $k^{th}$ intermediate state of
spin-parity $J^{\pi}$ has been used. Here $\vert\textrm{QRPA} \rangle$ is the QRPA
ground state and the operator $a_p^{\dagger}$ creates a proton quasiparticle on the
single-particle orbital $p=(n_p,l_p,j_p)$, where $n$ is the principal, $l$ the orbital
angular-momentum and $j$ the total angular-momentum quantum number. The operator $a_p$
is the corresponding annihilation operator and a similar definition applies for the
neutrons $n$. The single-particle orbitals are obtained by solving the Schr\"odinger
equation for a spherical Woods-Saxon mean-field potential \cite{Suhonen2007}.
By using this wave function one can obtain the transition densities
\begin{align}
( 0^{+}_{\rm gs} \vert \vert  [ c^{\dag}_{n'}\tilde{c}_{p'}]_J \vert
\vert J^{\pi}_{k_{1}} ) &=
\sqrt{[J]}\left[v_{n'}u_{p'}X^{J^{\pi}k_1}_{p'n'} +
u_{n'}v_{p'}Y^{J^{\pi}k_1}_{p'n'}\right] \ , \label{eq:density1} \\
( J^{\pi}_{k_{2}} \vert \vert  [ c^{\dag}_{n}\tilde{c}_{p}]_J \vert \vert 0^{+}_{i}) &=
\sqrt{[J]}\left[\tilde{u}_{n}\tilde{v}_{p}\tilde{X}^{J^{\pi}k_2}_{pn} +
\tilde{v}_{n}\tilde{u}_{p}\tilde{Y}^{J^{\pi}k_2}_{pn}\right] \, ,
\label{eq:density2}
\end{align}
where $v$ ($\tilde{v}$) and $u$ ($\tilde{u}$) correspond to the BCS occupation
and vacancy amplitudes of the final (initial) even-even nucleus. The
amplitudes $X$ and $Y$ ($\tilde{X}$ and $\tilde{Y}$) come from the pnQRPA
calculation starting from the final (initial) nucleus of the double-beta decay.
Here the initial and final states of the near-resonant $0\nu2$EC decay are assumed to be the
ground states of the even-even mother and daughter nuclei.

The $n^{th}$ excited $\omega_f=I^{\pi}_n$ state, where $I$ is the angular momentum of the
state, in the even-even daughter
nucleus is described in the QRPA formalism, and the corresponding wave function
can be presented as (see \cite{Suhonen2007})
\begin{align} \label{eq:ccQRPA-ph}
\vert I^{\pi}_n \,M_I\rangle &= Q^{\dagger}(I^{\pi}_n,M_I)\vert\textrm{QRPA} \rangle \nonumber \\
&= \sum_{a\le b}\big[X_{ab}^{\omega_f}A^{\dagger}_{ab}(IM_I) -
Y_{ab}^{\omega_f}\tilde{A}_{ab}(IM_I)\big]\vert\textrm{QRPA} \rangle \,,
\end{align}
where the normalized two-quasiparticle operators are defined as
\begin{align} \label{eq:A-dagger}
&A^{\dagger}_{ab}(IM_I) = \mathcal{N}_{ab}(I)\left[a^{\dagger}_aa^{\dagger}_b\right]_{IM_I} \,, \\
&\mathcal{N}_{ab}(I) = \frac{\sqrt{1+\delta_{ab}(-1)^I}}{1+\delta_{ab}}
\end{align}
for any state of angular momentum $I$ in the even-even nucleus. We denote here
$\tilde{A}_{ab}(IM_I)\equiv (-1)^{I+M_I}A_{ab}(I,-M_I)$. Here $a$ and $b$ denote the
quantum numbers of a single-particle orbital in a spherical nuclear mean field,
including the number of nodes $n$ (principal quantum number), the orbital
($l$) and total ($j$) angular momenta.
It should be noted that here the summation over $a\le b$ guarantees that there is no double
counting of two-quasiparticle configurations and this with the normalized operators
(\ref{eq:A-dagger}) guarantees that the wave function is properly normalized
with the normalization condition \cite{Suhonen2007}
\begin{equation} \label{eq:norm}
\sum_{a\le b}\Big(\big\vert X_{ab}^{\omega_f}\big\vert^2 -
\big\vert Y_{ab}^{\omega_f}\big\vert^2\Big) =1 \,.
\end{equation}
The creation operator $Q^{\dagger}(I^{\pi}_n,M_I)$ of (\ref{eq:ccQRPA-ph}) is usually called
the creation operator for a QRPA phonon.

For calculational convenience it is preferable to go from the restricted sum of
(\ref{eq:ccQRPA-ph}) to a non-restricted (free) one by introducing the correspondence
\begin{eqnarray} \label{eq:one-ph}
&\vert I^{\pi}_n \,M_I\rangle = \bar{Q}^{\dagger}(I^{\pi}_n,M_I)\vert\textrm{QRPA} \rangle
\nonumber \\
&= \sum_{ab}\big[\bar{X}_{ab}^{\omega_f}\bar{A}^{\dagger}_{ab}(IM_I) -
\bar{Y}_{ab}^{\omega_f}\tilde{\bar{A}}_{ab}(IM_I)\big]\vert\textrm{QRPA} \rangle \,,
\end{eqnarray}
where the barred two-quasiparticle operators are the ones of (\ref{eq:A-dagger}) without
the normalizer $\mathcal{N}_{ab}(I)$. Then the normalization condition becomes
\begin{equation} \label{eq:norm-bar}
\sum_{ab}\Big(\big\vert \bar{X}_{ab}^{\omega_f}\big\vert^2 -
\big\vert \bar{Y}_{ab}^{\omega_f}\big\vert^2\Big) = \frac{1}{2} \,.
\end{equation}
At the same time the two kinds of $X$ and $Y$ amplitudes are related by
\begin{equation} \label{eq:transfo}
\bar{X}_{ab}^{\omega_f} = \frac{\sqrt{1+\delta_{ab}}}{2}X_{ab}^{\omega_f} \ ;\
\bar{Y}_{ab}^{\omega_f} = \frac{\sqrt{1+\delta_{ab}}}{2}Y_{ab}^{\omega_f}\,, \ a\le b \,,
\end{equation}
for any $\omega_f = I^{\pi}_n$. The barred amplitudes are symmetrized ones and possess
the convenient symmetry relations (to generate amplitudes with $a>b$):
\begin{equation} \label{eq:symmetry}
\bar{X}_{ba}^{\omega_f} = (-1)^{j_a+j_b+J+1}\bar{X}_{ab}^{\omega_f} \ ;\
\bar{Y}_{ba}^{\omega_f} = (-1)^{j_a+j_b+J+1}\bar{Y}_{ab}^{\omega_f} \,.
\end{equation}

In the MCM, originally introduced in \cite{Suhonen1993},
the one-body transition densities
corresponding to a transition from the intermediate $\vert\omega\,M\rangle$ state
(\ref{eq:pnQRPA-st}) of the odd-odd nucleus to the final one-phonon
state (\ref{eq:one-ph}) of the even-even daughter nucleus are calculated
by first writing the transition
densities as ground-state-averaged multiple commutators and then applying the
quasi-boson approximation \cite{Suhonen2007} by replacing the QRPA vacuum by the
BCS vacuum when taking the ground-state average. The averaged multiple commutators
then become
\begin{align} \label{eq:one-ph-td}
&\langle \omega_f \,M_I \vert \beta^+_{L\mu}(np)\vert \omega  \,M\rangle \nonumber \\
&\approx \langle\textrm{BCS}\vert \big\lbrack \lbrack \bar{Q}(\omega_f,M_I),
\beta^+_{L\mu}(np)\rbrack ,
q^{\dagger}(\omega,M)\big\rbrack \vert\textrm{BCS}\rangle \,,
\end{align}
where $\vert\textrm{BCS}\rangle$ is the BCS ground state and
we have denoted the $\beta^+$ type of EC operator by
\begin{equation}
\beta^+_{L\mu}(np) \equiv \left[c^{\dagger}_n\tilde{c}_p\right]_{L\mu} \,,
\end{equation}\label{eq:trdens}
with $c^{\dagger}_n$ creating a neutron on orbital $n$ and $\tilde{c}_p$ annihilating
a proton on orbital $p$. Using the Wigner-Eckart theorem \cite{Suhonen2007}
one can convert (\ref{eq:one-ph-td}) to the reduced transition
density:
\begin{align} \label{eq:trdens-one}
&(\omega_f\vert\vert \left[c^{\dagger}_n\tilde{c}_p\right]_{L} \vert\vert\omega ) =
2\sqrt{[I][L][J]}(-1)^{I+L} \nonumber \\
&\times \sum_{n'}(-1)^{j_p+j_{n'}}\Big(\bar{X}^{\omega_f}_{n'n}X^{\omega}_{n'p}u_nu_p
-\bar{Y}^{\omega_f}_{n'n}Y^{\omega}_{n'p}v_nv_p\Big)
\left\{ \begin{array}{ccc} J & L & I \\ j_{n} & j_{n'} & j_{p}\end{array} \right\}
\nonumber \\
&+ 2\sqrt{[I][L][J]}(-1)^{I+J} \nonumber \\
&\times \sum_{p'}(-1)^{j_n+j_{p}}\Big(-\bar{X}^{\omega_f}_{p'p}X^{\omega}_{np'}v_nv_p
+\bar{Y}^{\omega_f}_{p'p}Y^{\omega}_{np'}u_nu_p\Big)
\left\{ \begin{array}{ccc} J & L & I \\ j_{p} & j_{p'} & j_{n}\end{array}
\right\} \,,
\end{align}
where the $v$ and $u$ factors are the occupation and vacancy amplitudes of the BCS
\cite{Suhonen2007}. The transition density (\ref{eq:trdens-one}) can be used to
compute the connection of the near-resonant $0\nu2$EC intermediate states to the final resonant
excited state $\omega_f$ of Eq.~(\ref{eq:one-ph}).

The MCM method has close connection with the boson-expansion method described in the papers of \textcite{Raduta:1991,Raduta:1991b,Raduta:1996}.

\subsubsection{Deformed quasiparticle random phase approximation \label{subsubsec:dQRPA}}

The near-resonant $0\nu2$EC NMEs for axially symmetric well-deformed nuclei can be calculated
in the adiabatic Bohr-Mottelson approximation in the intrinsic coordinate system
of a rotating nucleus. The nuclear excitations are characterized by the
parity $\pi$ and the quantum number $K$ which is associated to the projection of
the total angular momentum $J$ of the nucleus onto the intrinsic symmetry axis.
Then the $k^{th}$ intrinsic state of projection-parity $K^{\pi}$, $\vert K^{\pi},\,k\rangle$,
can be generated by the deformed QRPA approach \cite{Fang2011} in a way analogous to
Eq.~(\ref{eq:pnQRPA-st}):
\begin{align} \label{eq:dQRPA-st}
\vert K^{\pi} \,k\rangle &= q^{\dagger}(K^{\pi},k)\vert 0^+_{\rm gs} \rangle \nonumber \\
&= \sum_{pn} \left(X_{pn,K}^{k}a_p^{\dagger}a_{\bar{n}}^{\dagger}-Y_{pn,K}^{k}
a_{\bar{p}}a_n\right)\vert 0^+_{\rm gs} \rangle \,,
\end{align}
where for the quasiparticle operators $a_{\bar{p}}$ ($a_{\bar{n}}$) the $\bar{p}$
($\bar{n}$) denotes the time-reversed proton (neutron) orbital. The quasiparticle
pairs in (\ref{eq:dQRPA-st}) obey the selection rules $\Omega_p-\Omega_n=K$
and $\pi_p\pi_n=\pi$, where the involved parities are those of the single-particle
orbitals and $\Omega$ denotes the projection of the total single-particle angular
momentum $j$ on the intrinsic symmetry axis. The state $\vert 0^+_{\rm gs} \rangle$
denotes here the vacuum of the deformed QRPA. The single-particle states are
obtained by solving the Schr\"odinger equation for a deformed axially symmetric
Woods-Saxon mean-field potential \cite{Yousef2009}. In the deformed QRPA approach
the deformed calculation is transformed to a spherical QRPA framework by decomposing
the deformed Woods-Saxon wave functions first into deformed harmonic-oscillator (HO)
wave functions and these, in turn, into spherical HO wave functions \cite{Yousef2009}.
This also enables the use of realistic one-boson-exchange nucleon-nucleon potentials
in the many-body calculations \cite{Yousef2009}.

The one-body transition densities (\ref{eq:density1}) and (\ref{eq:density2}) of the
spherical QRPA are now replaced by the corresponding transition densities of the deformed
QRPA:
\begin{align}
\langle 0^{+}_{\rm gs} \vert c^{\dag}_{n'}c_{p'}]_J \vert K^{\pi}\, k_1 \rangle &=
v_{n'}u_{p'}X^{k_1}_{p'n',K^{\pi}} + u_{n'}v_{p'}Y^{k_1}_{p'n',K^{\pi}} \ , \label{eq:ddensity1} \\
\langle K^{\pi}\, k_2 \vert c^{\dag}_{n}\tilde{c}_{p} \vert 0^{+}_{i} \rangle &=
\tilde{u}_{n}\tilde{v}_{p}\tilde{X}^{k_2}_{pn,K^{\pi}} +
\tilde{v}_{n}\tilde{u}_{p}\tilde{Y}^{k_2}_{pn,K^{\pi}} \, .
\label{eq:ddensity2}
\end{align}
These transition densities are the ones used to compute the near-resonant $0\nu2$EC NMEs of the
decays of $^{152}$Gd, $^{164}$Er and $^{180}$W in the paper of \textcite{Fang2012}.

\subsubsection{Microscopic interacting boson model \label{subsubsec:IBM-2}}

The interacting boson model (IBM) is a theory framework based on $s$ and $d$
bosons which correspond to collective nucleon pairs coupled to angular momenta and
parities $0^+$ and $2^+$, respectively. An extension of the IBM is the microscopic
IBM (IBM-2) where the protons and neutrons form separate proton and neutron bosons.
The IBM-2 is in a way a phenomenological version of the nuclear shell model,
containing the seniority aspect and the
restriction to one magic shell in terms of the single-particle model space. The
Hamiltonian and the transition operators are constructed from the $s$ and $d$ bosons
as lowest-order boson expansions with coupling coefficients to be determined by
fits to experimental data on low-lying energy levels and E2 $\gamma$ transitions
associated with the $s$ and $d$ bosons, but the fitting does not use the spin or
isovector data available from e.g. the $\beta$ decays. One can also
relate the bosons to the underlying fermion model space
through a mapping procedure \cite{Otsuka1978,Otsuka1996}.

The microscopic IBM can be extended to include higher-multipole bosons, like
$g$ bosons, as well. Further extension concerns the description of odd-$A$ nuclei
by the use of the microscopic interacting boson-fermion model (IBFM-2)
\cite{Iachello1991}. The IBM concept can also be used to
describe odd-odd nuclei by using the interacting boson-fermion-fermion model (IBFFM)
and its proton-neutron variant, the proton-neutron IBFFM (IBFFM-2) \cite{Brant1988}.
Here problems arise from the interactions between the bosons and the one or
two extra fermions in the Hamiltonian, and from the transition operators containing
a large number of phenomenological parameters to be determined in some meaningful way.
While IBM-2 has been used quite much to calculate the $0\nu2\beta$ properties
of nuclei the IBFFM-2 has not. The IBM-2 calculations have to be done using the closure
approximation since it does not contain the spin-isospin degree of freedom
needed to access the intermediate odd-odd nucleus of the $0\nu2\beta$ decay,
in particular in the context of the near-resonant $0\nu2$EC decays.

\subsubsection{Energy-density functional method \label{subsubsec:EDF}}

The energy-density functional method (EDF) is a mean-field-based method that uses
closure approximation to compute the near-resonant $0\nu2$EC NMEs, and thus is well suited for $0\nu2$EC
transitions between two ground states, like in the cases of the near-resonant $0\nu2$EC decays of
$^{152}$Gd, $^{164}$Er and $^{180}$W treated in the paper of \textcite{Rodriguez2012}. In this theory
framework \cite{Rodriguez2010} density functionals based on the Gogny D1S
functional \cite{Berger1984} and D1M \cite{Goriely2009} in large single-particle bases
(11 major oscillator shells) are used. Both the particle-number
and angular-momentum projections are performed before the variation
for the mother and daughter nuclei, and configuration mixing is taken into account
using the generating coordinate method (GCM) \cite{Ring1980}. Hence, in the EDF the
initial and final ground states can be written as
\begin{align}
\vert 0^{+}_{\rm gs} \rangle = \sum_{\beta_2}g_{\beta_2}P^{J=0}P^NP^Z\vert\Phi_{\beta_2}\rangle\,,
\end{align}
where $P^{N}$ ($P^Z$) is the projection operator for a given neutron (proton) number and
$P^{J=0}$ is the projection operator for zero total angular momentum. The intrinsic
axially symmetric Hartree-Fock-Bogoliubov wave functions $\vert\Phi_{\beta_2}\rangle$ are
solutions to the variation equations after particle-number-projection constrained
to a given value of the axial quadrupole deformation $\beta_2$. The shape-mixing
coefficients $g_{\beta_2}$ are found by solving the Hill-Wheeler-Griffin
equation \cite{Ring1980}.

%\subsection{Simple estimation of the order of magnitude of the NMEs \label{subsec:simple}}
%\subsection{Estimations of the order of magnitude of the nuclear matrix elements \label{subsec:simple}}
%\subsection{Order of magnitude estimates of nuclear matrix elements \label{subsec:simple}}

\subsection{Decays of nuclides with the calculated nuclear matrix elements \label{subsec:simple}}

%\subsection{The decay probability of neutrinoless 2EC processes \label{subsec:simple}}

The order of magnitude of the NME (\ref{eq:ECEC-T}) can be estimated by constructing
a generic single-quasiparticle type NME (qp-NME) describing the conversion of a proton
pair to a neutron pair at the nuclear proton and neutron Fermi surfaces. This NME
picks the essential features of the transition since the most action is concentrated
at the Fermi surfaces. The detailed quasiparticle properties at the Fermi surfaces
can be obtained from a BCS calculation using the Woods-Saxon mean-field single-particle
energies \cite{Bohr1969}. In this simple estimation the collective effects are not
taken into account. These collective effects can be very important for $0\nu2$EC
transitions to the lowest-lying $0^+$ or $2^+$ states. However, for the
$J_f=0^+,2^+$ states at energies satisfying the resonance condition of the near-resonant $0\nu2$EC
decay the collective effects are not so important. In fact, at around these energies the
many-body wave functions can vary strongly from one state to the next, sometimes causing
coherent enhancements or incoherent cancellations. A qp-NME is
a kind of average between these two extremes and thus suitable for the role of a
generic NME in this case.

A plausible simplification of the NME (\ref{eq:NMEs}) is to consider the conversion
of an angular-momentum-zero-coupled proton pair to an angular-momentum-$J_f$-coupled
neutron pair at the nuclear Fermi surface. The zero-coupled proton pairs are the
most important contributors to the NMEs of the ordinary 0$\nu$2$\beta$ decay
\cite{Hyvarinen2015}, so that this is a good simplifying approximation. Considering the
$1^+$ type of intermediate states as the typical ones and taking $J_1=1$ for simplicity,
leads to the following simplified expression for the NME (\ref{eq:NMEs}):
\begin{align} \label{eq:simp1}
M_{K}^{\rm 2EC}(0^+_i\to J_f^+) & \approx \sqrt{\frac{3}{[J_f][j_p]}}(-1)^{j_p+j_n+1}
\left\{ \begin{array}{ccc} 1 & j_{n} & j_{p} \\
j_{n} & 1 & J_f \end{array} \right\} \\
 & \times ( nn:J_f \vert\vert h(r_{12})
\big\lbrack\mbox{\boldmath{$\sigma$}}_{1}\mbox{\boldmath{$\sigma$}}_{2}
\big\rbrack_{J_f} \vert\vert pp:0 )
( J^{+}_{f} \vert \vert  \big[ c^{\dag}_{n}
\tilde{c}_{p}\big]_1 \vert \vert 1^{+} )( 1^{+} \vert \vert  \big[ c^{\dag}_{n}
\tilde{c}_{p}\big]_1 \vert \vert 0^{+}_{i}) \,,
\end{align}
where the neutrino potential can be simplified to a Coulomb type of potential
\begin{equation}\label{eq:coulomb}
h(r_{12})=\frac{2R_A}{\pi}\frac{1}{r_{12}}
\end{equation}
by taking just
the leading %AA
contribution \cite{Hyvarinen2015} to the potential and approximating
the difference of the intermediate energy and the average of the parent and daughter
masses as zero, which is a rather good approximation for the ground state of the
intermediate nucleus. Here $R_A=1.2A^{1/3}\,\textrm{fm}$ is the nuclear radius for the nucleus of mass $A$.
In order to proceed further one has to convert the two-body NME to the center-of-mass
and relative coordinates for the computation of the associated radial integral
of the simplified neutrino potential (\ref{eq:coulomb}). This can be achieved by the use
of the Moshinsky brackets $M_{\lambda}$, first introduced by \textcite{Moshinsky1959} (see
\cite{Suhonen2012a} for more details).

Implementing the Moshinsky brackets and working out the angular-momentum algebra
results in a rather simple compact expression for the two-body NME in Eq.~(\ref{eq:simp1}):
\begin{align}\label{eq:simp2}
( nn:J_f \vert\vert h(r_{12})
\big\lbrack\mbox{\boldmath{$\sigma$}}_{1}\mbox{\boldmath{$\sigma$}}_{2}
\big\rbrack_{J_f} \vert\vert pp:0 ) & =
6\sqrt{[J_f][j_p]} [j_n]\sum_{S=0,1}G_{J_f}^{pn}(S) \nonumber \\
&\times \sum_{nn'lNL}M_S(n'lNL;n_nl_nn_nl_n)
M_S(nlNL;n_pl_pn_pl_p)I_{n'nl} \,,
\end{align}
where $n_n$ and $l_n$ are the principal and orbital angular-momentum quantum numbers
for the orbital occupied by the final neutrons and $n_p$ and $l_p$ are the
corresponding quantum numbers for the initial protons. The quantities
$M_S(n'lNL;n_nl_nn_nl_n)=\langle n'l,NL,S\vert n_nl_n,n_nl_n,S\rangle$ are the
Moshinsky brackets and the sum over the quantum numbers
$n$, $n'$ and $l$ refers to a sum over the principal and orbital angular-momentum
quantum numbers of the relative motion, and $N$ and $L$ symbolize the principal and
orbital angular-momentum quantum numbers associated with the center-of-mass coordinate.
The sum over $S$ denotes a sum over the possible total spins.
The geometric factor can be simplified to
\begin{equation}\label{eq:geofac}
G_{J_f}^{pn}(S) = [S] \sum_{S'} [S'](-1)^{S'+J_f+l_p+j_p+1/2}
\left\{ \begin{array}{ccc} l_p & l_p & S \\
{\scriptstyle \frac{1}{2}} & {\scriptstyle \frac{1}{2}} & j_p \end{array} \right\}
\left\{ \begin{array}{ccc} l_n & {\scriptstyle \frac{1}{2}} & j_n \\
l_n & {\scriptstyle \frac{1}{2}} & j_n \\ S & S' & J_f \end{array} \right\}
\left\{ \begin{array}{ccc} {\scriptstyle \frac{1}{2}} & {\scriptstyle \frac{1}{2}} & 1 \\
{\scriptstyle \frac{1}{2}} & {\scriptstyle \frac{1}{2}} & 1 \\
S' & S & J_f \end{array} \right\},
\end{equation}
and the Coulomb-type integral reads
\begin{equation}\label{eq:int}
I_{n'nl}=\int_0^{\infty} g_{n'l}(r)h(r)g_{nl}(r)r^2dr \,,
\end{equation}
where $g_{nl}(r)$ are the radial functions of the three-dimensional harmonic oscillator
and $h(r)$ is the simplified neutrino potential (\ref{eq:coulomb}).

The one-body transition densities involved in the expression (\ref{eq:simp1}) can be
obtained from Eqs.~(\ref{eq:density2}) and (\ref{eq:trdens-one}). In the quasiparticle
approximation the $Y$ amplitudes vanish and for the involved quasiparticle
transitions the $X$ factors are set to unity. Then one finds:
\begin{equation}\label{eq:densities}
(J^{+}_{f} \vert \vert  \big[ c^{\dag}_{n}
\tilde{c}_{p}\big]_1 \vert \vert 1^{+} )( 1^{+} \vert \vert  \big[ c^{\dag}_{n}
\tilde{c}_{p}\big]_1 \vert \vert 0^{+}_{i}) \approx 3\sqrt{{6}{[J_f]}} (-1)^{J_f+1}
\left\{ \begin{array}{ccc} 1 & 1 & J_f \\ {\scriptstyle \frac{9}{2}} &
{\scriptstyle \frac{9}{2}} & {\scriptstyle \frac{11}{2}} \end{array} \right\}
u_nu_p\tilde{u}_n\tilde{v}_p \,,
\end{equation}
where the occupation and vacancy amplitudes are obtained from BCS calculations in
the involved nuclei. The qp-NMEs calculated by using the simplified formalism of
Eqs.~(\ref{eq:simp1})--(\ref{eq:geofac}) are displayed in Table~\ref{tab:NME}.
In the $A=148-154$ region the proton-to-neutron single-quasiparticle transition is
$\pi 0h_{11/2}\to\nu 0h_{9/2}$ and for $A=194,202$ the transition is
$\pi 0h_{11/2}\to\nu 0i_{13/2}$.

\subsection{Section summary}

To conclude, Table~\ref{tab:NME} shows that the magnitudes of the computed $0v$2EC NMEs for
different theory frameworks can vary quite strongly for nuclei with $A\ge 152$.
These nuclei are deformed and thus rather challenging from the nuclear-structure
point of view. For these nuclei it is preferable to apply a nuclear-theory
framework which naturally contains the deformation degree of freedom, namely the
IBM-2, dQRPA, and EDF frameworks. However, as seen in Table~\ref{tab:NME}, the computed
NMEs show that there are big differences between the results obtained in these
different computational formalisms. The reason for these differences is not obvious
and is already well recognized in the case of the $0\nu 2\beta^-$ NMEs, as clearly
shown in the recent NME compilation of (Engel,2017). Since none of these theory
frameworks can systematically access the uncertainties of the calculations it is
hard to make a judicious choice between the different NMEs in terms of reliability.
This conclusion is valid for both the $0\nu 2\beta^-$ and $0\nu$2EC decay processes.
Only further studies and comparisons between these theory frameworks could shed
light on this rather disturbing situation and lead the way towards consistent
values of the NMEs for deformed heavy nuclei.

Another conspicuous feature of Table~\ref{tab:NME} is that the nuclear shell model (NSM),
standardly  used to compute the $0\nu 2\beta^-$ NMEs, does not contribute to the
calculations of the $0\nu$2EC NMEs. The reason for this is twofold: on one hand, for
the nuclei $^{152}$Gd, $^{156}$Dy, $^{164}$Er, and $^{180}$W the $2\nu$EC transition is
ground-state-to-ground-state and thus accessible, in principle, to the NSM.
Unfortunately, these nuclei are heavy (very) deformed nuclei and the NSM simply does
not have the necessary single-particle valence space in order to treat these
decays. On the other hand, for the lighter, nearly spherical nuclei, the NSM is
easier to install in terms of single-particle spaces, but the fact that the resonant
states in the daughter nuclei are highly excited excludes a reasonable description
of the corresponding wave functions by the NSM.

%\newpage

\section{Status of experimental searches}
% update by FAD & VIT on 30.04.2019; 21.04.2020; 25.05.2020
%%%%%%%%%%%%%%%%%%%%%%%%%%%%%%%%%%%%%%%%%%%%%%%%%%%%%%%%%%%
%%%%%%%%%%%%%%%%%%%%%%%%%%%%%%%%%%%%%%%%%%%%%%%%%%%%%%%%%%%
%%%%%%%%%%%%%%%%%%%%%%%%%%%%%%%%%%%%%%%%%%%%%%%%%%%%%%%%%%%
\setcounter{equation}{0}

\subsection{Experimental studies of 2EC processes}

The efforts of experimentalists were mainly concentrated on the
search for neutrinoless double-beta decay with emission ~of two
~electrons ~($2 \beta^-$) ~where limits ~on the half-lives ~of
$T_{1/2}>10^{24}-10^{26}$ yr were obtained\footnote{We refer
readers to the reviews of
\textcite{Tretyak:1995,Tretyak:2002,Elliott:2012,Giuliani:2012,Saakyan:2013,Cremonesi:2014,
Gomes:2015,
Sarazin:2015,Pas:2015,Bilenky:2015,Delloro:2016,Vergados:2016hso,Bar18}
and the ~recent experimental ~results ~from
\textcite{Ago18,NEMO-3,CUORE,EXO-200,Gando:2016,Azz18}.}. ~The most

\newpage

\begin{table}[H]
%\resizebox{\textwidth}{!}
%\renewcommand{\arraystretch}{2.5}
\scriptsize
\centering
\caption{Experimental half-life limits of neutrinoless 2EC for
transitions to the ground state (denoted as ``g.s.'') or to the
excited level of the daughter nuclide with possible resonant
enhancement. The mass differences between the mother and the
daughter atoms, $Q = M_{A,Z} - M_{A,Z-2}$, are taken from
the paper of \textcite{Wang:2017}; $\iota$ is the isotopic abundance of the
nuclide of interest in the natural isotopic compositions of the
elements \cite{Meija:2016}. To check the resonance enhancement
condition, the degeneracy parameter $\Delta = Q - E^* -
\epsilon^*_{\alpha \beta}$ is shown, where $E^*  = M^{*}_{A,Z-2} -
M_{A,Z-2}$ is the excitation energy of the daughter nuclide and
$\epsilon^*_{\alpha \beta} = M^{**}_{A,Z - 2} - M^{*}_{A,Z - 2}$
is the excitation energy of the atomic shell with the electron
vacancies $\alpha$ and $\beta$ in the $K$, $L$, $M$ or $N$ orbits.
The energies $E^*$ and the values of $J^\pi$ of the excited
nuclide levels are taken from the database of Brookhaven National
Laboratory (http://www.nndc.bnl.gov/ensdf/). The experimental
limits of the $^{54}$Fe$\rightarrow$$^{54}$Cr decay are at 68\%
confidence level (C.L.), in other cases at 90 \% C.L. The
de-excitation width of the electron shell of the daughter nuclides
$\Gamma_f = \Gamma_{\alpha} + \Gamma_{\beta}$ (see
\textcite{CAMP01}) is shown in column 6 (orbits are indicated in the
brackets). The resonance parameter $R_f = \Gamma_f/(\Delta^2 +
\Gamma_f^2/4)$ normalized on the value for the $0\nu$2EC decay
$^{54}$Fe $\rightarrow$ $^{54}$Cr (g.s. to g.s.) is given in
column 7.}
\label{tab:res-exp}
\begin{center}
\resizebox{0.9999\textwidth}{!}
{\begin{tabular}{|c|c|c|c|l|c|c|}
\hline
\hline
Transition                          & Decay channel,                &       &       &          &  &     \\
$Q$ (keV)                           & Level of daughter             &     $\Delta$ (keV)        &     Expt. limit               & ~~~~Experimental technique (Ref., Year)               & $\Gamma_f$ (eV)   & $R_f$ \\
$\iota$ (\%)                        & nuclei (keV)                  & ~                     &             (yr)             &                & ~ & ~ \\
\hline
\hline
$^{36}$Ar$\rightarrow$$^{36}$S     & $KL$, $0^+$ g.s.              & 427.65(19)         & $\geq3.6\times10^{21}$ & HPGe $\gamma$ spectrometry \cite{Agostini:2016}   & 1.04 ($KK$) & 1.2 \\
432.59(19)                         & ~                             & ~                     & ~                      & ~                                                 & & \\
0.3336(210)                        & ~                             & ~                     & ~                      & ~                                                 & & \\
\hline

$^{40}$Ca$\rightarrow$$^{40}$Ar    & 2EC, $0^+$ g.s.              & 187.10(2)             & $\geq1.4\times10^{22}$  & CaWO$_4$ scint. bolometer \cite{Angloher:2016}    & 1.32 ($KK$) & 8 \\
193.51(2)                          & ~                             & ~                     & ~                      & ~                                                 & & \\
96.941(156)                        & ~                             & ~                     & ~                      & ~                                                 & & \\
\hline

$^{50}$Cr$\rightarrow$$^{50}$Ti    & ~                             & 1159.7(5)          & --                     & ~                                                 & 1.78 ($KK$) & 0.3 \\
1169.6(5)                          & ~                             & ~                     & ~                      & ~                                                 & &  \\
4.345(13)                          & ~                             & ~                     & ~                      & ~                                                 & & \\
\hline

$^{54}$Fe$\rightarrow$$^{54}$Cr    & $KK$, $0^+$ g.s.              & 668.3(4)           & $\geq4.4\times10^{20}$ & HPGe $\gamma$ spectrometry \cite{Bikit:1998}      & 2.04 ($KK$) & 1 \\
680.3(4)                           & $KL$, $0^+$ g.s.               & ~                     & $\geq4.1\times10^{20}$ & HPGe $\gamma$ spectrometry \cite{Bikit:1998}      & & \\
5.845(105)                         & $LL$, $0^+$ g.s.               & ~                     & $\geq5.0\times10^{20}$ & HPGe $\gamma$ spectrometry \cite{Bikit:1998}      & & \\
\hline

$^{58}$Ni$\rightarrow$$^{58}$Fe    & $KL$, $0^+$ g.s.               & ~             & $\geq4.1\times10^{22}$ & HPGe $\gamma$ spectrometry \cite{Rukhadze:2018}   & ~ & ~ \\
1926.4(3)                          & $KK$, $2^+$ 1674.731(6)        & 237.4(3)                     & ~                      & ~                                                 & 2.38 ($KK$) & 9 \\
68.0769(190)                       & ~                             & ~                     & ~                      & ~                                                 & & \\
\hline

$^{64}$Zn$\rightarrow$$^{64}$Ni    & 2EC, $0^+$ g.s.              & 1075.6(7)      & $\geq3.2\times10^{20}$  & ZnWO$_4$ scintillator \cite{Belli:2011b}          & ~   & ~ \\
1094.9(7) & ~ & ~ & ~ & ~ & ~ & ~ \\
49.17(75) & ~ & ~ & ~ & ~ & ~ & ~ \\

\hline
$^{74}$Se$\rightarrow$$^{74}$Ge    & $KK$,  $0^+$ g.s.             &                       & $\geq6.2\times10^{18}$ & HPGe $\gamma$ spectrometry \cite{Bar07a}          & & \\
1209.24(1)                         & $KL$,  $0^+$ g.s.             &                       & $\geq6.4\times10^{18}$ & HPGe $\gamma$ spectrometry \cite{Bar07a}          & & \\
0.86(3)                            & ~                             &                       & $\geq9.6\times10^{18}$ & HPGe $\gamma$ spectrometry \cite{Lehnert:2016b}   & & \\
                                   & $LL$,  $0^+$ g.s.             &                       & $\geq4.1\times10^{18}$ & HPGe $\gamma$ spectrometry \cite{Bar07a}          & & \\
                                   & ~                             &                       & $\geq5.8\times10^{18}$ & HPGe $\gamma$ spectrometry \cite{Lehnert:2016b}   & & \\
                                   & $LL$,  $2^+$ 1204.205(7)      & $(2.21-2.60)\pm0.01$    & $\geq5.5\times10^{18}$ & HPGe $\gamma$ spectrometry \cite{Bar07a}          & 7.6 ($L_1L_1$) & $3.4\times10^{5}$ \\
                                   & ~                             & ~                     & $\geq4.3\times10^{19}$\footnotemark[1] & HPGe $\gamma$ spectrometry \cite{Fre11}         & & \\
                                   & ~                             & ~                     & $\geq1.5\times10^{19}$\footnotemark[2] & HPGe $\gamma$ spectrometry \cite{Jeskovsky:2015} & & \\
                                   & ~                             & ~                     & $\geq7.0\times10^{18}$ & HPGe $\gamma$ spectrometry \cite{Lehnert:2016b}   & & \\
\hline

$^{78}$Kr$\rightarrow$$^{78}$Se    & $KK$, $0^+$ g.s.              & ~                     & $\geq5.5\times10^{21}$ & Proportional counter                              & & \\
2847.67(26)                        & $LL$, $(2^+)$ 2838.49(7)      & $(5.88-6.32)\pm0.26$     & $\geq5.4\times10^{21}$ & filled with enriched                              & 7.6 ($L_1L_1$) & $4.8\times10^{4}$ \\
0.355(3)                           &                               & ~                     &                        & $^{78}$Kr (99.8\%) \cite{Gavrilyuk:2013}          & & \\
\hline

$^{84}$Sr$\rightarrow$$^{84}$Kr    & $KK$, $0^+$ g.s.              & ~                     & $\geq6.0\times 10^{16}$ & HPGe $\gamma$ spectrometry \cite{Belli:2012a}    & & \\
1789.8(12)                         & $KL$, $0^+$ g.s.              & ~                     & $\geq1.9\times 10^{16}$ & HPGe $\gamma$ spectrometry \cite{Belli:2012a}    & & \\
0.56(2)                            & $LL$, $0^+$ g.s.              & ~                     & $\geq5.9\times 10^{16}$ & HPGe $\gamma$ spectrometry \cite{Belli:2012a}    & & \\
 ~                                  & $KK$, $2^+$ 881.615(3)        & 879.5(12)           &                       &                                                   & 5.4 ($KK$)& 1.5 \\
\hline

$^{92}$Mo$\rightarrow$$^{92}$Zr    & $KK$, $0^+$ g.s.               & ~                     & $\geq6.8\times 10^{19}$ & CaMoO$_4$ scintillator in coincidence            & & \\
1650.45(19)                        & $KK$, $0^+$ 1382.77(7)         & 231.68(19)            & ~                      & with HPGe detector \cite{Kan13}                   & 7.66 ($KK$) & $31$ \\
14.649(106)                        & $KK$, $4^+$ 1495.46(5)         & 118.99(19)               & ~                      &                                                   & 7.66 ($KK$) &  $118$ \\
\hline

$^{96}$Ru$\rightarrow$$^{96}$Mo    & $KK$ $0^+$ g.s.               & ~                      & $\geq1.0\times 10^{21}$ & HPGe $\gamma$ spectrometry \cite{Belli:2013b}    & ~ & ~ \\
2714.50(12)                        & $KL$ $0^+$ g.s.               & ~                     & $\geq2.3\times 10^{20}$ & HPGe $\gamma$ spectrometry \cite{Belli:2013b}    & & \\
5.54(14)                           & $LL$ $0^+$ g.s.               & ~                     & $\geq2.3\times 10^{20}$ & HPGe $\gamma$ spectrometry \cite{Belli:2013b}    & & \\
~                                  & $KL$, $2^+$ 2700.21(6)        & $-(8.23-8.58)\pm0.13$   & $\geq2.0\times 10^{20}$ & HPGe $\gamma$ spectrometry \cite{Belli:2013b}    & 8.32 ($KL_1$) & $2.5\times10^{4}$ \\
~                                  & $LL$, $2^+$ 2700.21(6)        & $(8.56-9.25)\pm0.13$           & ~ & ~    & 7.6 ($L_1L_1$) & $2.3\times10^{4}$ \\
~                                  & $LL$, 2712.68(10)             & $-(3.22-3.91)\pm0.16$   & $\geq3.6\times 10^{20}$ & HPGe $\gamma$ spectrometry \cite{Belli:2013b}    & 5.63 $L_1L_2$ & $9.0\times10^{4}$ \\
\hline

$^{102}$Pd$\rightarrow$$^{102}$Ru  & $KK$, $2^+$ 1103.047(13)        & 56.0(4)               & ~                      & ~                                                 & 10.7 ($KK$) & 738 \\
1203.3(4)                          & ~                             & ~                     & ~                      & ~                                                 & & \\
1.02(1)                            & ~                             & ~                     & ~                      & ~                                                 & & \\
\hline

$^{106}$Cd$\rightarrow$$^{106}$Pd  & $KK$, $0^+$ g.s.              & ~                     & $\geq1.0\times 10^{21}$ & Enr. $^{106}$CdWO$_4$ scintillator \cite{Belli:2012} & & \\
2775.39(10)                        & $KL$, $0^+$ g.s.              & ~                     & $\geq1.3\times 10^{21}$ & Enr. $^{106}$CdWO$_4$ scintillator\footnotemark[3] \cite{Belli:2016} & & \\
1.245(22)                          & $LL$, $0^+$ g.s.              & ~                     & $\geq1.0\times 10^{21}$ & Enr. $^{106}$CdWO$_4$ scintillator \cite{Belli:2012} & & \\
 ~                                  & $KK$, $0^+$ 2624.40(5)         & $102.29\pm0.11$         & ~                     &                                           & 12.5 ($KK$) & 260 \\
 ~                                  & $KK$, $(1)^+$ 2705.30(8)         & $21.39\pm0.13$         & ~                     &                                           & 12.5 ($KK$) & $5.9\times10^{3}$ \\
~                                  & $KK$, 2717.59(21)             & $9.10\pm0.23$           & $\geq1.1\times 10^{21}$ & Enr. $^{106}$CdWO$_4$ scintillator\footnotemark[3] \cite{Belli:2016} & 12.5 ($KK$) & $3.3\times10^{4}$ \\
~                                  & $KL$, $4^+$ 2741.0(5)         & $(6.4-6.9)\pm0.5$     & $\geq9.5\times 10^{20}$ & Enr. $^{106}$CdWO$_4$ scintillator \cite{Belli:2012} & 10.2 ($KL_1$) & $5.3\times10^{4}$ \\
~                                  & ~                             & ~                     & $\geq1.7\times 10^{20}$ & HPGe $\gamma$ spectrometry of                    & & \\
~                                  & ~                             & ~                     & ~                       & enriched $^{106}$Cd \cite{Rukhadze:2011b}        & & \\
~                                  & $KL$, $2,3^-$ 2748.2(4)       & $-(0.8-0.3)\pm0.4$    & $\geq1.4\times 10^{21}$ & Enr. $^{106}$CdWO$_4$ scintillator\footnotemark[3] \cite{Belli:2016} & 8.3 ($KL_3$) & $1.6\times10^{7}$ \\
\hline

$^{108}$Cd$\rightarrow$$^{108}$Pd  & 2EC, $0^+$ g.s.              & 223.1(8)              & $\geq1.0\times 10^{18}$ & CdWO$_4$ scintillator \cite{Belli:2008a}          & 12.5 ($KK$) & 55 \\
271.8(8)                           & ~                             & ~                     & ~                      & ~                                                 & & \\
0.888(11)                          & ~                             & ~                     & ~                      & ~                                                 & & \\
\hline

$^{112}$Sn$\rightarrow$$^{112}$Cd  & $KK$, $0^+$ 1870.96(5)        & $-4.62\pm0.17$          & $\geq1.3\times 10^{21}$ & HPGe $\gamma$ spectrometry of enriched           & 14.6 ($KK$) & $1.5\times10^{5}$ \\
1919.80(16)                        & $KK$, $0^+$ g.s.              & ~                     & $\geq1.1\times 10^{21}$ & $^{112}$Sn \cite{Barabash:2011}                  & & \\
0.97(1)                            & $KL$, $0^+$ g.s.              & ~                     & $\geq8.2\times 10^{20}$ & ~                                                & & \\
~                                  & $LL$, $0^+$ g.s.              & ~                     & $\geq6.4\times 10^{20}$ & ~                                                & & \\
\hline

$^{120}$Te$\rightarrow$$^{120}$Sn  & $KK$, $0^+$ g.s.              & ~                     & $\geq6.0\times 10^{17}$ & HPGe $\gamma$ spectrometry \cite{Barabash:2007}  & & \\
1730(3)                            & $KL$, $0^+$ g.s.              & ~                     & $\geq3.9\times 10^{17}$ & HPGe $\gamma$ spectrometry \cite{Barabash:2007}  & & \\
0.09(1)                            & $LL$, $0^+$ g.s.              & ~                     & $\geq2.9\times 10^{17}$ & HPGe $\gamma$ spectrometry \cite{Barabash:2007}  & & \\
     ~                              & $KK$, $2^+$ 1171.265(15)      & $500\pm3$         & ~                     & ~                                             & 17.1 ($KK$) & 15 \\
\hline

$^{124}$Xe$\rightarrow$$^{124}$Te  & $KK$, $0^+\div4^+$ 2790.41(9) & $10.3\pm2.2$         & --                     & ~                                                 & 19.8 ($KK$) & $4.1\times10^{4}$ \\
2863.9(22)                         & $KK$, $2^+$ 2808.66(8)        & $-8.0\pm2.2$          & --                     & ~                                                 & 19.8 ($KK$) & $6.8\times10^{4}$ \\
%0.095(5)                           & $KK$, $2^+$ 2817.48(11)       & $-16.8\pm2.2$    & --                     & ~                                                 & 19.8 ($KK$) & $7.0\times10^{-8}$\\
0.095(5)                             & $KL_1$, $2^+$ 2817.48(11)       & $10.1\pm2.2$    & --                     & ~                                                 & 12.1 ($KL_1$) & $2.6\times10^{4}$\\
~                                  & $L_1L_1$, 2853.2(6)               & $1.2\pm2.3$     & --                     & ~                                                 & 4.4 ($L_1L_1$) & $6.4\times10^{5}$ \\
\hline
\hline
\end{tabular}}
\end{center}
\end{table}

\begin{table}[H]
\scriptsize
\centering
\caption{Continued from Table \ref{tab:res-exp}.}
\label{tab:res-exp-2}
\begin{center}
\resizebox{0.9999\textwidth}{!}
{\begin{tabular}{|c|c|c|c|l|c|c|}
\hline
\hline
Transition                          & Decay channel,                &       &       &          &  &     \\
$Q$ (keV)                           & Level of daughter             &     $\Delta$ (keV)        &     Expt. limit               & ~~~~Experimental technique (Ref., Year)               & $\Gamma_f$ (eV)   & $R_f$ \\
$\iota$ (\%)                        & nuclei (keV)                  & ~                     &             (yr)             &                & ~ & ~ \\
\hline
\hline
$^{126}$Xe$\rightarrow$$^{126}$Te  & 2EC, $0^+$ g.s.                & 854(4)              & ~                      & ~                                                 & 19.8 ($KK$) & 6 \\
918(4)                          & $KK$, $2^+$ 666.352(10)           & 188(4)                & ~                      & ~                                                    & 19.8 ($KK$) & 122 \\
0.089(3)                           & ~                             & ~                      & ~                      & ~                                                 & & \\
\hline

$^{130}$Ba$\rightarrow$$^{130}$Xe   & 2EC, $0^+$ g.s.                & ~            & $\geq2.8\times 10^{21}$ & Geochemical \cite{MECH01}\footnotemark[4]  &  & ~ \\
2618.9(26)                          &         & ~                    & $\geq7.4\times 10^{20}$ & Geochemical \cite{Pujol:2009}\footnotemark[4]                                            & & \\
0.11(1)                             & $KK$, $0^+$ 2017.06(16)        & $532.7\pm2.6$             & ~                         & ~                                              & 23.0 ($KK$) & 18 \\
~                                   & $KK$, 1,2 2502.207(25)        & $47.6\pm2.6$             & ~                         & ~                                              & 23.0 ($KK$) & $2.2\times10^{3}$ \\
~                                   & $KK$, 2544.43(8)               & $5.3\pm2.6$           & ~ & ~                       & 23.0 ($KK$) & $1.5\times10^{5}$\\
~                                  & $L_1L_2$, 2608.426(19)             & $-0.1\pm2.6$ & ~ & ~                       &  5.1 ($L_1L_2$)& $1.6\times10^{8}$ \\
\hline

$^{132}$Ba$\rightarrow$$^{132}$Xe  & 2EC, $0^+$ g.s.                & 774.8(11)               & $\geq2.5\times 10^{21}$ & Geochemical \cite{MECH01}\footnotemark[5]                       & 23.0 ($KK$) & 8 \\
843.9(11)                          & $KK$, $2^+$ 667.715(2)         & $107.1\pm1.1$                 & ~                        & ~                                               & 23.0 ($KK$)& 436 \\
0.10(1)                            & ~                             & ~                     & ~                        & ~                                               & & \\
\hline

$^{136}$Ce$\rightarrow$$^{136}$Ba  & $KK$, $0^+$ g.s.              & ~                     & $\geq2.1\times 10^{18}$ & HPGe $\gamma$ spectrometry \cite{Belli:2017}     & &  \\
2378.55(27)                        & $KL$, $0^+$ g.s.              & ~                     & $\geq3.4\times 10^{18}$ & HPGe $\gamma$ spectrometry \cite{Belli:2017}     & &  \\
0.186(2)                           & $LL$, $0^+$ g.s.              & ~                     & $\geq8.4\times 10^{18}$ & HPGe $\gamma$ spectrometry \cite{Belli:2017}     & &  \\
~                                  & $KK$, $0^+$ 2141.38(3)        & $162.29\pm0.27$         & $\geq4.2\times 10^{18}$ & HPGe $\gamma$ spectrometry \cite{Belli:2017}     & 26.4 ($KK$) & 218 \\
~                                  & 2EC, $0^+$ 2315.26(7)        & $-11.59\pm0.28$         & $\geq2.5\times 10^{18}$ & HPGe $\gamma$ spectrometry \cite{Belli:2017}     & 26.4 ($KK$) & $4.3\times10^{4}$  \\
\hline

$^{138}$Ce$\rightarrow$$^{138}$Ba  & $KK$, $0^+$ g.s.              & 616(5)             & $\geq5.5\times 10^{17}$ & HPGe $\gamma$ spectrometry \cite{Belli:2014}     & 26.4 ($KK$) & 15 \\
691(5)                             & $KL$, $0^+$ g.s.              & ~                     & $\geq8.3\times 10^{17}$ & HPGe $\gamma$ spectrometry \cite{Belli:2017}     & &  \\
0.251(2)                           & $LL$, $0^+$ g.s.              & ~                     & $\geq4.2\times 10^{18}$ & HPGe $\gamma$ spectrometry \cite{Belli:2017}     & &  \\
\hline

$^{144}$Sm$\rightarrow$$^{144}$Nd   & $KK$, $0^+$ g.s.              & ~                     & $\geq4.4\times 10^{19}$ & HPGe $\gamma$ spectrometry \cite{Belli:2019a}                              & & \\
1782.4(8)                           & $KL$, $0^+$ g.s.              & ~                     & $\geq1.7\times 10^{19}$ & HPGe $\gamma$ spectrometry \cite{Belli:2019a}                              & & \\
3.08(4)                             & $LL$, $0^+$ g.s.              & ~                     & $\geq1.4\times 10^{19}$ & HPGe $\gamma$ spectrometry \cite{Belli:2019a}                              & & \\
~                                  & $KK$, $4^+$ 1314.669(13)       & $380.6\pm0.8$             & ~                         & ~                                     & 34.8 ($KK$) & 52 \\
~                                   & $KK$, $2^+$ 1560.920(13)      & $134.3\pm0.8$             & ~                         & ~                                     & 34.8 ($KK$) & 419 \\
\hline

$^{152}$Gd$\rightarrow$$^{152}$Sm  & all transitions               & 1.11(18)              & $\geq6.0\times 10^{8}$ & Analysis of average                               & 23.1 ($KL_1$) & $4.1\times10^{6}$ \\
55.69(18)                          & $KL$, $0^+$ g.s.              & $(1.1-2.1)\pm0.2$     & $\geq6.0\times 10^{8}$ & parent-daughter abundances                        & & \\
0.20(3)                            & ~                             & ~                     & ~                      & \cite{Nozzoli:2018}                               & & \\
\hline

$^{156}$Dy$\rightarrow$$^{156}$Gd  & $KK$, $0^+$ g.s.              & ~                     & $\geq2.2\times 10^{16}$ & HPGe $\gamma$ spectrometry \cite{BELL11} & &  \\
2005.95(10)                        & $KL$, $0^+$ g.s.              & ~                     & $\geq1.7\times 10^{16}$ & HPGe $\gamma$ spectrometry \cite{BELL11} & &  \\
0.056(3)                           & $LL$, $0^+$ g.s.              & ~                     & $\geq1.7\times 10^{16}$ & HPGe $\gamma$ spectrometry \cite{BELL11} & &  \\
~                                  & $KK$, 2$^+$ 1914.835(5)       & $-9.33\pm0.10$          & $\geq1.1\times 10^{16}$ & HPGe $\gamma$ spectrometry \cite{BELL11} & 44.8 ($KK$) & $1.1\times10^{5}$ \\
~                                  & $KL_1$, 1$^-$ 1946.344(10)    & $0.99\pm0.10$           & $\geq1.0\times 10^{18}$ & HPGe $\gamma$ {sp. of enr.} $^{156}$Dy \cite{Fin15} & 26.2 ($KL_1$) & $5.8\times10^{6}$  \\
~                                  & $KL_1$, 0$^-$ 1952.400(6)     & $-5.06\pm0.10$          & $\geq2.2\times 10^{17}$ & HPGe $\gamma$ {sp. of enr.} $^{156}$Dy \cite{Fin15} & 26.2 ($KL_1$) & $2.2\times10^{5}$ \\
~                                  & $LL_1$, 0$^+$ 1988.5(2)       & $0.7\pm0.22$           & $\geq9.5\times 10^{17}$ & HPGe $\gamma$ {sp. of enr.} $^{156}$Dy \cite{Fin15} & 7.6 ($L_1L_1$) & $3.4\times10^{6}$ \\
~                                  & $LL_3$, 2$^+$ 2003.749(5)     & $-12.28\pm0.10$         & $\geq6.7\times 10^{16}$ & HPGe $\gamma$ {sp. of enr.} $^{156}$Dy \cite{Fin15} & 7.4 ($L_3L_3$) & $1.1\times10^{4}$ \\
\hline

$^{158}$Dy$\rightarrow$$^{158}$Gd  & $KK$, $0^+$ g.s.              & 181.7(24)             & $\geq4.2\times 10^{16}$ & HPGe $\gamma$ spectrometry \cite{BELL11} & 44.8 ($KK$) & 295 \\
282.2(24)                          & $KK$, $2^+$ 79.5143(15)          & $102.2\pm2.4$         & $\geq2.6\times 10^{14}$ & HPGe $\gamma$ spectrometry \cite{BELL11} & 44.8 ($KK$) & 932 \\
0.095(3)                           & $L_1L_1$, 4$^+$ 261.4580(16)    & $4.0\pm2.4$           & $\geq3.2\times 10^{16}$ & HPGe $\gamma$ spectrometry \cite{BELL11} & 7.6 ($L_1L_1$) & $1.0\times10^{5}$ \\
\hline

$^{162}$Er$\rightarrow$$^{162}$Dy  & $KK$, $0^+$ g.s.              & ~                     & $\geq1.0\times 10^{18}$ & HPGe $\gamma$ spectrometry \cite{Belli:2018}  & & \\
1846.96(30)                        & $KL$, $0^+$ g.s.              & ~                     & $\geq9.6\times 10^{17}$ & HPGe $\gamma$ spectrometry \cite{Belli:2018}  & & \\
0.139(5)                           & $LL$, $0^+$ g.s.              & ~                     & $\geq1.3\times 10^{18}$ & HPGe $\gamma$ spectrometry \cite{Belli:2018} & &  \\
~                                  & $KK$, $3^-$ 1738.999(4)       & $0.4\pm0.3$           & $\geq5.5\times 10^{8}$ & Analysis of average parent-daughter & 50.6 ($KK$) & $7.4\times10^{7}$ \\
~                                  & ~                              &  ~            & ~                     & abundances \cite{Nozzoli:2018}   & ~ & ~ \\
~                                  & $KL$, $2^+$ 1782.68(9)        & $(1.4-2.7)\pm0.3$     & $\geq5.0\times 10^{17}$ & HPGe $\gamma$ spectrometry \cite{Belli:2018} & 29.6 ($KL_1$) & $3.1\times10^{6}$ \\
~                                  & $LL$, $4^-$ 1826.753(4)       & $(2.1-4.6)\pm0.3$     & $\geq5.5\times 10^{8}$ & Analysis of average parent-daughter  & 8.6 ($L_1L_1$) & $4.2\times10^{5}$ \\
~                                  & ~      & ~     & ~ & abundances \cite{Nozzoli:2018} & &  \\
\hline

$^{164}$Er$\rightarrow$$^{164}$Dy  & all transitions               & ~                     & $\geq1.0\times 10^{9}$ & Analysis of average parent-daughter & ~ & ~ \\
25.08(11)                          & $L_1L_1$, $0^+$ g.s.              & $6.99\pm0.11$     & ~                      & abundances \cite{Nozzoli:2018}\footnotemark[6] & 8.6 ($L_1L_1$) & $3.8\times10^{4}$ \\
1.601(3)                           & ~                             & ~                     & ~                      & ~                                     & & \\
\hline

$^{168}$Yb$\rightarrow$$^{168}$Er   & $KK$, $0^+$ g.s.              & ~                     & $\geq7.3\times 10^{17}$ & HPGe $\gamma$ spectrometry \cite{Belli:2019b} & &   \\
1409.27(25)                         & $KL$, $0^+$ g.s.              & ~                     & $\geq6.9\times 10^{17}$ & HPGe $\gamma$ spectrometry \cite{Belli:2019b} & &  \\
0.123(3)                            & $LL$, $0^+$ g.s.              & ~                     & $\geq1.2\times 10^{18}$ & HPGe $\gamma$ spectrometry \cite{Belli:2019b} & &  \\
 ~                                  & 2EC, $0^+$ 1217.169(14)      & $77.13\pm0.25$         & $\geq1.5\times 10^{18}$ & HPGe $\gamma$ spectrometry \cite{Belli:2019b} & 57.0 ($KK$) & $2.1\times10^{3}$  \\
% ~                                  & $KL_1$, $1^-$ 1358.899(5)      & $-16.87\pm0.25$         & ~                         & ~                                 & 33.2 ($KL_1$) & $1.2\times10^{-7}$  \\
 ~                                  & $M_1M_1$ $(2)^-$ 1403.7357(23) & $1.12\pm0.25$         & $\geq1.9\times 10^{18}$ & HPGe $\gamma$ spectrometry \cite{Belli:2019b} & 27.2 ($M_1M_1$)& $4.7\times10^{6}$ \\
\hline

$^{174}$Hf$\rightarrow$$^{174}$Yb   & $KK$, $0^+$ g.s.      & ~                      & $\geq5.8\times 10^{17}$ & HPGe $\gamma$ spectrometry \cite{Danevich:2020} & 64.0 ($KK$) &  15 \\
1100.0(23)                          & $KL$, $0^+$ g.s.       & ~                     & $\geq1.9\times 10^{18}$ & HPGe $\gamma$ spectrometry \cite{Danevich:2020} & ~ & ~ \\
0.16(12)                            & $LL$, $0^+$ g.s.       & ~                     & $\geq7.8\times 10^{17}$ & HPGe $\gamma$ spectrometry \cite{Danevich:2020} & ~ & ~ \\
 ~                                  & $KK$, $2^+$ 76.471(1) & $900.9\pm2.3$         & $\geq7.1\times 10^{17}$ & HPGe $\gamma$ spectrometry \cite{Danevich:2020}  & 64.0 ($KK$) &  17\\
  ~                                  & $KL$, $2^+$ 76.471(1) &                      & $\geq6.2\times 10^{17}$ & HPGe $\gamma$ spectrometry \cite{Danevich:2020}  & ~ & ~ \\
 ~                                  & $LL$, $2^+$ 76.471(1) &                      & $\geq7.2\times 10^{17}$ & HPGe $\gamma$ spectrometry \cite{Danevich:2020}  & ~ & ~ \\
\hline

$^{180}$W$\rightarrow$$^{180}$Hf   & $KK$, $0^+$ g.s.              & $12.53\pm0.28$         & $\geq9.4\times 10^{18}$ & CaWO$_4$ scint. bolometer \cite{Angloher:2016} & 71.8 ($KK$) & $9.9\times10^{4}$ \\
143.23(28)                         & ~                              & ~                     & $\geq1.3\times 10^{18}$ & ZnWO$_4$ scintillator \cite{Belli:2011b} & &  \\
0.12(1)                            & $L_1L_1$, $2^+$ 93.3240(20)     & $27.36\pm0.28$          & ~                       & ~                                        & 11.4 ($L_1L_1$) & $3.3\times10^{3}$ \\
\hline

$^{184}$Os$\rightarrow$$^{184}$W   & $KK$, $0^+$ g.s.              & ~                     & $\geq2.0\times 10^{17}$ & HPGe $\gamma$ spectrometry \cite{Belli:2013a} & &  \\
1452.8(7)                          & $KL$, $0^+$ g.s.              & ~                     & $\geq1.3\times 10^{17}$ & HPGe $\gamma$ spectrometry \cite{Belli:2013a} & &  \\
0.02(2)                            & $LL$, $0^+$ g.s.              & ~                     & $\geq1.4\times 10^{17}$ & HPGe $\gamma$ spectrometry \cite{Belli:2013a} & &  \\
~                                  & 2EC, $0^+$ 1002.49(4)    & $311.3\pm0.7$          & $\geq3.5\times 10^{17}$ & HPGe $\gamma$ spectrometry \cite{Belli:2013a} & 80.2 ($KK$) & 180 \\
~                                  & $KK$, $(0)^+$ 1322.152(22)    & $-8.4\pm0.7$          & $\geq2.8\times 10^{16}$ & HPGe $\gamma$ spectrometry \cite{Belli:2013a} & 80.2 ($KK$) & $2.5\times10^{5}$ \\
~                                  & $KL$, $2^+$ 1386.296(13)      & $-(13.2-15.1)\pm0.7$  & $\geq6.7\times 10^{16}$ & HPGe $\gamma$ spectrometry \cite{Belli:2013a} & 46.4 ($KL_1$) & $4.4\times10^{4}$ \\
~                                  & $LL$, $2^+$ 1431.02(5)        & $(-2.4 - +1.4)\pm0.7$ & $\geq8.2\times 10^{16}$ & HPGe $\gamma$ spectrometry \cite{Belli:2013a} & 9.6 ($L_2L_3$) & $2.4\times10^{9}$ \\
\hline

$^{190}$Pt$\rightarrow$$^{190}$Os  & $KK$, $0^+$ g.s.              & ~                     & $\geq5.7\times 10^{15}$ & HPGe $\gamma$ spectrometry \cite{Belli:2011d} & &  \\
1401.3(4)                          & $KL$, $0^+$ g.s.              & ~                     & $\geq1.7\times 10^{16}$ & HPGe $\gamma$ spectrometry \cite{Belli:2011d} & &  \\
0.012(2)                           & $LL$, $0^+$ g.s.              & ~                     & $\geq3.1\times 10^{16}$ & HPGe $\gamma$ spectrometry \cite{Belli:2011d} & &  \\
~                                  & $KN$, $1,2$ 1326.9(5)         & $(-0.1 - +0.3)\pm0.6$  & ~                     & ~ & $52.3$ ($KN_1$) & $7.0\times10^{8}$ \\
~                                  & $LM$, $(0,1,2)^+$ 1382.4(2)   & $(2.9-6.1)\pm0.5$    & $\geq2.9\times 10^{16}$ & HPGe $\gamma$ spectrometry \cite{Belli:2011d} & 21.9 ($L_1M_1$) & $5.7\times10^{5}$ \\
~                                  & $L_3M_3$, $3^-$ 1387.00(2)        & $1.0\pm0.4$ & ~                      & ~ & 12.6 ($L_3M_3$) & $2.9\times10^{6}$\\
\hline

$^{196}$Hg$\rightarrow$$^{196}$Pt  & $KK$, $0^+$ g.s.              & $661.8\pm3.0$              & $\geq2.5\times 10^{18}$ & Ge(Li) $\gamma$ spectrometry \cite{Bukhner:1990} & 99 ($KK$) & 49 \\
818.6(30)                          & $KL$, $2^+$ 688.693(5)     & $37.6\pm3.0$             & ~                      & ~ & 58.3 ($KL_1$) & $8.9\times10^{3}$ \\
0.15(1)                            & ~                             & ~                      & ~                      & ~ & ~ & ~ \\
\hline
\hline
\end{tabular}}
\end{center}
\end{table}

%\begin{flushleft}
 \footnotetext[1]{The result $4.3\times10^{19}$ yr of \textcite{Fre11} for $^{74}$Se was corrected to $1.4\times10^{18}$ yr by \textcite{Jeskovsky:2015} and reestimated as $3.9\times10^{16}$ yr by \textcite{Lehnert:2016b}. See details in these works.}
 \footnotetext[2]{The result $1.5\times10^{19}$ yr of \textcite{Jeskovsky:2015} for $^{74}$Se was reestimated as $6.3\times10^{17}$ yr by \textcite{Lehnert:2016b}.}
 \footnotetext[3]{In coincidence with HPGe detectors.}
 \footnotetext[4]{The limit is valid for all possible $2\beta$-decay modes in $^{130}$Ba.}
 \footnotetext[5]{The limit is valid for all possible 2EC-decay modes in $^{132}$Ba.}
 \footnotetext[6]{The limit is valid for all possible 2EC-decay modes in $^{164}$Er.}
%\end{flushleft}

\hspace{-3.3mm}sensitive $0\nu2\beta^-$ experiments provide limits on the
effective Majorana mass of the electron neutrino on the level of $
|m_{\beta\beta}| < 0.1 - 0.7$ eV. The uncertainties in
$m_{\beta\beta}$ are related to the uncertainties inherent in the
nuclear structure model calculations of NMEs and the in-medium
modifications of the axial-vector coupling $g_A$.

\begin{figure}[htbp]
\begin{center}
\resizebox{0.38\textwidth}{!}{\includegraphics{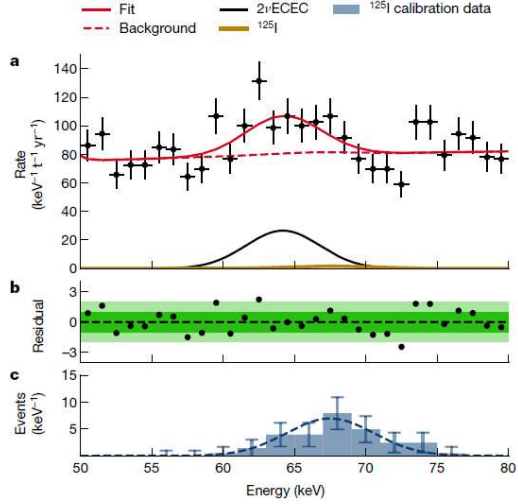}}
\caption{The result of the XENON1t data analysis in the region of
interest for $2\nu$2EC in $^{124}$Xe: (a) $2\nu$2EC decay peak
with area 126 events, obtained by the fit of the experimental
data, (b) residuals for the experimental data best fit, (c) a
histogram of the $^{125}$I peak produced by neutron activation
after the detector calibrations with an external $^{241}$AmBe
neutron source. The peak area in panel (a) corresponds to the
half-life of $^{124}$Xe $1.8 \times 10^{22}$ years. For detail see
the original work of \textcite{XENON:2019}. Reprinted with permission
from \textcite{XENON:2019}.} \label{fig:XENON}
\end{center}
\end{figure}

The sensitivity of the experiments in search for double-beta-plus
processes such as double-electron capture (2EC), electron capture
with positron emission (EC$\beta^+$) and double-positron emission
($2\beta^+$) is substantially lower (see reviews
\cite{Tretyak:1995,Tretyak:2002} and the references in
Tables~\ref{tab:res-exp} and \ref{tab:res-exp-2}). The most
sensitive experiments give limits on double-beta-plus processes on
the level of $\lim T_{1/2}\sim10^{21}-10^{22}$ yr (for $^{36}$Ar,
$^{40}$Ca, $^{58}$Ni, $^{64}$Zn, $^{78}$Kr, $^{96}$Ru, $^{106}$Cd,
$^{112}$Sn, $^{120}$Te, $^{124}$Xe, $^{126}$Xe, $^{130}$Ba,
$^{132}$Ba). It should be noted that until recently there were
only indications even for the allowed two neutrino double electron
capture. However, in 2019 the XENON collaboration claimed
observation of two-neutrino double electron capture in $^{124}$Xe
with the half-life of $(1.8 \pm 0.5) \times 10^{22}$ years
\cite{XENON:2019}. The result of the XENON1t data analysis in the
region of interest for $2\nu$2EC in $^{124}$Xe is shown in Fig.
\ref{fig:XENON}. Double beta decay of $^{130}$Ba was measured by
the geochemical method; half-life is $T_{1/2}=(2.2\pm
0.5)\times10^{21}$ yr obtained by \textcite{MECH01}, and
$T_{1/2}=(6.0\pm 1.1)\times10^{20}$ yr by \textcite{Pujol:2009}. An
indication of the $2\nu$2EC in $^{78}$Kr with the half-life of
$T_{1/2} = [9.2^{+5.5}_{-2.6}(\rm{stat}) \pm 1.3(\rm{syst})]
\times 10^{21}$ yr was obtained in $^{78}$Kr
\cite{Gavrilyuk:2013}. The indications of barium 2EC decay should
be confirmed in direct counting experiments, while the $^{78}$Kr
and $^{124}$Xe results also need to be confirmed with higher
statistics in independent experiments.

In Tables \ref{tab:res-exp} and \ref{tab:res-exp-2}, the experimental limits on
neutrinoless 2EC half-lives are presented. The data is given for
the transitions for which the probability is expected to be
greatest: either to the ground state or to the excited levels of
the daughter nuclide with possible resonant enhancement of the
decay rate. The de-excitation width of the electron shell of the
daughter nuclides $\Gamma_f= \Gamma_{\alpha}+ \Gamma_{\beta}$ and
the resonance parameter $R_f = \Gamma_f/(\Delta^2 + \Gamma_f^2/4)$
(normalized on the $R_f$ value for $^{54}$Fe) are also included in
the tables to indicate the status of the transitions as potential
resonances.

\begin{figure}[htbp]
\begin{center}
\resizebox{0.38\textwidth}{!}{\includegraphics{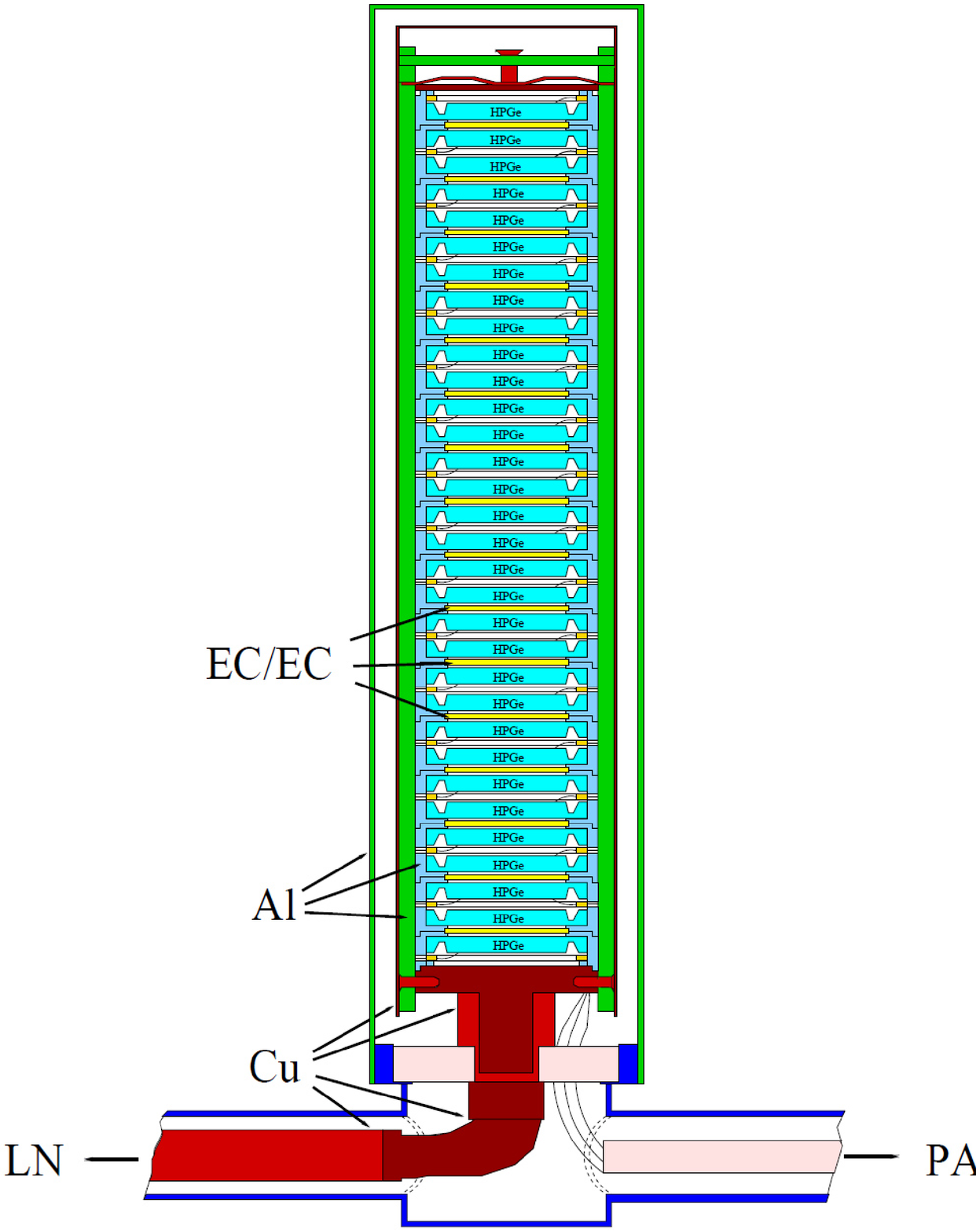}}
\resizebox{0.38\textwidth}{!}{\includegraphics{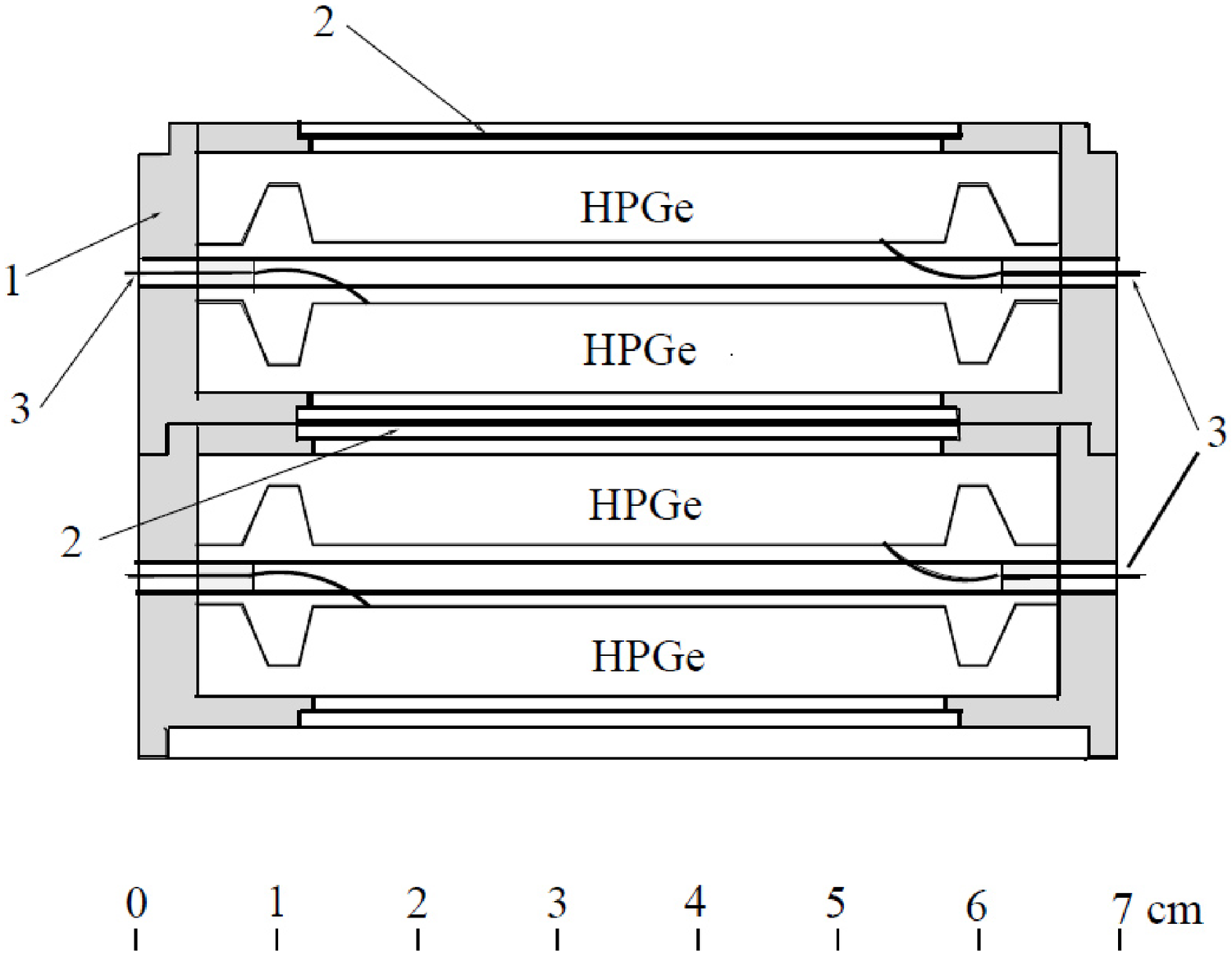}}
\vspace{3mm} \caption{(Color online) Left: Schematic view of the
TGV-2 detector. HPGe - planar-type Ge detectors, EC/EC - enriched
$^{106}$Cd foils, Al - construction details made from Al-Si alloy,
Cu - construction details made from copper, LN - liquid nitrogen,
PA - preamplifiers. Right: Section view of the stack of HPGe
detectors. 1 - cylindrical holders for the detectors, 2 -
$^{106}$Cd foils, 3 - electric contacts
(bronze wires in teflon insulators). Reprinted with permission
from \textcite{Rukhadze:2006}.} \label{fig:tgv}
\end{center}
\end{figure}

The experiments to investigate double-beta-plus processes can be
divided into three groups depending on the experimental technique:
geochemical investigations (the only examples are the searches for
$2\beta$ decay of barium \cite{MECH01,Pujol:2009,Mes17}), $\gamma$
spectrometry (the search for $\gamma$ quanta expected in the decay
processes by using $\gamma$ spectrometers, mainly HPGe detectors) and
the calorimetric approach (potentially $2\beta$ active
nuclei are incorporated into a detector). For instance,
investigations of $^{106}$Cd were realized by using the last two
methods: the search for $\gamma$ quanta from enriched $^{106}$Cd
samples with HPGe
\cite{Rukhadze:2011b,Rukhadze:2011a,Barabash:1996} or NaI(Tl)
detectors \cite{Belli:1999} ($\gamma$ spectrometry); and with the
help of a CdTe cryogenic bolometer \cite{Ito:1997}, CdZnTe diodes
\cite{Dawson:2009}, and a cadmium tungstate crystal scintillator enriched in $^{106}$Cd
($^{106}$CdWO$_4$) \cite{Belli:2012} (calorimetric approach).
Fig.~\ref{fig:tgv} presents a $\gamma$ spectrometry approach used
in the TGV-2 experiment.

\begin{figure}[htbp]
\begin{center}
\resizebox{0.38\textwidth}{!}{\includegraphics{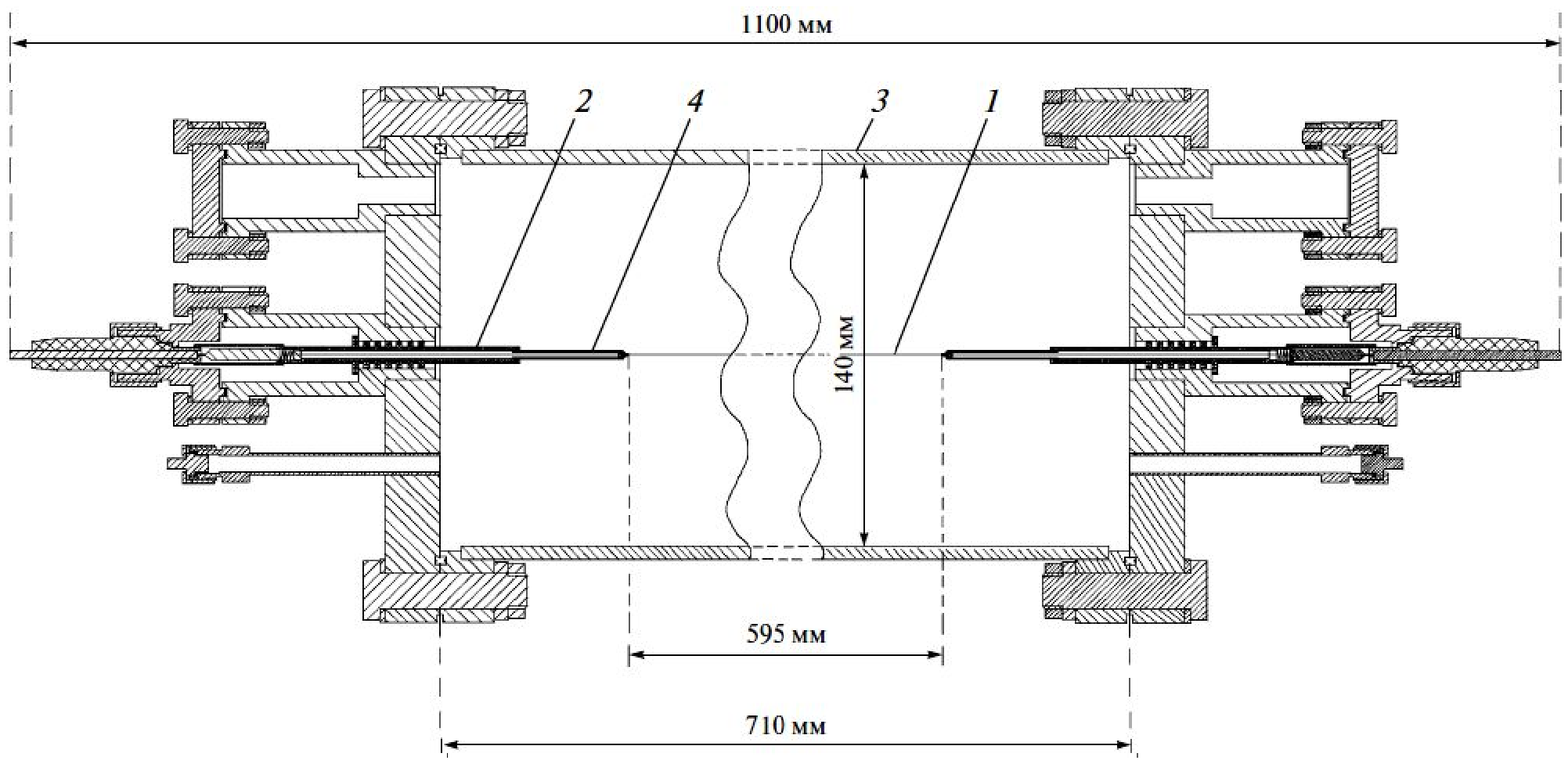}}
\resizebox{0.38\textwidth}{!}{\includegraphics{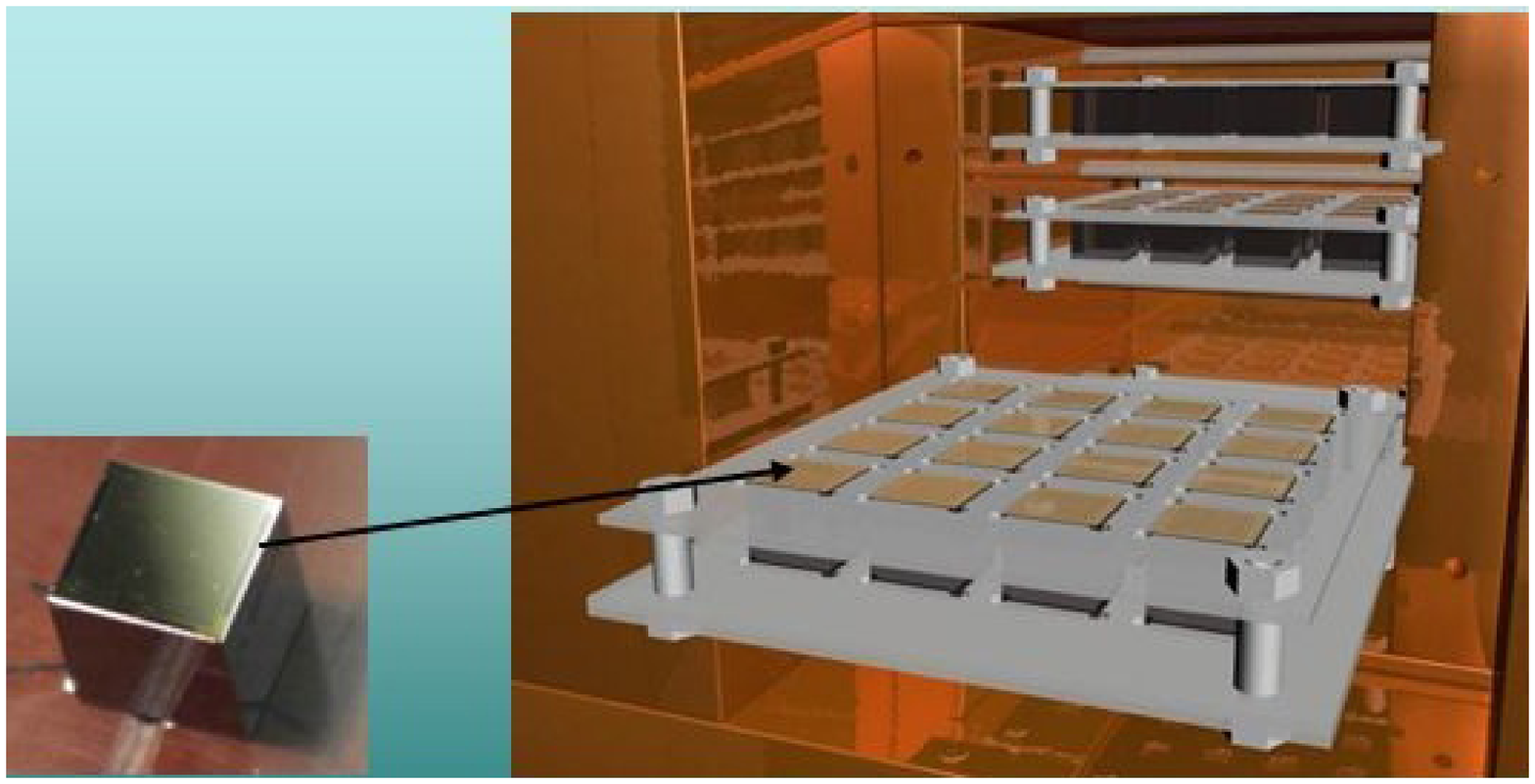}}
\vspace{3mm} \caption{(Color online) Left: Low counting proportional chamber filled with enriched $^{78}$Kr used in the experiment \cite{Gavrilyuk:2013}. 1 - Anode wire, 2 - Insulator, 3 - Cathode, 4 - Copper tubes. Reprinted with permission from \textcite{Gav10}. Right: The concept of the COBRA CdZnTe detectors array. Figure courtesy of K. Zuber.} \label{fig:prop-COBRA}
\end{center}
\end{figure}

The proportional chamber filled with enriched $^{78}$Kr used in
the experiment by \textcite{Gavrilyuk:2013} is an example of the
calorimetric approach (see Fig. \ref{fig:prop-COBRA}, left).
CdZnTe diodes were applied by the COBRA collaboration to
investigate double beta processes in Zn, Cd and Te isotopes (see
Fig. \ref{fig:prop-COBRA}, right). Another case are scintillation
calorimetric experiments, as for instance the experiment to search
for $2\beta$ decay of $^{106}$Cd with the help of the
$^{106}$CdWO$_4$ crystal scintillator \cite{Belli:2012}. The
second stage of the experiment with the $^{106}$CdWO$_4$
scintillation detector operated in coincidence with four HPGe
$\gamma$ detectors \cite{Belli:2016}, represents a combination of
the last two approaches.

\begin{figure}[htbp]
\begin{center}
\resizebox{0.618\textwidth}{!}{\includegraphics{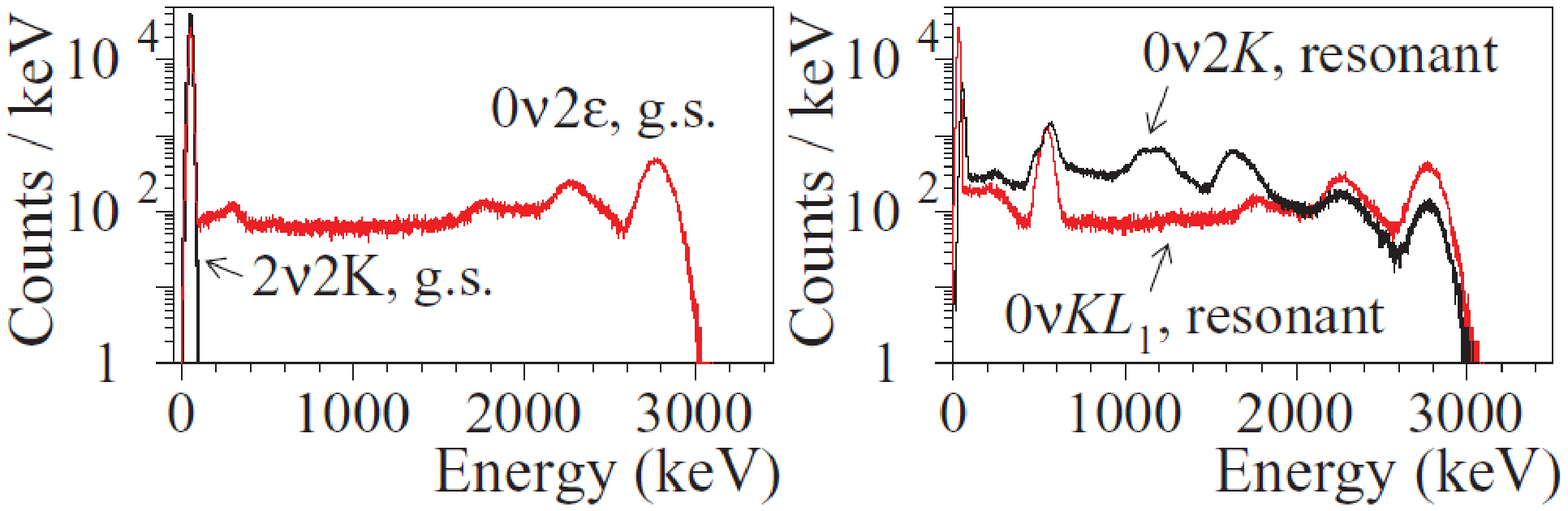}}
\vspace{5mm}\caption{(Color online) Monte Carlo simulated energy
spectra of $2\nu KK$, $0\nu$2EC decays of $^{106}$Cd to the ground
state of $^{106}$Pd, and neutrinoless resonant transitions of
$^{106}$Cd to excited levels in $^{106}$Pd. Reprinted with
permission from \textcite{Belli:2012}.}
 \label{fig:sim}
\end{center}
\end{figure}

An advantage of the $\gamma$ spectrometry method is the high energy
resolution of HPGe detectors (on the level of a few keV). However,
the typical detection efficiency of HPGe $\gamma$ detectors is rather
low: on the level of fractions of percent to several percent
depending on the decay mode and the set-up
configuration\footnote{See, e.g.,
\textcite{Belli:2013b,Barabash:2011,Barabash:2007,BELL11,Belli:2013a,Belli:2011d,Rukhadze:2011a}.}.
The detection efficiency of the calorimetric method is much
higher. Depending on the decay mode, energy threshold, detector
volume and composition, the detection efficiency can approach to
almost 100\%\footnote{See, e.g.,
\textcite{Angloher:2016,Belli:2011b,Gavrilyuk:2013,Belli:2012,Andreotti:2011,Mei:2014,Kiel:2003}.}.
In addition, the calorimetric method allows to distinguish between
different modes and channels of $2\beta$ processes. Examples of
Monte Carlo simulated responses of the $^{106}$CdWO$_4$ detector
to different channels and modes of $2\beta$ decay of
$^{106}$Cd are presented in Fig. \ref{fig:sim}.

\nopagebreak

\nopagebreak

\nopagebreak

\nopagebreak

\begin{figure}[htbp]
%hspace{-20mm} \vspace{-35mm}
\begin{center}
\resizebox{0.38\textwidth}{!}{\includegraphics{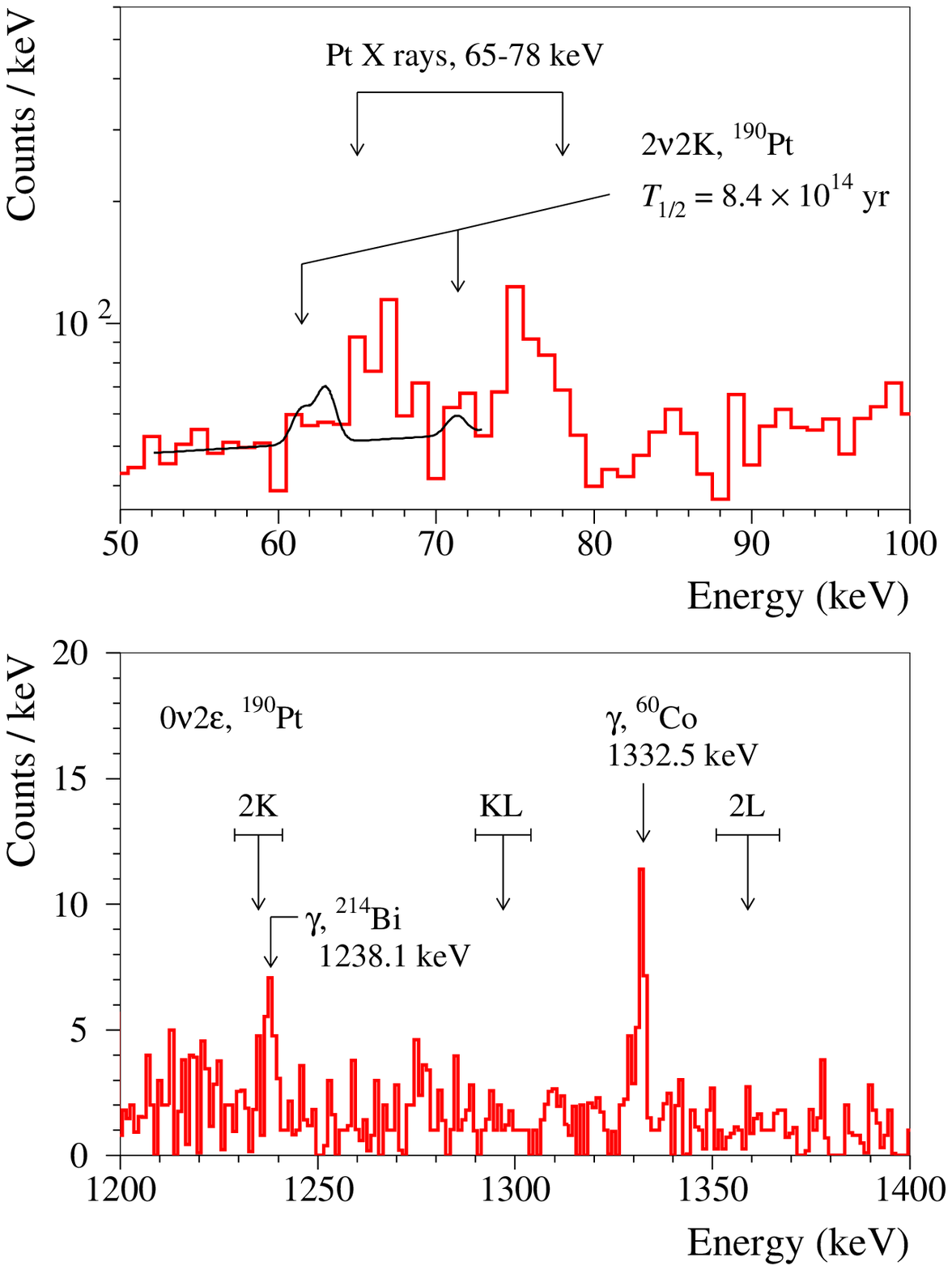}}
%vspace{4cm} \hspace{-4cm}
\caption{(Color online) (Upper panel) Low-energy part of the
spectrum accumulated by a HPGe detector with a platinum sample
over 1815 h. The distribution expected for the $2\nu KK$ decay of
$^{190}$Pt with a half-life of $8.4\times 10^{14}$ yr excluded at
90\% C.L. is shown by a solid line. (Lower panel) Part of the
energy spectrum accumulated with the platinum sample in the energy
region where peaks from the $0\nu$2EC processes in $^{190}$Pt
($KK, KL$ and $LL$) to the ground state of $^{190}$Os are
expected. The error bars take into account the uncertainty of the
$Q$-value of 2EC in $^{190}$Pt available at the time when the
experiment was realized ($\pm 6$ keV) and different energies of
the $L$ X-rays of osmium. Reprinted with permission from
\textcite{Belli:2011d}.} \label{fig:2n-0n}
\end{center}
\end{figure}

However, the $\gamma$ spectrometry approach can clearly distinguish
the $2\nu$ and $0\nu$ modes of the 2EC decay, too. The $2\nu$2EC
decay to the ground state of the daughter nucleus will give a
cascade of X-rays and Auger electrons with the individual energies
scaling up to the maximal energy of the $K$ X-rays. A combination
of X-ray peaks is expected in the decay (as it is shown in Fig.
\ref{fig:2n-0n} (top) for $^{190}$Pt); X-rays from
other atomic shells ($L, M, ...$) can be emitted in the $2\nu$2EC,
too. However, detection of such small energies by, e.g., HPGe
$\gamma$ spectrometry is problematic given that even in the case
of one of the heaviest 2EC nuclides $^{190}$Pt the energy of $L$ X
rays varies within 9 -- 13 keV.

In the process of $0\nu$2EC decay, in addition to X-rays, it is
expected that one or more inner-bremsstrahlung photons are emitted
carrying off the total decay energy, which in the $2\nu$ process
is taken by the neutrinos \cite{ros65}. The energy of the $\gamma$
quanta is expected to be equal to $E_{\gamma}=Q -
\epsilon^*_{\alpha\beta}$, where $\epsilon^*_{\alpha\beta}$ is the
excitation energy of the atomic shell with two vacancies $\alpha$
and $\beta$ of the daughter nucleus. Therefore, the expected
energies of the quanta for the $0\nu$2EC decay to the g.s.
of the daughter nucleus are much higher than those in the
$2\nu$2EC decay (see Fig. \ref{fig:2n-0n} (bottom)).

There are several reasons for the lower sensitivity of the
$2\beta^+$ experiments (in comparison to the $2\beta^-$ ones).
First of all, the development of experimental techniques and the
scale of the experiments are rather modest. E.g., the amount of
enriched isotopes utilized does not exceed the level of tens of
grams, while tens and even hundreds of kilograms of isotopically
enriched materials are already used in $2\beta^-$ experiments.
Then one should take into account the in general much lower
abundance of $2\beta^+$ active isotopes ($\iota$) in the natural
isotopic compositions of elements which is usually lower than 1\%.
There are only 6 nuclei, namely $^{40}$Ca ($\iota=96.941\%$),
$^{54}$Fe (5.845\%), $^{58}$Ni (68.0769\%), $^{64}$Zn (49.17\%),
$^{92}$Mo (14.649\%) and $^{96}$Ru (5.54\%) from the full list of
thirty four double-beta-plus candidates with $\iota$ values
greater than 5\% \cite{Tretyak:2002,Meija:2016}. For comparison,
from the full list of thirty five potentially $2\beta ^{-}$
decaying nuclides, only 5 candidates have $\iota$ below 5\%.
It is interesting to note that, starting from $^{74}$Se, the list
of $2\beta ^{+}$/EC$\beta ^{+}/$2EC nuclei practically coincides
with the list of the so-called bypassed (or p-) nuclei
\cite{fra86} which cannot be created in usual r- and s- processes
of nucleosynthesis by successive neutron captures, and whose
abundances, as a result, are suppressed in comparison to those of
r- and s- nuclei.

Finally, the released energy in the $2\beta ^{+}$ and
EC$\beta^{+}$ decays is lower than that in the $2\beta ^{-}$
decays which results in a lower probability for such processes due
to the small phase space factors. For $0\nu2\beta^-$ decay
processes, because of the relation between half-life
$T_{1/2}^{0\nu }$, phase space factor $G_{0\nu }$, NME
($M^{0\nu}$) and $m_{\beta \beta}$: $1/T_{1/2}^{0\nu} = G_{0\nu }
| M^{0\nu } | ^{2} | m_{\beta\beta}|^{2}$, the lower value of
$G_{0\nu}$ for $2\beta^+$/EC$\beta ^{+}$ processes results in a
weaker limit on the effective neutrino mass (or other parameters),
even if the same experimental $T_{1/2}$ limit is reached as for
the $2\beta ^{-}$ decay. To undergo $2\beta ^{+}$ decay, the
energy release should exceed four electron masses, and only 6
nuclei from the full list of 34 have enough energy for this:
$^{78}$Kr ($Q=2848$ keV), $^{96}$Ru (2715 keV), $^{106}$Cd (2775
keV), $^{124}$Xe (2864 keV), $^{130}$Ba (2619 keV), and $^{136}$Ce
(2379 keV). All these nuclides together with 16 additional
nuclides: $^{50}$Cr, $^{58}$Ni, $^{64}$Zn, $^{74}$Se, $^{84}$Sr,
$^{92}$Mo, $^{102}$Pd, $^{112}$Sn, $^{120}$Te, $^{144}$Sm,
$^{156}$Dy, $^{162}$Er, $^{168}$Yb, $^{174}$Hf, $^{184}$Os, and
$^{190}$Pt, undergo EC$\beta ^{+}$ decays for which the released
energy should exceed $2m_{e} + \epsilon^*_{\alpha}$, where
$\epsilon^*_{\alpha}$ is the binding energy of the captured
electron. And all of these 22 nuclides, together with the 12
remaining ones: $^{36}$Ar, $^{40}$Ca, $^{54}$Fe, $^{108}$Cd,
$^{126}$Xe, $^{132}$Ba, $^{138}$Ce, $^{152}$Gd, $^{158}$Dy,
$^{164}$Er, $^{180}$W, and $^{196}$Hg, undergo 2EC processes. The
lower energy release also results in a lower experimental
sensitivity, since the background counting rate typically
decreases with energy.
In the following we overview the current status of experimental
searches of the $2\beta ^{+}$, EC$\beta ^{+}$ and 2EC processes.

%---------------------------------------------------------------------

\textbf{$^{36}$Ar}. The first result for $0\nu KL$ capture in
$^{36}$Ar, $T_{1/2}^{0\nu} \geq 1.9\times10^{18}$ yr, was obtained \cite{Chk08} in the course of the R\&D investigations in
the GERDA experiment in search for $0\nu2\beta^-$ decay of
$^{76}$Ge using naked HP$^{76}$Ge detectors in liquid argon. It
was improved by 3 orders of magnitude with the data from Phase I
of the GERDA experiment in underground conditions of the Gran
Sasso underground laboratory (LNGS, depth of 3600 m w.e.) where
89.2 t of liquid argon were used as the coolant medium and shield,
$T_{1/2}^{0\nu}(KL)>3.6\times10^{21}$ yr at 90\% C.L. \cite{Agostini:2016}.

%---------------------------------------------------------------------

\textbf{$^{40}$Ca}. Limits for 2EC decay in $^{40}$Ca were set in
1999 with the help of two 370 g low radioactive CaF$_{2}$(Eu)
crystal scintillators at the LNGS over
$\simeq104$ d \cite{Bel99a}. The low energy threshold of 4 keV
allowed to also set a limit on the $2\nu$ process, when an
energy release in the detector of only 6.4 keV is expected. The
achieved $T_{1/2}$ limits (at the level of $10^{21}$ yr) were
recently improved slightly
 with CaWO$_4$ scintillating bolometers
(total exposure of 730 kg~d) \cite{Angloher:2016} used in the
CRESST-II dark matter experiment.

%---------------------------------------------------------------------

\textbf{$^{50}$Cr}. The first limit on EC$\beta^+$ decay of
$^{50}$Cr was set with photographic emulsions to record tracks of
$\beta$ particles \cite{Fre52}. The measurements underground at a
depth of 570 m gave a half-life limit on the level of $10^{14}$
yr. It was improved in 1985 with two HPGe detectors of 110
cm$^{3}$ volume each, searching in coincidence for two 511 keV
$\gamma$ quanta after annihilation of the emitted positron;
measurements of a 148 g Cr sample for 163 h resulted in the limit
$T_{1/2}^{0\nu +2\nu }>1.8\times 10^{17}$ yr \cite{Nor85}. The
best current EC$\beta^+$ sensitivity ($T_{1/2}>1.3\times 10^{18}$
yr) was achieved by \textcite{Bik03} with a HPGe detector which
measured, in coincidence with a NaI(Tl), a 209 g CrO$_{3}$ sample
for 720 h.

%---------------------------------------------------------------------

\textbf{$^{54}$Fe}. The only known limits on double-electron
capture in $^{54}$Fe were set in the measurements of \textcite{Bikit:1998}
with the help of a HPGe detector placed in the centre of an iron
cube with an inner volume of 1 m$^{3}$ and a wall thickness of 25
cm. The search for $\gamma$-rays emitted in the 2EC decay for
$\simeq6700$ h gave $T_{1/2}$ limits on the level of
$(4.1-5.0)\times 10^{20}$ yr depending on the decay channel ($KK$,
$KL$ or $LL$).

%---------------------------------------------------------------------

\textbf{$^{58}$Ni}. The first limit on EC$\beta ^{+}$ decay of $%
^{58}$Ni of $\simeq 10^{17}$ yr was reported in an
experiment with photoemulsion plates \cite{Fre52}. A level of
$\lim T_{1/2}\sim10^{19}$ yr was reached in an experiment with a
Ge(Li) detector and a 2.1 kg Ni sample for 187 h \cite{Bel82}, and
in measurements with a 1.6 kg Ni sample using two NaI(Tl)
scintillators in coincidence for 100 h \cite{NORM84}. The
sensitivity was improved by one order of magnitude in underground
measurements at the Baksan Neutrino Observatory of 660 m w.e.
depth with two NaI(Tl) scintillators and a
1.9 kg Ni sample \cite{Vas93}. Further improvement of the
sensitivity to the neutrinoless 2EC process of
$T_{1/2}>2.1\times10^{21}$ yr was achieved by using a 7.3
kg Ni sample measured for 1400 h with a low background HPGe
detector at the Felsenkeller underground laboratory (110 m w.e.)
\cite{Lehnert:2016a}.
The best to date sensitivity was reported by \textcite{Rukhadze:2018}
where a large volume $\approx600$ cm$^3$ HPGe detector and a 21.7
kg Ni sample installed at the Modane underground laboratory (4800 m w.e.)
are used to search for
$\gamma$ quanta expected in the 2EC and EC$\beta^+$ processes in
$^{58}$Ni. A preliminary limit on the $0\nu$2EC g.s. to
g.s. transition in $^{58}$Ni is $T_{1/2}>4.1\times10^{22}$ yr.

%---------------------------------------------------------------------

\textbf{$^{64}$Zn}. Searches for $^{64}$Zn EC$\beta ^{+}$ activity started in 1952
when a Zn sample was measured with photoemulsions \cite{Fre52}.
However, no useful limits could be extracted from these data,
because the emitted positrons have a quite low energy of $\le 73$
keV. In 1953, the characteristic X-rays resulting from 2EC decay
were searched for with proportional counters. The determined limit
was on the level of $10^{16}$ yr \cite{ber53}. In 1985, EC$\beta
^{+}$ decay of $^{64}$Zn was searched for with two HPGe detectors
with volume 110 cm$^{3}$ each looking for 511 keV $\gamma$-rays in
coincidence from a 228 g Zn sample. A limit $T_{1/2}^{0\nu +2\nu
}>2.3\times 10^{18}$ yr was set after 161 h of measurements
\cite{Nor85}.

In 1995, an indication on a positive EC$\beta ^{+}$ effect was
claimed in measurements (at the Earth's surface)
using a 350 g Zn sample with HPGe and
NaI(Tl) detectors in coincidence. After nearly 400 h of data
taking an excess of counts in the 511 keV annihilation peak
was observed that corresponds to a half-life  $T_{1/2}^{0\nu
+2\nu }=(1.1\pm 0.9)\times 10^{19}$ yr \cite{Bik95}.
This claim was disproved in two experiments:

(1) Search for double-beta-plus processes in $^{64}$Zn with
emission of $\gamma$ quanta by using a HPGe detector (with volume
456 cm$^{3}$) and a CsI(Tl) scintillator ($\simeq400$
cm$^{3}$) in coincidence. The measurements of a 460 g Zn sample
for 375 h at the Cheong Pyung underground laboratory (1000 m w.e.)
gave a limit $T_{1/2}^{0\nu +2\nu }>1.3\times 10^{20}$ yr
\cite{Kim:2007};

(2) A scintillation detector with a 117 g ZnWO$_{4}$ crystal
scintillator was utilized in an experiment at the LNGS for 1902 h.
EC$\beta^+$ decay was not observed, the limits $T_{1/2}^{0\nu
}>2.2\times 10^{20}$ yr and $T_{1/2}^{2\nu }>2.1\times 10^{20}$ yr
were set \cite{Bel07}. The result was further improved with a
larger 0.7 kg detector
 to $T_{1/2}^{0\nu +2\nu }\geq9\times10^{20}$ \cite{Belli:2011b}.

The 2EC processes in $^{64}$Zn were searched for in the following
experiments:

(1) half-life limits on the 2EC decay of $^{64}$Zn on level of
$10^{17}$ yr were set with the help of a Cd$_{0.9}$Zn$_{0.1}$Te
semiconductor detector with mass 2.89 g over 1117 h of data taking
at the LNGS  \cite{Kiel:2003}; the sensitivity was moderately
improved in the next stage of the experiment with four
Cd$_{0.9}$Zn$_{0.1}$Te crystals (6.5 g each) \cite{blo07};
advancements of the experiment sensitivity are expected with the
currently running set-up with 64 Cd$_{0.9}$Zn$_{0.1}$Te detectors
%with a total mass of 380 g
\cite{Ebe16a}.

(2) ZnWO$_{4}$ crystal scintillator, mass of 4.5 g, 429 h of data
collection in the Solotvina underground laboratory (1000 m w.e.),
half-life limits on the level of $10^{18}$ yr \cite{dan05};

(3) ZnWO$_{4}$ scintillators with mass up to 0.7 kg at LNGS
\cite{Belli:2011b}; the strongest to date
%half-life
limits on the 2EC decay of $^{64}$Zn were set:
$T^{2\nu}_{1/2}\geq1.1\times10^{19}$ yr and
$T^{0\nu}_{1/2}\geq3.2\times10^{20}$ yr.

%---------------------------------------------------------------------

\textbf{$^{74}$Se}. The first search for EC$\beta ^{+}$ processes
in $^{74}$Se was performed in measurements of a 563 g Se sample
for 437 h with a 400 cm$^{3}$ HPGe detector and resulted in
$T_{1/2}$ limits on the level of $10^{18}-10^{19}$ yr
\cite{Bar07a}, depending on the decay mode ($2\nu$ or $0\nu$) and
excited level of the daughter nucleus. The next search for a
potentially resonant $LL$ capture of $^{74}$Se to the 1204 keV
level of $^{74}$Ge was realized in the experiment \cite{Fre11} by
HPGe $\gamma$ spectrometry of a 3 kg Se sample with a limit
$T_{1/2}\geq4.3\times10^{19}$ yr. However, this value was
corrected to $>1.4\times10^{18}$ yr by \textcite{Jeskovsky:2015} and
re-evaluated further to $>3.9\times10^{16}$ yr by
\textcite{Lehnert:2016b} for the reason that the method of setting the
limit was not reliable enough. In the work by \textcite{Lehnert:2016b}
also the limit \cite{Jeskovsky:2015} for the $0\nu$2EC transition
of $^{74}$Se to the 1204 keV excited level of $^{74}$Ge,
$T_{1/2}\geq1.5\times10^{19}$ yr, obtained with the help of HPGe
$\gamma$ spectrometry in coincidence with NaI(Tl) scintillation
counter of the same selenium sample as in the experiment
\cite{Fre11}, was re-evaluated to $T_{1/2}\geq6.3\times10^{17}$ yr
for the same reason of a not quite correct interpretation of the
experiment sensitivity as a half-life limit.
%Authors of work
\textcite{Lehnert:2016b}, after measurements of a $\approx2.5$ kg
selenium sample by using HPGe $\gamma$ detector located in the
Felsenkeller Underground Laboratory, gave the strongest to date
limit $T_{1/2}\geq7.0\times10^{18}$ yr for the $0\nu LL$
transition of $^{74}$Se to the 1204.2 keV excited state of
$^{74}$Ge.

%---------------------------------------------------------------------

\textbf{$^{78}$Kr}. $^{78}$Kr is one of the six nuclides whose
$Q$-values allow all three channels of double beta-plus-decays:
2EC, EC$\beta^+$ and $2\beta^+$. The first experiment to search
for EC$\beta^+$ and $2\beta^+$ decay of $^{78}$Kr was performed in
1994 with a high pressure ionization chamber placed within an
array of 6 large volume ($\simeq3000$ cm$^{3}$ each) NaI(Tl)
scintillators. The ionization chamber contained 35 lt of enriched
(to 94.08\%) $^{78}$Kr. The data were collected in the Canfranc
Tunnel Laboratory (675 m w.e.) for 4435 h; derived $T_{1/2}$
limits for $2\beta^{+}$ and EC$\beta^{+}$ processes remain among
the best reached for double-beta-plus nuclei ($\simeq10^{21}$ yr)
\cite{Sae94}.

Experiments \cite{gav00,Gav11} at the Baksan Neutrino Observatory
(BNO, 4900 m w.e.) with enriched $^{78}$Kr lead to limits at the
level of $10^{20}-10^{21}$ yr. An indication on the $2\nu
KK$ capture with $T_{1/2}=9.2^{+5.7}_{-2.9}\times10^{21}$ yr was
obtained with a large proportional
counter (49 lt) filled by gas enriched in $^{78}$Kr to 99.81\%
\cite{Gavrilyuk:2013} (recently updated to
$1.9^{+1.3}_{-0.8}\times10^{22}$ yr \cite{Rat17}). Limits for
other modes of decay were also set, in particular, a
possible resonant $0\nu KK$ capture to the $2^+$
2838 keV level of $^{78}$Se was limited as
$T_{1/2}\geq5.4\times10^{21}$ yr.

%---------------------------------------------------------------------

\textbf{$^{84}$Sr}. The first limit ($\simeq 10^{13}$ yr)
on EC$\beta ^{+}$ decay of $^{84}$Sr was derived by
\textcite{Tretyak:1995} by analysis of the photoemulsion
experiment \cite{Fre52} with corrections on the
decay energy and the natural abundance of the isotope.
Recently, 2EC and EC$\beta ^{+}$ processes in
$^{84}$Sr has been investigated at a level of
$T_{1/2}\sim 10^{15}-10^{16}$ yr
with a small (6.6 g) sample of SrI$_2$(Eu) crystal
scintillator measured over 706 h at LNGS by a 468 cm$^3$ HPGe
detector \cite{Belli:2012a}.
Furthermore, R\&D with a 54 g SrF$_2$ scintillating bolometer
\cite{Dia13} allowed to set limits on few decay modes to
$T_{1/2}\sim 10^{15}-10^{16}$ yr.

%---------------------------------------------------------------------

\textbf{$^{92}$Mo}. A limit on EC$\beta ^{+}$ decay of $^{92}$Mo
can be extracted from a photoemulsion experiment
\cite{Fre52} as $\simeq $10$^{15}$ yr. In 1955, double positron
tracks emitted from a Mo foil were searched for with a Wilson
cloud chamber \cite{win55}. The determined limit of $4\times
10^{18}$ yr should be discarded, because, as we know now, the
energy released in the decay is not sufficient for
emission of two positrons.
Search for 2EC transition of $^{92}$Mo to excited levels of
$^{92}$Zr was realized in 1982 in measurements on the Earth's
surface using a 1.82 kg Mo sample with a 130 cm$^{3}$ Ge(Li)
detector for 161 h, with limits
on the level of 10$^{18}$ yr \cite{Bel82}. In 1985, a
324 g Mo sample was sandwiched between two HPGe detectors
(110 cm$^{3}$ each) looking for coincidence of 511 keV
$\gamma $ quanta; $T_{1/2}$ limits of $\simeq $10$^{17}$ yr were
set \cite{Nor85}. In 1987, positrons emitted from thin Mo foil in
EC$\beta ^{+}$ decay of $^{92}$Mo were searched for with a time
projection chamber; after 3.05 h of data taking, limits were set
as $10^{17}-10^{18}$ yr \cite{ell87}. The best known limits on the double beta decay of $^{92}$Mo on the
level of $10^{21}$ yr were set in 1995--1997 in underground
measurements on 2.5 kg monocrystal Mo rods with a 400 cm$^{3}$
HPGe detector in the Modane underground laboratory for 1316 h
\cite{aun95,Bar97}.

The search for EC$\beta^+$ decay of $^{92}$Mo was also pursued in
the YangYang underground laboratory at a depth of 700 m with a 277
g CaMoO$_4$ crystal scintillator surrounded by 14 big CsI
scintillators which served as a passive and active shielding
\cite{Lee11}; the achieved limit of $T_{1/2}\geq2.3\times10^{20}$
yr is slightly better than the result of \textcite{Bar97}
($T_{1/2}\geq1.9\times10^{20}$ yr). The sensitivity of the
experiment \cite{Kan13}, where a 411 g CaMoO$_4$ crystal was used
in coincidence with a HPGe detector, was on the level of
$10^{19}-10^{20}$ yr.

%---------------------------------------------------------------------

\textbf{$^{96}$Ru}. In fact, $^{96}$Ru is the only nuclide from
the list of 6 potential $2\beta ^{+}$ decaying nuclides which has
an appreciable natural isotopic abundance of $\iota = 5.54\%$.
Notwithstanding these favourable features for experimental
investigation, the first search was performed only in 1985 were a
Ru sample of 50 g was measured with two HPGe detectors (110
cm$^{3}$) in coincidence for 178 h. Determined limits on $2\beta
^{+}$ and EC$\beta ^{+}$ decays to the ground state and excited
levels of $^{96}$Mo were in the range of 10$^{16}$ yr
\cite{Nor85}. These limits were improved to the level of $\simeq
10^{19}$ yr by \textcite{Belli:2009a,Andreotti:2012}. The best results
on the level of $10^{20}-10^{21}$ yr were obtained at LNGS in measurements with a highly purified Ru sample (mass of 720 g)
and four HPGe detectors $\simeq 225$ cm$^3$ each for 5479 h
\cite{Belli:2013b}.

%---------------------------------------------------------------------

\textbf{$^{102}$Pd}. First $T_{1/2}$ limits on 2EC processes in
$^{102}$Pd were obtained in the experiment at the Felsenkeller
underground laboratory, where a 802 g Pd sample was measured with
a HPGe detector; the results were on the level of $10^{18}$ yr
\cite{Lehnert:2011}. They were improved in measurements with two
sandwiched HPGe detectors at the HADES underground laboratory (500
m w.e.) \cite{Lehnert:2013}. Further measurements were performed
at the LNGS in a set-up with four HPGe detectors ($\simeq 225$
cm$^3$ each) for 2093 h; joint analysis of all three measurements
pushed the $T_{1/2}$ limits to the level of $10^{19}$ yr
\cite{Lehnert:2016c}.

%---------------------------------------------------------------------

\textbf{$^{106}$Cd}. The nuclide $^{106}$Cd is among the most
investigated $2\beta ^{+}$ nuclides with a long history of
experimental studies. Limits on $2\beta ^{+}$ and EC$\beta ^{+}$
decays on the level of 10$^{15}$ yr can be extracted from the
measurements using a Cd sample with photographic emulsions \cite
{Fre52}. Searches for positrons emitted in $2\beta ^{+}$ decay
were performed in 1955 with a Wilson cloud chamber in the magnetic
field and a Cd foil with a mass of 30 g; the experiment resulted
in a limit of $\simeq $10$^{16}$ yr \cite{win55}. In 1984, in
measurements of a 153 g Cd sample for 72 h with two NaI(Tl) in
coincidence, limits on the level of $\simeq 10^{17}$ yr were
determined for $2\beta ^{+}$, EC$\beta ^{+}$ and 2EC processes
\cite{NORM84}.

An experiment with a $^{116}$CdWO$_{4}$ crystal scintillator (15
cm$^{3}$, enriched in $^{116}$Cd to 83\%) in the Solotvina
underground laboratory was aimed at the investigation of
$2\beta^-$ decay of $^{116}$Cd. However, the data collected over
2982 h were also used to set limits for other nuclides, in
particular for $^{106}$Cd on the level of 10$^{17}$ to 10$^{19}$
yr, depending on the decay mode \cite{geo95}. A ``source =
detector'' approach and a low energy threshold allowed to set a
limit on $2\nu KK$ capture ($\simeq $10$^{17}$ yr) for the first
time.

An external Cd foil (331 g) was measured at the Modane underground
laboratory with a 120 cm$^{3}$ HPGe detector for 1137 h, aiming at
the detection of $\gamma$ quanta from annihilation of positrons
and from de-excitation of the daughter $^{106}$Pd nucleus if the
excited levels are populated \cite{Barabash:1996}. Half-life
limits on the level of 10$^{18}$--10$^{19}$ yr were set. Also in
1996, a large (1.046 kg) CdWO$_{4}$ scintillator with natural Cd
composition was measured at the LNGS for 6701 h \cite{dan96a}.
Limits on the half-life for $2\beta ^{+}$ and EC$\beta ^{+}$
decays were on the level of $\simeq $10$^{19}$ yr for $0\nu $, and
$\simeq $10$^{17}$ yr for $2\nu $ processes. A small (0.5 g) CdTe
bolometer operating at a temperature of 10 mK was tested for 72 h
by \textcite{Ito:1997}. The derived limit on $0\nu$EC$\beta^{+}$ decay
of $^{106}$Cd was 1.4$\times 10^{16}$ yr, lower than those reached
with more traditional techniques at that time, particularly due to
the small size of the CdTe bolometer.

Cd samples of natural composition with 1.25\% of $^{106}$Cd
were used in all the above experiments (in $^{116}$CdWO$_{4}$
scintillator \cite{geo95} $^{106}$Cd was even depleted).
On contrary, a cadmium sample with mass of 154 g enriched by $^{106}$Cd in 68\%
was used in the experiment \cite{Belli:1999}.
Measurements at LNGS
with two low background NaI(Tl) scintillators for 4321 h enabled
to reach a level of the half-life sensitivity of more than
10$^{20}$ yr for $2\beta ^{+}$, EC$\beta ^{+}$ and 2EC processes
accompanied by emission of $\gamma $ quanta. In 2003, the long-term (14183 h) experiment in the Solotvina
underground laboratory with three enriched $^{116}$CdWO$_{4}$
scintillators (total mass of 330 g) was finished \cite{Dan03}.
Together with measurements for 433 h with a
non-enriched CdWO$_{4}$ crystal of 454 g, sensitivity
for $^{106}$Cd was improved by approximately one order of magnitude
in comparison with \textcite{geo95}.

A radiopure cadmium
tungstate crystal scintillator with a mass of 216 g produced from
cadmium enriched in $^{106}$Cd to 66\% ($^{106}$CdWO$_4$) was
grown in \cite{Belli:2010}. Measurements at the LNGS
gave limits on $2\beta$ processes on
the level of $10^{20}$ yr \cite{Belli:2012}. In the second stage,
the $^{106}$CdWO$_4$ scintillator was installed between four HPGe
detectors ($\simeq 225$ cm$^3$ each) to
improve the sensitivity to decay modes accompanied by $\gamma$
quanta; this resulted in limits of $T_{1/2} \geq 10^{20}-10^{21}$
yr \cite{Belli:2016}. In the
third stage, the $^{106}$CdWO$_4$ detector was
running in coincidence with two big CdWO$_4$ crystal scintillators
in a close geometry to increase the detection efficiency of
$\gamma$ quanta emitted from the $^{106}$CdWO$_4$ crystal. The
sensitivity of the experiment is approaching the theoretical
predictions for the $2\nu$EC$\beta^+$ decay of $^{106}$Cd (that
are in the interval $10^{21}-10^{22}$ yr):
$T_{1/2}\geq4\times10^{21}$ yr \cite{Polischuk:2019}.

At present, there are two other running experiments probing the
double-beta decays of $^{106}$Cd: COBRA and TGV-2. The COBRA
experiment started at the LNGS with two small semiconductor
detectors: Cd$_{0.9}$Zn$_{0.1}$Te (mass of $\simeq $3 g), and CdTe
($\simeq $6 g) \cite{Kiel:2003}. CdZnTe crystals are used in the
current stage \cite{Ebe13,Ebe16b}. Results for $^{106}$Cd
$T_{1/2}$ limits are on the level of $10^{18}$ yr.
The main aim of the TGV-2 experiment (located at the Modane
underground laboratory) is the search for $2\nu KK$ decay of
$^{106}$Cd (this channel has the lowest expected half-life). In
the experiment, 32 planar HPGe detectors are used with a total
sensitive volume $\approx400$ cm$^3$ (see Fig.~\ref{fig:tgv}). The Cd foils
were enriched in
$^{106}$Cd to $60-75$\% in the first stage of the experiment
\cite{Rukhadze:2011b,Rukhadze:2011a,Rukhadze:2006}; now 23.2 g of
Cd enriched to 99.57\% of $^{106}$Cd are used \cite{Ruk16}. After
$\simeq 8200$ h of data taking, limits are on
the level of $10^{20}$ yr.

%---------------------------------------------------------------------

\textbf{$^{108}$Cd}. First limits
(10$^{16}$--10$^{17}$ yr) on 2EC decay of $^{108}$Cd were
determined in 1995, with a $^{116}$CdWO$_{4}$
scintillator in the Solotvina underground laboratory
\cite{geo95}. They were improved later to more than 10$^{17}$ yr
with higher statistics collected (14183 h) and with data of the larger (454 g)
CdWO$_{4}$ scintillator with natural Cd composition
\cite{Dan03,dan96b}. The best limits on the 2EC decay
($\simeq 10^{18}$ yr) were obtained at LNGS
in the COBRA experiment \cite{Kiel:2003} and with a
CdWO$_4$ crystal scintillator of 434 g measured for 2758 h in the
low-background DAMA/R\&D set-up \cite{Belli:2008a}.

%-----------------

\textbf{$^{112}$Sn}. The first limit on the 0$\nu$EC$\beta ^{+}$
decay of $^{112}$Sn on the level of $\simeq 10^{13}$ yr was
derived by \textcite{Tretyak:1995} from the measurements of Sn samples with photographic emulsions \cite{Fre52}. In 2007 bounds
on EC$\beta ^{+}$ processes in $^{112} $Sn were improved to
$\simeq $10$^{18}$ yr in
an R\&D to develop tin-loaded scintillators for $2\beta $ experiments
with $^{112}$Sn and $^{124}$Sn. A 1 lt sample of tetrabutyl-tin
(C$_{4}$H$_{9}$)$_{4}$Sn with the Sn concentration of 34\% was
measured for 140 h with a 456 cm$^{3}$ HPGe detector in the Cheong
Pyung underground laboratory \cite{Kim:2007}.
In another experiment, a 72 cm$^{3}$ Ge detector was used for
measurements of a 1.24 kg Sn sample on the Earth's surface for 831
h \cite{Dawson:2008a}. For 2EC processes with population of the
ground and excited states of $^{112}$Cd, limits were on the level
of $10^{18}$ yr.

In a series of measurements by %different groups
\textcite{Dawson:2008b,Bar08,Kid08,Bar09} performed mostly underground
with HPGe detectors and external Sn sources (with natural
composition and enriched in $^{112}$Sn) limits on the
level of $10^{19}-10^{20}$ yr were obtained. Limits
in the range of $(0.1-1.6)\times 10^{21}$ yr were achieved with a 100 g Sn sample
enriched in $^{112}$Sn to 94.32\% measured in the Modane
underground laboratory with a 380 cm$^3$ HPGe detector for 3175 h
\cite{Barabash:2011}.

%---------------------------------------------------------------------

\textbf{$^{120}$Te}. The first limit on $0\nu$EC$\beta^{+}$ decay
of $^{120}$Te ($\simeq 10^{12}$ yr) was determined by
\textcite{Tretyak:1995} on the basis of the measurements of Te samples
with photoemulsions \cite{Fre52}. It was improved to $\simeq
10^{16}$ yr by the COBRA collaboration with small semiconductor
detectors Cd$_{0.9}$Zn$_{0.1}$Te ($\simeq 3$ g, 1117 h of data
collection) and CdTe ($\simeq 6$ g, 1645 h) in the LNGS
\cite{Kiel:2003}. Limits on $KK$ captures to the ground state and
the first excited level of $^{120}$Sn were set on the level of
10$^{16}$ yr \cite{Kiel:2003}. Somewhat improved results can be
found in \cite{blo07,Daw09}. The 2EC limits were further improved
to $10^{17}-10^{18}$ yr by measuring 1 kg of natural TeO$_{2}$
powder with a 400 cm$^{3}$ HPGe detector for 475 h at the Modane
underground laboratory \cite{Barabash:2007}. The best results for
the EC$\beta^+$ mode were obtained in the CUORICINO/CUORE-0
experiments at LNGS with TeO$_2$ bolometers:
$T_{1/2}>2.7\times10^{21}$ yr for $0\nu$EC$\beta^+$ \cite{Ald18} and
$T_{1/2}>7.6\times10^{19}$ yr for $2\nu$EC$\beta^+$
\cite{Andreotti:2011}.

%---------------------------------------------------------------------

\textbf{$^{124}$Xe}. $^{124}$Xe has the highest available energy
which could be released in $2\beta ^{+}$ processes (2864 keV). In
the first experiment with a gridded ionization chamber (volume of
3.66 lt), filled with Xe and installed at the Baksan Neutrino
Observatory 850 m w.e. underground, the Xe gas of natural
composition and Xe sample enriched in $^{136}$Xe to 93\% were
measured over 710 h and 685 h, respectively. Limits on $2\beta
^{+}$ and EC$\beta ^{+}$ decays of $^{124}$Xe were set in the
range of $10^{14}-10^{18}$ yr, depending on the decay mode
\cite{Bar89}. A limit on $2\nu KK$ decay was set by using a multiwire wall-less
proportional counter (fiducial volume of 4.44 lt, working pressure
of 4.8 atm), also installed at the Baksan Neutrino Observatory but
at a larger depth of 4700 m w.e. After the measurements for
$\simeq 1600$ h with Xe samples of different isotopic composition,
a limit on $2\nu KK$ decay of $^{124}$Xe was set as $T_{1/2}\geq
1.1\times10^{17}$ yr \cite{Gav98}.

Recent developments and the start of the operation of massive dark
matter detectors based on Xe allowed to extract limits on $2\nu
KK$ decay of $^{124}$Xe: $T_{1/2}>2.1\times10^{22}$ yr from 832 kg
of liquid Xe at the Kamioka Observatory, Japan (2700 m w.e.)
measured for 19200 h \cite{Abe18}, and $T_{1/2}>6.5\times10^{20}$
yr from the XENON100 TPC with 62 kg of liquid Xe measured at LNGS
for 5390 h \cite{Apr17}. In both these experiments natural Xe was
used, with a $^{124}$Xe abundance of $\iota= 0.095\%$. The result
from the Baksan Neutrino Observatory, Russia (4900 m w.e.)
obtained with 52 g of Xe enriched in $^{124}$Xe to 21\% after 1800
h of measurement with a copper proportional counter is also known:
$T_{1/2}>2.5\times10^{21}$ yr \cite{Gav15c}. The 2$\nu$2EC in
$^{124}$Xe with a half-life of $T_{1/2}=(1.8 \pm 0.5)
\times 10^{22}$ yr was finally observed in 2019 by using the XENON1T
dark-matter detector \cite{XENON:2019}.

%---------------------------------------------------------------------

{\bf $^{126}$Xe}. The limit on $^{126}$Xe $2\nu KK$ decay
was derived in the XMASS experiment:
$T_{1/2}>1.9\times10^{22}$ yr \cite{Abe18}.

%---------------------------------------------------------------------

\textbf{$^{130}$Ba}. The first limits were derived on the level of
10$^{11}$ yr for the $2\beta ^{+}$ decay and 10$^{12}$ yr for the
EC$\beta ^{+}$ decay \cite{Tretyak:1995} from the experiment with
photographic emulsions \cite{Fre52}. In 1996, by re-analyzing the
old geochemical measurements \cite{SRIN76} with respect to the
amount of the daughter nuclide $^{130}$Xe accumulated in a
BaSO$_{4}$ sample during geological time, the limit of
$T_{1/2}>4\times 10^{21}$ yr was set for all modes of $^{130}$Ba
decay \cite{Bar96a}. It should be noted that an indication on the
effect with $T_{1/2}=2.1^{+3.0}_{-0.8}\times10^{21}$ yr was
obtained for another sample \cite{Bar96a}. A claim on positive
observation of the $^{130}$Ba decay with the half-life
$T_{1/2}=(2.16\pm0.52)\times 10^{21}$ yr (any decay channel and
mode) was reported by \textcite{MECH01} by analysis of $^{130}$Xe
excess in a BaSO$_{4}$ sample. It should be stressed that the
$T_{1/2}$ value is consistent with the theoretical estimates for
the $2\nu KK$ decay of $^{130}$Ba. The excess of $^{130}$Xe was
also found in another geochemical experiment \cite{Pujol:2009} but
the obtained half-life was smaller: $T_{1/2}=(6.0\pm
1.1)\times10^{20}$ yr. In the recent analysis by \textcite{Mes17}, this
disagreement was explained by the contribution from cosmogenics,
and the result of \textcite{MECH01} was considered as more reliable.

As for the direct experiments with $^{130}$Ba, only one experiment
was performed in 2004. A BaF$_{2}$ crystal scintillator with a
mass of 3615 g was measured in coincidence with two NaI(Tl)
detectors for 4253 h at LNGS
\cite{Cer04}. Derived limits on 2$\beta ^{+}$ and EC$\beta ^{+}$
decays to the ground state and few excited levels of $^{130}$Xe
are on the level of 10$^{17}$ yr, far from what is needed to check
the possible positive claim of \textcite{MECH01,Pujol:2009}.

%---------------------------------------------------------------------

\textbf{$^{132}$Ba}. The first limit for 2EC in $^{132}$Ba
$T_{1/2}>3.0\times10^{20}$ yr was set by \textcite{Bar96a} from the
re-analysis of the geochemical data on excess of $^{132}$Xe in
BaSO$_{4}$ samples \cite{SRIN76}. An excess of $^{132}$Xe was observed in the most ``promising'' barite sample in the already mentioned geochemical measurements \cite{MECH01} (in other four samples the $^{132}$Xe excess was too large), leading to the $^{132}$Ba half-life
 of $T_{1/2}=(1.3\pm 0.9)\times 10^{21}$ yr. However, for
$^{132}$Ba the authors preferred more cautiously to give a limit of
$T_{1/2}>2.2\times 10^{21}$ yr.

%---------------------------------------------------------------------

\textbf{$^{136}$Ce}. First limits on $2\beta^+$ decay of
$^{136}$Ce were obtained with the help of two CeF$_{3}$ crystal
scintillators (mass of 75 g and 345 g) measured for 88 h and 693
h, respectively, in the low background set-up at LNGS. In
particular, a sensitivity to the $0\nu 2\beta ^{+}$ decay was on
the level of $T_{1/2}>10^{18}$ yr \cite{Ber97}.

Limits for EC$\beta ^{+}$ and 2EC processes in $^{136}$Ce were
derived from  the long-term (13949 h) measurements with a 635 g
Gd$_{2}$SiO$_{5}$(Ce) crystal scintillator in the Solotvina
underground laboratory; the results for $T_{1/2}$ limits were on
the level of $10^{13}-10^{16}$ yr, depending on the decay mode
\cite{Dan01}. The limit on $2\nu KK$ capture was later improved
from $\simeq $10$^{13}$ yr to $\simeq $10$^{16}$ yr in
measurements with a CeF$_{3}$ scintillation detector (49 g) for
2142 h at LNGS \cite{Bel03}. A small (6.9 g) CeCl$_3$ scintillator
was measured at LNGS for 1638 h also in the ``source = detector''
approach resulting in half-life limits on the level of
$10^{16}-10^{17}$ yr \cite{Bel11a}.

Measurements of Ce-containing materials as external targets with
HPGe detectors at LNGS (CeCl$_3$ 6.9 g, HPGe 244 cm$^3$, 1280 h)
\cite{BELL09} and CeO$_2$ 732 g, HPGe 465 cm$^3$, 1900 h
\cite{Belli:2014}) lead to limits of $\simeq 10^{17}-10^{18}$ yr
for modes accompanied by emission of $\gamma$ quanta. The last
sample was additionally purified (Th contents in a 627 g sample
was reduced by a factor 60) and remeasured for 2299 h with the
same detector. This lead to an improvement of $T_{1/2}$ limits to
the level of $>10^{18}$ yr \cite{Belli:2017}.

%---------------------------------------------------------------------

\textbf{$^{138}$Ce}. In the experiments described above
\cite{Belli:2014,BELL09,Dan01,Bel03,Bel11a}, the 2EC decays in
$^{138}$Ce were searched for too; the to-date best limits on the
neutrinoless $KK$, $KL$ and $LL$ decays are on the level of
$(4.0-5.5)\times10^{17}$ yr \cite{Belli:2017}.

%---------------------------------------------------------------------

{\bf $^{144}$Sm}. The first limit on $2\beta$ decay of $^{144}$Sm
($\simeq 8\times10^8$ yr) was obtained recently by
\textcite{Nozzoli:2018} by analysing the average abundances of
parent-daughter nuclei in the Earth's crust. The first counting
experiment to search for double-beta processes in $^{144}$Sm by
low-background HPGe $\gamma$ spectrometry was performed at the
LNGS by using a highly purified samarium oxide sample, with the
limits on different channels and modes of 2EC and
EC$\beta^+$ decays on the level of
$T_{1/2}\geq(0.13-1.3)\times10^{20}$ yr \cite{Belli:2019a}.

%---------------------------------------------------------------------

{\bf $^{152}$Gd}.
The first limit
on 2EC in $^{152}$Gd was obtained similarly to $^{144}$Sm
as $T_{1/2} > 6\times10^8$ yr
\cite{Nozzoli:2018}.

%---------------------------------------------------------------------

{\bf $^{156}$Dy}. First searches for double beta processes in
$^{156}$Dy were realized at the LNGS by measurements of a 322 g
sample of dysprosium oxide Dy$_2$O$_3$ of 99.98\% purity grade
with a HPGe detector (244 cm$^3$) for 2512 h \cite{BELL11}. The
obtained limits were on the level of $\simeq 10^{14}-10^{16}$ yr,
depending on the decay mode. In the work of \textcite{Fin15} two enriched Dy$_2$O$_3$ sources (803
mg, enriched in $^{156}$Dy to 21.58\%, and 344 mg, 20.9\%) were
measured at the Kimballton underground research facility, USA
(1450 m w.e.) with two HPGe detectors in coincidence. The
$T_{1/2}$ limits were improved to the level of $\simeq
10^{17}-10^{18}$ yr.

%---------------------------------------------------------------------

{\bf $^{158}$Dy}. The limits on 2EC processes in $^{158}$Dy on the
level of $T_{1/2}
> (0.35-1.0)\times10^{15}$ yr ($2\nu$ mode) and $T_{1/2}
> (0.026-4.2)\times10^{16}$ yr ($0\nu$) were obtained at LNGS in the experiment
by \textcite{BELL11} described above.

%---------------------------------------------------------------------

{\bf $^{162}$Er}. The first limit for $^{162}$Er
$T_{1/2}>5.5\times10^8$ yr was obtained by
\textcite{Nozzoli:2018} similarly to $^{144}$Sm (valid for all decay modes).
Much better results close to $10^{18}$ yr for specific
transitions were obtained in measurements of a
highly-purified Er$_2$O$_3$ sample (326 g) with a HPGe detector (465
cm$^3$) for 1934 h at LNGS \cite{Belli:2018}.

%---------------------------------------------------------------------

{\bf $^{164}$Er}. The first limit of $T_{1/2}>1.0\times10^9$ yr
was obtained by\textcite{Nozzoli:2018} (valid for
all 2EC transitions).

%---------------------------------------------------------------------

{\bf $^{168}$Yb}. The first limit of $T_{1/2}>5.7\times10^8$ yr
was obtained by \textcite{Nozzoli:2018}. A much higher sensitivity for different modes and
channels of the decay ($\lim T_{1/2} \sim
10^{14}-10^{18}$ yr) was reached in a low-background experiment
with a sample of highly purified ytterbium oxide measured by a
low-background HPGe detector \cite{Belli:2019b}.

%---------------------------------------------------------------------

{\bf $^{174}$Hf}. The first limit $T_{1/2}>5.8\times10^8$ yr was
obtained by \textcite{Nozzoli:2018}.
The first counting
experiment to search for 2EC and EC$\beta^+$ decay of $^{174}$Hf
was realized using a high-pure 180 g sample of hafnium and the
ultra low-background HPGe-detector system located 225 m
underground at the HADES laboratory. After 75 days of
data taking, limits were set on the level of
$T_{1/2}>10^{16}-10^{18}$ yr \cite{Danevich:2020}.

%---------------------------------------------------------------------

\textbf{$^{180}$W}. First experimental limits on 2EC decays of
$^{180}$W were derived in 1995 from measurements with an enriched
$^{116}$CdWO$_{4}$ crystal scintillator (15 cm$^{3}$) in the
Solotvina underground laboratory for 2982 h \cite{geo95}. They
were upgraded (to $\simeq 10^{16}$ yr) with three $^{116}$CdWO$_{4}$ crystals
(total mass of 330 g) and higher statistics (14183 h), as well as
with measurements for 433 h of a non-enriched CdWO$_{4}$ crystal
(454 g) \cite{Dan03,dan96b}. The limits were further improved in
measurements at LNGS with ZnWO$_4$ scintillators to $\simeq
10^{18}$ yr \cite{Belli:2011b,Bel08c}, and with CaWO$_4$
scintillating bolometers to $\simeq 10^{19}$ yr
\cite{Angloher:2016}.

%---------------------------------------------------------------------

\textbf{$^{184}$Os}. The first limit on EC$\beta ^{+}$ decay of
$^{184}$Os ($\simeq 10^{10}$ yr) was derived by \textcite{Tretyak:1995} from data of the
experiment with photoemulsions \cite{Fre52}. Searches for 2EC and EC$\beta^+$ decays of
$^{184}$Os (including possible resonant $0\nu$2EC transitions)
were performed by \textcite{Belli:2013a} in measurements at the LNGS
with a 173 g ultra-pure Os sample and a 465 cm$^3$ HPGe detector
for 2741 h; this gave limits of $T_{1/2} \simeq 10^{16}-10^{17}$
yr.

%---------------------------------------------------------------------

\textbf{$^{190}$Pt}. Once again, the first limit on EC$\beta ^{+}$
decay of $^{190}$Pt was set from a reanalysis of the photoemulsion
experiment \cite{Fre52} by \textcite{Tretyak:1995} with $T_{1/2} \geq
10^{11}$ yr. Recently 2EC decays of $^{190}$Pt with emission of
$\gamma$ quanta were searched for at LNGS with a 468 cm$^3$ HPGe
detector and a 42 g platinum sample \cite{Belli:2011d} for 1815 h;
half-life limits were on the level of $10^{16}$ yr.

%---------------------------------------------------------------------

\textbf{$^{196}$Hg}. First experimental limits ($\simeq 10^{17}$
yr) on 2EC in $^{196}$Hg were obtained with a Ge(Li) detector (35
cm$^{3}$) which was surrounded by a massive (320 kg) Hg shield
(containing 480 g of $^{196}$Hg) in the Solotvina underground
laboratory for 1478 h \cite{zde86}. The sensitivity was further
improved by one order of magnitude in 1990 with the same Hg
shielding but with a larger HPGe detector (165 cm$^{3}$) over 1109
h of data taking \cite{Bukhner:1990}.

%---------------------------------------------------------------------

\subsection{Prospects for possible future experiments}

There are currently no large-scale projects to search for
$0\nu$2EC. However, further research (corrections of atomic mass values,
discovery of new excited levels) may lead to the identification of
nuclides with favorable conditions for the resonant process. Also detection of $0\nu$2EC process, along with observation of $0\nu2\beta^-$ decay in different nuclei (see discussion, e.g, in \cite{Giuliani:2018}), will be requested in a case of positive evidence of $0\nu2\beta^-$ decay in one nuclide. Furthermore, the light neutrino exchange mechanism,
being the most favourable one, is not a only possible. The investigations of EC$\beta^+$ and $2\beta^+$  decay
might be one of the tools to identify the
mechanism of the decay, whether it is mediated by light neutrino
mass or right-handed currents admixture in weak interaction
\cite{Hirsch:1994}. Here we try to give reader an imagination whether high sensitivity $0\nu$2EC experiments are in principle possible.

One can estimate a sensitivity of possible large scale 2EC
experiments assuming the performance of the most advanced
$2\beta^-$ experiments. For instance both the GERDA and
Majorana $^{76}$Ge $2\beta$ experiments achieve a
background level of $\sim 10^{-3}$ counts/(yr keV kg) in the region of interest $\sim2$ MeV
\cite{Ago18,Ell16}. Assuming the use of 100 kg of a highly
radiopure isotopically enriched material, HPGe detectors with mass of 100
kg, an energy resolution (full width at half of maximum) of 3 keV,
a detection efficiency of $\sim 5\%$, and a measurement time of 5
yr, one can get an estimation of the experimental sensitivity on
the level of $\lim T_{1/2}\sim(4-9)\times 10^{25}$ yr. Moreover, there are plans for a ton-scale
experiment with HPGe detectors \cite{LEGEND} that could further
improve the sensitivity.

The bolometric detectors, thanks to their excellent energy
resolution and possibility to realize the calorimetric approach
with detectors of different chemical formula, look like a very
promising approach in the search for resonant 2EC. Using the level of background estimated by
the CUORE collaboration $\sim 0.004$ counts/(yr keV kg)
\cite{Alessandria:2012}, a mass of isotope of 100 kg (embedded in
highly radiopure crystal scintillators \cite{rad-cont-scint}) and
a detection efficiency around 10\% (e.g., in a peak near
$Q_{\beta\beta}$ of $^{106}$Cd), one can get an estimation of the
experimental sensitivity on the level of $\lim T_{1/2}\sim4\times
10^{25}$ yr for a 5 years experiment.

It should be stressed one more advantage of possible large scale $0\nu$2EC experiments
in comparison to the $0\nu2\beta^-$ searches. In most of the $0\nu$2EC processes closest to resonant ones (see Tables
\ref{tab:table18.1} and XIII), the energy of $\gamma$ quanta expected in the
decays is many times larger than the energy of the X-ray quanta
emitted in the allowed $2\nu$2EC. For instance, in the case of $0\nu$2EC decay of
$^{156}$Dy to the $0^+$ 1988.5 keV excited level of $^{156}$Gd
$\gamma$ quanta with energy 1899.5 keV are expected, while energy of
the X-ray quanta in the $2\nu LL$ process in $^{156}$Dy should be
within $7.2-8.4$ keV (the binding energies on the $L_1$, $L_2$ and
$L_3$ shells of gadolinium atom). Thus, the background due to the
$2\nu$2EC mode in an experiment to search for $0\nu$2EC decay to
excited level(s) with energy much higher than the energy of X-ray
quanta emitted in the $2\nu$2EC decay will never play a role in
practice in contrast to the $0\nu2\beta^-$ experiments where
background caused by the $2\nu$ mode becomes dominant due to poor
energy resolution (scintillation and some gaseous detectors)
\cite{Tretyak:1995} or time resolution (low-temperature
bolometers) \cite{Chernyak:2012}. However, for the cases of the
g.s. to g.s. resonant transitions (e.g., in $^{152}$Gd), a very
small energy release ($Q_{2\mathrm{EC}}(^{152}$Gd$)=55.7$ keV)
makes separation between the effect searched for and the allowed
$2\nu KL$ process practically rather problematic.

\subsection{Neutrinoless 2EC with radioactive nuclides}

Since there is no clealry identified ``ideal nucleus'' to find a
resonant 2EC among stable (or long-lived) nuclei, it makes sense
to consider experiments with radioactive nuclei. Using a few kg of
a radioactive material looks challenging but maybe possible taking
into account previous experience with e.g. the $^{51}$Cr source
for the GALLEX solar neutrino experiment \cite{Cri96} or the
already started BEST sterile neutrino experiment
\cite{Barinov:2019}. Possibilities of large amounts of radioactive
isotopes production for 2EC experiments are discussed in
Sec.~VII.C.

The most realistic approach to search for $0\nu$2EC decay of
radioactive isotopes is $\gamma$ spectrometry. For this purpose
one could use a low background HPGe detector array and samples of
radioactive material as a source. The isotope should be producible
in an amount of tens of kg, should not cause $\gamma$ background
in the region of interest, the half-life of the isotope relative
to the ordinary decay channel should be long enough to carry out
several years of measurements and to avoid thermal destruction of
the sample. In the most favorable case the main decay channel of
the isotope is low energy $\beta$ (EC) or $\alpha$ decay without
$\gamma$-ray emission. The analysis performed by
\textcite{Tretyak:2005} has shown that, despite the very large energy
of the $2\beta^-$ decay of some radioactive nuclei (and therefore
much faster $0\nu2\beta^-$ decay), realization of a high
sensitivity experiment will be limited mainly by heat release in
the detector. However, while the problem is important in the
calorimetry approach requested by $2\beta^-$ experiments, it
becomes less troublesome in the search for $0\nu$2EC processes by
$\gamma$ spectrometry. Nevertheless, taking into account current
rather problematic possibilities for production of long-lived
radionuclides in large amounts and experimental challenges in
ultra-low background measurements with radioactive samples, this
approach is still far from practical realization.

\subsection{Section summary}

To conclude, the highest up-to-date sensitivity to the $0\nu$2EC
decay at the level of $\lim T_{1/2} \sim 10^{21} - 10^{22}$ yr is
achieved using gaseous ($^{78}$Kr), scintillation ($^{106}$Cd),
low-temperature bolometric ($^{40}$Ca) detectors, $\gamma$
spectrometry with HPGe detectors ($^{36}$Ar, $^{58}$Ni, $^{96}$Ru,
$^{112}$Sn), and geochemical method ($^{130}$Ba, $^{132}$Ba).
Observation of the allowed two neutrino mode of 2EC decay is
claimed only for three nuclides: $^{78}$Kr, $^{124}$Xe and
$^{130}$Ba. However, the $2\nu$2EC processes in $^{78}$Kr and
$^{124}$Xe were detected in a single experiment for each nuclide.
Both the claims should be confirmed in independent investigations.
As for the geochemical result for $^{130}$Ba, it needs to be
approved by detection in direct counting experiments. The
sensitivity to the $0\nu$2EC experiments is $3-4$ orders of
magnitude lower than that for the most sensitive $0\nu2\beta^-$
experiments. The main reasons are in general lower sensitivity of
the $0\nu$2EC experiments to the absolute neutrino mass (with a
similar half-life sensitivity), in most of the cases very low
isotopic concentration and limited capabilities of the isotopes of
interest enrichment, a more complicated $0\nu$2EC radiative effect
signature (which results in a lower detection efficiency to the
$0\nu$2EC process), a typically smaller energy release than those
for $0\nu2\beta^-$ decays, while the smaller energy of the effect
searched for complicates suppression of the radioactive
background.

However, the situation might be changed in a case of resonant
enhancement of the $0\nu$2EC. In this case $\gamma$ spectrometry
of external isotopically enriched source with the help of HPGe
diodes, low-temperature bolometers look the most promising
detection techniques for possible large scale experiments to
search for resonant $0\nu$2EC. In addition to the stable nuclide
candidates, searches for the $0\nu$2EC process might be considered
in radioactive nuclides with a reasonably long half-life and
relevant decay mode which will not interfere with the $0\nu$2EC
effect. The most realistic approach in this case could be $\gamma$
spectrometry with HPGe detectors. However, realization of a high
sensitivity experiment with radioactive isotope will be limited by
practical difficulties of its production in large amounts.

%%%%%%%%%%%%%%%%%%%%%%%%%%%%%%%%%%%%%%%%%%%%%%%%%%%%%%
\section{Precise determination of 2EC decay energies}
%%%%%%%%%%%%%%%%%%%%%%%%%%%%%%%%%%%%%%%%%%%%%%%%%%%%%%
\label{Precise determination of $Q$ values}
\setcounter{equation}{0}
\textcolor{black}{The probability of 0$\nu$2EC is in general substantially lower than that of 0$\nu$2$\beta$ decay. The only process that can increase the probability of 0$\nu$2EC and thus make it attractive for its experimental search is its resonant enhancement.
Assuming a resonantly enhanced 0$\nu$2EC-transition is found, its use in experiments on the determination of the neutrino type can provide some advantages compared to the investigation of neutrinoless double $\beta^{-}$-decay. First, there might be a variety of excited nuclear states with different low spin values and different parities in one nuclide to which the double-electron-capture transition can be resonantly enhanced resulting in relatively short partial half-lives. Second, no essential reaction induced background from the two-neutrino mode is expected in the case of ground-state-to-ground-state transitions.
For two nuclides, connected by transition between the nuclear ground states only, the neutrinoless mode dominates since the two-neutrino transition is strongly suppressed by phase space: no energy is left for the neutrinos to carry away. Thus, in the past ten years experimenters focused on the search for nuclides in which such resonantly enhanced 0$\nu$2EC can take place.}

\subsection{Basics of high-precision Penning-trap mass spectrometry}

In order to determine the degree of resonant enhancement of the 0$\nu$2EC-transitions, %given in Tables 1 and 2,
their $Q$-values must be measured with an uncertainty of approximately 100 eV.
The $Q$-value is the mass difference of the initial and final states of the transition
\begin{equation}
Q/c^2=M_i-M_f=M_f\cdot (\frac{M_i}{M_f}-1)=M_f\cdot (R-1),
\end{equation}
where $M_i$ and $M_f$  are the atomic masses of the initial and final states, respectively, and $R=M_i/M_f$  their ratio. An uncertainty $\delta Q$ of the $Q$-value determination is given by
\begin{equation}
\delta Q=Q\cdot \sqrt{ \left( \frac{\delta M_f}{M_f} \right)^2+\left(\frac{\delta R}{R-1}\right)^2}\cong Q\cdot \frac{\delta R}{R-1}; R>1,
\end{equation}
where $\delta M_f$ and $\delta R$ are the uncertainties of the mass of the final state and the mass ratio, respectively. Thus, a determination of the $Q$-value with an uncertainty of a hundred eV implies the measurement of the mass ratio $R$ with an uncertainty of approximately $10^{-9}$.

\begin{figure*} [tbh]
\begin{center}
\includegraphics[width=0.35\textwidth]{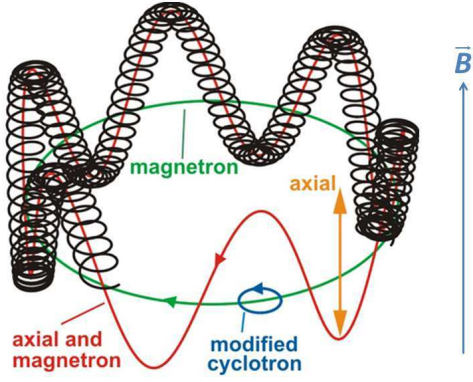}
\caption{\label{fig1} (color online) Motion of an ion in a Penning trap. It can be considered to be a combination of three independent eigenmotions, cyclotron and magnetron motions in the plane orthogonal to the magnetic field lines, and axial motion along the magnetic field line.}
\end{center}
\end{figure*}

The only technique which is capable of providing such a low uncertainty for a large variety of stable and radioactive nuclides is nowadays considered to be high-precision Penning-trap mass-spectrometry (PTMS) \cite{Blaum13, Myers13}. This technique is superior in achievable sensitivity, accuracy and resolving power to all other mass-measurement methods due to the very idea which forms the basis of a Penning trap: one confines a single ion with mass $M$ and electrical charge $q$ to a minute volume by the superposition of extremely stable and strong static homogeneous magnetic field $B$ and weak static quadrupole electric potential. In such a field configuration a charged particle performs a quite complex periodic motion, which is considered to be a product of three circular motions with very stable eigenfrequencies, namely, cyclotron,  magnetron and axial motions with the frequencies $\nu_+$, $\nu_-$ and $\nu_z$, respectively (see Fig.~\ref{fig1}) \cite{Brown86}).

A certain combination of these eigenfrequencies yields the so-called free cyclotron frequency
\begin{equation}
\nu_c=\frac{1}{2\pi}\cdot \frac{q}{M}\cdot B,
\end{equation}
i.e., the frequency of an ion with charge-to-mass ratio $q/M$ in a homogeneous magnetic field $B$. A determination of the mass of a charged particle via a measurement of its free cyclotron frequency, the most precisely measurable quantity in physics, is a remarkable trick which sets the Penning trap
beyond any other mass-measurement technique.
The measurement of the mass ratio $R$ reduces to the measurement of the ratio $R=\nu_c^f/\nu_c^i$  of the cyclotron frequencies $\nu_c^f$ and $\nu_c^i$ of the final and initial states of the transition, respectively.

\begin{figure*} [tbh]
\begin{center}
\includegraphics[width=0.4\textwidth]{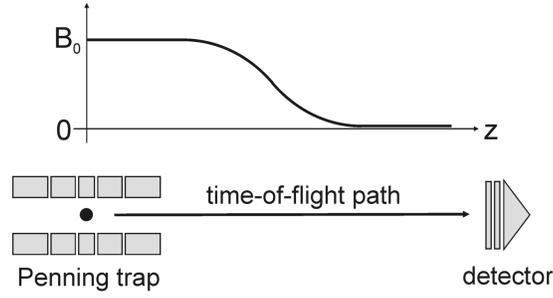}
\caption{\label{fig2} Basic principle of the ToF-ICR technique. The ions are ejected from the trap and sent to the detector through a strong gradient of the magnetic field. The time of flight of the ions depends on their orbital magnetic moment in the trap and is a function of the ion's cyclotron frequency. For details see text.}
\end{center}
\end{figure*}

The majority of the Penning-trap facilities are built for mass-measurements on radioactive nuclides and employ for the measurement of the cyclotron frequency the so-called Time-of-Flight Ion-Cyclotron-Resonance technique (ToF-ICR) \cite{Graeff80, Koenig95}. A key component of this method is a micro-channel-plate (MCP) detector, which is placed on the axis of the Penning trap (direction $\vec{z}$, see Fig. 17) in a region with very low magnetic field. The MCP detector serves as a counter for single ions. The cyclotron frequency is determined from the measurement of the time of flight (ToF) of the ion passing the strong gradient of the magnetic field between the Penning trap and the MCP detector (Fig.~\ref{fig2}).

An ion in a magnetic-field gradient is subject to a force which acts in direction $\vec{z}$ and is given by
\begin{equation}
\vec{F}=-\frac{E_r}{B_0}\cdot \frac{\partial B}{\partial z}\cdot \vec{z},
\end{equation}
where $E_r$ is the ion's radial kinetic energy and $B_0$ is the magnetic field in the trap.
The ion's time of flight between the trap located at zero position and the detector at $z_{det}$ can thus be determined by
\begin{equation}
T(E_r)=\int_0^{z_{det}}dz\cdot \sqrt{\frac{M}{2(E_0-qV(z)-\mu(E_r)B(z))}},
\end{equation}
where $E_0$ is the total initial energy of the ion, $V(z)$ and
$B(z)$ are the electric and magnetic fields along the ion's path
between the trap and MCP detector, and $\mu(E_r)$ is the ion's
orbital magnetic moment in the Penning trap.

\begin{figure*} [tbh]
\begin{center}
\includegraphics[width=0.6\textwidth]{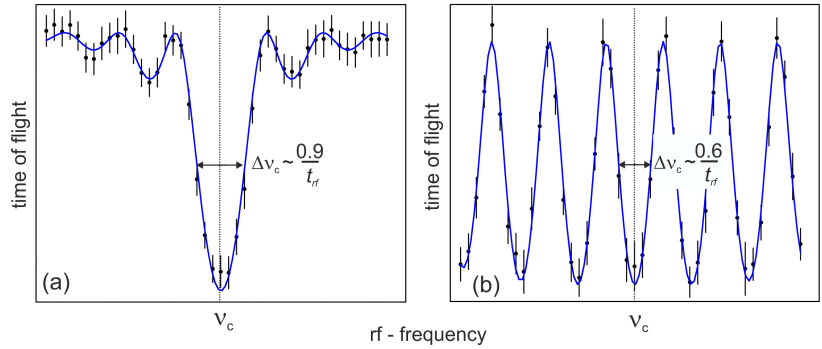}
\caption{\label{fig3} The line shape of the time of flight vs. the (a) one-pulse and (b) two-pulse (Ramsey) rf-field of duration $T_{rf}$, respectively. For details see text.}
\end{center}
\end{figure*}

The orbital magnetic moment of the ion in the trap and hence its time of flight can be manipulated by applying a quadrupolar radiofrequency field (rf-field) of certain temporal profile at a frequency near the ion's cyclotron frequency  in the trap. By varying the frequency of the rf-field of duration $T_{rf}$, one obtains the time of flight vs frequency of the rf-field as shown, e.g., in Fig.~\ref{fig3}(a) for a one-pulse rf-field \cite{Koenig95} and in Fig.~\ref{fig3}(b) for a two-pulse rf-field \cite{Kretzschmar2007,George2007,George07}.

The two-pulse method (two-pulse Ramsey) is usually the method of choice since it provides the highest precision for the determination of the cyclotron frequency $\nu_c$ due to a large number of very pronounced periodic minima in the time-of-flight line shape.
The cyclotron frequency of singly charged ions of mass 100 u in a magnetic field of 7 T is approximately 10$^6$ Hz. A half-an-hour measurement of the cyclotron frequency with the rf-pulse duration of 2 s and two-pulse Ramsey configuration allows a determination of the cyclotron frequency with a relative uncertainty of about $5\cdot10^{-9}$ to $10^{-8}$. The final achievable uncertainty of the frequency-ratio determination is usually defined by the instability of the magnetic field in time \cite{Droese11} and amounts to approximately $10^{-9}$ in the experiments which employ the ToF-ICR detection technique.

Recently, a novel Phase-Image Ion-Cyclotron-Resonance (PI-ICR) technique has been invented \cite{Eliseev2013,Eliseev2014} for on-line facilities like SHIPTRAP. In this method the measurement of the ion-cyclotron frequency is based on the projection of the ion position in the trap onto a position-sensitive detector. This allows one to monitor the time evolution of the ion  motion and thus to measure the trap-motion frequencies of the ion with subsequent determination of the ion free cyclotron frequency.
Compared to the conventional ToF-ICR technique, the PI-ICR offers a gain in precision and resolving power of approximately 5 and 50, respectively. This has made it feasible to carry out  measurements of mass ratios of long-lived nuclides with an uncertainty of just a few ten eV at on-line Penning-trap facilities \cite{Nesterenko2014,SEliseev2015,Jonas2019}.

\subsection{Decay energies of 2EC transitions in virtually stable nuclides}

Since 12 out of 19 most promising nuclide pairs from Tables~\ref{tab:ground_states} and \ref{tab:excited_states}
have been addressed with SHIPTRAP \cite{Block07} and  three pairs have been investigated with the mass-spectrometers JYFLTRAP \cite{Kol04}, TRIGATRAP \cite{Ket08} and the FSU Penning trap \cite{Red05}, which are in many aspects similar to SHIPTRAP, the experiments on the determination of the $Q$-values are described here by the example of the SHIPTRAP mass spectrometer.

\begin{figure*} [tbh]
\begin{center}
\includegraphics[angle = 0,width=0.7\textwidth]{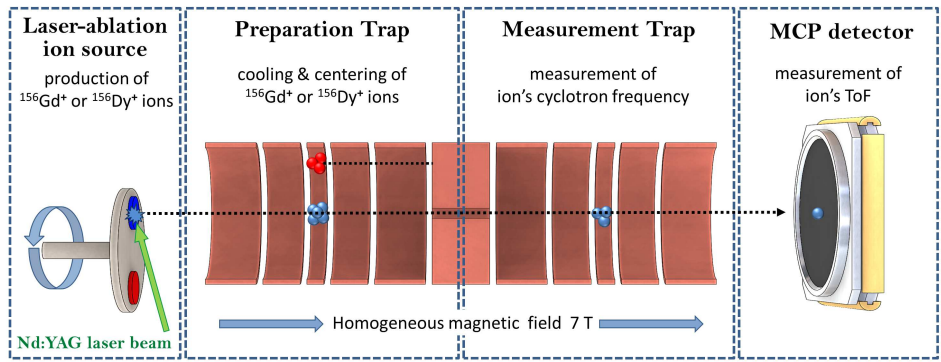}
\caption{\label{fig4} Sketch of the off-line SHIPTRAP facility. The ions produced with a laser ablation are guided by ion transport optics towards the Penning traps. After the mass-selective cooling and centering of the ions in the preparation trap, they are sent into the measurement trap, where their cyclotron frequency is measured with the ToF-ICR technique. For details see text.}
\end{center}
\end{figure*}

The SHIPTRAP facility has been built for experiments on transuranium nuclides produced in fusion-evaporation reactions at GSI, Darmstadt. A detailed description of the entire facility can be found in \cite{Block07}. Here, only off-line SHIPTRAP - the part relevant to the measurements of the $Q$-values - is described (Fig.~\ref{fig4}).

The nuclides of interest are virtually stable and can be purchased in sufficient amounts in different chemical forms. For a production of singly charged ions of these nuclides a laser-ablation ion source was used \cite{Ankur07}. For this, a few milligrams
of the nuclide of interest were shaped into a 5$\times$5 mm$^2$ solid target on a rotatable holder. These targets were then irradiated with  short laser pulses. The frequency-doubled Nd:YAG laser (532 nm) has a pulse duration of 3 - 5 ns, a pulse energy of 4 - 12 mJ and a diameter of the laser beam on the target of less than 1 mm. The material is ionized by laser induced desorption, fragmentation and ionization. A series of electrostatic electrodes transport the ions from the ion source towards the Penning-trap mass spectrometer.

The Penning-trap mass spectrometer has two cylindrical Penning traps - the preparation trap (PT) and measurement trap (MT) - placed in a magnetic field of 7 T created by a superconducting magnet. The PT separates the ions of interest from unwanted ions by employing the mass-selective buffer gas cooling technique \cite{Savard91}. From the PT only the ions of interest pass into the MT where their cyclotron frequency is measured with the ToF-ICR technique described above.

\begin{figure*} [b]
\begin{center}
\includegraphics[width=0.64\textwidth]{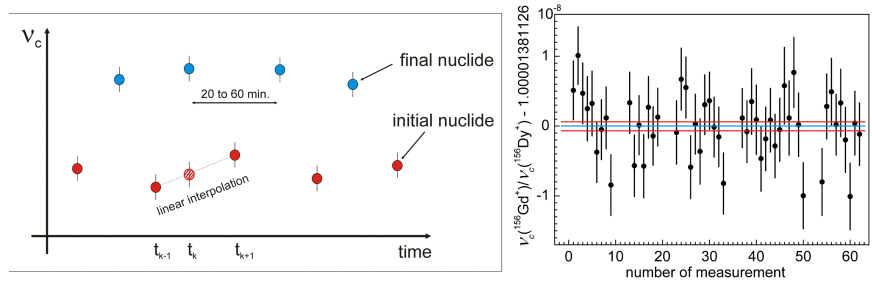}
\caption{\label{fig5} (Colour online) Illustration of the principle of an alternate measurement of the cyclotron frequencies of two nuclides (left) (for details see text), which results in a determination of their ratio (right) (by the example of the frequencies-ratio measurements of $^{156}$Gd and $^{156}$Dy). In the left figure the filled squares and filled circles represent the single measurements of the frequencies of the mother and daughter nuclides, respectively. The hatched circle is the value obtained from the linear interpolation to the time $t_k$ of the frequencies measured at times $t_{k-1}$ and $t_{k+1}$. In the right plot the central and upper/lower lines represent the weighted mean frequency ratio and the error of the determination of the weighted mean frequency ratio, respectively.}
\end{center}
\end{figure*}

The ratio of the cyclotron frequencies of the initial and final nuclides of the transition is obtained by a measurement of the two cyclotron frequencies alternately as schematically depicted in Fig.~\ref{fig5} (left).

In this case, the ratio which corresponds, e.g., to measurement time $t_k$ (see Fig.~\ref{fig5} (left)) is made up of the frequency of the mother nuclide measured at time $t_k$ and the frequency of the daughter nuclide which is obtained by linear interpolation of the frequencies of the daughter nuclide measured at times $t_{k-1}$ and $t_{k+1}$. The measurement time of one ratio-point usually does not exceed 1 hour and is measured with an uncertainty of better than $10^{-8}$. Thus, a  measurement campaign of a few-days results in a determination of the frequency ratio with the required uncertainty of about $10^{-9}$, as shown in Fig.~\ref{fig5} (right) by the example of double-electron capture in $^{156}$Dy \cite{ELIS11Dy}. The corresponding uncertainty of the $Q$-value is about 100 eV.
The results obtained in this measurement campaign are summarized in the review of \textcite{Eliseev2012} and included in Tables \ref{tab:ground_states} and \ref{tab:excited_states}.
\textcolor{black}{The choice of nuclide pairs for measurements was made from the analysis of data for all stable nuclides, which from the energy balance can undergo double electron capture. This data set is presented in Tables \ref{tab:res-exp} and \ref{tab:res-exp-2}. There, in addition to energy, are presented other parameters that allow one to judge the proximity of a particular level to the resonant state, leading to an increase in the probability of double capture. These parameters include the energy gap $\Delta$ and the width $\Gamma$ of the states of electrons undergoing capture, on which the resonance gain factor $R$ depends. The most promising cases were selected for the experiments, the results of which are given in Tables \ref{tab:ground_states} and \ref{tab:excited_states}.}
\begin{table}[tbh]
\caption{
Parameters of the three most promising 2EC transitions between nuclear ground states which $Q$-values have been precisely measured with SHIPTRAP.
}
\label{tab:ground_states}
\centering
\scriptsize

\begin{tabular}{cccc}
\hline\hline
 2EC transition  &  $^{152}$Gd $\rightarrow$ $^{152}$Sm \cite{ELIS11Gd}   &   $^{164}$Er $\rightarrow$ $^{164}$Dy \cite{Eliseev2012}    &   $^{180}$W $\rightarrow$ $^{180}$Hf \cite{DROE12}   \\
\hline
electron orbitals        & $KL_1$      &  $L_1L_1$   &  $KK$       \\
$Q_{\mathrm{2EC}}$ (old) / keV          & 54.6(35)   &  23.3(39)  &  144.4(45) \\
$\Delta$ (old) / keV     & 0.27(350)  &  5.05(390)  &  12.4(45)   \\
$Q_{\mathrm{2EC}}$ (new) / keV          & 55.70(18)   &  25.07(12)  &  143.20(27) \\
$\Delta$ (new) / keV     & 0.83(18)    &  6.82(12)   &  11.24(27)  \\
\hline\hline
\end{tabular}
\end{table}

\begin{table}[tbh]
\caption{
2EC transitions to nuclear excited states which are of interest in the search for the 0$\nu$2EC process and for which $Q_{\mathrm{2EC}}$-values have been precisely measured with Penning traps.
}
\label{tab:excited_states}
\centering
\scriptsize

\begin{tabular}{ccccccc}
\hline\hline
 0$\nu$2EC transition                      &  $E^*$ / keV   &   $J_f^{\pi}$   &   electron orbitals &  $Q_{\mathrm{2EC}}$ / keV    & $\Delta$ / keV  & Ref.  \\
\hline
$^{74}$Se $\rightarrow$ $^{74}$Ge    &  1204.205(7)  &   2$^+$     &   $L_2L_3$          &  1209.169(49) & 2.50(5)         &  \textcite{Kolhinen2010}     \\
                                     &               &             &                     &  1209.240(7)  & 2.57(1)         &  \cite{74Se2}     \\
$^{78}$Kr $\rightarrow$ $^{78}$Se    &  2838.49(7)  &   2$^+$     &   $L_1M_4$          &  2847.75(27)  & 7.55(28)       &  \textcite{BUST13}     \\
                                     &               &             &   $L_3M_2$          &               & 7.66(28)       &                    \\
$^{96}$Ru $\rightarrow$ $^{96}$Mo    &  2700.21(6)   &   2$^+$     &   $L_2L_2$          &  2714.51(13)  & 9.05(14)        &  \textcite{96Ru162Er168Yb}      \\
$^{102}$Pd $\rightarrow$ $^{102}$Ru  &  1103.047(13) &   2$^+$     &   $KL_3$            &  1203.27(36)  & 75.26(36)       &  \textcite{102Pd106Cd}     \\
$^{106}$Cd $\rightarrow$ $^{106}$Pd  &  2748.2(4)    &   (2,3)$^-$ &   $KL_3$            &  2775.39(10)  & 0.33(41)       &  \textcite{102Pd106Cd}      \\
$^{112}$Sn $\rightarrow$ $^{112}$Cd  &  1871.00(19)  &   0$^+$     &   $KK$              &  1919.82(16)  & 4.50(25)       &  \textcite{Rahaman2009}      \\
$^{120}$Te $\rightarrow$ $^{120}$Sn  &       ?       &     ?       &      ?              & 1714.81(1.25) &    ?            &  \textcite{SCIE09}     \\
$^{124}$Xe $\rightarrow$ $^{124}$Te  &  2790.41(9)   & 0$^+$-4$^+$ &   $KK$              &  2856.82(13)  & 1.96(16)        &  \textcite{124Xe130Ba}      \\
$^{130}$Ba $\rightarrow$ $^{130}$Xe  &  2544.43(8)   & 0$^+$       &   $KK$              &  2623.71(26)  & 10.15(26)       &  \textcite{124Xe130Ba}      \\
$^{136}$Ce $\rightarrow$ $^{136}$Ba  &  2315.32(7)   & 0$^+$       &   $KK$              &  2378.53(27)  & 11.67(28)      &  \textcite{136Ce}      \\
$^{144}$Sm $\rightarrow$ $^{144}$Nd  & 1560.920(13)  &   2$^+$     &   $KL_3$            & 1782.59(87)   & 171.89(87)      &  \textcite{102Pd106Cd}     \\
$^{156}$Dy $\rightarrow$ $^{156}$Gd  &  1946.375(6)  &   1$^-$     &   $KL_1$            &  2005.95(10)  & 0.75(10)        &  \textcite{ELIS11Dy}      \\
                                     &  1952.385(7)  &   0$^-$     &   $KM_1$            &               & 1.37(10)        &                    \\
                                     &  1988.5(2)    &   0$^+$     &   $L_1L_1$          &               & 0.54(24)        &                   \\
                                     &  2003.749(5)  &   2$^+$     &   $M_1N_3$          &               & 0.04(10)        &                   \\
$^{162}$Er $\rightarrow$ $^{162}$Dy  &  1782.68(9)   &   2$^+$     &   $KL_3$            &  1846.95(30)  & 2.69(30)        &  \textcite{96Ru162Er168Yb}     \\
$^{168}$Yb $\rightarrow$ $^{168}$Er  & 1403.7357(23) &   2$^-$     &   $M_2M_2$          &  1409.27(25)  & 1.52(25)        &  \textcite{96Ru162Er168Yb}      \\
$^{184}$Os $\rightarrow$ $^{184}$W   & 1322.152(22)  &   0$^+$     &   $KK$              &  1453.68(58)  & 8.89(58)       &  \textcite{184Os}      \\
$^{190}$Pt $\rightarrow$ $^{190}$Os  &  1326.9(5)    &   1,2       &   $KN_1$            &  1401.57(47)  & 0.14(69)        &  \textcite{190Pt}      \\
                                     &               &             &   $KN_2$            &               & 0.25(69)        &                    \\
                                     &               &             &   $KN_3$            &               & 0.34(69)        &                   \\
                                     &               &             &   $KN_4$            &               & 0.51(69)        &                    \\
\hline\hline
\end{tabular}
\end{table}

%\subsection{Prospects for decay energies of 2EC transitions in radioactive nuclides}
\subsection{Prospects for measurements of decay energies in radioactive nuclides}

The use of radioactive nuclides in the search for 2EC transitions was first proposed by \textcite{BerNov}. Of particular interest were nuclides which are situated very far from the valley of beta-stability. They can undergo double-electron capture with a much higher decay energy than the nuclides along the valley of beta-stability. Since the probability of double-electron capture exhibits a very strong dependence on the $Q_{\mathrm{2EC}}$-value, the half-lives of such nuclides with regard to double-electron capture can be short enough to be measured. Similar to the virtually stable nuclides, also here one could search for resonantly enhanced 0$\nu$2EC by measuring the $Q_{\mathrm{2EC}}$-values of promising 2EC transitions.

The main decay mode of such nuclides must not hamper the search for double-electron capture. In other words, it should be either alpha-decay to a nuclear ground state or electron capture with a low $Q$-value.
Furthermore, from practical point of view, the half-lives of such nuclides should be at least a few years. The most promising near-resonant nuclides which fulfill these criteria are listed in Table \ref{tab:radioactive_2EC}; their specific $0\nu$2EC data are given in Table~XIV.

\begin{table}[tbh]
\caption{
Radioactive nuclides with half-lives $T_{1/2} > 1$~yr, which might be of interest for $0\nu$2EC. $Q_{\alpha}$ and $T_{1/2}^{\alpha}$ are the $Q$ value and half-life of the $\alpha$-decay mode, respectively. The $Q_{\mathrm{2EC}}$, $Q_{\mathrm{EC}}$ and $Q_{\alpha}$ values and $T_{1/2}^{\mathrm{EC}}$ are from \textcite{Wang:2017}, $T_{1/2}$ are from from the database of Brookhaven National
Laboratory (http://www.nndc.bnl.gov/ensdf/).
%The half-lives are in years, $Q$ values are in keV.
}
\label{tab:radioactive_2EC}
\centering
\scriptsize

%\resizebox{\textwidth}{!}
{
\begin{tabular}{cccccccc}
\hline\hline
Mother & $T_{1/2}$ & $Q_{\mathrm{2EC}}$ & $T_{1/2}^{\mathrm{2EC}\ast )}$ & $Q_{\mathrm{EC}}$ & $T_{1/2}^{\mathrm{EC}}$ & $Q_{\alpha }$ & $T_{1/2}^{\alpha }$ \\
nuclide & yr & keV & yr & keV & yr & keV & yr \\
\hline
$^{148}$Gd & $71.1\pm 1.2$ & 3066.9(9) & $10^{25-33}$ & 30(10) & 10$^{3}$ & 3271 & 70 \\
$^{150}$Gd & $(1.79\pm 0.08)\times 10^{6}$ & 1287(6) & $10^{23-33}$ & - & - & 2807 & $(3.00\pm 0.15)\times 10^{6}$ \\
$^{154}$Dy & $(3.0\pm 1.5)\times 10^{6}$ & 3312(7) & $10^{23-30}$ & - & - & 2945 & 3$\times 10^{6}$ \\
$^{194}$Hg & $444\pm 77$ & 2576(3) & $10^{25-31}$ & 28(4) & 10$^{3}$ & 2698 & $\sim 10^{20\ast )}$ \\
$^{202}$Pb & $(5.25\pm 0.28)\times 10^{4}$ & 1405(4) & $10^{22-32}$ & 40(4) & 10$^{2}$ & 2589 & $\sim 10^{23\ast\ast )}$ \\
\hline\hline
\end{tabular}
}
\begin{flushleft}
\vspace{-3mm}
~~~~~~~~~~~~~~~~~~~~~~~~~~~~~~~~~~~~~~~~$^{*)}$ According to Sec.~VIII.\\
~~~~~~~~~~~~~~~~~~~~~~~~~~~~~~~~~~~~~~~$^{**)}$ Values estimated with the Geiger-Nuttal approach \cite{Geiger}.
\end{flushleft}
\end{table}

Provided a resonantly enhanced transition has been found in these nuclides by means of Penning-trap mass spectrometry, a suitable mechanism will have to be found for a production of macroquantities of these nuclides. Spalation/fragmentation reactions with high-energy particles and  a fusion with heavy ions do not allow the production of a quantity which would be sufficient for a large-scaled experiment in search of 0$\nu$2EC (see Table \ref{tab:production_of_radioactive_nuclides1}).
\begin{table}[tbh]
\caption{
Estimated yields for the production of the radioactive nuclides of interest for the search for neutrinoless 2EC in spalation/fragmentation reactions with high-energy particles and fusion with heavy ions.
}
\label{tab:production_of_radioactive_nuclides1}
\centering
\scriptsize

\begin{tabular}{ccccc}
\hline\hline
 Mother                       &  ISOLDE/CERN $^{*)}$                      &   yield / atoms            &  yield / atoms                      &  max. counts                             \\
   nuclide                    &  counts / ions/$\mu$C      &   (spallation, 3 months)   &  (fusion with $^{48}$Ca, 3 months)  &  $\mid\lambda$Y$\mid_{2EC}$ / y$^{-1}$    \\
\hline
$^{148}$Gd                           &  6$\times10^8$                     &   $\approx 10^{16}$        &   too small                         &                                           \\
$^{150}$Gd                           &  4$\times10^8$                     &   $\approx 10^{16}$        &   $\approx 10^{13}$                 &         $\approx 10^{-7}$                 \\
$^{154}$Dy                           &  8$\times10^9$                     &   $\approx 10^{17}$        &   $\approx 10^{13}$                 &         $\approx 10^{-6}$                 \\
$^{194}$Hg                           &  7$\times10^9$                     &   $\approx 10^{17}$        &   $\approx 10^{13}$                 &                                           \\
$^{202}$Pb                           &  $10^7$                            &   $\approx 10^{14}$        &   too small                         &         $\approx 10^{-9}$                 \\
\hline\hline
\end{tabular}
\begin{flushleft}
\vspace{-3mm}
~~~~~~~~~~~~~~~~~~~~~~~~~~~~~~~$^{*)}$ http://isolde.web.cern.ch
\end{flushleft}
\end{table}
A more promising production mechanism is an irradiation of samples with neutrons in a reactor. Unfortunately, only relatively short-lived nuclides can be produced in this manner (see Table \ref{tab:production_of_radioactive_nuclides2}).
To calculate the production yields, the mass of the irradiated sample was taken equal to 1 mole, and the neutron flux was
10$^{15}$ n/cm$^2$/s.
Thus, in principle a production of sufficient amount of $^{169}$Yb and $^{175}$Hf is feasible.
However, since these nuclides are too short-lived, their handling might be problematic.\\
\begin{table}[tbh]
\caption{
Parameters of relatively short-lived nuclides with near-resonant transitions for the search for 0$\nu$2EC,
which can be produced in a reactor.
The $Q_{\mathrm{EC}}$ and $Q_{\mathrm{2EC}}$-values are in keV units \cite{Wang:2017};
$E^*$ in keV units and $J_i^{\pi}$ are from the database of Brookhaven National Laboratory (http://www.nndc.bnl.gov/ensdf/).
}
\label{tab:production_of_radioactive_nuclides2}
\centering
\scriptsize
%\resizebox{\textwidth}{!}
{\begin{tabular}{ccccccc}
\hline\hline
nuclide     &  $J^{\pi}_i$  &   $T_{1/2}^{\mathrm{exp}}$   &  $Q_{\mathrm{EC}}$   &  $Q_{\mathrm{2EC}}$  & $E^*$, possible relevant levels           & $\mid\lambda$Y$\mid_{\mathrm{EC}}$ / y$^{-1}$        \\
            &             &                      &                          &                          & in daughter nuclide   &  (3 months in reactor)                       \\
\hline
$^{93}$Mo   &  5/2$^+$       &   $(4.0 \pm 0.8)\times 10^3$ y    &   405.8~(15)              &  315.0~(5)                & 266.8~(3/2$^+$)                        &  $10^{-3}$                                   \\
$^{113}$Sn  &  1/2$^+$       &   $115.09 \pm 0.03$ d              &   1039.0~(16)             &  715.2~(16)               & 680.52~(3/2$^+$), 708.57~(5/2$^+$)         &  3$\times10^{-3}$                            \\
$^{145}$Sm  &  7/2$^-$       &   $340 \pm 3$ d              &   616.2~(25)              &  780.6~(9)               & 748.275~(9/2$^-$)                      &  $10^{-2}$                                   \\
$^{169}$Yb  &  7/2$^-$       &   $32.018\pm 0.005$ d               &   897.6~(11)             &  545.5~(3)                &  ?                                 &  10                                          \\
$^{175}$Hf  &  5/2$^{(-)}$       &   $70 \pm 2$ d               &   683.9~(20)              &  213.9~(23)              &  ?                                 &  4                                           \\
\hline\hline
\end{tabular}}
\end{table}

%%%%%%%%%%%%%%%%%%%%%%%%%%%%%%%%%%%%%%%%%%%%%%%%%%%%%%%%%%%
%%%%%%%%%%%%%%%%%%%%%%%%%%%%%%%%%%%%%%%%%%%%%%%%%%%%%%%%%%%
%%%%%%%%%%%%%%%%%%%%%%%%%%%%%%%%%%%%%%%%%%%%%%%%%%%%%%%%%%%
\section{Normalized half-lives of near-resonant nuclides}
%%%%%%%%%%%%%%%%%%%%%%%%%%%%%%%%%%%%%%%%%%%%%%%%%%%%%%%%%%%
%%%%%%%%%%%%%%%%%%%%%%%%%%%%%%%%%%%%%%%%%%%%%%%%%%%%%%%%%%%
%%%%%%%%%%%%%%%%%%%%%%%%%%%%%%%%%%%%%%%%%%%%%%%%%%%%%%%%%%%
\setcounter{equation}{0}

In this section, estimates of the $0\nu2$EC half-lives of nuclei closest to the resonance condition are presented.
Over the past ten years, great progress has been made due to accurate measurements of $Q_{\mathrm{2EC}}$-values,
which made it possible to clarify whether the resonance condition for the prospective nuclides is satisfied (see Sec.~VII).
The $Q_{\mathrm{2EC}}$ values of the identified earlier prospective nuclides have now been measured.

%%%%%%%%%%%%%%%%%%%%%%%%%%%%%%%%%%%%%%%%%%%%%%%%%%%%%%%%%%%%%%%%%%%%%%%%%%%%%%%%%%%%%%%%%%%%%%%%%%%%
%\begin{landscape}
%\begin{turnpage}

\begin{table}[]
\vspace{-4mm}
\caption{The $ 0 \nu $2EC processes closest to the resonant ones.
The normalized half-lives $\tilde{T}_{1/2}$ take newly measured $Q_{\mathrm{2EC}}$ values %of Tables III and IV
into account. The 1st column reports the natural isotopic abundance ($\iota$)
of the parent nuclides. The spin and parity of the daughter nuclides are given in column 3. If the spin or parity are unknown, their suggested or assumed values are given in round or square brackets, respectively. Column 4 reports the excitation energies of daughter nuclides together with the experimental errors. Column 5 lists the degeneracy parameter of the two atoms $\Delta = M^{**}_{A,Z-2} - M_{A,Z}$, including the excitation energy of the electron shell; the errors indicate the experimental uncertainty in the $Q_{\mathrm{2EC}}$ values. The quantum numbers of the electron vacancies $\alpha$ and $\beta$ are given in the next two columns, where $ n $ is the principal quantum number, $ j $ is the total angular momentum, and $ l $ is the orbital angular momentum. Columns 8, 9 and 10 enumerate the energies of the vacancies  $\epsilon_{\alpha}^*$ and  $\epsilon_{\beta}^*$ and the energy shift $\Delta\epsilon_{\alpha\beta}^*$ due to the Coulomb interaction, relativistic and collective electron shell effects. Column 11 presents the widths of the excited electron shells. The minimum and maximum normalized half-lives (in years) corresponding to the 99\% C.L. interval \textcolor{black}{determined by the uncertainty in the degeneracy parameter $\Delta$}
are presented in the last two columns. The masses, energies and widths are given in keV.}
\label{tab:table18.1}
\centering
\renewcommand{\arraystretch}{1.1}
\addtolength{\tabcolsep}{1pt}
\scriptsize
%\tiny
\vspace{2mm} \resizebox{\textwidth}{!}
{\begin{tabular}{|c|c|c|c|r|c|c|r|r|r|r|c|c|c|c|c|c|} \hline
\hline $\iota$  \T \B &   Transition  & $J^{\pi}_f$ & $M^*_{A,Z-2}
- M_{A,Z - 2}$ & $M^{**}_{A,Z-2} - M_{A,Z}$ & $(n 2j l)_{\alpha}$
& $(n 2j l)_{\beta}$ & $\epsilon_{\alpha}^*\;\;$ &
$\epsilon_{\beta}^*\;\;$ & $\Delta\epsilon_{\alpha\beta}^*$ &
$\Gamma_{\alpha \beta}\;\;\;\;\;\;$  & $\tilde{T}_{1/2}^{\min}$ &
$\tilde{T}_{1/2}^{\max}$ \\
\hline \hline
5.52\% \T                                 &$^{ 96}_{ 44}$Ru$ \to ^{ 96}_{ 42}$Mo$^{**}$  & 0$^{+ }$    &2742  $\pm$  1      & 28.1  $\pm$   0.13   &  310
   &  410    &  0.50    &  0.06    &  0.02    &$9.5\times 10^{  -3}$
&$1.0\times 10^{37}$       &$1.3\times 10^{37}$      \\
\hline
1.25\%   \T                               &$^{106}_{~48}$Cd$ \to ^{106}_{~46}$Pd$^{**}$  &   $[0^+]$   &2737   $\pm$   1      & 11.0  $\pm$   0.10  &  110
   &  110    & 24.35    & 24.35    &  0.72    &$1.3\times 10^{  -2}$
&$8\times 10^{31}$       &$2\times 10^{32}$     \\
 \T  \B    &    &             &                  & -10.3  $\pm$   0.10    &  110    &
210    & 24.35    &  3.60    &  0.18    &$1.0\times 10^{  -2}$   &
$3\times 10^{32}$       &$8\times 10^{32}$    \\
\hline
0.095\%  \T                               &$^{124}_{~54}$Xe$ \to ^{124}_{~52}$Te$^{**}$  & [$0^+$]   &2853.2    $\pm$   0.6
&  2.5  $\pm$   0.12    &  210    &  310   &  4.94    &  1.01    &  0.06    &$1.2\times 10^{  -2}$
 &$5\times 10^{31}$       &$1\times 10^{33}$      \\
 \T  \B                                &
       &             &                 &  1.6  $\pm$   0.12    &  210    &  410   &  4.94    &  0.17    &  0.02    &$4.6\times 10^{  -3}$
 &$2\times 10^{29}$       &$3\times 10^{ 35}$      \\
\hline
0.185\%  \T  \B                             &$^{136}_{~58}$Ce$ \to ^{136}_{~56}$Ba$^{**}$  & 0$^{+ }$    &2315.32   $\pm$   0.07~$^{*)}$ & 12.6  $\pm$  0.27
                                                          &  110    &  110    & 37.44    & 37.44    &  0.93    &$2.6\times 10^{  -2}$
&$2\times 10^{31}$       &$2\times 10^{31}$      \\
                                        &                                           &  [0$^{+}$]  &2349.5   $\pm$   0.5  & 46.7  $\pm$  0.27    &  110    &  110    & 37.44    & 37.44    &  0.93    &$2.6\times 10^{  -2}$    &$2\times 10^{32}$       &$2\times 10^{32}$      \\
                                        &                                           &             &                  & 14.6  $\pm$  0.27    &  110    &  210    & 37.44    &  5.99    &  0.23    &$1.5\times 10^{  -2}$    &$1\times 10^{32}$       &$2\times 10^{32}$      \\
                                        &                                           &             &                  &  9.7  $\pm$  0.27   &  110    &  310    & 37.44    &  1.29    &  0.08    &$2.4\times 10^{  -2}$    &$2\times 10^{32}$       &$3\times 10^{32}$      \\
                                        &                                           & $(1^+,2^+)$ &2392.1  $\pm$  0.6
& 20.9  $\pm$  0.27    &  210    &  310    &  5.99    &  1.29    &  0.05    &$1.3\times 10^{  -2}$    &$3\times 10^{33}$       &$5\times 10^{33}$      \\
                                        &                                           & $(1^+,2^+)$ &2399.87   $\pm$  0.05
& 64.9$ \pm$ 0.27   &  110    &  210    & 37.44    &  5.99    &  0.19    &$1.5\times 10^{  -2}$    &$1\times 10^{33}$       &$1\times 10^{33}$      \\
\hline \T
0.20\%                                  &$^{152}_{~64}$Gd$ \to ^{152}_{~62}$Sm$^{*} $  & 0$^{+ }$    &   0
& 39.0 $\pm$  0.18    &  110    &  110  & 46.83    & 46.83    &  1.09    &$4.0\times 10^{  -2}$    &$5\times 10^{31}$       &$5\times 10^{31}$      \\
                                        &                                           &             &                  & -0.9  $\pm$  0.18   &  110    &  210    & 46.83    &  7.74    &  0.28    &$2.3\times 10^{  -2}$    &$3\times 10^{28}$       &$3\times 10^{29}$      \\
                                        &                                           &             &                  & -1.3  $\pm$  0.18    &  110    &  211    & 46.83    &  7.31    &  0.33    &$2.3\times 10^{  -2}$    &$1\times 10^{31}$       &$5\times 10^{31}$      \\
                                        &                                           &             &                  & -7.1 $\pm$  0.18    &  110    &  310    & 46.83    &  1.72    &  0.09    &$3.2\times 10^{  -2}$    &$3\times 10^{31}$       &$3\times 10^{31}$      \\
                                        &                                           &             &                  & -8.5 $\pm$  0.18    &  110    &  410    & 46.83    &  0.35    &  0.04    &$2.4\times 10^{  -2}$    &$2\times 10^{32}$       &$3\times 10^{32}$      \\
\hline \T
0.06\%                                  &$^{156}_{~66}$Dy$ \to ^{156}_{~64}$Gd$^{**}$  & 1$^{- }$    &1946.375  $\pm$  0.006
& -1.1  $\pm$   0.10    &  110    &  211    & 50.24    &  7.93    &  0.33    &$2.6\times 10^{  -2}$    &$5\times 10^{29}$       &$2\times 10^{30}$      \\
                                        &                                           &             &                  & -7.6  $\pm$    0.10  &  110    &  311    & 50.24    &  1.69    &  0.10    &$2.8\times 10^{  -2}$    &$2\times 10^{32}$       &$2\times 10^{32}$      \\
                                        &                                           & 0$^{- }$    &1952.385  $\pm$  0.007
&  5.3  $\pm$   0.10  &  110    &  210    & 50.24    &  8.38    &  0.29    &$2.6\times 10^{  -2}$    &$8\times 10^{31}$       &$1\times 10^{32}$      \\
                                        &                                           &             &                  & -1.4  $\pm$   0.10  &  110    &  310    & 50.24    &  1.88    &  0.10    &$3.5\times 10^{  -2}$    &$1\times 10^{31}$       &$3\times 10^{31}$      \\
                                        &                                           & 1$^{- }$    &1962.037  $\pm$  0.012
& 14.5   $\pm$   0.10   &  110    &  211    & 50.24    &  7.93    &  0.33    &$2.6\times 10^{  -2}$    &$2\times 10^{32}$       &$2\times 10^{32}$      \\
                                        &                                           &             &                  &  8.1  $\pm$   0.10  &  110    &  311    & 50.24    &  1.69    &  0.10    &$2.8\times 10^{  -2}$    &$2\times 10^{32}$       &$2\times 10^{32}$      \\
                                        &                                           & 1$^{+ }$    &1965.950  $\pm$  0.004
& 18.8  $\pm$   0.10  &  110    &  210    & 50.24    &  8.38    &  0.23    &$2.6\times 10^{  -2}$    &$1\times 10^{31}$       &$1\times 10^{31}$      \\
                                        &                                           &             &                  & 12.2  $\pm$   0.10  &  110    &  310    & 50.24    &  1.88    &  0.08    &$3.5\times 10^{  -2}$    &$2\times 10^{31}$       &$2\times 10^{31}$      \\
                                        &                                           &             &                  & 10.6  $\pm$   0.10  &  110    &  410    & 50.24    &  0.38    &  0.03    &$2.7\times 10^{  -2}$    &$8\times 10^{31}$       &$8\times 10^{31}$      \\
                                        &                                           &  [$0^+$]    &1970.2   $\pm$  0.8
& 65.8$\pm$  0.10    &  110    &  110    & 50.24    & 50.24    &  1.14    &$4.5\times 10^{  -2}$    &$8\times 10^{31}$       &$8\times 10^{31}$      \\
                                        &                                           &             &                  & 23.1$\pm$  0.10   &  110    &  210    & 50.24    &  8.38    &  0.29    &$2.6\times 10^{  -2}$    &$5\times 10^{31}$       &$8\times 10^{31}$      \\
                                        &                                           &             &                  & 16.4$\pm$  0.10   &  110    &  310    & 50.24    &  1.88    &  0.10    &$3.5\times 10^{  -2}$    &$8\times 10^{31}$       &$1\times 10^{32}$      \\
                                        &                                           & 0$^{+ }$    &1988.5   $\pm$   0.2~$^{*)}$ & 84.1  $\pm$   0.10  &  110    &  110    & 50.24    & 50.24    &  1.14    &$4.5\times 10^{  -2}$    &$1\times 10^{32}$       &$1\times 10^{32}$      \\
                                        &                                           &             &                  & 41.4  $\pm$   0.10    &  110    &  210    & 50.24    &  8.38    &  0.29    &$2.6\times 10^{  -2}$    &$2\times 10^{32}$       &$2\times 10^{32}$      \\
                                        &                                           &             &                  & -0.6  $\pm$   0.10    &  210    &  210    &  8.38    &  8.38    &  0.19    &$7.6\times 10^{  -3}$    &$8\times 10^{25}$       &$8\times 10^{30}$      \\
                                        &                                           &             &                  & -1.1  $\pm$   0.10   &  210    &  211    &  8.38    &  7.93    &  0.14    &$7.7\times 10^{  -3}$    &$8\times 10^{31}$       &$8\times 10^{32}$      \\
                                        &                                           & 0$^{+ }$    &2026.664   $\pm$   0.006   & 79.5 $\pm$ 0.10  &  110    &  210    & 50.24    &  8.38    &  0.23    &$2.6\times 10^{  -2}$    &$2\times 10^{32}$       &$2\times 10^{32}$      \\
\hline \T
0.14\%                                  &$^{168}_{~70}$Yb$ \to ^{168}_{~68}$Er$^{**}$  & 1$^{- }$    &1358.898  $\pm$  0.005
& 16.7 $\pm$  0.25    &  110    &  211    & 57.49    &  9.26    &  0.36    &$3.3\times 10^{  -2}$    &$8\times 10^{31}$       &$1\times 10^{32}$      \\
                                        &                                           &             &                  &  9.2  $\pm$  0.25   &  110    &  311    & 57.49    &  2.01    &  0.11    &$3.5\times 10^{  -2}$    &$8\times 10^{31}$       &$1\times 10^{32}$      \\
                                        &                                           &             &                  &  7.5 $\pm$  0.25  &  110    &  411    & 57.49    &  0.37    &  0.04    &$3.4\times 10^{  -2}$    &$2\times 10^{32}$       &$3\times 10^{32}$      \\
                                        &                                            & 0$^{+ }$    &1422.10  $\pm$  0.03 &129.0  $\pm$  0.25  &  110    &  110    & 57.49    & 57.49    &  1.25    &$5.7\times 10^{  -2}$    &$1\times 10^{32}$       &$2\times 10^{32}$ \\
\hline \T
0.13\%                                  &$^{180}_{~74}$W$ \to ^{180}_{~72}$Hf$^*$  & 0$^{+ }$    &         0
&-11.2 $\pm$ 0.27   &  110    &  110    & 65.35    & 65.35    &  1.36    &$7.2\times 10^{  -2}$    &$5\times 10^{ 29}$       &$5\times 10^{ 29}$      \\
                                        &                                           &             &                  &-66.2 $\pm$ 0.27    &  110    &  210    & 65.35    & 11.27    &  0.36    &$4.2\times 10^{  -2}$    &$1\times 10^{32}$       &$1\times 10^{32}$      \\
\hline \T
0.02\%                                  &$^{184}_{~76}$Os$ \to ^{184}_{~74}$W$^{**}$ & $(0)^{+ }$  &1322.152 $\pm$ 0.022~$^{*)}$
&  8.8 $\pm$ 0.58   &  110    &  110    & 69.53    & 69.53    &  1.42    &$8.0\times 10^{  -2}$    &$2\times 10^{29}$       &$3\times 10^{29}$      \\
                                        &                                           &             &                  &-49.6 $\pm$ 0.58   &  110    &  210    & 69.53    & 12.10    &  0.37    &$4.6\times 10^{  -2}$    &$3\times 10^{31}$       &$5\times 10^{31}$      \\
\hline \T
0.014\%                                 &$^{190}_{~78}$Pt$ \to ^{190}_{~76}$Os$^{**}$  &$(0,1,2)^+$  &1382.4  $\pm$ 0.2~$^{*)}$
&  7.0 $ \pm$ 0.47 &  210    &  210    & 12.97    & 12.97    &  0.24    &$1.4\times 10^{  -2}$    &$2\times 10^{31}$       &$33\times 10^{31}$      \\
                                        &                                           &             &                  & -3.0 $\pm$ 0.47  &  210    &  310    & 12.97    &  3.05    &  0.11    &$2.2\times 10^{  -2}$    &$2\times 10^{30}$       &$1\times 10^{31}$      \\
                                        &                                           &             &                  & -3.3 $\pm$ 0.47  &  210    &  311    & 12.97    &  2.79    &  0.09    &$1.6\times 10^{  -2}$    &$2\times 10^{32}$       &$1\times 10^{33}$      \\
                                        &                                           &             &                  & -5.5 $\pm$ 0.47 &  210    &  410    & 12.97    &  0.65    &  0.05    &$1.5\times 10^{  -2}$    &$8\times 10^{31}$       &$2\times 10^{32}$      \\ \hline \hline
 \end{tabular}}
\begin{flushleft}
$^{*)}$~Decay channels with the known NMEs listed in Table~\ref{tab:NME}.
\end{flushleft}
\end{table}

%\end{turnpage}

\subsection{Decays of virtually stable nuclides}

New estimates of the half-lives, taking the recent $Q$-value measurements into account, are presented in Tables~\ref{tab:table18.1} and \ref{table83}.
In view of the considerable variance of the NME values for various nuclides (Table~\ref{tab:NME})
and to not mix distinct physical effects, the normalized half-lives $\tilde{T}_{1/2}$ with NME = 3 are provided.
\textcolor{black}{In the following we focus mainly on the light Marjorana neutrino exchange mechanism of Fig.~3~(a).}

Astrophysical restrictions on the sum of the diagonal neutrino masses obtained by the Planck Collaboration
from the study of cosmic microwave background anisotropies yield $\sum m_{\nu} < 120$~meV \cite{Aghanim:2018}.
The best restrictions on the effective electron neutrino Majorana mass in double-beta decay experiments
are obtained by the KamLAND-Zen Collaboration \cite{Gando:2016}:
$|m_{\beta\beta}| < (61 - 165)$~meV for commonly used NMEs and the unquenched axial-vector coupling $g_A = 1.27$.
Exotic scalar interactions modify the mass of neutrinos in nuclear matter \cite{Kovalenko:2014},
so the effective electron neutrino Majorana mass in $0\nu2\beta^-$ decay and $0\nu$2EC can differ
from that derived from astrophysical data and tritium beta decay.

\textcolor{black}{
The decay half-life is determined by the decay width: $T_{1/2}^{0\nu\mathrm{2EC}}= {\ln 2}/{\Gamma _{i}}$,
where $\Gamma _{i}$ is given by Eq. (\ref{G1}), with $M_{i} = M_{A,Z}$ and $M_{f} = M_{A,Z-2}^{**}$;
$\Gamma _{f} \equiv \Gamma _{\alpha \beta }$ is the decay width of the daughter atom with vacancies of electrons
in the states $\alpha $ and $\beta$. The amplitude $V_{\alpha \beta }$ entering $\Gamma _{i}$ is
defined by Eq.~(\ref{eq:ECECRate-1}); for the light Majorana neutrino exchange mechanism,
$V_{\alpha \beta }$ simplifies to Eq.~(\ref{LNVP2}). }

\textcolor{black}{
Assuming the light Majorana neutrino exchange mechanism is dominant,
$T_{1/2}^{0\nu\mathrm{2EC}}$ scales with powers of the overlap factor $K_{Z}$, the neutrino mass $m_{\beta \beta }$,
the axial-vector coupling $g_{A}$, and the nuclear matrix element $M_\mathrm{2EC}$.
The decay half-life can be written as follows:
\begin{equation} \label{T12abs}
T_{1/2}^{0\nu\mathrm{2EC}} = K_Z^{-2}\left(\frac{1.27}{g_A}\right)^4 \left(\frac{100~\mathrm{meV}}{|m_{\beta\beta}|}\right)^2 \left(\frac{3}{|M_{\mathrm{2EC}}|}\right)^2 \tilde{T}_{1/2}.
\end{equation}
The normalized half-life $\tilde{T}_{1/2}$ does not depend on
$K_Z$, $m_{\beta\beta}$, $g_A$, and $M^{2EC}$.
Various schemes for calculating $K_Z$ are discussed by \textcite{Krivoruchenko:2019b}.
The nucleon and non-nucleon spin-isospin correlations and the renormalization effects of the axial-vector coupling $g_A$
are discussed in by \textcite{Suhonen2019}.
%$T_{1/2}$ are given in Table~\ref{tab:NME} for nuclides with
%the known NMEs in the conjunction with $K_Z$ of Eq.~(\ref{OTA}), $m_{\beta \beta }=100$ meV, and $g_{A}=1.27$.
%Tables XIV, XVI, and XVII report $\tilde{T}_{1/2}$ for nuclides with the known and unknown NMEs.
}

\begin{table}[]
\centering
\renewcommand{\arraystretch}{1.1}
%\addtolength{\tabcolsep}{-1pt}
\scriptsize
\caption{The normalized half-lives of $0\nu2$EC processes not included in Table~\ref{tab:table18.1},
but for which precise $Q_{\mathrm{2EC}}$ values are known, and/or experimental constraints on the $0\nu2$EC decay half-lives exist.
Transitions with $\tilde{T}_{1/2}^{\min} > 10^{35}$ y are not shown.
An asterisk with a parenthesis $^{*)}$ indicates channels with the known NMEs listed in Table~\ref{tab:NME}.
The other notations are the same as in Table~\ref{tab:table18.1}. }
\label{table83}
\resizebox{\textwidth}{!}
{\begin{tabular}{|c|c|c|c|r|c|c|r|r|r|r|c|c|c|c|c|c|}
\hline \hline
$\iota$                            &   Transition                              & $J^{\pi}_f$ & $M^*_{A,Z-2} - M_{A,Z - 2}$ & $M^{**}_{A,Z-2} - M_{A,Z}$ & $(n 2j l)_{\alpha}$ & $(n 2j l)_{\beta}$ & $\epsilon_{\alpha}^*\;\;$ & $\epsilon_{\beta}^*\;\;$ & $\Delta\epsilon_{\alpha\beta}^*$ & $\Gamma_{\alpha\beta}~~~~$  & $\tilde{T}_{1/2}^{\min}$ & $\tilde{T}_{1/2}^{\max}$ \\
\hline
\hline
5.52\%                                  &$^{ 96}_{ 44}$Ru$ \to ^{ 96}_{ 42}$Mo$^{**}$  & [0$^{+ }$]    &2712.68 $\pm$ 0.10~$^{*)}$
                                                                                                                         &  1.6 $\pm$  0.13    &  210    &  310    &  2.87    &  0.50    &  0.05    &$1.0\times 10^{  -2}$    &$5\times 10^{32}$       &$2\times 10^{33}$      \\
                                        &                                              & 0$^{+ }$    &2742 $\pm$ 1& 68.1 $\pm$   0.13    &  110    &  110    & 20.00    & 20.00    &  0.64    &$9.0\times 10^{  -3}$    &$1\times 10^{34}$       &$1\times 10^{34}$      \\
\hline
1.25\%                                  &$^{106}_{~48}$Cd$ \to ^{106}_{~46}$Pd$^{**}$  & [0$^{+ }$]    &2717.59 $\pm$ 0.21~$^{*)}$& -8.4 $\pm$  0.10    &  110    &  110    & 24.35    & 24.35    &  0.72    &$1.3\times 10^{  -2}$    &$5\times 10^{31}$       &$8\times 10^{31}$      \\
                                        &                                              & 2,3$^{- }$    &2748.2 $\pm$ 0.4&  0.5 $\pm$  0.10    &  110    &  231    & 24.35    &  3.17    &  0.18    &$8.3\times 10^{  -3}$    &$2\times 10^{31}$       &$2\times 10^{36}$      \\
\hline
0.97\%                                  &$^{112}_{~50}$Sn$ \to ^{112}_{~48}$Cd$^{**}$  & 0$^{+ }$    &1871.00 $\pm$ 0.19&  5.4 $\pm$  0.16    &  110    &  110    & 26.71    & 26.71    &  0.76    &$1.5\times 10^{  -2}$    &$1\times 10^{31}$       &$2\times 10^{31}$      \\
\hline
0.11\%                                  &$^{130}_{~56}$Ba$ \to ^{130}_{~54}$Xe$^*$  & (1$^{+ }$)  &2533.4  $\pm$ 0.3  & -50.2 $\pm$ 0.29          &  110    &  210    & 34.56    &  5.45    &  0.18    &$1.4\times 10^{  -2}$    &$8\times 10^{32}$       &$8\times 10^{32}$      \\
                                        &                                            & (1$^{+ }$)  &2544.43 $\pm$ 0.08 & -39.1 $\pm$ 0.29          &  110    &  210    & 34.56    &  5.45    &  0.18    &$1.4\times 10^{  -2}$    &$5\times 10^{32}$       &$5\times 10^{32}$      \\
                                        &                                            & (1$^{+ }$)  &2628.360$\pm$ 0.001& 44.8  $\pm$ 0.29          &  110    &  210    & 34.56    &  5.45    &  0.18    &$1.4\times 10^{  -2}$    &$8\times 10^{32}$       &$8\times 10^{32}$      \\
                                        &                                            & (1$^{+ }$)  &2637.50 $\pm$ 0.05 & 53.9  $\pm$ 0.29          &  110    &  210    & 34.56    &  5.45    &  0.18    &$1.4\times 10^{  -2}$    &$1\times 10^{33}$       &$1\times 10^{33}$      \\
\hline
0.06\%                                  &$^{156}_{~66}$Dy$ \to ^{156}_{~64}$Gd$^{**}$  & [0$^{+ }$]    &1804 $\pm$ 7 & -100.4 $\pm$  0.10    &  110    &  110    & 50.24    & 50.24    &  1.14    &$4.5\times 10^{  -2}$    &$1\times 10^{32}$       &$3\times 10^{32}$      \\
                                        &                                            & 0$^{+ }$    &1851.239 $\pm$ 0.007~$^{*)}$&-53.1 $\pm$  0.10    &  110    &  110    & 50.24    & 50.24    &  1.14    &$4.5\times 10^{  -2}$    &$5\times 10^{31}$       &$5\times 10^{31}$      \\
 \hline
0.14\%              &$^{162}_{~68}$Er$ \to ^{162}_{~66}$Dy$^{**}$  & 0$^{+ }$    &1666.27 $\pm$ 0.20 &-72.0 $\pm$  0.30    &  110    &  110    & 53.79    & 53.79    &  1.19    &$5.1\times 10^{  -2}$    &$5\times 10^{31}$       &$8\times 10^{31}$      \\
                                       &                                             & 1$^{+ }$    &1745.716 $\pm$ 0.007 & -38.2 $\pm$  0.30   &  110    &  210    & 53.79    &  9.05    &  0.25    &$3.0\times 10^{  -2}$    &$3\times 10^{31}$       &$3\times 10^{31}$      \\
\hline
1.601\%                                        &$^{164}_{~68}$Er$ \to ^{164}_{~66}$Dy$^{*}$  & 0$^{+ }$    & 0           & 38.0 $\pm$  0.12   &  110    &  210    & 53.79    &  9.05    &  0.31    &$3.0\times 10^{  -2}$    &$1\times 10^{32}$       &$1\times 10^{32}$      \\
\hline
0.014\%                                 &$^{190}_{~78}$Pt$ \to ^{190}_{~76}$Os$^{**}$  & 1,2    &1326.9 $\pm$ 0.5~$^{*)}$   & -0.1 $\pm$ 0.47   &  110    &  410    & 73.87    &  0.65    &  0.07    &$5.2\times 10^{  -2}$    &$3\times 10^{25}$       &$2\times 10^{29}$      \\
                                        &                                              &             &                  & -0.2  $\pm$ 0.47  &  110    &  411    & 73.87    &  0.55    &  0.04    &$5.1\times 10^{  -2}$    &$5\times 10^{26}$       &$3\times 10^{30}$      \\
\hline \hline
\end{tabular}}
\end{table}

The value $ \Delta \epsilon^*_ {\alpha \beta} $ shown in Tables~\ref{tab:table18.1} - \ref{table83} %\ref{tab:table18.2}
is a correction to the two-hole excitation energy (\ref{east}).
In the lowest approximation, $ \Delta \epsilon^*_ {\alpha \beta} $ is the Coulomb interaction energy of the holes.
For the estimates reported here, the values of $\Delta \epsilon^*_{\alpha \beta}$ are determined
empirically from the Auger electron spectroscopy, as described in Sec.~IV.B. If no experimental data are available,
$\Delta \epsilon^*_{\alpha \beta}$ are determined from calculations using the G{\lowercase{\scshape RASP}}2K package.
The calculations are performed for noble gas atoms with the simplest electron shell structure.
For the remaining atoms with identical quantum numbers of the holes,
the interaction energy is obtained via interpolation with a power function $\epsilon^*_{\alpha \beta} \sim aZ^b$.
Tables~\ref{tab:table18.1} and \ref{table83} %\ref{tab:table18.2}
report the minimum and maximum normalized values of the half-lives.
The confidence interval of the 99\%
probability is determined by the uncertainty in the degeneracy parameter of
the parent and daughter atoms, $ M^{**}_{A,Z-2} - M_{A,Z} \pm 2.6 \sigma $,
where $\sigma^2 = (\Delta Q)^2 + (\Delta E^*)^2$ and $\Delta Q$ and
$\Delta E^*$ are the errors in the $Q_{\mathrm{2EC}}$ value and the excitation energy $E^{*} =  M^*_{A,Z-2} - M_{A,Z - 2}$ of
the daughter nuclide, respectively. If this interval includes zero, then at the 99\% C.L.,
the resonance is not excluded, so that $\tilde{T}_{1/2}^{\min}$ gives the unitary limit.

In addition to the perspective nuclear pairs discussed by \textcite{KRIV11}, the $Q_{\mathrm{2EC}}$ values of other nuclide pairs were also measured.
Experimental limits on the half-lives of the $0\nu2$EC decays of some other nuclei were previously known or established in recent years.
These additional cases are analyzed on the same grounds; the normalized half-life estimates are presented in Table~\ref{table83}. %\ref{table83}.
Decays with a minimum normalized half-life of more than $10^{34}$ years are not listed.
Tables~\ref{tab:table18.1} and \ref{table83}
thus report all interesting cases of the atoms for which $Q_{\mathrm{2EC}}$ values were measured
and/or the experimental limits on the $0\nu2$EC half-lives are available.

In the near-resonance region, one finds a group of excited levels of $^{130}_{\;\,54}$Xe$^*$ with unknown quantum numbers.
We provide estimates for four excitation levels of $^{130}_{\;\,54}$Xe$^*$ with energies
$2533.40$,
$2544.43$,
$2628.36$, and
$2637.50$ keV of the assumed spin-parity $1^+$.
According to our estimates, these decays all appear to be non-resonant. The upper and lower limits of the normalized half-lives
coincide with the accuracy under consideration because of small errors in the excitation energy of the daughter nucleus
and the precise $Q_{\mathrm{2EC}}$ value of the pair. The same remark applies to the decays
$^{136}\mathrm{Ce} \to {^{136}\mathrm{Ba}^{**}}(2399.9 )$,
$^{156}\mathrm{Dy} \to {^{156}\mathrm{Gd}^{**}}(1851.24)$,
$^{162}\mathrm{Er} \to {^{162}\mathrm{Dy}^{**}}(1745.72)$,
$^{180}\mathrm{W}  \to {^{180}\mathrm{Hf}^{* }}$,
the decay $^{156}\mathrm{Dy} \to {^{156}\mathrm{Gd}^{**}}(1988.5 )$ accompanied by the $KK$ or $KL_1$ capture,
and to some other decays.

The decay of $^{152}$Gd is not resonant at 99\% C.L.,
but its probability remains sufficiently high, with a normalized half-life of $3 \times10^{28} - 3 \times10^{29}$ years.
The case of $^{156}$Dy, which decays into an excited state of gadolinium with energy $1988.5$ keV,
is noteworthy. The error in the $Q_{\mathrm{2EC}}$ value of this pair is 100 eV;
the resonance capture of electrons from the $L_1 L_1$ state with a normalized half-life of $ 8 \times 10^{25}$ years
cannot be excluded at 99\% C.L.
The NME for decay into the excited level $J^{\pi} = 0^+$ and the energy of $ 1851.239 $ keV according to \textcite{Kotila2014} is equal to 0.35 (Table~\ref{tab:NME}).
Taking the same NME value for the level $ 1988.5$ keV, one finds the half-life of $ 1.1 \times 10^{28} $ years,
which is even less than the half-life of $^{152}$Gd.
For other cases considered in Tables~\ref{tab:table18.1} and \ref{table83}, %\ref{tab:table18.2},
the minimum values of the normalized half-life are above $ 10^{29} $ years.

\subsection{Decays of long-lived radionuclides}

The radioactive elements are challenging to address experimentally;
if one chooses the longest-living ones, then the experimental difficulties
in working with such substances may be minimized.
We performed an analysis of weakly radioactive elements with lifetimes of longer
than a year using the Brookhaven National Laboratory database
\footnote{Center for Nuclear Studies, Department of Physics, The George Washington University. \\Data Analysis Center: http://www.nndc.bnl.gov/.}.
From the point of view of resonant capture,
two isotopes of gadolinium and isotopes of mercury and lead are of interest.
The half-life estimates for these elements are presented in Table~XIV. %\ref{tab:table18.2}.
The $Q_{\mathrm{2EC}}$ values of the pairs are not well-known; thus, column five reports two errors
in the masses of the parent and daughter nuclei. Column four shows the error in the excitation energy of the daughter nuclide.
The first column gives the half-life of the radioactive parent nuclei with respect to the dominant decay.
At the present level of knowledge of the parameters of the long-lived radionuclides,
any of the channels listed in Table~XIV. %\ref{tab:table18.2} can be exactly resonant.

\begin{table}[]
\label{tab:table18.2}
\centering
\renewcommand{\arraystretch}{1.1}
%\addtolength{\tabcolsep}{-1pt}
\scriptsize
\caption{The near-resonant $ 0 \nu $2EC for long-lived radionuclides.
The first column reports the half-life of the parent nuclide.
An asterisk with a parenthesis $^{*)}$ indicates channels with the known NMEs listed in Table~\ref{tab:NME}.
The fifth column gives the degeneracy parameter,
taking into account the excitation energy of the electron shell of the daughter in the ground state.
The errors indicate uncertainties in measuring masses of the parent and daughter nuclides, respectively.
The other notations are the same as in Table~\ref{tab:table18.1}.
}
\resizebox{\textwidth}{!}
{\begin{tabular}{|c|c|c|c|r|c|c|r|r|r|r|c|c|c|c|c|c|}
\hline \hline
$T_{1/2}$                            &   Transition                              & $J^{\pi}_f$ & $M^*_{A,Z-2} - M_{A,Z - 2}$ & $M^{**}_{A,Z-2} - M_{A,Z}$ & $(n 2j l)_{\alpha}$ & $(n 2j l)_{\beta}$ & $\epsilon_{\alpha}^*\;\;$ & $\epsilon_{\beta}^*\;\;$ & $\Delta\epsilon_{\alpha\beta}^*$ & $\Gamma_{\alpha\beta}~~~~$  & $\tilde{T}_{1/2}^{\min}$ & $\tilde{T}_{1/2}^{\max}$ \\
\hline
\hline
$71.1 \pm 1.2$ y                        &$^{148}_{~64}$Gd$ \to ^{148}_{~62}$Sm$^*$  & [0$^{+ }$]    &3004 $\pm$ 3~$^{*)}$ & -7.5 $\pm$ 2.8 $\pm$ 2.4    &  110    &  210    & 46.83    &  7.74    &  0.28    &$2.3\times 10^{  -2}$    &$2\times 10^{25}$       &$8\times 10^{31}$      \\
                                        &                                           & [1$^{+ }$]    &                 & -7.6 $\pm$ 2.8 $\pm$ 2.4    &  110    &  210    & 46.83    &  7.74    &  0.22    &$2.3\times 10^{  -2}$    &$8\times 10^{24}$       &$8\times 10^{31}$      \\
$71.1 \pm 1.2$ y                        &$^{148}_{~64}$Gd$ \to ^{148}_{~62}$Sm$^{**}$  & [0$^{+ }$]    &3022 $\pm$ 3 & 10.5 $\pm$  2.8 $\pm$  2.4    &  110    &  210    & 46.83    &  7.74    &  0.28    &$2.3\times 10^{  -2}$    &$2\times 10^{25}$       &$1\times 10^{32}$      \\
$(1.79 \pm 0.08) \times 10^{6}$ y                   &$^{150}_{~64}$Gd$ \to ^{150}_{~62}$Sm$^{**}$  & 0$^{+ }$    &1255.40 $\pm$ 0.22~$^{*)}$ & 15.4 $\pm$  6.3 $\pm$  2.4    &  110    &  310    & 46.83    &  1.72    &  0.09    &$3.2\times 10^{  -2}$    &$2\times 10^{26}$       &$8\times 10^{32}$      \\
$(3.0 \pm 1.5) \times 10^{6}$ y                   &$^{154}_{~66}$Dy$ \to ^{154}_{~64}$Gd$^{**}$  & 1,2$^{+ }$    &3264.42 $\pm$ 0.21~$^{*)}$ &  8.2 $\pm$  7.6 $\pm$  2.5    &  110    &  210    & 50.24    &  8.38    &  0.23    &$2.6\times 10^{  -2}$    &$5\times 10^{24}$       &$3\times 10^{31}$      \\
$444 \pm 77$ y                                 &$^{194}_{~80}$Hg$ \to ^{194}_{~78}$Pt$^{**}$  & (0$^{+ }$,1$^{+ }$,2$^{+ }$)    &2450 $\pm$ 5~$^{*)}$ &-27.6 $\pm$ 12.5 $\pm$  0.9    &  110    &  210    & 78.39    & 13.88    &  0.31    &$5.8\times 10^{  -2}$    &$2\times 10^{24}$       &$1\times 10^{31}$      \\
                                         &                                            & (0$^{+ }$,1$^{+ }$,2$^{+ }$)  &2472 $\pm$ 5~$^{*)}$ & -5.6 $\pm$ 12.5 $\pm$  0.9    &  110    &  210    & 78.39    & 13.88    &  0.31    &$5.8\times 10^{  -2}$    &$2\times 10^{24}$       &$5\times 10^{30}$      \\
                                         &                                             & [1$^{+ }$]    &2500 $\pm$ 10~$^{*)}$ & 22.4 $\pm$ 12.5 $\pm$  0.9    &  110    &  210    & 78.39    & 13.88    &  0.31    &$5.8\times 10^{  -2}$    &$2\times 10^{24}$       &$1\times 10^{31}$      \\
$(5.25 \pm 0.28) \times 10^4$ y                     &$^{202}_{~82}$Pb$ \to ^{202}_{~80}$Hg$^{**}$  & 0$^{+ }$    &1411.37 $\pm$ 0.12~$^{*)}$ & 17.6 $\pm$  8.2 $\pm$  0.6    &  210    &  310    & 14.84    &  3.56    &  0.12    &$2.6\times 10^{  -2}$    &$2\times 10^{25}$       &$2\times 10^{32}$      \\
                                        \hline \hline
\end{tabular}}
\end{table}

%%%%%%%%%%%%%%%%%%%%%%%%%%%%%%%%%%%%%%%%%%%%%%%%%%%%%%%%%%%
%%%%%%%%%%%%%%%%%%%%%%%%%%%%%%%%%%%%%%%%%%%%%%%%%%%%%%%%%%%
%%%%%%%%%%%%%%%%%%%%%%%%%%%%%%%%%%%%%%%%%%%%%%%%%%%%%%%%%%%
\section{Conclusions}
%%%%%%%%%%%%%%%%%%%%%%%%%%%%%%%%%%%%%%%%%%%%%%%%%%%%%%%%%%%
%%%%%%%%%%%%%%%%%%%%%%%%%%%%%%%%%%%%%%%%%%%%%%%%%%%%%%%%%%%
%%%%%%%%%%%%%%%%%%%%%%%%%%%%%%%%%%%%%%%%%%%%%%%%%%%%%%%%%%%
\setcounter{equation}{0}

Neutrino physics is a field of science characterized by a wealth of ideas, a multitude of unsolved problems and mysteries that intrigue the imagination.
Despite the more than half a century that has passed since the discovery of this particle,
it remains the least understood among the fermions of the Standard Model. Great hopes are placed on the study of neutrinoless double-beta ($0\nu2\beta^-$) decay able to shed light on the type of neutrinos (Dirac or Majorana) and on the total lepton number conservation. However, numerous experimental attempts to observe $0\nu2\beta^-$ decay have been unsuccessful so far.
%Great hopes are placed on the study of double-beta decay, the neutrinoless mode of which is able to shed light on the type of neutrinos (Dirac or Majorana) and on the total lepton number conservation. However, numerous experimental attempts to observe neutrinoless double-beta decay have been unsuccessful so far.

In the steady stream of efforts devoted to the study of $2\beta^-$ decay,
the reverse process of two-electron capture (2EC) was in the shadows until recently.
In this review, a comprehensive characterization of this process is presented, with a main focus on its neutrinoless mode,
which can give the same information about the properties of the neutrino as the $0\nu2\beta^-$ decay.
All aspects, both theoretical and experimental, revealing in detail unusual physical phenomena
associated with the manifestations of the properties of mysterious neutrinos, are considered.

To determine the possibility of observing the neutrinoless 2EC process in real experimental conditions,
the lifetimes of nuclides in which the resonant conditions can occur were calculated.
For realistic estimates it is necessary to describe properties of electron shells of atoms
and 2EC nuclear matrix elements.

%%%%%%%%%%%%% atomic shell

Atomic electron shells are involved in the $0\nu2$EC process through the overlap of the electron
wave functions with the nucleus and the excitation energy of electron shells.
The short-distance electron wave functions are well defined in the framework of the multielectron Dirac-Hartree-Fock schemes.
%The decaying atom $(A,Z)$ remains electrically neutral, and the electron shell goes into an excited state with two vacancies in low orbits.
%Two additional electrons inherited from the parent atom occupy outer shells.
Valence electrons are bound by several eV, so the excitation energy of the shells is close to the double ionization potential.
Experimental progress in atomic spectroscopy made it possible to study multiple ionization processes.
%The bulk of information comes from Auger spectroscopy, from which the double ionization potentials can be extracted.
%Some of the combinations of quantum numbers of the electron vacancies are not
%available for the analysis through Auger spectroscopy. In such cases,
%we derived the excitation spectra with the help of the multiconfiguration Dirac-Hartree-Fock package G{\lowercase{\scshape RASP}}2K.
Analysis of theoretical uncertainties and comparison with the Auger spectroscopic data leads to the conclusion that
the excitation energies of $0\nu2$EC can be determined with an accuracy of 60 eV or better for heavy atoms,
%The uncertainties decrease with decreasing $Z$, and for light atoms, the accuracy can be estimated at about 10 eV,
which is comparable to the accuracy in measuring the atomic masses in Penning traps and the de-excitation width of electron shells.
An improvement of theoretical schemes or direct measurements of the excitation energy of electron shells with quantum numbers
relevant for the $0\nu2$EC process %in multiple ionization of atoms
is of high interest.

%%%%%%%%%%%%% NME

The nuclear theory frameworks used to evaluate the nuclear wave functions
involved in the nuclear matrix elements of the resonant neutrinoless 2EC transitions
cover theories based on the quasiparticle random-phase approximation (both spherical
and deformed QRPA), boson mapping (the microscopic interacting boson model,
IBM-2) or modern energy-density functionals, EDF. The latter two base
the matrix-element computations on the closure approximation and only the
QRPA-based theories avoid the use of closure. On the other hand, the IBM-2 and
EDF flexibly take into account the deformation degree of freedom, as also
to a certain extent the deformed QRPA does.

%Calculation of the nuclear matrix elements involved in the resonant neutrinoless 2EC processes is not an easy task.
The most straightforward are the matrix elements
involved in the ground-state-to-ground-state captures in the deformed nuclei
$^{152}$Gd, $^{156}$Dy, $^{164}$Er and $^{180}$W. %Here only the ground-state nuclear wave functions are involved.
For these nuclides the different
theory frameworks (QRPA, IBM-2, EDF) give consistent results within a factor
of $2-3$ for the values of the nuclear matrix elements.
Theoretical treatment of the excited nuclear resonant states,
in particular those with high excitation energies, is a challenge. The problem is the identification
of the theoretical state that corresponds to the resonant experimental state
of a certain spin-parity, $J^{\pi}$, in particular when there are several
experimental and theoretical states close to the resonant state. The intrinsic
properties of these close-lying states can vary quite strongly from one state to the
next so that selecting the proper state is essential for a reliable prediction
of the 2EC nuclear matrix element. This effect is magnified in deformed
nuclei, with possible coexisting structures %of more or less deformation
at around similar excitation energies.

At the quark-lepton level, the underlying physical LNV  mechanisms of the $0\nu2$EC, $0\nu$EC$\beta^+$, $0\nu 2\beta^+$, and $0\nu 2\beta^-$  processes are essentially the same. In the Standard Model represented by the sector of renormalized dimensions-4 interactions the total number of leptons $L$ is conserved. The corresponding  $\Delta L =2$ contributions can appear via non-renormalizable effective operators  of higher dimensions. We specified all the operators up to dimension 9 and discussed their possible high-scale origin from renormalizable theories. We presented, with some detail, three popular large-scale scenarios beyond the Standard Model.
The conventional mechanism of Majorana neutrino exchange is highlighted by the fact that the corresponding operator has minimal dimension $d = 5$.

The effective electron neutrino Majorana mass $m_{\beta\beta}$ determines the amplitudes of $0\nu2 \beta^-$ decay and the $0\nu2 $EC process.
We reported the lower and upper half-life limits of $0\nu2$EC for near-resonant nuclides with the known nuclear matrix elements
for the effective electron neutrino Majorana mass of $|m_{\beta\beta}| = 100$ meV.% in Table~XIII.
%Depending on the underlying physics, the contributions of the above-mentioned $d>5$ effective operators can also dominate the $0\nu 2$EC process.
%Since there is currently no information on the corresponding nuclear matrix elements of the higher dimension operators,
%we restricted ourselves in Sec.~VIII by the estimates of the normalized half-lives for the near-resonant virtually stable and long-lived radionuclides.

%%% Experiment

The experimental sensitivity to the $0\nu$2EC process is currently
lower as compared to that of the $0\nu2\beta^-$ decay. The
strongest $0\nu$2EC half-life limits are at the level of
$T_{1/2} \sim 10^{21}-10^{22}$ yr, while the $0\nu2\beta^-$
experiments have already achieved the sensitivity level of
$\lim T_{1/2} \sim 10^{24}-10^{26}$ yr. The highest up-to-date sensitivity to the $0\nu$2EC process is
achieved using quite diverse experimental techniques: gaseous
($^{78}$Kr), scintillation ($^{106}$Cd) and cryogenic
scintillating bolometric detectors ($^{40}$Ca), HPGe $\gamma$
spectrometry ($^{36}$Ar, $^{58}$Ni, $^{96}$Ru, $^{112}$Sn), and
geochemical methods ($^{130}$Ba, $^{132}$Ba).
%The sensitivity of the experiments can be significantly improved by
%increasing the amount of isotopes of interest (in particular, through the utilization of enriched materials),
%by increasing the detection efficiency, reducing the background and providing the highest possible energy resolution.

The prospects for finding $0\nu$2EC become more favorable
if a resonance effect in double-electron capture occurs - a phenomenon that is peculiar only to $0\nu$2EC. The resonant $0\nu$2EC effect is expected to be clearly identified thanks to the high accuracy of the $\gamma$ quanta energies expected in the decay, while background due to the neutrino accompanied decay (X-ray with energies up to several tens of keV)  will never play a role in practice (in contrast to the $0\nu2\beta^-$ experiments where background caused by the $2\nu$ mode becomes dominant due to poor energy or time resolution). The experimental sensitivity can be significantly improved by increasing the amount of isotopes of interest and utilization of enriched materials, by increasing the detection efficiency, reducing the background and providing the highest possible energy resolution. HPGe detectors and low temperature bolometers appear to be the most suitable detection techniques
for $0\nu$2EC experiments with a sensitivity of $T_{1/2}\sim  10^{25}-10^{26} $yr. Moreover, the complicated signature of the resonant
effect can be a definite advantage, since the energies of the $\gamma$ quanta expected in most decays are tabulated
and usually known with a very high accuracy, which will ensure reliable identification of the effect.

The resonance in 2EC is associated with the degeneracy of mother and intermediate daughter atomic states
that can in principle only be fulfilled in 2EC provided the process is neutrinoless.
These conditions cannot occur in the 2$\beta^-$ decays. However, an insufficiently accurate knowledge of
the atomic mass differences between the mother and daughter states has blocked the ascertainment of this possibility so far.
%The atomic mass differences have until recently only been known
%to a precision many orders of magnitude lower than the widths of the intermediate resonant states being on the order of about 10 eV.

The development of Penning-Trap Mass Spectrometry (PTMS) has radically altered this situation.
PTMS is superior to all other known methods of mass spectrometry. It has successfully been used
for the determination of mass differences of nuclides with unprecedented low uncertainties down to the eV level.
Mass differences for nineteen such pairs %of atoms
connected via 2EC and listed in Tables~VII and VIII
have been measured by PTMS, and new measurements are still planned for other pairs.

%%%%

The refined analysis shows that at $90\%$ C.L.
none of the stable isotopes are exactly resonant when accompanied by electron capture from favorable states $ns_{1/2}$ and $np_{1/2}$.
Yet, at $99\%$ C.L.
it is impossible to exclude the exactly resonant character of the $^{190}_{~78}$Pt$ \to ^{190}_{~76}$Os$^{**}$ decay
to the excited state 1326.9 $\pm$ 0.5 keV of the daughter.
For the vanishing degeneracy parameter and $J^{\pi} = 1^+$, the half-life appears to be $3.3 \times 10^{26}$ years.
A more accurate knowledge of the $Q_{\mathrm{2EC}}$-value, the excitation energy and $J^{\pi}$ of $^{190}_{~76}$Os$^{**}$ would certainly be desirable.
The $^{156}_{~66}$Dy$ \to ^{156}_{~64}$Gd$^{**}$ decay to the excited state of $1988.5 \pm 0.2$ keV also demonstrates the proximity to the resonance.
The decay half-life $1.1\times10^{28}$ years is within the 99\%
confidence interval.
For the ground-to-ground state transitions, the $^{152}_{~64}$Gd $\to ^{152}_{~62}$Sm$^*$
decay with a lower half-life  limit of $7\times10^{27}$ years is the most encouraging case.
The long-lived radionuclides listed in Table~XIV can be resonant at $90\%$ C.L., however,
production of these nuclides in significant amounts is technically complex.

%%%%%%%%%%%%%%%%%%%%

The data suggests that with further refinement of the parameters,
the nuclides under discussion could become competitive with nuclides that decay through the $ 0\nu2\beta^-$ channel,
for which the lower half-life limit is already set at %the level of
$\sim 10^{26}$ years.
This level of sensitivity is also achievable for 2EC processes.
Moreover, since not all excited states of nuclides are known experimentally, there can also exist other, not yet identified near-resonant nuclides.
Special demands are placed on the nuclear spectroscopy for the search of new relevant
excited nuclear states and/or exact determination of the spin and parity values for them.

%%%%%%%%%%%%%%%%%%%%

The search for nuclides satisfying the resonance condition in the 2EC process
requires new experimental and theoretical efforts.
The experimental side should focus on improving the accuracy of mass measurements by ion traps,
the search for new excited states of nuclides in the resonance region,
the determination of their excitation energies and quantum numbers, and
the determination of the excitation energy of the electron shells using atomic spectroscopy methods.
The theory should be aimed at refining theoretical schemes for calculating the 2EC half-life.
Further progress requires the joint efforts of theorists in atomic, nuclear and particle physics,
as well as the development and implementation of advanced technologies that are already on the horizon.
The double-electron capture process can prove to be an important player in the world beyond the Standard Model.

\section*{Acknowledgements}

The authors appreciate the valuable discussions with M.~Block, Amand~Faessler, D.~Frekers, T.~Eronen, V.~Shabaev, F.~{\v S}imkovic, and
I.~Tupitsyn on topics related to 2EC physics.
This work is supported in part by the Max-Planck-Society.
K.B. and S.E. acknowledge support by the European Research Council (ERC) under the European Union's Horizon 2020 research and innovation programme under Grant Agreement No. 832848-FunI.
F.A.D. and V.I.T. were supported in part by the Program of the National Academy of Sciences of Ukraine
``Fundamental research on high-energy physics and nuclear physics (International Cooperation)''.
S.K. acknowledges the support of the ANID Fondecyt (Chile) Grant  No. 1190845 and the CONICYT (Chile) PIA/Basal FB0821.
M.I.K. is supported by the RFBR Grant No.~18-02-00733 (Russia) and the Alexander von Humboldt-Stiftung (Germany).
Yu.N.N. is supported by the Extreme Matter Institute EMMI (GSI Darmstadt).
J.S. was partly supported by the Academy of Finland under the Academy project No. 318043.
F.A.D., M.I.K., Yu.N.N. and V.I.T. wish to acknowledge the kind hospitality at Max-Planck-Institut f\"ur Kernphysik
in Heidelberg, where a part of this work has been done.

%\newpage

\section*{Abbreviations, symbols, and units of measure}

\begin{table}[H]
\label{tabresexp}
{\begin{tabular}{llll}
$A$              & {mass number } & {m w.e.}                  &    {meters of water equivalent}\\
{CC}             & {Charged Current}                          & {NME}            & {Nuclear  Matrix  Element}\\
{C.L.}           & {confidence level}                         & {PI-ICR}         & {Phase  Image  Ion  Cyclotron  Resonance}\\
%{CVC}            & {Conservation of Vector Current }\\
{DEIP}           & {Double-electron Ionization Potential}     & {PMNS}           & {Pontecorvo-Maki-Nakagawa-Sakata  mixing  matrix}\\
$E^*$            & {excitation energy of nucleus}             & {PTMS}           & {Penning  Trap  Mass  Spectrometry}\\
{EDF}            & {Energy Density Functional}                & {PT}             & {Preparation  Trap}\\
{2EC}            & {double-electron capture}                  & $Q$ {value}      & {mass  difference between neutral parent and daughter atoms}\\
{FWHM}           & {Full Width at Half Maximum}               & {QCD}            & {Quantum  Chromodynamics}\\
{GCM}            & {Generating Coordinate Method}             & {QED}            & {Quantum  Electrodynamics}\\
{GT}             & {Gamow-Teller beta-decay type}             & {QRPA}           & {Quasiparticle  Random  Phase  Approximation}\\
{g.s.}           & {ground state}                             & {RGE}            & {Renormalization Group Equation}\\
{HPGe}           & {Hyper-Pure Germanium detector}            & {SEIP}           & {Single-electron  Ionization  Potential}\\
{HO}             & {Harmonic Oscillator}                      & {SM}             & {Standard  Model}\\
{IBM}            & {Interacting Boson Model}                  & {SSB}            & {Spontaneous  Symmetry  Breaking}\\
{IBM-2}          & {Microscopic IBM}                          & {SUSY}           & {Supersymmetric  model}\\
{IBFM-2}         & {Microscopic Interacting Boson Fermion Model} & RPV              & $R$-parity violating \\
{IBFFM-2}        & {Proton-neutron IBFFM}                     & {ToF-ICR}        & {Time-of-Flight  Ion  Cyclotron  Resonance}\\
{IBFFM}          & {Interacting Boson-Fermion-Fermion Model}  & $T_{1/2}$        & {radioactive  decay  half-life}\\
$K,L,M$, etc.      & {orbitals of atomic electrons}             & $Z$              & {atomic  number}\\
{LNV}            & {Lepton Number Violation}                  & $0\nu${2EC}           & {neutrinoless double-electron  capture}\\
{LRSM}           & {Left-Right Symmetric Model}               & $0\nu2\beta^-$        & {neutrinoless  double-beta decay}\\
{LQ}             & {Leptoquarks}                              & $\Gamma$              & {electromagnetic decay width}\\
$M$              & {neutral atom  mass  value}                & $\Delta$              & {the  degeneracy  parameter = $ Q$ minus total excitation}\\
                 &                                            &                       &  energies of nucleus and electron shells \\
{MCP}            & {Micro-Channel  Plate}                     & $\epsilon^*$          & {excitation energy of electron shell}\\
{MCM}            & {Multiple-Commutator  Model}               & $\nu$                 & {neutrino}\\
{MT}             & {Measurement  Trap}                        & $\nu_i$               & {ion  frequencies  in  the  trap} \\
\end{tabular}}
\end{table}

%\newpage

\bibliography{biblio}

\end{document}